\documentclass[prb,draft,amsfonts,amssymb,amsmath,showpacs]{revtex4}
\begin{document}

\title{Dynamical Mass Generations and Collective Excitations in the 
(Supersymmetric-)Nambu$-$Jona-Lasinio Model and 
a Gauge Theory with Left-Right-Asymmetric Majorana Mass Terms}
\author{Tadafumi Ohsaku}
\date{\today}

\newcommand{\bmx}{\mbox{\boldmath $x$}}
\newcommand{\bmp}{\mbox{\boldmath $p$}}
\newcommand{\bmk}{\mbox{\boldmath $k$}}
\newcommand{\kfey}{\ooalign{\hfil/\hfil\crcr$k$}}
\newcommand{\pfey}{\ooalign{\hfil/\hfil\crcr$p$}}
\newcommand{\Pfey}{\ooalign{\hfil/\hfil\crcr$P$}}
\newcommand{\partfey}{\ooalign{\hfil/\hfil\crcr$\partial$}}
\newcommand{\Dfey}{\ooalign{\hfil/\hfil\crcr$D$}}
\newcommand{\Dcalfey}{\ooalign{\hfil/\hfil\crcr${\cal D}$}}
\newcommand{\afey}{\ooalign{\hfil/\hfil\crcr$A$}}
\newcommand{\gfey}{\ooalign{\hfil/\hfil\crcr$G$}}
\newcommand{\hfey}{\ooalign{\hfil/\hfil\crcr$h$}}

\begin{abstract}

The structure of effective potential surface of the Nambu$-$Jona-Lasinio ( NJL ) model 
with right-left asymmetric Majorana mass terms 
( corresponds to the single-flavor type-II seesaw situation of neutrino ) is investigated.
After the dynamical generation of Dirac mass, 
two collective modes appear similar to the case of ordinary NJL model,
and the phase mode ( phason ), 
which corresponds to majoron or pion at vanishing Majorana mass parameter(s),
has an excitation mass.
The mechanism of generation of phason as a pseudo Nambu-Goldstone boson is examined by mathematical manner,
summarized into a set of theorems
( claims as the generalized Nambu-Goldstone theorem ).
The mass of phason is also evaluated in a supersymmetric version of the NJL-type model,
and phason mass takes the order of that of axion commonly accepted today.  
An $SU(2_{c})$-gauge model is constructed for the context of neutrino seesaw mechanism, 
and the Schwinger-Dyson equation of dynamical mass functions is examined.
Several physical implications such as decay modes of phason, a non-linear sigma model for phason are given.
It is proposed that the method/result of this paper can be applied to an understanding 
on the origin of the Kobayashi-Maskawa matrix.
The mathematical structure of the generalized Nambu-Goldstone ( GNG ) theorem is examined from 
various viewpoints, and several physical/mathematical perspectives are given.

\end{abstract}


\maketitle

$\mathfrak{Lange \,\, lieb \,\, ich \,\, dich \,\, schon, \,\, m\ddot{o}chte \,\, dich, \,\, mir \,\, zu \,\, Lust, ...}$
( $\mathfrak{Friedrich\,\, H\ddot{o}lderlin, \,\, "Heidelberg"}$ )

\vspace{1mm}

$\mathfrak{Zwei \,\, Dinge \,\, erf\ddot{u}llen \,\, das \,\, Gem\ddot{u}t \,\, mit \,\, immer \,\, neuer \,\, und \,\, zunehmender \,\, Bewunderung \,\, und \,\, Ehrfurcht,}$ 

$\mathfrak{je \,\, \ddot{o}fter \,\, und \,\, anhaltender \,\, sich \,\, das \,\, Nachdenken \,\, damit \,\, besch\ddot{a}ftigt:}$ 

$\mathfrak{Der \,\, bestirnte \,\, Himmel \,\, \ddot{u}ber \,\, mir, \,\, und \,\, das \,\, moralische \,\, Gesetz \,\, in \,\, mir.}$

( $\mathfrak{I. \,\, Kant, \,\, Kritik \,\, der \,\, praktischen \,\, Vernunft \,\, (1788)}$ )

\section{Introduction}

In spite of non-renormalizability, 
four-fermion interaction models
~[1,4,11,12,21,24,28,36,48,54,79,93,94,95,103,]
~[111,112,137,141,146,160,161,172,173,174,175,176,177,178,211,217,221,226,234,237,244,245] 
are quite useful in condensed matter, nuclei and hadrons, 
and particle physics:
They seem to have infinite-number of applications.
For example, quantitative evaluations of physical quantities in asymptotic-free gauge theory such 
as quantum chromodynamics ( QCD ) are usually not an easy task, 
not only by its strong-coupling nature of the infrared region, 
but also the impossibility to use the Slavnov-Taylor identity in a nonperturbative calculation: 
We have to employ approximations to obtain an integration kernel and the running gauge coupling
which has an infrared ( IR ) divergence, cannot avoid gauge dependence, 
reduce the theory to an Abelianized version ( so-called "QCD-like" model ),
even in a calculation of a self-consistent mass function in a Schwinger-Dyson ( SD ) equation
~[3,7,13,64,69,70,93,95,124,138,139,142,147,168,181,199,217,234,244,245].
( Though various numerical results of SD equation show that 
it gives good agreement with experimental measurements. ) 
On the contrary, in a four-fermion interaction model, 
we have to introduce a cutoff, 
and cannot evaluate a physical quantity under a renormalization-group invariant manner.
Thus, a prediction of a four-fermion model is less confidential 
compared with that of an asymptotic-free gauge theory.
However, it is much easier to examine, for example, 
collective modes associated with a spontaneous symmetry breaking,
than by a gauge theory:
In a gauge theory, one should employ a Bethe-Salpeter formalism for examinations of collective modes,
and it is usually difficult and complicated to solve
~[93,124,138,147,199,245]. 
In fact, applying method of global analysis to a four-fermion model
is relatively much easier than a gauge theory:
We can say a four-fermion model is mathematically more "well-defined" than a gauge theory,
and a gauge theory has too many freedoms which should be fixed before doing a global analysis of it.
Therefore, it is a natural attitude to examine dynamical symmetry breaking
by a four-fermion model first, and for the next step, one should investigate a gauge model.

\vspace{2mm}

The Standard Model ( SM ) accurately describes physical phenomena of energy scales accessible 
by accelerators until now, and it is still our starting point of scientific investigations
on particle phenomenology, hadrons/nuclei, atomic physics, astrophysics, and cosmology
~[61,229].
The main problems of the SM, their importances 
${\it established}$ ${\it by}$ ${\it experimental}$ ${\it facts}$, 
we have not yet found ultimate solutions/understandings are, 
(1) origins of masses in quarks and charged leptons,
(2) origins of flavor mixings,
(3) CP-violations,
(4) neutrino masses,
(5) candidates of dark sector in the SM
~[5,9,26,35,45,61,80,83,97,101,105,109,110,113,131,134,152,167,192,212,213,233].
The subject which we consider in this paper will relate to ${\it all}$ of these problems, 
with an emphasis on ${\it neutrino}$ ${\it seesaw}$ ${\it mechanism}$.
For example, a CP-violation may cause, 
a difference of life times of particle and anti-particle, a particle-number non-conservation,
breakdown of time-reversal invariance if a system/theory keeps CPT and Lorentz invariances. 
If CPT is conserved, the flavor fraction in mass eigenstates in neutrino oscillation 
must be the same for neutrinos and anti-neutrinos,
at least in the framework of the usual neutrino oscillation theory~[183]. 
We might obtain another picture/interpretation 
if we get a radical/essential modification for modern neutrino oscillation paradigm 
for explaining various experimental facts.
In experimental examinations, neutrinos seem to have no magnetic moment~[61] 
and they are electrically neutral, while the discussion on the type of mass of neutrino 
( Dirac mass or Majorana mass ) has not yet meet a unique conclusion until now~[110,152].
If neutrino is Dirac particle, they can have magnetic moments,
$\mu^{D}_{\nu}=3eG_{F}m_{\nu}/(8\sqrt{2}\pi^{2})$, 
induced by corrections of weak interaction~[68], 
while it has no diagonal electric/magnetic moments when it is Majorana.
Hence an examination on neutrino magnetic moment can provide us 
to distinguish neutrinos as Dirac or Majorana~[210].
If neutrinos have magnetic moment, some astrophysical processes inside stars,
especially neutron star cooling, might be affected~[89].
In fact, a supernova will release 90 percent of its energy by neutrino emission~[214].
Neutrinoless double $\beta$ decay can say something about whether neutrinos are Dirac or Majorana~[9,227].
Neutrinos can be regarded as relic particles of dynamics/evolution of early Universe~[202].
By these various interesting issues and recent remarkable advances in experiments,
surely neutrino physics is now "the key" to understand physics 
from "beyond the SM" to astro-particle physics and cosmology
~[6,9,14,15,32,65,80,83,152,167,183].
Moreover, a paper, which claims that neutrino emissions from supernovae
can become a probe of gravitational wave detector, is published recently~[182].
Now, we have arrived at the age that 
neutrinos are probes for investigations on various problems of fundamental physics.

\vspace{2mm}

The top-quark condensation model of the SM is one of candidates toward "beyond the SM", 
and it stands a quite close place to the SM~[12,48,95,111,137,146,244].
In the top-condensation model, Bardeen, Hill and Lindner employed the method of renormalization-group
with a compositeness condition ( set from/below a GUT ( grand unified theory ) scale ) 
as a boundary condition of the RG equation~[12],
while Miransky, Tanabashi and Yamawaki used the Schwinger-Dyson approach supplemented with 
the Pagels-Stokar formula~[146]. 
( It was proved that these two methods are equivalent at the large-$N$ limit~[48,244]. 
Hence, both of these approaches examined the first derivative of the SM effective potential
with respect to Higgs VEV ( vacuum expectation value ),
and thus these methods cannot be applied to the case of first-order phase transition. 
For a full examination, we need a total/global profile of the effective potential,
and it is an interesting issue for us, 
especially in the Minimal Supersymmetric Standard Model ( MSSM )~[28,29,30,36]. )

\vspace{2mm}

Observations of neutrino oscillation give us the evidence that neutrinos have masses,
at least two of three generations~[212,213]. 
A very economical explanation for neutrino tiny mass is the seesaw mechanism~[76,145,151,246]:
In the mechanism, neutrinos have both Dirac and right-handed Majorana mass terms
and a Nambu-Goldstone ( NG ) boson ( majoron~[34] ) will arise from the Higgs sector 
associated with giving a VEV for a lepton-number-violating Majorana mass term of neutrino.
Hence it is an interesting question that what will happen 
if a fermion-field acquires a Dirac mass dynamically under 
having both left- and right- handed Majorana mass terms~[178].  
Recently, Antusch et al. gave 
a theory of dynamical seesaw mechanism of neutrinos 
which should relate to the top-condensation model~[4].
In such a dynamical model, 
the theory will be reduced to a Ginzburg-Landau ( a low-energy effective ) bosonic model,
and an examination of collective fields of a Ginzburg-Landau model are usually done by
their energy-momentum relations, global structure of energy profile, and mode-mode couplings~[155].
Sometimes, a mode-mode coupling reduces a Higgs VEV.
An investigation on possible relations between majoron and leptogenesis, dark energy, was given by Ref.~[96].
In Ref.~[127], decay processes of right-handed neutrinos caused by interactions with majorons 
are examined, and it is argued that they would contribute to lepton asymmetry.
In Ref.~[15], majoron-mediated neutrino decay processes and its implication to astrophysical, 
supernova neutrino emission are considered. 
Majoron and axion are related to ( coming from ) 
a phase degree of freedom of a mass parameter of a field theory.
If majoron is very light, it could be found through the Raffelt-Stodolsky mechanism~[197] 
which is usually discussed in axion-photon mixing~[185].
The dark energy seems to dominate dynamics of the accelerated expansion of the Universe,
and there are several models of mass-varying ( depends on environment ) neutrinos
for explanations of dark energy~[22,106].

\vspace{2mm}

Recently, a paper on seesaw mechanism of neutrino mass in 
the SUSY Nambu$-$Jona-Lasinio ( NLJ ) model supplemented by Majorana mass terms are given by the author~[178].
In the model, dynamical generation of a Dirac mass term, 
and a relation between phase degrees of freedom of Dirac, left- and right-handed Majorana mass parameters
were discussed. 
Kobayashi and Maskawa found a phase of the quark mixing matrix 
which cannot be absorbed by any field redefinition in a six-flavor theory~[113]. 
It is a well-known fact that the Kobayashi-Maskawa phase is too weak to keep lepton/baryon asymmetries.
The purpose of this paper is to give an investigation on dynamical mass functions,
and the mass phases, they cannot be absorbed by a field redefinition, as collective modes,
with the emphasis on the neutrino seesaw mechanism.

\vspace{2mm}

This paper is organized as follows: 
In Sec. II, 
a rather long "Prelude" of this work, which intends to summarize 
the essential part of our result, will be given, 
and then we introduce an NJL type four-fermion model with left-right asymmetric Majorana mass terms,
evaluate the one-loop effective potential of it, and examine the first and second derivatives
with respect to collective fields. 
We will find that the ordinary Nambu-Goldstone ( NG ) theorem is modified 
in a dynamical symmetry breaking with an explicit symmetry breaking parameter, 
and then we investigate a generalization of the NG theorem.
A supersymmetric extension of the result of Sec. II is given in Sec. III,
can be regarded as an "Intermezzo" by a reader.  
In Sec. IV, we construct a gauge model for neutrino seesaw mechanism,
and examine an SD equation for Dirac and Majorana masses.
Several physical implications of our theory is given in Sec. V.
An investigation of the mathematical structure of our 
generalized Nambu-Goldstone ( GNG ) theorem is given in Sec. VI.
We will see how the GNG theorem contacts with various theories in physics/mathematics.
Finally, we give our conclusion of this works with futher investigations in Sec. VII.

\vspace{2mm}

Throughout this paper, 
we use the conventions $g_{\mu\nu}={\rm diag}(1,-1,-1,-1)$ and 
$\gamma_{5}=i\gamma^{0}\gamma^{1}\gamma^{2}\gamma^{3}$.
The chiral projectors are defined as follows:
\begin{eqnarray}
P_{+} &\equiv& \frac{1}{2}(1+\gamma_{5}), \quad P_{-} \equiv \frac{1}{2}(1-\gamma_{5}). 
\end{eqnarray}

\section{The Nambu$-$Jona-Lasinio Model with Left-Right Asymmetric Majorana Mass Terms}

\subsection{Preliminary}

We start our investigation by a re-expression of a well-known result in the Nambu$-$Jona-Lasinio ( NJL ) model.
The Lagrangian is
\begin{eqnarray}
{\cal L}^{NJL} &=& \bar{\psi}(i\partfey-|m^{(0)}|)\psi + G\bigl[(\bar{\psi}\psi)^{2}+(\bar{\psi}i\gamma_{5}\psi)^{2}\bigr]   
= -\frac{1}{2G}|M_{dyn}|^{2} + \bar{\psi}\bigl( i\partfey - |m^{(0)}| - M_{dyn}P_{+} - M^{\dagger}_{dyn}P_{-}  \bigr)\psi,     \\
M_{dyn} &=& |M_{dyn}|e^{i\theta_{\chi}}, \qquad ( \theta_{\chi}\in {\bf R}^{1} ).
\end{eqnarray} 
Note that the way of entering the auxiliary field $M_{dyn}$ is slightly different 
from the orthodox definition given in literature~[93].
We have written the pion degree of freedom by a linear combination of phase $\theta_{\chi}$.
The $U(1)$ phase $\theta_{\chi}$ cannot be absorbed by any field redefinition, 
since ${\cal L}^{NJL}$ has bare mass $|m^{(0)}|$.
This Langangian gives the following mass eigenvalue:
\begin{eqnarray}
M &=& \sqrt{|m^{(0)}|^{2}+|M_{dyn}|^{2}+2|m^{(0)}||M_{dyn}|\cos\theta_{\chi}}.
\end{eqnarray}
Note that $M$ is a periodic function of $\theta_{\chi}$.
Hence, the one-loop effective potential, 
its first derivatives with respect to $|M_{dyn}|$ and $\theta_{\chi}$ also become periodic:
\begin{eqnarray}
V_{eff} &=& \frac{|M_{dyn}|^{2}}{2G} - \frac{1}{16\pi^{2}}\Bigl[ 
\Lambda^{2}M^{2} + \Lambda^{4}\ln\Bigl( 1 + \frac{M^{2}}{\Lambda^{2}} \Bigr)
- M^{4}\ln\Bigl( 1 + \frac{\Lambda^{2}}{M^{2}} \Bigr) \Bigr],   \\
\frac{\partial V_{eff}}{\partial |M_{dyn}|} &=& \frac{|M_{dyn}|}{G} 
- \frac{|M_{dyn}|+|m^{(0)}|\cos\theta_{\chi}}{4\pi^{2}}
\Bigl[ 
\Lambda^{2}-M^{2}\ln\Bigl( 1 +\frac{\Lambda^{2}}{M^{2}} \Bigr) 
\Bigr],   \\
\frac{\partial V_{eff}}{\partial\theta_{\chi}} &=&
\frac{|M_{dyn}||m^{(0)}|\sin\theta_{\chi}}{4\pi^{2}}
\Bigl[ 
\Lambda^{2}-M^{2}\ln\Bigl( 1 +\frac{\Lambda^{2}}{M^{2}} \Bigr) 
\Bigr].
\end{eqnarray}
Thus, one finds $\theta_{\chi}=2n\pi$ ( $n\in {\bf Z}$ ) are minima and stable.
From the definition of the fluctuation of $\theta_{\chi}$ around $2n\pi$ by 
$\tilde{\theta}_{\chi}\equiv \delta_{\theta_{\chi}}(M_{dyn})=|M_{dyn}|\delta\theta_{\chi}$,
the product of a wavefunction renormalization constant and a square of mass of fluctuation 
$\tilde{\theta}_{\chi}$ becomes the formula of pion mass known in literature~[103]:
\begin{eqnarray}
Z^{-1}_{\theta_{\chi}}(m_{\theta_{\chi}})^{2} 
&=& \frac{1}{2!}\frac{1}{|M_{dyn}|^{2}}\frac{\partial^{2}V_{eff}}{\partial\theta^{2}_{\chi}}\Big|_{\theta_{\chi}=2n\pi, \frac{\partial V_{eff}}{\partial |M_{dyn}|}=0}    \nonumber \\ 
&=& \frac{1}{2G}\frac{|m^{(0)}|}{|m^{(0)}|+|M_{dyn}|} 
\sim \frac{\Lambda^{2}}{8\pi^{2}}\frac{|m^{(0)}|}{|m^{(0)}|+|M_{dyn}|}.
\end{eqnarray}

\vspace{2mm}

Next, we consider a more generic case.
Let $\hat{g}$ be an element of a neighborhood of the identity $\hat{1}$ of a compact Lie group, 
and let $\psi\to \hat{g}\psi$ be a global gauge transformation.
Let $i\widehat{\Dfey}$ be a Dirac operator properly defined for a problem.
In a non-Abelian case, 
we consider a Lie group defined by an exponential mapping $\hat{g}=\exp(i\theta^{\alpha}T_{\alpha})$, 
and the corresponding Lie algbera $[T_{\alpha},T_{\beta}]=if_{\alpha\beta\gamma}T_{\gamma}$
which gives a tangent space at a unity $\hat{1}$,
and $\{\theta^{\alpha}\}$ denote the local coordinates of the first kind.
The local coordinate system is fixed at the moment.
Then, if $[\widehat{M}^{(0)},T_{i}]\ne 0$, ( $i=1,\cdots,n\le N$, $N$: dimension of the Lie group ), 
$\hat{g}$ of the following Lagrangian
\begin{eqnarray}
{\cal L}^{g-NJL} &=& -\frac{1}{2G}\bigl(\widehat{M}_{dyn}\bigr)^{2} +\bar{\psi}(i\widehat{\Dfey} - \widehat{M}^{(0)} - \hat{g}\widehat{M}_{dyn}\hat{g}^{-1})\psi
\end{eqnarray}
cannot be absorbed by any field redefinition.
In that case,
\begin{eqnarray}
0 &=& \frac{\partial\Gamma[\widetilde{M}]}{\partial\theta_{i}} = 
-(2G)^{-1}\widehat{M}_{dyn}\frac{\partial\widehat{M}_{dyn}}{\partial\theta_{i}} 
-(2G)^{-1}\frac{\partial\widehat{M}_{dyn}}{\partial\theta_{i}}\widehat{M}_{dyn} 
-i{\rm Tr}\frac{1}{i\widehat{\Dfey}-\widetilde{M}}\frac{\partial\widetilde{M}}{\partial\theta_{i}},   \\
\widetilde{M} &=& \widehat{M}^{(0)} + \hat{g}\widehat{M}_{dyn}\hat{g}^{-1}, 
\end{eqnarray}
give the stationary condition of the action in the parameter space $\{\theta_{i}\}$. 
( Note that the action $\Gamma[\widetilde{M}]$ given from ${\cal L}^{g-NJL}$ is a real-analytic function 
of the variable defined by the matrix $\widetilde{M}$. )
The periodicity of the gap equation is coming from, mainly,
$\frac{\partial\widetilde{M}}{\partial\theta_{i}}$,
and this fact is clearer if we consider a zeta-function regularization scheme for the model
( in this case, a zeta function will get the periodicity~[50] ).
The product of squares of mass eigenvalues $\lambda_{j}$ ( $j=1,\cdots, l$ ) are obtained by
\begin{eqnarray}
\lambda^{2}_{1}(\theta_{i})\times \cdots\times \lambda^{2}_{l}(\theta_{i}) &=& {\rm det}\widetilde{M}.
\end{eqnarray}
The mass eigenvalue $M$ of the chiral $U(1)$ case (4) can be obtained by the same way, 
with applying chiral projectors.  
The periodicity of roots arises from the characteristic polynomial ${\rm det}(p^{2}+\widetilde{M}^{2})$
which reflects a symmetry and its breakdown.
Our ultimate interest would be projected onto the determinant ${\rm det}(i\partfey+\afey+\widetilde{M})$,
because it may show an inconsistency between the gauge principle and the mass parameters $\widetilde{M}$.
The following scheme can be considered: 
(1) Matrix $\widetilde{M}$ is defined on a linear space, 
and usually it takes a Lie-algebra-valued form,
(2) via a functor between the matrix space of $\widetilde{M}$ and 
the space of ${\rm det}\widetilde{M}$,
we obtain a set of roots of ${\rm det}\widetilde{M}$,
(3) a root, a continuous function of $\{\theta_{i}\}$, 
and its derivatives will show a periodicity.
Roots of the first derivative give a discrete set for defining stable points of the theory.

\vspace{2mm}

For example, we will consider an $SU(2_{f})$-flavor case with its explicit breaking
$SU(2_{f})\to U(1)$.
Thus, we set
\begin{eqnarray}
\widehat{M}^{(0)} &\equiv& \left(
\begin{array}{cc}
|m^{(0)}_{1}| & 0 \\
0 & |m^{(0)}_{2}| 
\end{array}
\right), \quad \widehat{M}_{dyn} \equiv \left(
\begin{array}{cc}
M_{dyn1}P_{+} + M^{\dagger}_{dyn1}P_{-} & 0    \\
0 & M_{dyn2}P_{+} + M^{\dagger}_{dyn2}P_{-} 
\end{array}
\right), 
\end{eqnarray}
with
\begin{eqnarray}
\hat{g} &\equiv& e^{i(\theta_{1}\tau_{1}+\theta_{2}\tau_{2})} = \left(
\begin{array}{cc}
\cos |\theta| & i\frac{\theta_{-}}{|\theta|}\sin|\theta| \\
i\frac{\theta_{+}}{|\theta|}\sin|\theta| & \cos|\theta|
\end{array}
\right),  \nonumber \\
|\theta| &\equiv& \sqrt{\theta^{2}_{1}+\theta^{2}_{2}}, \quad \theta_{\pm} \equiv \theta_{1}\pm i\theta_{2},
\quad M_{dyn1} \equiv |M_{dyn1}|e^{i\theta_{\chi}}, 
\quad M_{dyn2} \equiv |M_{dyn2}|e^{i\theta_{\chi}}.
\end{eqnarray}
Here, we have taken the common phase for the dynamical chiral mass parameters.
( Thus, $\theta_{\chi}$ is still a pseudo-NG boson by setting $|m^{(0)}_{1}|=0$. )
$(\theta_{1},\theta_{2})$ defines a local coordinate system of $\hat{g}$ around the unity $\hat{1}$.
$\hat{g}$ causes a mixing of the iso-doublet.
Note that $\hat{g}$ in the above expression must not contain the symmetric generator ${\tau}_{3}$
for a consistency,
and $e^{i\theta_{3}\tau_{3}}$ should be multiplied such as 
$e^{i\theta_{3}\tau_{3}}(\widehat{M}^{(0)} + \hat{g}\widehat{M}_{dyn}\hat{g}^{-1})e^{-i\theta_{3}\tau_{3}}$
as a gauge transformation, can be extended as a local gauge at $\widehat{M}_{dyn}=0$.
Since $[\hat{g},e^{i\theta_{3}\tau_{3}}]\ne 0$, 
there is a coupling between the modes $\theta_{3}$ and $(\theta_{1},\theta_{2})$
due to the fact that the breakdown $SU(2_{f})\to U(1_{f})$ does not give a coset topology,
and the $\theta_{3}$-mode may be absorbed after the dynamical generation of $\widehat{M}_{dyn}$
by the Higgs mechanism.  
After a diagonalization of the mass matrix 
$\widehat{M}^{(0)} + \hat{g}\widehat{M}_{dyn}\hat{g}^{-1}$, 
we get the mass eigenvalues as follows:
\begin{eqnarray}
\lambda_{\pm}(\theta_{\chi},\theta_{1},\theta_{2}) &\equiv& \sqrt{\frac{A}{2}\pm\frac{1}{2}\sqrt{A^{2}-4B}},
\end{eqnarray}
where,
\begin{eqnarray}
A &\equiv& |c_{1}|^{2}+|c_{2}|^{2}  + 2|c_{3}|^{2}|c_{4}|^{2},   \\
B &\equiv& |c_{1}|^{2}|c_{2}|^{2} + |c_{3}|^{4}|c_{4}|^{4}
- |c_{4}|^{2}\bigl( c_{1}c_{2}(c^{\dagger}_{3})^{2} + (c_{1}c_{2})^{\dagger}(c_{3})^{2} \bigr), 
\end{eqnarray}
and
\begin{eqnarray}
c_{1} &\equiv& |m^{(0)}_{1}| 
+ |M_{dyn1}|e^{i\theta_{\chi}}\cos^{2}|\theta|
+ |M_{dyn2}|e^{i\theta_{\chi}}\sin^{2}|\theta|,   \\
c_{2} &\equiv& |m^{(0)}_{2}| 
+ |M_{dyn2}|e^{i\theta_{\chi}}\cos^{2}|\theta|
+ |M_{dyn1}|e^{i\theta_{\chi}}\sin^{2}|\theta|,   \\
c_{3} &\equiv& \bigl( |M_{dyn1}| - |M_{dyn2}| \bigr)e^{i\theta_{\chi}},  \\
c_{4} &\equiv& -i\frac{\theta_{-}}{|\theta|}\cos|\theta|\sin|\theta|. 
\end{eqnarray}
Note that the mass eigenvalues are given in terms of three phases $\theta_{\chi}$ ( chiral ), 
$\theta_{1}$ and $\theta_{2}$ ( flavor ).
Since the mass spectra $\lambda_{\pm}$ depend on $(\theta_{1},\theta_{2})$ via the form 
$|\theta|=\sqrt{\theta^{2}_{1}+\theta^{2}_{2}}$ ( "radial" coordinate ), 
they depend symmetrically on $(\theta_{1},\theta_{2})$ 
due to the isotropy of the two-dimensional pseudo-NG space:
The periodicity of the $SU(2_{f})$ flavor space of the effective potential is determined by $|\theta|$.
This might be a confusing fact, because the mass eigenvalues of fermions $\lambda_{\pm}$
split toward the direction of the two-dimensional flavor space, no degeneracy in the flavor space
( while, Kramers degeneracies in the spinor space exist ),
and the generators $\tau_{1}$ and $\tau_{2}$ are completely broken.
Therefore, here we will emphasize the fact that the pseudo-NG manifold has the $SO(2)$-symmetry in 
the coordinate system of the first kind $(\theta_{1},\theta_{2})$,
and their mass eigenvalues are degenerate:
Namely, {\it fermions have different mass eigenvalues, while pseudo-NG bosons are degenerated}. 
If we consider the case where all of the generators are broken,
then $|\theta|=\sqrt{\theta^{2}_{1}+\theta^{2}_{2}+\theta^{2}_{3}}$ will arise 
from the Casimir invariant of $SU(2_{f})$. 
We will write $\hat{g}$ explicitly:
\begin{eqnarray}
\hat{g} &\equiv& e^{i(\theta_{1}\tau_{1}+\theta_{2}\tau_{2}+\theta_{3}\tau_{3})}
= \left(
\begin{array}{cc}
\cos |\theta| + i\frac{\theta_{3}}{|\theta|}\sin|\theta| & i\frac{\theta_{-}}{|\theta|}\sin|\theta| \\
i\frac{\theta_{+}}{|\theta|}\sin|\theta| & \cos|\theta|-i\frac{\theta_{3}}{|\theta|}\sin|\theta|
\end{array}
\right), \quad |\theta| = \sqrt{\theta^{2}_{1}+\theta^{2}_{2}+\theta^{2}_{3}}.
\end{eqnarray}
An explicit calculation of eigenvalues of $\widetilde{M}$ in this case is a little bit complicated,
though we can find some characters of the case where all of generators of $SU(2_{f})$ are broken:
Since there is no key to distinguish $\theta_{1}$ and $\theta_{2}$ in $\hat{g}$ 
of the expression given in (22), still the $U(1)$ symmetry of $(\theta_{1},\theta_{2})$-space holds. 
Therefore, we conclude that the $U(1)$ symmetry of $(\theta_{1},\theta_{2})$ in the diagonal breaking (13) 
is not a result of the remaining Cartan subalgera $\tau_{3}$.
If we diagonalize it as 
$U\widehat{M}^{(0)}U^{-1}+(U\hat{g}U^{-1})(U\widehat{M}_{dyn}U^{-1})(U\hat{g}^{-1}U^{-1})$,
$\hat{g}'\equiv U\hat{g}U^{-1}\ne \hat{g}$ in general.
Our definition $|\theta|^{2}-\theta^{2}_{1}-\theta^{2}_{2}=c$ shows
a Lorentz symmetry $O(2,1)$, and the light-like region $c=0$ is physical:
The diagonal breaking of $SU(2_{f})$ contains an $SL(2,{\bf R})$ symmetry. 
In the case of diagonal breaking (13), 
trigonometric functions of $(\theta_{1},\theta_{2})$ inside $\lambda_{\pm}$
are factorized with those of $\theta_{\chi}$
( not like $\cos(\theta^{2}_{\chi}+\theta^{2}_{1})$ ), 
and this fact is essential to clarify the geometric structure of 
the pseudo-NG manifold.
It is noteworthy to recall that a two-dimensional non-commutative Lie algebra is
isomorphic with an affine mapping $x\to\alpha x + \beta$ defined over ${\bf R}$~[166].
It is well-known that ${\bf H}\simeq SL(2,{\bf R})/SO(2)
\simeq \{ \left(
\begin{array}{cc}
a & b \\
0 & a^{-1}
\end{array}
\right), a>0, b\in {\bf R} \}$
( $SO(2)$ is an isotropy subgroup )
and this matrix is obtained by an exponential mapping of the two-dimensional non-commutative Lie algebra.
At the degeneracy condition of the {\it fermion} mass spectra in the two-dimensional flavor space,
they will recover the same spectra of the chiral symmetry breaking given in (4).
Our observation should be included into a complexification 
$\hat{g}=\exp(iz_{1}\tau_{1}+iz_{2}\tau_{2}+iz_{3}\tau_{3})\in SL(2,{\bf C})$
in which a group action of it is complex-analytic
( $z'_{a}=f_{a}(z_{1},z_{2},z_{3})$, $a=1,2,3$ and $f_{a}$ are holomorphic ),
and our result given above is a "section" ( not in the sense of fibre bundle! ) of the complex manifold.
( Note that both $SU(N)$ and $SL(N,{\bf R})$ are subgroups of $SL(N,{\bf C})$,
and there is a one-to-one correspondence between a finite-dimensional irreducible representation
of $SU(N)$ and that of $SL(N,{\bf R})$.
If $G$ is a connected compact Lie group, there is a connected complex Lie group $G_{\bf C}$
and $G\subset G_{\bf C}$.
In this case, a homogeneous space $G_{\bf C}/G$ is simply connected,
namely the fundamental group is trivial. )
It is known fact that, due to Baily-Borel,
$X/\Gamma$ ( $X$; a bounded open connected subset of ${\bf C}^{n}$ 
with a fixed point under an involution $z\to -z$,
$\Gamma$; a congruence subgroup )
is embedded into a Shimura variety $X^{*}/\Gamma$.
Moreover, $\widehat{M}_{dyn}$ of the diagonal breaking can be interpreted as a maximal torus of $U(N_{f})$,
namely ${\rm diag}(e^{it_{1}},\cdots,e^{it_{N_{f}}})$ 
( $t_{1},\cdots,t_{N_{f}}\in{\bf R}$ ),
and our $\widehat{M}_{dyn}$ can be expressed by a complexification of the maximal torus,
${\rm diag}(e^{iz_{1}},\cdots,e^{iz_{N_{f}}})$
( $z_{1},\cdots,z_{N_{f}}\in D\subset{\bf C}$ ).
In this case, for example, one can use a domain of $z_{l}$ in ${\rm det}\widetilde{M}(z_{l},\bar{z}_{l})$
( $l=1,\cdots,N_{f}$ ) as 
an Osgood space $\bigotimes^{N_{f}}_{l=1}\widehat{\bf C}_{l}$~[164] or, 
a poly-upper-halves $\bigotimes^{N_{f}}_{l=1}{\bf H}_{l}$
( the total matrix $\widetilde{M}$ containes both holomorphic and anti-holomorphic parts, 
while $\widehat{M}^{(0)}$ {\it can} be chosen as real, $M_{dyn1}$ and $M_{dyn2}$ are holomorphic ).
Now, $(SL(2,{\bf C}))^{N_{f}}$ acts on $\widehat{M}_{dyn}$.
The periodicity of eigenvalues $\lambda_{l}(\theta_{i})$ ( not $\lambda^{2}_{l}(\theta_{i})$,
and $\lambda_{l}(z_{i})$ are given by infinite-order serieses of $z_{a},\bar{z}_{a}$ ),
is obtained by taking values of $z_{a},\bar{z}_{a}$ on the real axis of 
$\bigotimes^{N_{f}}_{l=1}D_{l}\subset\bigotimes^{N_{f}}_{l=1}{\bf C}_{l}$.
Note that any element $\hat{g}$ of Lie group $G$ is given by
$\bigcup_{\hat{g}\in G}\hat{g}\hat{T}\hat{g}^{-1}$
where $\hat{T}$ is a maximal torus, and thus
$\hat{g}\widehat{M}_{dyn}\hat{g}^{-1}$ is a group element ( a linear combination of elements ) of $G$.
We should mention that a complexification of a connected compact semisimple Lie group is uniquely determined,
and its fundamental group $\pi_{1}$ is a finite Abelian.  
Even though it is difficult to derive more generic expression of mass eigenvalues in a compact Lie group,
we find that the periodicity is mainly coming from $\hat{g}$ such that:
\begin{eqnarray}
\hat{g} &\equiv& e^{\vartheta\left(
\begin{array}{cc}
a & -b \\
b & a 
\end{array}
\right)} = \left(
\begin{array}{cc}
e^{\vartheta a}\cos \vartheta b & -e^{\vartheta a}\sin \vartheta b \\
e^{\vartheta a}\sin \vartheta b & e^{\vartheta a}\cos \vartheta b \\
\end{array}
\right).
\end{eqnarray}
Thus, from this expression which is obtained by a resummation of taylor series of $\hat{g}$, 
we recognize the fact that it is impossible to generate a periodicity 
toward a direction of Grassmann coordinates
by a Lie supergroup of a $Z_{2}$-grading superalgebra by the mechanism we have discussed.

\vspace{2mm}

Next, we will examine an $SO(3_{f})$ 
( the set of its group elements is doubly-connected over a sphere,
locally isomorphic to $SU(2)$ )
case to establish the periodicity.
By defining 
\begin{eqnarray}
A &\equiv& \left(
\begin{array}{ccc}
0 & -\rho_{3} & \rho_{2}  \\
\rho_{3} & 0 & -\rho_{1}  \\
-\rho_{2} & \rho_{1} & 0
\end{array}
\right), \quad  \widehat{M}^{(0)} \equiv \left(
\begin{array}{ccc}
0 & 0 & 0 \\
0 & 0 & 0 \\
0 & 0 & |m^{(0)}_{3}|
\end{array}
\right), \quad
\widehat{M}_{dyn} \equiv \left(
\begin{array}{ccc}
0 & 0 & 0 \\
0 & 0 & 0 \\
0 & 0 & M_{dyn3}
\end{array}
\right),  \nonumber \\
\hat{g} &\equiv& e^{A}, \quad
\widetilde{M} \equiv 
\widehat{M}^{(0)} + \hat{g}(\widehat{M}_{dyn}P_{+}+\widehat{M}^{\dagger}_{dyn}P_{-})\hat{g}^{-1},
\end{eqnarray}
( here, we take the mass function on a complex number even though $SO(3)$ is real )
and utilizing the Rodrigues formula in the case of $SO(3)$,
\begin{eqnarray}
e^{A} &\equiv& 1_{3\times 3} + \frac{\sin|\rho|}{|\rho|}A + \frac{1-\cos|\rho|}{|\rho|^{2}}A^{2},   \\
|\rho| &\equiv& \sqrt{\rho^{2}_{1}+\rho^{2}_{2}+\rho^{2}_{3}},  
\end{eqnarray}
one finds the mass matrix in the following form:
\begin{eqnarray}
\widehat{M}^{SO(3)}_{dyn} &\equiv& \widehat{M}_{dyn}P_{+}+\widehat{M}^{\dagger}_{dyn}P_{-},   \\
M^{SO(3)}_{dyn3} &\equiv& M_{dyn3}P_{+}+M^{\dagger}_{dyn3}P_{-},   \\
\widehat{M}^{(0)} + \hat{g}\widehat{M}^{SO(3)}_{dyn}\hat{g}^{-1}
&=& \left(
\begin{array}{ccc}
\frac{\sin^{2}|\rho|}{|\rho|^{2}}\rho^{2}_{2}M^{SO(3)}_{dyn3}   & 
-\frac{\sin^{2}|\rho|}{|\rho|^{2}}\rho_{1}\rho_{2}M^{SO(3)}_{dyn3}   & 
\frac{\cos|\rho|\sin|\rho|}{|\rho|}\rho_{2}M^{SO(3)}_{dyn3}   \\
-\frac{\sin^{2}|\rho|}{|\rho|^{2}}\rho_{1}\rho_{2}M^{SO(3)}_{dyn3}   & 
\frac{\sin^{2}|\rho|}{|\rho|^{2}}\rho^{2}_{1}M^{SO(3)}_{dyn3}   & 
-\frac{\cos|\rho|\sin|\rho|}{|\rho|}\rho_{1}M^{SO(3)}_{dyn3}   \\
\frac{\cos|\rho|\sin|\rho|}{|\rho|}\rho_{2}M^{SO(3)}_{dyn3}   & 
-\frac{\cos|\rho|\sin|\rho|}{|\rho|}\rho_{1}M^{SO(3)}_{dyn3}   & 
|m^{(0)}_{3}|+\cos^{2}|\rho|M^{SO(3)}_{dyn3}    
\end{array}
\right), 
\end{eqnarray}
( by setting $\rho_{3}=0$ ).
The periodicity of the secular equation of fermion determinant 
$0={\rm det}[\pfey-(\widehat{M}^{(0)}+\hat{g}\widehat{M}^{SO(3)}_{dyn}\hat{g}^{-1})]$
is obvious, coming from the "periodicity" of exponential mapping from the algebra to the group. 
An interesting fact is that the mass parameters of the entries of the matrix of the determinant vanish
at $\rho_{1}=\rho_{2}=0$, except $|m^{(0)}_{3}|+\cos^{2}|\rho|M^{SO(3)}_{dyn3}$ remains.
Namely, the entries vanish at $\rho_{1}=\rho_{2}=0$ give massive pseudo-NG modes,
and those entries should take small values.
The Pauli principle is satisfied, 
and the result indicates that it seems easy to generate a mass hierarchy
( via a possible higher-order radiative collection or so ).

\vspace{2mm}

Now, we mention on the functional structure of the mass function (4)
form the viewpoint of theory of harmonic mapping.
It is a known fact that ( the Weierstrass minimal surface )~[46], 
when $h$ is a harmonic mapping of a domain $D$ of a two-dimensional manifold,
the following function
\begin{eqnarray}
F &\equiv& |\partial_{x}h|^{2} - |\partial_{y}h|^{2} -2i\langle \partial_{x}h,\partial_{y}h \rangle
\end{eqnarray} 
is holomorphic on $z=x+iy\in D$.
We will obtain the square of (4) from (30) by the following correspondences,
with an analytic continuation to ${\bf H}$:
\begin{eqnarray}
|m^{(0)}| \leftrightarrow |\partial_{x}h|, \quad |M_{dyn}| \leftrightarrow i|\partial_{y}h|.
\end{eqnarray}
The anti-holomorphic counterpart of it can also be obtained by an analytic continuation to $\overline{\bf H}$.
This fact is interesting because $M^{2}$, the square of (4), 
can be regarded as a holomorphic function of $z\in D\subset{\bf C}$, 
it defines a harmonic mapping of two-dimensional space implicitly,
and the chiral degree of freedom $\theta_{\chi}$ ( pion ) is given by 
an angle between two vectors $\partial_{x}h$ and $\partial_{y}h$. 
Since this holomorphicity is always kept in the mass eigenvalues of ${\rm det}\widetilde{M}$, 
it should be satisfied in a path integration of mass functions:
A path integration of the auxiliary fields in the NJL-type model 
traces inside a larger space which contains a subspace of the holomorphicity
( there is a loss of information after a diagonalization of a matrix,
and $|m^{(0)}|+M_{dyn}P_{+}+M^{\dagger}_{dyn}P_{-}$
contains three variables while this is a matrix of 16-dimensional linear space ).
We should emphasize the fact that this is only the case in the chiral symmetry
( due to functional structures of dispersions ),
and other vector/axialvector-type masses, and of course, Cooper pairing masses also,
cannot have this character:
Pion seems not only physically but also mathematically special.
Because $h$ is harmonic, we can implicitly consider the action as the back ground of the mass function (30):
\begin{eqnarray}
S[h] &\equiv& \int \partial_{\mu}h^{l}\partial^{\mu}h^{l},
\end{eqnarray} 
where $l=1,\cdots,N$, and we have replicated the function $h$, 
and the degree of freedom could be interpreted as "flavor" of the theory.
One can consider not only a possible functional form of $h$
but also a possible deformation of $h$ in its functional space
( an infinite dimensional Banach/Hilbert space ).
If we regard $h^{l}$ as independent quantities, 
\begin{eqnarray}
S[h] &=& \int d^{2}z \sqrt{-g}g_{\mu\nu}g_{lm}\partial_{\mu}h^{l}\partial_{\nu}h^{m}
\end{eqnarray}
is obtained. 
This is an artificial prescription, 
and restrictions on possible forms/types on $g_{\mu\nu}$ and $g_{lm}$ may exist, 
though a Ricci flow~[66,188] ( an RG-flow with a cutoff regularization ) obtained from $S[h]$
reflects some relations/differences of harmonic functions as a "generating function" of mass function.
( Note that, mathematically, a Ricci flow can be obtained in any Riemannian manifold of any dimensions. )
Furthermore, it might be possible to extend/generalize the viewpoint
of harmonic mapping/function to those of harmonic form,
namely a Laplacian and Hodge theorem over a complex manifold,
to get a new interpretation on mass functions
or to extend the notion of mass functions.

\vspace{2mm}

Before closing this subsection, 
we will make a comment on a complex structure of the dynamical mass $M_{dyn}$.
In fact, $|m^{(0)}|\to |m^{(0)}|+M_{dyn}P_{+}+M^{\dagger}_{dyn}P_{-}$ can be interpreted 
as a "chiral/antichiral" affine mapping.
The generation of the phase degree of freedom $\theta_{\chi}$ is included 
by the following (quasi)conformal mapping~[98] $|M_{dyn}|\to |M_{dyn}|e^{i(\alpha z + \beta\bar{z})}$ 
and its complex conjugate ( $z, \alpha, \beta \in {\bf C}$ ). 
A Lagrangian with the following mass and its mapping
\begin{eqnarray}
|m^{(0)}| + zP_{+} + \bar{z}P_{-} \to |m^{(0)}| + wP_{+} + \bar{w}P_{-},  \qquad
w = f(z,\bar{z}), \quad \bar{w} = \bar{f}(\bar{z},z)
\end{eqnarray}
still keeps its Hermiticity. 
By a (quasi)conformal mapping of a complex mass parameter $z (\bar{z})\to w$, 
we can handle variations $f(z,\bar{z})\to f(z+\delta z,\bar{z}+\delta\bar{z})$ 
with respect to the amplitude $|M_{dyn}|$ and 
the phase $\theta_{\chi}$ of the $U(1)$ case in a unified manner, 
and can construct first and second variations of the theory more mathematically. 
( It is a kind of complexification $e^{i\theta}\to e^{i z}$, $\theta\in {\bf R}$, 
$z\in \widehat{\bf C}$ or ${\bf C}$ or ${\bf H}$. )
Since a quasi-conformal mapping induces a deformation of complex structure,
there are Riemann surfaces and their Teichm\"{u}ller/moduli~[98] spaces 
as the background of complex mass functions,
though ${\bf C}$ or the Riemann sphere $\widehat{\bf C}$ 
may physically appropriate domain for a complex mass parameter $\widetilde{M}$
( in the one-dimensional case ). 
On the other hand, there is a discussion on a convergence property and invariance of measure 
of the path integration of auxiliary fields, 
which may give a restriction to a form of integration domain 
( especially in an interacting bosonic sector )~[72].   
For example, an affine mapping $\theta_{\chi}\to a\theta_{\chi}+b$ 
( a scalling and a change of the origin, $a,b\in {\bf R}$ )
generated by a matrix of $GL(2,{\bf R})$
gives essentially the same mass spectra of (4),
namely an absolute value of $\theta_{\chi}$ 
( and also a value of period, i.e., $\pi$ or $2\pi$, ... ) has no physical meaning,
and this is a kind of gauge degree of freedom,
should be removed from a path integration measure ${\cal D}\theta_{\chi}$.
Hence, a physical angle will be defined by $\theta_{\chi}/{\rm affine}$,
and the physical Hilbert space is 
${\cal V}_{phys-\theta_{\chi}}\equiv {\cal V}_{\theta_{\chi}}/{\rm affine}$.  
( The affine transformation does not give a symmetry of the system, simply a redefinition of parameters. )
Linear fractional transformations on Riemann surfaces are classified into mainly three categories:
(1) parabolic type, $f(z)=z+b$ ( $b\in {\bf C}$, $b\ne 0$ ),
(2) elliptic type, $f(z)=e^{i\theta}z$ ( $\theta\in {\bf R}$, $\theta\ne 2n\pi$ ), 
(3) hyperbolic type, $f(z)=\lambda z$ ( $\lambda > 0$, $\lambda\ne 1$ ). 
Any variation of the mass function will be obtained by those mappings.
It is well-known fact that the automorphism ${\rm Aut}({\bf H})$
of the upper-half plane, $f(z)=(az+b)/(cz+d)$,
( $a,b,c,d\in{\bf R}$, $ad-bc=1$ ) is completely classified
into these three types, uniquely classified by examining where fixed points of those transformations
locate on ${\bf C}$. 
A Teichm\"{u}ller space is introduced via 
a Fuchsian model $\Gamma$ and a Fricke coordinate system.
( If $F$ is a fundamental region of $\Gamma$, 
then $\gamma(F)\cap F= \phi$, $\gamma\in\Gamma$, 
and ${\bf H}=\cup_{\gamma\in\Gamma}\gamma(\overline{F})$, $\overline{F}$ is a closure of $F$.
A canonical system of generators of fundamental group $\pi_{1}$ is given by
elements of ${\rm Aut}({\bf H})$,
and a set of parameters of linear transformations of ${\rm Aut}({\bf H})$
defines a Fricke coordinates which is an ${\bf R}^{6g-6}$ where $g$ denotes a genus of Riemann surface. )
It should be emphasized that the complex structure we have discussed
exists not only in the chiral mass, but also in vector/axialvector-type mass functions.
Hence, this kind of examinaion is generically meaningful for mass functions and their quantum 
fluctuations/perturbations, 
especially when a theory has an explicit+dynamical symmetry breaking mechanism.
The mass function, a Riemann surface, is an implicit function of model parameters of the Lagrangian
\begin{eqnarray}
M_{dyn} &=& z = z(G,\Lambda,|m^{(0)}|)
\end{eqnarray}
and then a variation with respect to these model parameters causes a "motion" of a point
over a Riemann surface.
We can go further into an abstract generalization of our results.
If we replace a Riemann surface, same as a mass function, 
to an abstract Riemann surface 
( a set of "place", and a place is an equivalence class of a discrete valuation 
of algebraic function field )
which appears in number theory and congruence zeta functions
( it will play a special role in the proof of Weil conjecture~[86] ),
then one could introduce a notion of Grothendieck-Galois theory~[58].
( For example, Riemann surfaces handled by the p-adic Teichm\"{u}ller theory and covering groups
might be inserted into an abstruct dynamical Lagrangian~[150]. )
We can summarize our result schematically:
\begin{eqnarray}
i\widehat{\Dfey} &\to& {\rm moduli \, space \, of \, connections},  \\
\widetilde{M} &\to& 
{\rm Teichmueller/moduli \, space \, of \, (holomorphic/anti-holomorphic) \, Riemann \, surfaces}.
\end{eqnarray}
Since a Lagrangian of non-linear sigma model is defined in terms of a "normalized" mass function,
then we could consider a Lagrangian over an abstract Riemann surface
( with a generalized derivative, to generalize topological field theory ). 
For a possible classification by liftings and universal covering groups of chiral mass case, 
we give the following chart:
\begin{eqnarray}
i\widehat{\Dcalfey} + \widetilde{\cal M} &\longrightarrow& {\rm det}(i\widehat{\Dcalfey} + \widetilde{\cal M})  \nonumber \\
\qquad \Bigg\downarrow   & & \qquad \Bigg\downarrow \nonumber \\
i\widehat{\Dfey} + \widetilde{M} &\longrightarrow& {\rm det}(i\widehat{\Dfey} + \widetilde{M}).
\end{eqnarray}
Here, the first column implies globally defined covering spaces possibly include discrete subgroups,
while the lower column are considered as quantities defined locally in specified senses.
The vertical morphisms contain local homeomorphisms defined in specified senses.

\subsection{The Lagrangian and the Propagator}

We introduce the following NJL-type four-fermion contact interaction Lagrangian 
with left- and right- handed Majorana mass terms:
\begin{eqnarray}
{\cal L}^{NJL+M} &=&  \xi^{\dagger}i\bar{\sigma}^{\mu}\partial_{\mu}\xi +\eta^{\dagger}i\sigma^{\mu}\partial_{\mu}\eta
-\frac{1}{2}\bigl( m_{R}\xi^{\dagger}i\sigma_{2}\xi^{*}-m^{\dagger}_{R}\xi^{T}i\sigma_{2}\xi \bigr)
-\frac{1}{2}\bigl( m_{L}\eta^{T}i\sigma_{2}\eta-m^{\dagger}_{L}\eta^{\dagger}i\sigma_{2}\eta^{*} \bigr) +G\eta^{\dagger}\xi\xi^{\dagger}\eta.
\end{eqnarray}
The Majorana mass terms break the following global $U(1)$ symmetries:
\begin{eqnarray}
\xi \to \xi' = e^{i\theta_{1}}\xi, \quad \eta \to \eta' = e^{i\theta_{2}}\eta.
\end{eqnarray}
Namely, both the gauge $U(1)_{V}$ and chiral $U(1)_{A}$ symmetries are explicitly broken 
by introducing non-vanishing mass parameters $m_{R}$ and $m_{L}$.
The coupling constant $G$ has its mass dimension as $[{\rm mass}]^{-2}$.
As usual, the four-fermion interaction is prepared to generate a dynamical Dirac mass term in our model.
Hence, it will be converted such as
\begin{eqnarray}
G\eta^{\dagger}\xi\xi^{\dagger}\eta &\to& -m_{D}\xi^{\dagger}\eta-m^{\dagger}_{D}\eta^{\dagger}\xi.
\end{eqnarray}
Here, we do not intend to introduce Majorana mass terms dynamically:
Usually, it is energetically unfavorable to generate 
a left-right asymmetric Majorana mass terms dynamically, 
and we will obtain a scalar Cooper-pair-type mass given by a linear combination 
of a right and a left handed Majorana mass terms. 
Through the method of auxiliary fields of composites, 
the generating functional of the theory becomes
\begin{eqnarray}
{\cal Z} &=& \int
{\cal D}m_{D}{\cal D}m^{\dagger}_{D}{\cal D}\Psi_{MN}{\cal D}\overline{\Psi_{MN}} \exp\Bigg[ i\int d^{4}x \Bigl( -\frac{|m_{D}|^{2}}{G}+\frac{1}{2}\overline{\Psi_{MN}}\Omega^{F}_{M}\Psi_{MN}  \Bigr) + ({\rm sources}) \Bigg].
\end{eqnarray}
Here, the matrix and fields in ${\cal Z}$ are defined as follows:
\begin{eqnarray}
\Omega^{F}_{M} &\equiv& \left(
\begin{array}{cc}
i\partfey -m^{\dagger}_{R}P_{+} -m_{R}P_{-} & -m^{\dagger}_{D}P_{+} -m_{D}P_{-}  \\
-m^{\dagger}_{D}P_{+} -m_{D}P_{-}  & i\partfey -m^{\dagger}_{L}P_{+} -m_{L}P_{-} 
\end{array}
\right), 
\end{eqnarray}
and
\begin{eqnarray}
\Psi_{MN} &\equiv& \left(
\begin{array}{c}
\psi_{MR} \\
\psi_{ML}
\end{array}
\right), \quad \overline{\Psi_{MN}} = (\overline{\psi_{MR}},\overline{\psi_{ML}}), \quad \psi_{MR} = \left(
\begin{array}{c}
\xi \\
i\sigma_{2}\xi^{*}
\end{array}
\right), \quad \psi_{ML} = \left(
\begin{array}{c}
-i\sigma_{2}\eta^{*} \\
\eta
\end{array}
\right). 
\end{eqnarray}
$\psi_{MR}$ and $\psi_{ML}$ are right- and left- handed Majorana fields, respectively.
$\Psi_{MN}$ can be called as a Majorana-Nambu-notation field~[178].
For our convenience, we write the global $U(1)$ transformation laws: 
\begin{eqnarray}
U(1)_{V}: & & 
\psi'_{MR} = e^{i\gamma_{5}\theta}\psi_{MR}, \quad
\psi'_{ML} = e^{-i\gamma_{5}\theta}\psi_{ML}, \quad 
\Psi'_{MN} = e^{i(\gamma_{5}\otimes\tau_{3})\theta}\Psi_{MN},  \quad 
\overline{\Psi'_{MN}} = \overline{\Psi_{MN}}e^{i(\gamma_{5}\otimes\tau_{3})\theta},   \\
U(1)_{A}: & & 
\psi'_{MR} = e^{i\gamma_{5}\theta}\psi_{MR}, \quad 
\psi'_{ML} = e^{i\gamma_{5}\theta}\psi_{ML}, \quad
\Psi'_{MN} = e^{i(\gamma_{5}\otimes 1)\theta}\Psi_{MN},  \quad
\overline{\Psi'_{MN}} = \overline{\Psi_{MN}}e^{i(\gamma_{5}\otimes 1)\theta}.
\end{eqnarray}

\vspace{2mm}

In momentum space, $\Omega^{F}_{M}$ will be inverted into 
( the method, see Ref.~[172] ):
\begin{eqnarray}
(\Omega^{F}_{M})^{-1} &\equiv& \frac{1}{D(k)}\left(
\begin{array}{cc}
G_{11}(k) & G_{12}(k) \\
G_{21}(k) & G_{22}(k)
\end{array}
\right),
\end{eqnarray}
where, 
\begin{eqnarray}
G_{11}(k) &\equiv& (k^{2}-|m_{L}|^{2})(\kfey+m_{R}P_{+}+m^{\dagger}_{R}P_{-}) -(m_{D}P_{+}+m^{\dagger}_{D}P_{-})(\kfey-m^{\dagger}_{L}P_{+}-m_{L}P_{-})(m_{D}P_{+}+m^{\dagger}_{D}P_{-}),  \\
G_{12}(k) &\equiv& (\kfey+m_{R}P_{+}+m^{\dagger}_{R}P_{-})(m^{\dagger}_{D}P_{+}+m_{D}P_{-})(\kfey+m_{L}P_{+}+m^{\dagger}_{L}P_{-})-|m_{D}|^{2}(m_{D}P_{+}+m^{\dagger}_{D}P_{-}),  \\
G_{21}(k) &\equiv& (\kfey+m_{L}P_{+}+m^{\dagger}_{L}P_{-})(m^{\dagger}_{D}P_{+}+m_{D}P_{-})(\kfey+m_{R}P_{+}+m^{\dagger}_{R}P_{-})-|m_{D}|^{2}(m_{D}P_{+}+m^{\dagger}_{D}P_{-}),  \\
G_{22}(k) &\equiv& (k^{2}-|m_{R}|^{2})(\kfey+m_{L}P_{+}+m^{\dagger}_{L}P_{-}) -(m_{D}P_{+}+m^{\dagger}_{D}P_{-})(\kfey-m^{\dagger}_{R}P_{+}-m_{R}P_{-})(m_{D}P_{+}+m^{\dagger}_{D}P_{-}), 
\end{eqnarray}
and the denominator is
\begin{eqnarray}
D(k) &=& [k^{2}-(M^{F}_{+})^{2}]^{2}[k^{2}-(M^{F}_{-})^{2}]^{2} = (k_{0}-E^{F}_{+}(\bmk))^{2}(k_{0}+E^{F}_{+}(\bmk))^{2}(k_{0}-E^{F}_{-}(\bmk))^{2}(k_{0}+E^{F}_{-}(\bmk))^{2}.
\end{eqnarray}
Here, the energy spectra will be obtained into the following forms~[178]:
\begin{eqnarray}
E^{F}_{\pm}(\bmk) &=& \sqrt{\bmk^{2}+(M^{F}_{\pm})^{2}},  \\
M^{F}_{\pm} &=& \sqrt{|m_{D}|^{2} + \frac{|m_{R}|^{2}+|m_{L}|^{2}}{2}\mp \frac{1}{2}\sqrt{ (|m_{R}|^{2}-|m_{L}|^{2})^{2} + 4|m_{D}|^{2}(|m_{R}|^{2}+|m_{L}|^{2}+2|m_{R}||m_{L}|\cos\Theta) } }, \\
\Theta &\equiv& \theta_{R} + \theta_{L} -2\theta_{D},
\end{eqnarray}
where, the definitions of phases of the mass parameters are given by
\begin{eqnarray}
m_{R} = |m_{R}|e^{i\theta_{R}}, \quad m_{L} = |m_{L}|e^{i\theta_{L}}, \quad m_{D} = |m_{D}|e^{i\theta_{D}}.
\end{eqnarray}
There are arbitralinesses of definitions of those phases,
\begin{eqnarray}
& & 
\theta_{R} \to \theta'_{R} = \alpha_{R}\theta_{R} + \beta_{R}, \quad
\theta_{L} \to \theta'_{L} = \alpha_{L}\theta_{L} + \beta_{L}, \quad
\theta_{D} \to \theta'_{D} = \alpha_{D}\theta_{D} + \beta_{D}, \\
& & \Theta \to \Theta' = \theta'_{R} + \theta'_{L} -2\theta'_{D},
\end{eqnarray}
where,
\begin{eqnarray}
\alpha_{R}, \alpha_{L}, \alpha_{D}, \beta_{R}, \beta_{L}, \beta_{D} \in {\bf R}.
\end{eqnarray}
Namely, $SL(2,{\bf R})$ acts on those phase spaces.
$\Theta$ in $M^{F}_{\pm}$ is a dimensionless quantity: Its mass dimension is $[{\rm mass}]^{0}$.
At first glance, the spectra $E^{F}_{\pm}(\bmk)$ seem to have the complicated structures, 
though the spectra do not break the proper orthochronous Lorentz symmetry $O_{+}(3,1)$.
Note that $\Omega^{F}_{M}$ is an $8\times 8$ matrix, 
each of the energy spectra has two-fold degeneracy.
From the method to obtain the propagator $(\Omega^{F}_{M})^{-1}$, 
we find the reasons of the degeneracy: 
(1) the theory keeps the Lorentz symmetry,
(2) we do not consider vector and axial-vector type dynamical masses in here.
( Especially, a vector mean field $l_{\mu}=g^{-1}\partial_{\mu}g$, $g=e^{i\gamma_{5}\alpha}$ 
causes an anomaly~[54]. )
For example, the mass eigenvalues under the type-I seesaw condition $|m_{L}|=0$ and $|m_{R}|\gg |m_{D}|$ 
of neutrino~[152] will be obtained from $M^{F}_{\pm}$ given above as 
\begin{eqnarray}
M^{F}_{+} &\approx& \frac{|m_{D}|^{2}}{|m_{R}|}, \quad M^{F}_{-} \approx |m_{R}|.
\end{eqnarray}
The situation with non-vanishing $m_{L}$ gives a type-II-like 
seesaw mechanism while $m_{L}=0$ will be called as a type-I seesaw of the single-flavor model.
The Pauli-G\"{u}rsey symmetry defined as a rotation of particle and anti-particle space,
$\psi' = a\psi + b\gamma_{5}\psi^{c}$,
$\overline{\psi'} = a^{*}\bar{\psi} - b^{*}\overline{\psi^{c}}\gamma_{5}$,
$|a|^{2}+|b|^{2} = 1$,
is broken due to the Dirac and Majorana mass terms while the kinetic term keeps it~[184].
This symmetry restricts the type of interaction in the "classical" theory of left-handed neutrino.

\vspace{2mm}

By imposing the self-consistency condition of the propagator, 
one finds two gap equations of $m_{D}$ in the following forms ( in the Euclidean region ):
\begin{eqnarray}
m_{D}\pm m^{\dagger}_{D} &=& G\int_{k}\frac{(m_{D}\pm m^{\dagger}_{D})(k^{2}+|m_{D}|^{2})-m^{\dagger}_{D}m_{R}m_{L}\mp m_{D}m^{\dagger}_{R}m^{\dagger}_{L}}{[k^{2}+(M^{F}_{+})^{2}][k^{2}+(M^{F}_{-})^{2}]}.
\end{eqnarray}
It seems impossible to determine both $|m_{D}|$ and $\theta_{D}$ by these equations.
Later, we will find that $\theta_{D}=0$ while $\theta_{R}+\theta_{L}=\Theta=(2n+1)\pi$ 
from the one-loop effective potential of the theory.

\vspace{2mm}

If we use the formalism of Dirac bispinors, the fermion matrix in ${\cal Z}$ will be replaced by 
\begin{eqnarray}
\Omega^{F}_{D} &\equiv& \left(
\begin{array}{cc}
i\partfey -m_{D}^{\dagger}P_{+} -m_{D}P_{-} & -m^{\dagger}_{L}CP_{+} -m_{R}CP_{-}  \\
-m^{\dagger}_{R}CP_{+}-m_{L}CP_{-}  & -C^{-1}(i\partfey)C +m^{\dagger}_{D}P_{+} +m_{D}P_{-} 
\end{array}
\right),   
\end{eqnarray}
with the following definitions of fields,
\begin{eqnarray}
\Psi_{DN} &\equiv& \left(
\begin{array}{c}
\psi_{D} \\
\overline{\psi_{D}}^{T}
\end{array}
\right), \quad \overline{\Psi_{DN}} = (\overline{\psi_{D}},\psi^{T}_{D}), \quad \psi_{D} = \left(
\begin{array}{c}
\xi \\
\eta
\end{array}
\right).
\end{eqnarray}
Here, $\Psi_{DN}$ is a Dirac-Nambu notation~[172].
Needless to say, the formalism of the Dirac fields is physically equivalent to that of the Majorana fields.

\subsection{The One-Loop Effective Potential}

We will employ the steepest descent approximation for the integrations of collective fields. 
We get the following effective action:
\begin{eqnarray}
\Gamma^{NJL+M}_{eff} &=& -\int d^{4}x \frac{|m_{D}|^{2}}{G} -\frac{i}{2}\ln{\rm Det}\Omega^{F}_{M}.
\end{eqnarray}
To examine the structure and the stationary condition of the effective potential 
$V^{NJL+M}_{eff}\equiv-\Gamma^{NJL+M}_{eff}/\int d^{4}x$, 
first we neglect the contribution of quantum fluctuation $\delta m_{D}$ of the collective field $m_{D}$.
Due to the existence of the phase $\Theta$ in our theory, 
collective excitations will obtain some effects 
coming from a nontrivial $\Theta$-dependent structure of the potential of our model.
Our model has its characteristic feature at this point.
By the four-dimensional covariant cutoff regularization for the momentum integration, 
the effective potential of the theory is found to be
\begin{eqnarray}
V^{NJL+M}_{eff} &=& \frac{|m_{D}|^{2}}{G} - \frac{1}{16\pi^{2}}\Bigg[ \Lambda^{2}(M^{F}_{+})^{2} + \Lambda^{2}(M^{F}_{-})^{2} + \Lambda^{4}\ln\Bigl(1+\frac{(M^{F}_{+})^{2}}{\Lambda^{2}}\Bigr)\Bigl(1+\frac{(M^{F}_{-})^{2}}{\Lambda^{2}}\Bigr)    \nonumber \\
& & -(M^{F}_{+})^{4}\ln\Bigl(1+\frac{\Lambda^{2}}{(M^{F}_{+})^{2}}\Bigr) -(M^{F}_{-})^{4}\ln\Bigl(1+\frac{\Lambda^{2}}{(M^{F}_{-})^{2}}\Bigr)  \Bigg].
\end{eqnarray}
We should mention that this potential is not normalized as $V^{NJL+M}_{eff}(|m_{D}|=0)=0$.
The first-derivative of $V^{NJL+M}_{eff}$ with respect to the collective field $|m_{D}|$ becomes 
\begin{eqnarray}
\frac{\partial V^{NJL+M}_{eff}}{\partial|m_{D}|} &=& 
\frac{2|m_{D}|}{G} - \frac{1}{8\pi^{2}}\Bigg[ \frac{\partial (M^{F}_{+})^{2}}{\partial|m_{D}|}F^{F}_{+} + \frac{\partial (M^{F}_{-})^{2}}{\partial|m_{D}|}F^{F}_{-} \Bigg], 
\end{eqnarray}
while, we also have to take into account the following derivative,
\begin{eqnarray}
\frac{\partial V^{NJL+M}_{eff}}{\partial\Theta} &=& 
- \frac{1}{8\pi^{2}}\Bigg[ \frac{\partial (M^{F}_{+})^{2}}{\partial\Theta}F^{F}_{+} + \frac{\partial (M^{F}_{-})^{2}}{\partial\Theta}F^{F}_{-} \Bigg], 
\end{eqnarray}
where,
\begin{eqnarray}
F^{F}_{\pm} &\equiv& \Lambda^{2}-(M^{F}_{\pm})^{2}\ln\Bigl( 1+\frac{\Lambda^{2}}{(M^{F}_{\pm})^{2}} \Bigr).
\end{eqnarray}
The global minimum of our model has to satisfy the vanishing conditions of both of these derivatives.
The derivatives become
\begin{eqnarray}
\frac{\partial (M^{F}_{\pm})^{2}}{\partial|m_{D}|} &=& 
2|m_{D}|\Bigg[ 1\mp\frac{|m_{R}|^{2}+|m_{L}|^{2}+2|m_{R}||m_{L}|\cos\Theta}{\sqrt{(|m_{R}|^{2}-|m_{L}|^{2})^{2}+4|m_{D}|^{2}(|m_{R}|^{2}+|m_{L}|^{2}+2|m_{R}||m_{L}|\cos\Theta)}} \Bigg],  \\
\frac{\partial (M^{F}_{\pm})^{2}}{\partial\Theta} &=& 
\pm\frac{2|m_{D}|^{2}|m_{R}||m_{L}|\sin\Theta}{\sqrt{(|m_{R}|^{2}-|m_{L}|^{2})^{2}+4|m_{D}|^{2}(|m_{R}|^{2}+|m_{L}|^{2}+2|m_{R}||m_{L}|\cos\Theta)}}.
\end{eqnarray}
Especially,
\begin{eqnarray}
\frac{\partial (M^{F}_{\pm})^{2}}{\partial|m_{D}|}\Big|_{\Theta=\pi} &=& 2|m_{D}|\Bigg( 1 \mp\frac{|m_{R}|-|m_{L}|}{\sqrt{(|m_{R}|+|m_{L}|)^{2}+4|m_{D}|^{2}}} \Bigg),   \\ 
\frac{\partial (M^{F}_{\pm})^{2}}{\partial\Theta}\Big|_{\Theta=\pi} &=& 0,
\end{eqnarray}
( assume $|m_{R}|>|m_{L}|$ ).
Here, mass dimensions of these derivatives 
$\frac{\partial (M^{F}_{\pm})^{2}}{\partial|m_{D}|}$ and $\frac{\partial (M^{F}_{\pm})^{2}}{\partial\Theta}$ 
are $[{\rm mass}]^{1}$ and $[{\rm mass}]^{2}$, respectively.
The first derivative $\frac{\partial V^{NJL+M}_{eff}}{\partial\Theta}$ has zero-points at $\Theta=0,\pi$,
and it is almost proportional to $-\sin\Theta$, always negative at $0<\Theta<\pi$. 
Therefore, a global minimum locates at a point on the line $\Theta=\pi$ of 
the two-dimensional effective potential surface $V^{NJL+M}_{eff}(|m_{D}|,\Theta)$:
We find $\Theta=0$ is always unstable, and the potential gives the lowest energy 
at $\Theta=\pi$ with fixed $|m_{D}|$.
This feature is always the case under various values of $|m_{R}|$ and $|m_{L}|$, 
both of the cases $|m_{R}|>|m_{L}|$ and $|m_{R}|<|m_{L}|$,
and this $\Theta$-dependence of $V^{NJL+M}_{eff}$ disappears when $|m_{R}|=0$ and/or $|m_{L}|=0$.
$V^{NJL+M}_{eff}(|m_{D}|,\Theta)$ is real and analytic except $|m_{D}|=\infty$.

\vspace{2mm}

If we derive the gap equation from the derivative $\frac{\partial V^{NJL+M}_{eff}}{\partial|m_{D}|}$ 
of the form before momentum integration of $V^{NJL+M}_{eff}$ is done, 
we obtain it in the following form ( in the Euclidean region ):
\begin{eqnarray}
|m_{D}| &=& G\int_{k} \frac{|m_{D}|(k^{2}+|m_{D}|^{2}-|m_{R}||m_{L}|\cos\Theta)}{[k^{2}+(M^{F}_{+})^{2}][k^{2}+(M^{F}_{-})^{2}]}.
\end{eqnarray}
Compared with (61), this is the "correct" gap equation and 
we should solve it with the condition $\Theta=(2n+1)\pi$ ( $n\in{\bf Z}$ ).
Equation (61) with $\theta_{D}=0$, $\theta_{R}+\theta_{L}=\Theta$ coincides with (73). 
Hence, we conclude that the theory at one-loop level chooses $\theta_{D}=0$ 
while $\theta_{R}+\theta_{L}=\Theta=(2n+1)\pi$.
Due to the CPT theorem of a Lorentz invariant theory,
we can use the time-reversed Lagrangian of (39) for examining 
whether the system dynamically breaks CP symmetry or not.  
Since phase of mass parameters cannot be absorbed at $m_{R}\ne m_{L}$, $m_{D}\ne 0$ in our model 
( and, the gauge symmetry is explicitly broken ), 
it might be possible that the invariance under the CP transformation
$\psi_{D} \to \gamma^{0}C\bar{\psi}^{T}_{D} = \gamma^{0}\psi^{c}_{D}$,
$\bar{\psi}_{D} \to -\psi^{T}_{D}C^{-1}\gamma^{0} = \overline{\psi^{c}_{D}}\gamma^{0}$,
( we use $C\equiv i\gamma^{2}\gamma^{0}$ ) 
is dynamically broken. 
However, each of the energy spectra obtained by (54) has two-fold Kramers degeneracy, 
and thus the theory should conserve CP ( namely, time-reversal ) symmetry.
( In other words, the mechanism of a CP violation in the SM may somewhat different with our one-flavor model,
such as the Kobayashi-Maskawa mechanism of 6 flavors ) 
We have found that $\theta_{D}=0$ and $\theta_{R}+\theta_{L}=\Theta=(2n+1)\pi$ will be chosen as the vacuum. 
We find when 
\begin{eqnarray}
\theta_{R} = \Bigl(2j+\frac{1}{2}\Bigr)\pi, \quad
\theta_{L} = \Bigl(2l+\frac{1}{2}\Bigr)\pi,
\qquad {\rm or} \qquad
\theta_{R} = \Bigl(2j+\frac{3}{2}\Bigr)\pi, \quad 
\theta_{L} = \Bigl(2l+\frac{3}{2}\Bigr)\pi,  
\end{eqnarray}
( $j,l\in{\bf Z}$ )
both of the Majorana mass terms are invariant under a T-transformation 
( we use the definition and the phase convention of time-reversal transformation 
as $\psi_{D}(x_{0})\to i\gamma^{1}\gamma^{3}\psi_{D}(-x_{0})$ 
with taking complex conjugations to $c$-numbers/matrices  ):
In this phase choice,
\begin{eqnarray}
m_{R} = -m^{\dagger}_{R}, \qquad m_{L} = -m^{\dagger}_{L},
\end{eqnarray}
and CP is conserved.
In summary, the theory chooses the vacuum as $\theta_{D}=0$, 
$\theta_{R}=\theta_{L}=\pi/2$ and CP is conserved at the vacuum state.
A similar situation will happen in theory of relativistic superconductivity~[172,173].  
In relativistic theory of superconductivity, 
the Lagrangian with spin-singlet scalar Cooper pairing is given by
\begin{eqnarray}
{\cal L}_{sc} &\equiv& \bar{\psi}_{D}(i\partfey-m^{\dagger}_{D}P_{+}-m_{D}P_{-}+\gamma^{0}\mu)\psi_{D} +\Delta^{*}_{S}\psi^{T}_{D}C\gamma_{5}\psi_{D} - \Delta_{S}\bar{\psi}_{D}\gamma_{5}C\bar{\psi}^{T}_{D}.
\end{eqnarray}
Here, $\mu$ is chemical potential.
The Majorana mass terms of (39) coincide with the Cooper-pair mass terms at 
\begin{eqnarray}
\Delta_{S} &=& -\frac{m_{R}}{2} = \frac{m^{\dagger}_{L}}{2}.
\end{eqnarray}
The Cooper-pair mass term is formally not symmetric under a CP-transformation in general,
and it becomes CP-even at $\Delta_{S}=-\Delta^{*}_{S}$
( and the Pauli principle also be satisfied at this condition ).
Because $\xi$ and $\eta$ have common mass in this case,
we cannot consider $\theta_{R}$ and $\theta_{L}$ independent with each other.
Thus, the condition $\Delta_{S}=-\Delta^{*}_{S}$ with $m_{D}=m^{\dagger}_{D}$ 
cannot be achieved simultaneously, in general. 
The phase degree of freedom of Cooper pair gives a vortex inside superconductor/superfluid,
and gives various interesting phenomena they have been observed by experiments.
We wish to emphasize that our theory chooses the CP ( time-reversal ) invariant vacua variationally, 
and $U(1)$ symmetries are broken from the beginning of the theory, 
while a $Z(N)$ symmetry ( a notation in physics ) arises in the one-loop effective potential. 
Therefore, we can say our theory does not have uncountable infinitely degenerate vacua 
usually appear in a spontaneous symmetry breakdown of a theory, 
while it has countable infinite number of vacua
( the cardinal number $\aleph_{0}$~[27] ).
It is interesting for us to find such a phenomenon in other physical system/situation,
and this phenomenon may be summarized into the following generic conjecture/theorem:

\vspace{2mm}

{\bf Conjecture.1}:

When a continuous and global symmetry of a compact Lie group in 
a Hermitian/unitary quantum field theory is broken explicitly by a parameter,
and if an order parameter which breaks the {\it same} symmetry develops,
infinitely countable degenerate vacua will arise from the effective potential of the theory.

\vspace{2mm}

Proof:

The Nambu-Goldstone theorem states that infinitely degenerate uncountable ( continuous ) vacua
will arise associated with a zero-mode ( a Nambu-Goldstone mode )
if a continuous symmetry ( inner automorphism ) is broken spontaneously in a quantum field theory.
While if the continuous symmetry is broken from the beginning of the theory
( in that case, the effective potential of the theory is not a class function of an action of
a Lie group $G$ which corresponds to an NG mode since 
$f(g\widetilde{M}g^{-1})\ne f(\widetilde{M})$, where $\forall g\in G$ ), 
the NG theorem cannot be applied.    
The Lagrangian, or, at least its effective potential 
of the theory must be single-valued under a variation
with respect to the parameter $\theta$ ( the direction of the continuous symmetry )
if the potential is real and has no singular point in coordinate $\theta$,
thus this criterion requests the theory to make its potential as a periodic form
in a domain $D$ ( $U(1)$-case, for example, $0\le \theta \le 2\pi$ ).
Inside this domain, the potential always has at least a maximum and a minimum,
since the continuous infinite degeneracy is lifted.
Thus, after the domain of the effective potential is appropriately extended 
( $U(1)$-case: $0\le \theta \le 2\pi\to -\infty < \theta < \infty$,
by making a correspondence between a function $F$ on a torus ${\bf T}$ and the real number field ${\bf R}$, 
i.e. $F( t \bmod 2\pi) = F(t)$ ( $t\in {\bf R}$ ) ), 
a countable infinitely degenerate vacua must arise in the theory.
We should mention that this proof is rigorous for the $U(1)$ case,
while we need more rigorous discussion for cases of non-Abelian compact Lie groups.

\vspace{2mm}

Here, we wish to emphasize that this conjecture/theorem gives a notion which will generalize
the Nambu-Goldstone theorem in particle physics.
( A breaking of rotational invariance gives a periodic potential.
For example, the potential of rotation of the earth is surely periodic...
A magnon excitation may have a mass when an external magnetic field is
applied to a magnet, and a rotational potential is periodic. )
Physically, this phenomenon coming from the explicit symmetry breaking 
can be interpreted as a "localization" in a field-configuration space.
We should emphasize the fact that a degeneracy as a usual definition of "ordering"
of a phase is not suitable for our generalized Nambu-Goldstone ( GNG ) theorem.

\vspace{2mm}

We will call the total mathematical structure of explicit+dynamical symmetry breaking 
as "generalized Nambu-Goldstone theorem", which is revealed in this paper.

\vspace{2mm}

The set of stable points in the $U(1)$ case forms a lattice, should be called as 
an "NG-lattice." If one considers a system of three $U(1)$ degrees of freedom,
then the lattice looks like an optical lattice obtained by interferences of laser lights.

\vspace{2mm}

The origin of the Cabibbo-Kobayashi-Maskawa ( CKM ) mixing matrix~[26,113] 
( and the Pontecorvo-Maki-Nakagawa-Sakata ( PMNS ) matrix~[134,192] ) 
may also be understood/found from the viewpoint of this theorem. 
( The work which interprets the Cabibbo angle as a pseudo-NG boson is,
H. Pagels, Phys. Rev. {\bf D11}, 1213 (1975). )
The CKM matrix is in fact, a periodic matrix function with 
four coordinates $\theta_{1}$, $\theta_{2}$, $\theta_{3}$ and $\phi$. 
A family ( or, a flavor ) symmetry in it can be taken from a compact Lie group,
like an $SU(3)$ family symmetry~[32]. 
A problem is that the determination of the CKM matrix 
in our nature is under the variation principle or not. 
The NG theorem deeply depends on the variation principle
( for example, established via a Ward-Takahashi identity ), 
while the CKM matrix seems to show an inconsistency/uncertainty 
between mass eigenstates/bases/fields ( obeys the variation principle ) 
and flavor/horizontal symmetries.
Namely, the question asks that the CKM/PMNS matrices are fields or fundamental constants.  
Again, we mention that the phase parameters of CKM/PMNS matrices will obey Euler-Lagrange equations
if they are physical.
It may be the case that a pseudo-NG manifold parametrized by a set of compact Lie group generators 
is bounded and thus it could be studied by a Morse-theory approach~[23,144]. 
( Morse theory assumes a function of a domain, given as a subset of a manifold, as bounded.
Note that a compact group is a bounded and closed set, 
thus the parameter space of the group is also a bounded and closed domain,
and a volume of the group manifold defined by a Haar measure becomes finite. )
For example, the Vafa-Witten theorem~[225] or an axion potential~[186,187,206,207]
may fall into a class of this theorem ( conjecture ). 
Some exceptional cases are,
(1) a global gauge symmetry of BRS generator~[123],
(2) supersymmetry~[238],
and both of them are defined by fermionic generators.
Beside BRS, we need the notion of supermanifold~[53,201], 
namely a pseudo-NG supermanifold for addressing this issue.

\vspace{2mm}

The $U(1)$ p-NG ( pseudo Nambu-Goldstone ) mode~[25,74,230] determines a curve 
by the following mapping $c: (e_{1},e_{2}) \to M$,
where $(e_{1},e_{2})$ denotes the end points of the curve, 
and $M$ is a differentiable manifold.
Moreover, this curve is always closed 
in $D\equiv 0\le\Theta\le 2\pi$
due to its $U(1)$-nature and defines
\begin{eqnarray}
c: S^{1} \to M.
\end{eqnarray}
In our $U(1)$ p-NG case, 
we find an imbedding of $S^{1}$ to a three dimensional manifold ${\bf R}^{3}$, 
namely $g:S^{1}\to g(S^{1})\in {\bf R}^{3}$.
For example, if we consider a $2\times 2$ matrix case,
\begin{eqnarray}
{\cal A} &\propto& \left(
\begin{array}{cc}
a + a^{\dagger} & i(a-a^{\dagger})  \\
-i(a-a^{\dagger}) & a+a^{\dagger} 
\end{array}
\right) = 2|a|\left(
\begin{array}{cc}
\cos\theta_{1} & \sin\theta_{1}  \\
-\sin\theta_{1} & \cos\theta_{1} 
\end{array}
\right).
\end{eqnarray}
This matrix defines a closed curve $S^{1}\in{\bf R}^{2}$.
In our context, we regard $\theta_{1}$ as a p-NG mode associated with a $U(1)$ symmetry breaking. 
It is known fact that a compactification of string usually gives several $U(1)$ degrees of freedom~[81,193].
This kind of matrix can be generated by an NJL-type four-fermion model 
which has off-diagonal interactions in flavor space without much difficulties. 
Our approach has a similarity with the flavon mechanism ( the Froggatt-Nielsen mechanism~[18,56,67,189] ), 
though the flavon mechanism intends to obtain a mass matrix with mass hierarchy 
simultaneously by particle interactions, 
while our approach considers only mixing angles at the moment.
Our method is not enough to obtain a prediction of a mixing angle,
though it is suggestive fact that the diagonal part is dominant in the CKM mixing matrix of quark sector, 
while ( almost/near ) maximal mixing is observed in the PMNS matrix of neutrino sector. 
For example, the PMNS matrix $U_{PMNS}$ is given by
\begin{eqnarray}
U_{PMNS} &\equiv& V_{PMNS}P_{PMNS},   \\
V_{PMNS} &\equiv& \left(
\begin{array}{ccc}
c_{12}c_{13} & s_{12}c_{13} & s_{13}e^{-i\delta}  \\
-s_{12}c_{23}-c_{12}s_{23}s_{13}e^{i\delta} & c_{12}c_{23}-s_{12}s_{23}s_{13}e^{i\delta} & s_{23}c_{13}  \\
s_{12}s_{23}-c_{12}c_{23}s_{13}e^{i\delta} & -c_{12}s_{23}-s_{12}c_{23}s_{13}e^{i\delta} & c_{23}c_{13}
\end{array}
\right),   \\
P_{PMNS} &\equiv& {\rm diag}(1,e^{i\alpha},e^{i(\beta+\delta)}),   \\
c_{12} &\equiv& \cos\theta_{12}, \quad s_{23} \equiv \sin\theta_{23},  \cdots.
\end{eqnarray}
$V_{PMNS}$ can be factorized into three rotational matrices 
( three $S^{1}$-curves, show a rotational asymmetry ) of $SO(3)$,
while $P_{PMNS}$ is in fact, a maximal torus of the $U(3)$-group.
( In case of type-II seesaw, the mixing matrix becomes $U_{f}=e^{i\phi}P_{f}\tilde{U}_{f}Q_{f}$,
where $P_{f}={\rm diag}(1,e^{i\alpha},e^{i\beta})$ and $Q_{f}={\rm diag}(1,e^{i\rho},e^{i\sigma})$~[200]. )
Since $SO(3)\subset SU(3)$, the PMNS matrix is given by a product of four representations of $U(3)$.
In more generic case of $N_{f}$-flavors, a mixing matrix may take the form of
a product of $N_{f}$ curves with a maximal torus of $U(N_{f})$:
${\bf T}_{N_{f}} \equiv {\rm diag}(e^{i\lambda_{1}},\cdots,e^{i\lambda_{N_{f}}})$.
Note that $U(N_{f})/{\bf T}_{N_{f}}$ is a flag manifold
( a compact complex manifold ),
and thus several methods of the Borel-Weil theory 
( any finite-dimensional irreducible representation of $U(N)$ is given as a section
of holomorphic line bundle over a flag manifold $U(N)/{\bf T}_{N}$ )
might be useful for our context. 
The Lagrange plane $\Lambda(N_{f})\equiv U(N_{f})/O(N_{f})$ 
( it is a compact manifold, 
and it is considered in a little Higgs model~[189,20]
which can be interpreted as a kind of unified theory ) may play a role 
when we consider the symmetry breaking sequence
such as $U(N_{f})\to O(N_{f})\to {\rm nothing}$. 
We should recall that a ${\it closed}$ ${\it subgroup}$ $H$ of a Lie group $G$ are always a Lie group. 
These speculations are useful to find a symmetry and its breaking scheme for constructing 
a dynamical model of CKM/PMNS matrices.

\vspace{2mm}

If a potential is an analytic function of $\theta_{1}$, $\theta_{2}$, $\theta_{3}$ and $\phi$, 
and it is periodic into these directions,
we can always take 1st and 2nd derivatives of them, 
and there is a stable point ( a non-degenerate critical point as a notion of Morse theory ), 
and the potential has low-energy excitations ( quantum fluctuations ) around the point.
To emphasize the particle nature of those excitations,
we will call fluctuations of $\theta_{1}$, $\theta_{2}$, $\theta_{3}$ as mixon 
( and their possible superpartners as mixino,
similar to the relation between axion and axino ) 
and that of $\phi$ as cpon ( and its superpartner as cpino ). 
Here, we emphasize that a mixing matrix is considered as physical degrees of freedom ( a particle picture ),
and this is independent from the question whether these degrees of freedom are p-NG or not,
in principle. 
Then it becomes remarkable when we consider a supersymmetric counterpart of a mixing matrix.  
Therefore, there might be a reaction processes of squarks mediated by a mixino,
and such a reaction might be suppressed under a certain mechanism.
Our framework will radically modify the physical mechanism of neutrino oscillation.
A kaon reaction might also be described by this framework.
We have arrived at the notion in which "fundamental constants" of the SM are VEVs, 
their fluctuations and (p-)NG modes of an underlying theory. 
We can utilize VEVs of dilaton/moduli of strings sometimes used in string phenomenology~[81,193].

\vspace{2mm}

At a critical point,
\begin{eqnarray}
\frac{\partial f}{\partial x_{1}}(p) &=& \cdots = \frac{\partial f}{\partial x_{n}}(p) = 0,  
\end{eqnarray}
is satisfied, and if the Hessian
\begin{eqnarray}
{\cal H} &\equiv& \Bigl( \frac{\partial^{2}f}{\partial x_{i}\partial x_{j}}(p) \Bigr)_{1\le i, j \le n} 
\end{eqnarray}
is regular $\det{\cal H}\ne 0$, the point is non-degenerate. 
If a critical point is non-degenerate, 
it is isolated and it has no other critical point in its neighborhood~[144].
Our $U(1)$ periodic potential is just this case.
Here, $f:M\to {\bf R}$ is a $C^{\infty}$-class function,
$x_{i}$ ( $i=1,\cdots,n$ ) denotes a local coordinates at $p\in M$.
It is well-known that,
when all of eigenvalues of Hessian are positive, the critical point is stable,
and this corresponds to the case where a theory has a p-NG mode, 
while eigenvalues of NG modes are zero
and the critical point is degenerate.
Negative eigenvalues indicate tachyonic modes.
In Morse theory, it is known fact that a manifold is homeomorphic to $S^{n}$ 
if the manifold has a Morse function $f$ which has only two non-degenerate critical points.
The $U(1)$ case we have discussed above is an example of this fact.
A function of a bounded domain which will have a minimum must satisfy the following
Palais-Smale condition~[144]:
\begin{eqnarray}
\inf_{S}\|df\| &=& 0, \quad ( S \subset M, \quad f: M \to {\bf R} ).
\end{eqnarray}
Here, $df$ is an appropriately defined differential of $f$.
Then, $f$ has a critical point inside the closure $\bar{S}$.
Especially, a continuous function on a compact set always have at least
a maximum and a minimum.
Note the fact that several homogeneous spaces gives $S^{m}$, such as 
$SO(n+1)/SO(n)=O(n+1)/O(n)=S^{n}$,
$SU(n+1)/SU(n)=U(n+1)/U(n)=S^{2n+1}$,
$O(n+1)/(O(1)\times O(n))=S^{n}/{\bf Z}_{2}$.
For more generic case,
real Grassmann manifolds
$G_{k,n}({\bf R})=O(n)/(O(k)\times O(n-k))$,
Stiefel manifolds 
$V(m,r)=SO(m)/SO(m-r)$ and a generalized flag manifold
$U(n)/(U(n_{1})\times\cdots\times U(n_{k}))$ 
( $n=n_{1} + \cdots + n_{k}$ )
may also be interesting from our context.
If a matrix ${\cal M}$ is regular, ${\rm det}{\cal M}\ne 0$, it will be diagonalized as
${\cal M}=(V^{(1)}(\theta_{i}))^{\dagger}\lambda_{\cal M}V^{(2)}(\theta_{i})$ ( $i=1,\cdots,n$ ), 
like the case of CKM/PMNS theory
( $\lambda_{\cal M}$: a diagonal matrix ).
Thus, the regular condition of ${\cal M}$ excludes an NG-mode,
an effective action $\Gamma_{eff}$ given by a polynomial or a convergent series 
of ${\cal M}$ is a periodic function of $\theta_{i}$,
and defines a mapping $\Gamma_{eff}:{\bf R}^{n}\to{\bf R}^{1}$.
Non-degenerate critical points of ${\bf R}^{n}$ for $\Gamma_{eff}$,
and their physical meaning will be obtained by the prescription discussed above.

\vspace{2mm}

The phase of dynamical mass of the $U(1)$-case discussed previously is, 
in fact, a Fourier component/basis $\exp[i(\xi t)]$ defined on a torus ${\bf T}$ 
or the real number field ${\bf R}$ ( both of them are commutative ) 
namely a one-dimensional irreducible unitary representation:
\begin{eqnarray}
f_{k}: {\bf T} \to {\bf C}^{\times}, \quad t \bmod 2\pi \to e^{ikt}.
\end{eqnarray}
An irreducible unitary representation of a possibly non-commutative compact Lie group $G$ 
is obtained by the following relation of non-commutative harmonic analysis via the Peter-Weyl theorem 
( Fourier analysis is sometimes called as commutative analysis )~[114]:
\begin{eqnarray}
\pi(f) &=& \int_{G} f(g)\pi(g) dg,
\end{eqnarray}
where $dg$ is the Haar measure of $G$ ( via a Maurer-Cartan form of $g$ ).
If our approach to the CKM/PMNS matrices is correct,
their matrix elements can be given by some matrix elements of $\pi(g)$.
Then, if the parameters $\theta_{i}$ ( $i=1,2,3$ ) of the CKM/PMNS matrices are physical, 
then one should examine a variation of an effective action $\Gamma[\pi(f)]$ 
with respect to matrix elements of $\pi(f)$. 
Hence,

\vspace{2mm}

${\bf Proposition. 2}$:

The CKM/PMNS matrices can be examined by methods of non-commutative harmonic analysis.

\vspace{2mm}

An interesting problem is to know how infinitely countable degenerate vacua 
will arise from infinitely uncountable degenerate vacua, by (non)commutative harmonic analysis.  
Moreover, one would need method of (non)commutative harmonic superanalysis 
with theory of Lie supergroup to investigate p-NG supermanifold 
( here, we do not restrict ourselves on phenomena of "spontaneous/dynamical SUSY breaking" )
and supersymmetric CKM/PMNS matrices
~[16,26,31,40,62,78,102,113,130,134,148,156,192,201,238,242,248]. 
For this purpose, we also need to invent theory of super-Banach/Hilbert space 
for a construction of Finsler superspace,
with a consideration on a variational calculas.

\vspace{2mm}

We wish to comment on topological ( global ) aspect of the periodic effective potential~[157].
In fact, our $U(1)$ periodic effective potential has a tangent bundle
$TM_{pNG}$ and its dual ( cotangent ) bundle.
Here, $M_{pNG}$ is the base manifold, i.e. $S^{1}$.
In our case, only a trivial bundle is physically acceptable.
However we can consider the following theorem for our discussion:

\vspace{2mm}

{\bf Theorem. 3:}

The $U(1)$ potential acquires a winding ( topological ) number $n\in{\bf Z}$ by the definition
$V(\Theta \bmod 2n\pi)=V(\Theta)$.

\vspace{2mm}

Note that the fundamental group of $S^{1}\sim U(1)$ 
( $U(1)$ is connected, but not simply ) is $\pi_{1}(S^{1})\simeq {\bf Z}$,
and one can choose $n$ as a negative integer, also for a congruent, of course.
To avoid a possible misunderstanding, 
we should mention that the ${\it manner}$ of the definition of winding number given above 
is essentialy different with that of gauge fields ( such as monopole, instanton ), 
because we have utilized the fact that the extension of domain of $\Theta$ 
can be given completely arbitrarily
( if we do not consider a path integration of the auxiliary fields ),
while the starting point of discussions on Dirac monopole and instanton
is given by a Chern-Weil characteristic polynomial
( a class function of a Lie group ).
Thus, we could say our manner of the extension of domain 
is similar to complex analysis on Riemann surfaces,
and it implicitly has a replication of the theory.
( See also, G. 't Hooft,
{\it On the Phase Transition towards Permanent Quark Confinement},
Nucl. Phys. {\bf B138}, 1 (1978). )
The theory acquires a connected domain, but now it is not simply connected.
We need the extension of domain of $\Theta$ for a discussion of transitions between valleys
( similar to a case of discussion on an axion potential ),
which will be given in Sec. V.  
It is well-known fact that any function which satisfies $f(z+1)=f(z)$ can be rewritten as
\begin{eqnarray}
f(z) = g(q), \quad q \equiv e^{2\pi iz}, \quad z \in {\bf H}.
\end{eqnarray}
Here ${\bf H}$ is the upper half plane.
This periodocity is given by a transformation which is a special case of the following 
elliptic modular group $SL(2,{\bf Z})$,
which is included in real M\"{o}bius transformations
${\rm Aut}({\bf H})$ ( $a,b,c,d\in{\bf R}$, $ad-bc=1$ )~[98]:
\begin{eqnarray} 
f(z') = f\Bigl( \frac{az+b}{cz+d} \Bigr), \quad
\left(
\begin{array}{cc}
a & b \\
c & d 
\end{array}
\right) \in SL(2,{\bf Z}), \quad (a,b,c,d\in{\bf Z},ad-bc=1).
\end{eqnarray}
$f(z+1)=f(z)$ is given by $a=b=d=1,c=0$, as a special case of parabolic type.
( The generators of the modular group is usually chosen as
$S:z\to z+1$ and $T:z\to -1/z$. ) 
Therefore one finds:

\vspace{2mm}

{\bf Theorem. 4:}

Our $U(1)$ periodic potential can be interpreted from the modular group,
by an analytic continuation of $\Theta/2\pi$ to the upper half plane ${\bf H}$
which is conformally equivalent with the open unit disc $\Delta\equiv |z|<1$.

\vspace{2mm}

It is well-known that an analytic continuation of a periodic function gives various interesting characters:
For example, an $L$-function ( a generalized zeta function ) $L(f,s)$ 
can be defined from a Fourier expansion $f(z)=\sum^{\infty}_{n=1}a_{n}e^{2\pi izn}$
such that $L(f,s)=\sum^{\infty}_{n=1}(a_{n}/n^{s})$
via a Mellin transformation.
In fact, $\cos\Theta$ ( $=\frac{1}{2}(e^{i\Theta}+e^{-i\Theta})$ ) or $\cos\theta_{\chi}$
are a special case of modular functions ( a cusp form )
\begin{eqnarray}
f(\tau) &=& \sum_{n\ge 1}\alpha_{n}q^{n}, \qquad q=e^{2\pi i\tau}.
\end{eqnarray}
We can divide $V_{eff}\sim\cos\Theta$ into holomorphic $e^{i\Theta}$ and anti-holomorhic $e^{-i\Theta}$ parts,
and then regard them as the first terms of Laurent series of a modular function.
In that case, $V_{eff}$ is given by a square-root of 
a product of holomorphic and anti-holomorphic modular functions,
similar to the case of definition of partition function of 3D quantum gravity~[135].  
Then, now we can obtain a viewpoint from conformal field theory ( AdS/CFT ).
In that case, a partition function will be given by
\begin{eqnarray}
Z(\tau) &=& {\rm Tr}e^{-2\pi i\tau L_{0}},
\end{eqnarray} 
where, the trace is taken over a Hilbert space of a representation space of Virasoro algebra
( $L_{0}$: a Virasoro generator ). 
These consideration has been employed in theory of three-dimensional quantum gravity~[135].
The sequence given by our $U(1)$ periodic effective potential with a "winding number" can be interpreted
by employing a suitable dilatation of their extended domains of $\theta_{\chi}$. 
Now, we arrive at the following theorem.

\vspace{2mm}

{\bf Theorem. 5:}

The periodicity of the gap equation $\frac{\partial V^{NJL+M}}{\partial\Theta}=0$ 
with a winding number $n\in {\bf Z}$ 
defines a cyclotomic equation/polynomial, which corresponds to a regular polygon of $n$ edges.

\vspace{2mm}

From this aspect, the pseudo-NG mode obtains a mathematical structure/implication in 
the Iwasawa theory ( algebraic number theory ),
and also in the Fermat's last theorem and the Shimura-Taniyama-Weil ( modularity ) theorem
( and ultimately, the Langlands program and Shimura varieties )
~[44,49,63,82,99,100,126,143,219,240,241]. 
We have found the fact that the gap equation with an extended domain
gives the correspondence 
\begin{eqnarray}
U(1) \leftrightarrow {\rm Gal}(\mathbb{Q}(\zeta_{n})/\mathbb{Q}) \simeq (\mathbb{Z}/n\mathbb{Z})^{\times},
\end{eqnarray}
( $\mathbb{Q}$: rational number field, 
$\mathbb{Q}(\zeta_{n})$: a cyclotomic field,
$\zeta_{n}$: n-th root of unity )
i.e., a correspondence between the $U(1)$-degree of freedom and the Galois group.
Here, we argue that now our effective potential is a representation space of the Galois group, 
and a non-linear sigma model derived from the theory also
should have the same mathematical implication, 
especially in its potential term.
One should know that our definition of the winding number defines an orientation,
while a Galois group does not have it.

\vspace{2mm}

We have mentioned that the periodicity of the effective potential along the pseudo-NG mode 
implies a localization/decoherence of wavefunction of the system. 
( Hence, now we obtain the notion of "topological localization".  
Such kind of ( topological ) localization might be very general, 
realizes in various systems from particle phenomenology to condensed matters. )
Our statement says that pion has these methematical structures:
Pion is a localization of $\theta_{\chi}$ 
in $0\le \theta_{\chi}\le 2\pi$ or in $-\infty < \theta_{\chi} < +\infty$.
Our gap equation gives a relation between the "continuous" and "discrete" vacua
( they do not connect smoothly with the vanishing limit of an explicit symmetry breaking parameter ),
thus it is interesting for us if the relation can be understood by the notion
of derived category of Grothendieck~[86,132]
( and also, the Pontryagin duality and the Tannaka-Krein duality ).
It might be possible that the notions of derived category could derive various topological natures
of particles and matters.

\vspace{2mm}

From periodicity of the $U(1)$ Lie-group actions,
zeros of the gap equation derived from our discussion are always separable ( no multiple root ),
and thus an appropriately normalized $V_{eff}$  
and its derivative $\frac{\partial V_{eff}}{\partial\theta}$ 
( periodicities of both $V_{eff}$, after a normalization, 
and $\frac{\partial V_{eff}}{\partial\theta}$ define
cyclotomic polynomials on the $U(1)$-circle ) 
do not share any root 
( from the central theorem of the Galois theory and algebraic field extensions,
and this fact clearly realize in the $U(1)$ case shown above ).  
Therefore, if $\theta=\pi/4$ is the VEV of a theory,
its effective potential would have the period $\pi/2$:
This may give a suggestion on construction/origin of a neutrino mixing matrix.
Hence, an effective potential with explicit symmetry breaking parameters
will show a correspondence between a compact Lie group $\widehat{G}$ 
and a finite ( discrete ) group $\widehat{\Gamma}$,
which implicitly acts to the effective potential,
through its gap equation.

\vspace{2mm}

An extension of our examination for a case of non-Abelian compact group is an interesting subject 
( for example, a flavor-$SU(N_{f})$ symmetry case ).
If an $SU(N_{f})$ symmetry is broken diagonally with a mass matrix 
${\rm diag}(|m^{(0)}_{1}|,\cdots,|m^{(0)}_{N_{f}}|)$ as $SU(N_{f})\to (U(1))^{N_{f}-1}$, 
generators of its Cartan subalgebra remains unbroken,
and the number of pseudo NG bosons become $N_{f}(N_{f}-1)$.
In the $SU(2_{f})$ case discussed above, the dependence of the fermion mass spectra 
( and thus, the effective potential )
on $|\theta|=\sqrt{\theta^{2}_{1}+\theta^{2}_{2}}$ 
disappear at the equivalence limit ( degeneracy condition of fermion masses ) of two-flavor degree of freedom.
At that limit, the Galois group of the $\theta_{\chi}$ degree of freedom will 
be explicitly recovered by the theory:
Hence we can say the $|\theta|$-dependence of the mass spectra (15)-(21) gives
a discrepancy/deformation from a cyclotomic field.
The periodicity of the mass spectra of the $SU(2_{f})$ case is given by both $\theta_{\chi}$
and $|\theta|=\sqrt{\theta^{2}_{1}+\theta^{2}_{2}}$, 
and the latter defines one-dimensional curves $\pm\sqrt{\theta^{2}_{1}+\theta^{2}_{2}}=n\pi+{\rm const}.$
as a set of stable points.
The curve $V(|\theta|,\theta_{1},\theta_{2})\equiv |\theta|^{2}-\theta^{2}_{1}-\theta^{2}_{2}$
is homogeneous $V(c|\theta|,c\theta_{1},c\theta_{2})=c^{6}V(|\theta|,\theta_{1},\theta_{2})$
and $V(|\theta|,\theta_{1},\theta_{2})=0$ 
gives a physical ( physically meaningful ) space.
Any coupling between $\theta_{a}$, $\theta_{b}$ and $\theta_{c}$ ( $a\ne b\ne c$ ) in $SU(N_{f})$ 
( $N_{f}\ge 3$ )
breaks the homogeneity.
Note that, in the $U(1)$-case, the set of stable spaces is given by zero-dimensional objects,
namely points. 
Because both $\theta_{\chi}$ and $|\theta|$ are real, 
the effective potential of $SU(2_{f})$ defines a surface of ${\bf R}^{2}$,
and we can introduce elliptic functions for understanding the periodicity of the pseudo-NG manifold in
the $SU(2_{f})$ case:
Even though the effective potential of $SU(2_{f})$ is real analytic on both $\theta_{\chi}$ and $|\theta|$
and has no pole inside the fundamental region,
and we need both $z=\theta_{\chi}+i|\theta|$ and its conjugate inside the effective potenial,
it should have some group theoretical aspects arise from double-periodicity of elliptic functions
( a "projection" from the effective potential to another function to understand a set of 
periodic stable points,
similar to our consideration on cyclotomic polynomials of the Iwasawa theory in the $U(1)$ potential ).
A homeomorphism $\pi:{\bf C}\to{\bf C}/\Gamma$ ( $\Gamma$: a lattice group )
gives a torus which can have infinite number of complex structures, 
and it is topologically equivalent with an elliptic curve 
$w^{2}=z(z-1)(z-\lambda)$:
A torus ${\bf C}/\Gamma$ is always biholomorphically equivalent with an elliptic curve,
and it should have a group structure of elliptic curves.
The moduli space of torus is given by $M_{1}\simeq {\bf H}/PSL(2,{\bf Z})$
where $PSL(2,{\bf Z})=SL(2,{\bf Z})/\{\pm 1\}$
( any element of $PSL(2,{\bf R})$ is an isometry of $\bf{H}$ ), 
and it is biholomorphically equivalent with ${\bf C}$.
Moreover, if one considers $(\theta_{\chi},|\theta|)\to (z_{1},z_{2})\in{\bf H}$,
the effective potential gives a special case ( a deformed form ) of Hilbert modular form.

\vspace{2mm}

Let us consider the case of $SU(N_{f})$. 
Unfortunately, getting the explicit expression of an exponential mapping of Lie group
is not an easy task,
except $SU(2)$ ( $2\times 2$ skew-Hermite matrices )
and $SO(3)$ ( $3\times 3$ skew-symmetric matrices, and it is known as the Rodrigues formula  )
and solvable/nilpotent Lie groups, 
and still there are several investigations on it~[125,128].
Even in the case of $SU(3)$, it will derive a long recursion relation which is not useful
for our purpose.
If a VEV is generated diagonally such like ${\rm diag}(M_{dyn1},\cdots,M_{dynN_{f}})$
with a set of explicit symmetry breaking parameters ${\rm diag}(|m^{(0)}_{1}|,\cdots,|m^{(0)}_{N_{f}}|)$,
the periodicity of mass spectra would be given by
\begin{eqnarray}
|\theta|^{2} &=& \theta^{2}_{1} + \cdots + \theta^{2}_{N_{f}(N_{f}-1)}.
\end{eqnarray}
This is reasonable, because local coordinates of unbroken generators must not enter into $V_{eff}$ 
and $|\theta|^{2}$ arises from 
$\sum_{a}\theta^{2}_{a}X^{2}_{a}=\sum_{i}\theta^{2}_{i}G^{2}_{i}-\sum_{l}\theta^{2}_{l}H^{2}_{l}$
of a taylor series of $\hat{g}=\exp[i\sum_{a}\theta_{a}X_{a}]$,
where $G_{i}$ are all of the generators of $SU(N_{f})$, 
$X_{a}$ are broken generators, and $H_{l}$ are symmetric generators.  
In fact, we can obtain a quadratic form by a diagonalization
$(\sum\theta_{a}X_{a})^{2}=\sum_{ab}\theta_{a}X_{a}X_{b}\theta_{b}=\sum_{a}\tilde{\theta}^{2}_{a}$
and then the ( block ) diagonal part of $\hat{g}$ becomes $\cos|\theta|$.
This diagonalization is hoped to correspond 
to a "good" choice of parametrization for a p-NG manifold.
It is naturally expected that the approximation becomes very well because
we can choose the local cordinates for defining exponential mapping small enough
for a description of low-energy excitations,
and the higher-order terms will be suppressed.  
Because the rank of Lie algebra of the group $SU(N_{f})$ becomes $N_{f}-1$, 
and thus the number of Casimir invariants also $N_{f}-1$
( there are two Casimir invariants, second ( quadratic form of generators, so-called "Hamiltonian" ) 
and third ( coming from anti-commutators of generators ) orders, in the Lie algebra of $SU(3)$ ),
there are couplings between generators inside an exponential mapping $\hat{g}$
and they give terms such like $\theta_{a}\theta_{b}\theta_{c}$, 
and thus a description on the phase dependence of $V_{eff}$ by $|\theta|$ is an approximation,
though this is useful to reveal a structure of pseudo-NG manifold
because the phases $\theta_{\chi}$ ( chiral ) and $|\theta|$ ( flavor )
are {\it perpendicular}, periods of modulations of phases are {\it independently} defined with each other. 
Moreover, entries of block-diagonal parts give main contributions to
$\widetilde{M}=\widehat{M}^{(0)}+\hat{g}\widehat{M}_{dyn}\hat{g}^{-1}$
in cases of diagonal breakings
( since we have our main interest on the neighborhood of $\hat{1}$ ).
It should be mentioned here that the local coordinates are defined on a smooth manifold $M_{SU(N_{f})}$
and this smooth structure reflects to a smoothness of a mass eigenvalue,
namely $\lambda_{j}$ in (12) are smooth functions on an open domain $U\subset M_{SU(N_{f})}$,
$\lambda_{j}\in {\cal O}_{{M}_{SU(N_{f})}}(U)$ 
( ${\cal O}_{{M}_{SU(N_{f})}}$: a sheaf of smooth functions ).
Since ${\tilde{\theta}_{i}}$ ( $i=1,\cdots,N_{f}(N_{f}-1)$ ) are real, 
an $SO(N_{f}(N_{f}-1))$ (quasi)symmetry in the coordinates of mass phases 
in the vicinity of the origin implicitly arises in the effective potenial,
and this is always the case for a diagonal breaking of an $SU(N_{f})$ model.
( It is noteworthy to mention that, 
for example in an $SU(3_{f})$ case,
we will find that a number of unbroken generators does not 
give this (quasi)symmetry of p-NG manifold by counting numbers of generators. ) 
Thus, the set of stable points in the NG sector at a fixed $\theta_{\chi}$ in this case 
is given by infinite-number of $N_{f}(N_{f}-1)-1$-dimensional quadratic hypersurfaces 
( a set of projective spaces placed "periodically" ) after an extension of the domain,
via an equation ( a quadratic form ) such that 
\begin{eqnarray}
|\theta|^{2} &=& \tilde{\theta}^{2}_{1} + \cdots + \tilde{\theta}^{2}_{N_{f}(N_{f}-1)} 
= (|\theta_{stable}|+n\pi)^{2}.
\end{eqnarray}
Note that $\pi$ acts as a generator of transcendental numbers in the algebraic curve given above,
and the curve is defined over $\mathbb{C}P^{N_{f}(N_{f}-1)}$.
In the quadratic form, $\pi$ can be removed by a normalization,
while any higher-order coupling makes it impossible.
This algebraic curve is included into 
$C_{1}\tilde{\theta}^{2}_{1} + \cdots + 
C_{N_{f}(N_{f}-1)}\tilde{\theta}^{2}_{N_{f}(N_{f}-1)}-C_{s}(|\theta_{stable}|+n\pi)^{2}=0$
with $C_{l},C_{s}\in{\bf Q}$, $l=1,\cdots,N_{f}(N_{f}-1)$,
as a special example. 
Of cource, one can finds the rotational symmetry is exact or not by employing
an isometry of $O(N_{f}(N_{f}-1))$,
and if the symmetry is not exact, there is a discrepancy from the quadratic form $|\theta|^{2}$.
For the same purpose, we can employ a conformal transformation of $O(N_{f}(N_{f}-1),1)$ to the light-cone
$\theta^{2}_{1}+\cdots +\theta^{2}_{N_{f}(N_{f}-1)}-|\theta|^{2}=0$.   
Again, $V_{eff}$ depends on $|\theta|$, 
and mass eigenvalues of pseudo-NG bosons are (quasi)degenerated. 
We can consider both tangent and cotangent bundles of the pseudo-NG manifold,
and they can be utilized to analyze a collective mode in the pseudo-NG manifold.
Therefore, we obtain an insight that 
a pseudo-NG manifold with $N_{b}$ broken generators of a large class of breaking schema
will give geometric objects they are topologically equivalent with $S^{N_{b}-1}$-spheres
( each of the sphere has two critical points though this fact does not relate 
with those of p-NG manifolds directly, of course ). 
Note that the NG manifold must always be ${\it real}$ analytic with all of the phase variables.
In general, there are couplings between $\theta_{\chi}$ and 
$(\tilde{\theta}_{1},\cdots,\tilde{\theta}_{N_{f}(N_{f}-1)})$ inside mass eigenvalues,
then usually an effective potential cannot be factorized such like $\cos\theta_{\chi}\cos^{2}|\theta|$
( this factorization is exact in $SU(2_{f})$ case ),
though $V_{eff}$ must be a scalar and isotropic toward $\theta_{a}$ ( $a=1,\cdots,N_{f}(N_{f}-1)$ ),
we can assume the form of $V_{eff}$
from an examination on periodicities toward $\theta_{\chi}$ and $|\theta|$ as an approximation such that:
\begin{eqnarray}
V_{eff} &=& V_{eff}(\theta_{\chi},|\theta|) 
\propto \cos\theta_{\chi} 
\bigl( \cos^{2}\sqrt{H_{ab}\theta_{a}\theta_{b} + F_{abc}\theta_{a}\theta_{b}\theta_{c} 
+ \cdots } \bigr) + {\rm const.}
\sim \cos\theta_{\chi} \bigl( \cos^{2}(|\tilde{\theta}| + \cdots )\bigr) + {\rm const}.
\end{eqnarray}
Here, $H_{ab}$ give a real-Hermitian $N_{f}(N_{f}-1)\times N_{f}(N_{f}-1)$ metric,
and $F_{abc}$ are completely symmetric tensors.
 ( Note that $\widetilde{M}$ is Hermitian, and $V_{eff}$ is real analytic on $\widetilde{M}$. )
A contour line of the NG sector of $V_{eff}$ is given by
\begin{eqnarray}
H_{ab}\theta_{a}\theta_{b} + F_{abc}\theta_{a}\theta_{b}\theta_{c} + \cdots &=& const.
\end{eqnarray}
Hence, the NG manifold is an $N_{f}(N_{f}-1)-1$-dimensional geometric object,
looks like a sphere in the neighbourhood of the unit 
( i.e., the origin of the coordinate system ),
while the higher-order terms will become relevant at the region far from the origin.
Since we will have our interest on the dynamical behavior of the system around the origin,
a low-energy excitation, then we say the approximation by a sphere is physically relevant.
More generically,
\begin{eqnarray}
V_{eff} &\sim& \cos F(\{\theta_{a}\}),
\end{eqnarray}
and the set of stable points are given by 
\begin{eqnarray}
F(\{\theta_{a}\}) = (2n+1)\pi, \quad n\in{\bf Z}.
\end{eqnarray}
Under the spherical apploximation, 
a set of global minima will be obtained by
$0 = \frac{\partial V_{eff}}{\partial\theta_{\chi}} = \frac{\partial V_{eff}}{\partial|\theta|}$
( strictly, 
$0 = \frac{\partial V_{eff}}{\partial\theta_{\chi}} 
= \frac{\partial V_{eff}}{\partial\theta_{1}} = \cdots 
= \frac{\partial V_{eff}}{\partial\theta_{N_{f}(N_{f}-1)}}$ ).
In this case, one finds
(1) cyclotomic points, they indicate stable points with a fixed $\theta_{\chi}$, toward $|\theta|$-direction,
(2) cyclotomic points toward $\theta_{\chi}$-direction with a fixed $|\theta|$,
(3) a hyperplane $(|\theta|+n\pi)^{2}=\sum^{N_{f}(N_{f}-1)}_{m=1}(\tilde{\theta}_{m})^{2}$ 
attached at each cyclotomic point of the $\theta_{\chi}$-direction. 
In this case, the structure of pseudo NG manifold is essentially described 
by two coordinates $(\theta_{\chi},|\theta|)$.
Furthermore, after an analytic continuation in the $SU(2_{f})$ case, we get
\begin{eqnarray}
V_{eff} &\to& \cos z_{1} \cos^{2} z_{2}, \qquad z_{1}, z_{2} \in {\bf H},
\end{eqnarray}
and then one can consider two $SL(2,{\bf Z})$ actions in an $F_{eff}$ as a projection of $V_{eff}$:
\begin{eqnarray}
& & F_{eff}(z_{1},z_{2}) = F_{eff}(z_{1}+1,z_{2}) = F_{eff}(z_{1},z_{2}+1) = F_{eff}(z_{1}+1,z_{2}+1),   \\
& & F_{eff}\Bigl( \frac{a_{1}z_{1}+b_{1}}{c_{1}z_{1}+d_{1}}, \frac{a_{2}z_{2}+b_{2}}{c_{2}z_{2}+d_{2}} \Bigr)
= F_{eff}(z'_{1},z'_{2}), \\
& & \left(
\begin{array}{cc}
a_{l} & b_{l} \\
c_{l} & d_{l} 
\end{array}
\right) \in SL(2,{\bf Z}), \quad a_{l},b_{l},c_{l},d_{l} \in {\bf Z}, \quad a_{l}d_{l}-b_{l}c_{l}=1, \quad l=1,2,
\end{eqnarray}
and $F_{eff}(z_{1},z_{2})$ will be regarded as a Hilbert-Blumenthal modular form~[218].
If the fermion field $\psi$ has several "flavor" ( gauge ) degrees of freedom,
more complicated geometric object will arise.
One should consider the following matrix:
\begin{eqnarray}
\widetilde{M} &=& \widehat{M}^{(0)} + 
(\hat{g}_{1}\otimes \hat{g}_{2}\otimes\cdots\otimes\hat{g}_{a})\widehat{M}_{dyn}
(\hat{g}^{-1}_{1}\otimes \hat{g}^{-1}_{2}\otimes\cdots\otimes\hat{g}^{-1}_{a}).
\end{eqnarray}
Here, we consider the case where all of the Lie groups are diagonally broken
and the Cartan subalgebras of them remain unbroken.
In this case,
\begin{eqnarray}
V_{eff} &=& V_{eff}(\theta_{\chi},|\theta_{1}|,\cdots,|\theta_{a}|)
\sim \cos\theta_{\chi}\Bigl(\cos^{2}|\theta_{1}|\times\cdots\times\cos^{2}|\theta_{a}| + \cdots \Bigr) + {\rm const}.
\end{eqnarray}
The factorized form ( factorization ) of $\cos^{2}|\theta_{1}|\times\cdots\times\cos^{2}|\theta_{a}|$
is "exact" when all of $\hat{g}_{a}$ belong to $SU(2_{f})$. 
Therefore, we obtain
(1) cyclotomic points toward every $|\theta_{m}|$-directions ( $m=1,\cdots,a$ ) with a fixed $\theta_{\chi}$,
(2) cyclotomic points toward $\theta_{\chi}$-direction with all $|\theta_{m}|$ fixed,
(3) hyperplanes 
\begin{eqnarray}
(|\theta_{1}|+n_{1}\pi)^{2} = \sum_{j_{1}}(\tilde{\theta}^{(1)}_{j_{1}})^{2}, \quad \cdots, \quad
(|\theta_{a}|+n_{a}\pi)^{2} = \sum_{j_{a}}(\tilde{\theta}^{(a)}_{j_{a}})^{2}.
\end{eqnarray}
Thus we can say several flavor degrees of freedom of $SU(2_{f})$
make corresponding Galois groups with a set of widing numbers.
We will name it as a Galois flavoration
( a flavoration of Galois theory ).

\vspace{2mm}

Even if there are couplings between $\theta_{a}$, $\theta_{b}$ and $\theta_{c}$ ( $a\ne b\ne c$ )
at $SU(N_{f})$ ( $N_{f}\ge 3$ ) and there is a deviation from the quadratic form $|\theta|^{2}$, 
we can introduce a perturbative approach, namely,
\begin{eqnarray}
\widetilde{M} &=& \widehat{M}^{(0)} + \hat{g}\widehat{M}_{dyn}\hat{g}^{-1}
= \widetilde{M}_{0}(q) + \kappa\widetilde{M}_{1},   \\
{\rm det}\widetilde{M} &=& {\rm det}\widetilde{M}_{0}(q)\times{\rm det}
\Bigl( 1+\kappa\frac{\widetilde{M}_{1}}{\widetilde{M}_{0}(q)} \Bigr).
\end{eqnarray}
Here, $\widetilde{M}_{0}(q)$ is a part given by $|\theta|^{2}$, 
and $\widetilde{M}_{1}$ is a contribution
coming from higher-order couplings ( non-linear terms ) between local coordinates.
By a suitable definition of norms of $\widetilde{M}_{0}(q)$ and $\widetilde{M}_{1}$,
one can estimate a convergence radius of the perturbative expansion~[107],
while it should be ( absolutely ) converged 
if the perturbation is appropriately defined and
if the exponential mapping of a Lie group converges absolutely
( the usual case ), in principle.
We can say the method as "a perturbation theory for algebraic varieties",
especially, a perturbation to $N$-dimensional sphere.
Then we meet an interesting problem for us, namely, to show how a perturbation 
( topologically/essentially ) modifies an $N$-dimensional sphere,
derives another topological object ( our insight is, "sometimes the case while sometimes not" ).
Of course, one can utilze the Serre's GAGA principle to examine a cohomology
of a formal power series ring.
Thus, Morse theory should be applied to this problem to find stable points.
In a realistic physical situation, there are couplings between phase parameters,
and thus a stable global minimum ( minima ) in mass phases of a flavor space depends on
the chiral mass phase. 
Once we have proposed that the CKM/PMNS matrices may be understood by 
the mechanism of explicit$+$dynamical symmetry breaking
by utilizing a non-commutative harmonic analysis.
Hence, mixing angles and CP phase(s) ( Dirac and/or Majorana ) may couple inside a potential in our mechanism,
and their VEVs depend with each other.
( An application of our mechanism to realistic CKM/PMNS matrices is involved,
and one needs a slight generalization of our framework. )
Now, we arive at the following conjecture:

\vspace{2mm}

{\bf Conjecture. 6}:

Consider a simply-connected compact simple Lie group, 
especially $SU(N)$ in the mechanism of explicit$+$dynamical symmetry breakdown.
When the number of broken generators ( at the Lorentz symmetric case ) of a diagonal breaking
( entries of a mass matrix is diagonal ) of a system is $N_{b}$,
( approximately ) an $SO(N_{b})$ rotational (quasi)symmetry in a local coordinate space arises,
reflects the fact that the mass eigenvalues of pseudo-NG bosons are (quasi-)degenerated.
$N_{b}-1$-dimensional geometric objects ( hyperplanes ) defined on the local coordinates
give a set of stable points of the system,
they are imbedded in the total parameter space of the Lie group.
The objects form a line toward the direction vertical to the phase $\theta_{\chi}$ of chiral condensates.

\vspace{2mm}

A dimension of a pseudo-NG manifold is determined under the usual manner 
( at least in a case of no Lorentz symmetry breaking )
since the notion of cosets or homogeneous spaces still survive:
The action of Lie group $G$ to the total manifold $M$ of the theory $G\times M\to M$ is transitive, 
gives a $G$-orbit, and such a manifold is always a coset $G/H$, 
where $H$ is a {\it closed} subgroup of $G$
( the manifold $G/H$ is Hausdorff if $H$ is closed ).
Thus a non-linear-realization for p-NG boson fields can be introduced.
One can consider a principal bundle $(G,G/H,f,H)$ also in our generalized Nambu-Goldstone theorem,
where $G$ is the total space, $G/H$ is the base manifold,
$f$ is the projection $G\to G/H$, and $H$ is the fiber. 
Note that theory of cosets or homogeneous spaces are useful for considerations on NG manifolds,
while it is neutral to determine a set of stable points in an NG manifold
in our generalized Nambu-Goldstone theorem.
For example, in the case of generalized flag manifold $U(n)/(U(n_{1})\times\cdots\times U(n_{k}))$,
the dimension of p-NG manifold is $n^{2}-\sum^{k}_{i=1}n^{2}_{i}$
while the set of periodic stable points becomes 
an $n^{2}-\sum^{k}_{i=1}n^{2}_{i}-1$-dimensional geometric object.
The effective potential must not have coordinates of unbroken generators of a Lie group,
and thus we do not have to consider whole space of a local coordinates of the Lie group
for our examination of geometry of the pseudo-NG manifold itself, in principle.
We should distinguish geometry of a group manifold of the beginning of a quantum field theory
with its pseudo-NG manifold:
In general, the universal covering group and its discrete subgroup will be introduced
to define the global structure and topological nature of a Lie group
( the fundamental group of a simply-connected compact simple Lie group is trivial ),
then a Lie group will be devided into a countable number of pieces. 
While, an exponential mapping $\hat{g}=\exp[i\theta_{a}T_{a}]$ inside a Lagrangian
is defined by a local coordinate system $\{\theta_{a}\}$ around the unity $\hat{1}$ of a Lie group, 
and the pseudo-NG manifold is obtained "locally" in that sense.
The method of extension of domain of variables of a pseudo-NG manifold is {\it another} story
from a topological nature of the Lie-group given in the beginning of a classical Lagrangian. 
If a Lie algebra has a homomorphism,
it will uniquely define a local homomorphism of neighborhood of the unity of the corresponding Lie group
by a lifting of the homomorphism,
though it is not sure that the homomorphism will be extended over the whole space of the Lie group.
If the Lie group is connected, the extension is unique.
We should examine a global aspect of the Lie group,
especially the fundamental group given by the universal covering group.
( Note if $M_{1}$ and $M_{2}$ are connected manifolds, then
$\pi_{1}(M_{1}\times M_{2})\simeq \pi_{1}(M_{1})\times \pi_{1}(M_{2})$.
Sometimes it is also the useful fact for us that $\pi_{1}(G/H)\simeq \pi_{1}(G)$ 
when $G$ is a connected Lie group and $H$ is a simply connected closed subgroup of $G$. )
Those situations are almost parallel with those of Riemann surfaces
where the notion of biholomorphic equivalences and discrete subgroups play the key-roles. 
In the $SU(2_{f})$ case, an {\it energy} surface of the NG sector 
can be parametrized as a two-dimensional coordinate system
$(0\le \theta_{\chi}\le 2\pi,0\le |\theta|\le \pi)$ as a "unit cell"
( we can say there is one redundancy in $(\theta_{1},\theta_{2})$ to describe 
variations of {\it potential} energy in the NG sector ),
and there might be several choices to extend a domain of those variables:
A trivial extension or a M\"{o}bius-band-like one
${\bf R}\times{\bf R}/{\bf Z}$
( a torus vs a Klein bottle ).
Each unit cell has a minimum ( or, minima ), 
and we gain a possibility to consider a transition or an interaction
between minima after a periodic extension of domain, 
similar to philosophy of discussions of axion potential.
Therefore, topological nature of several extended domains 
can be examined by tangent/cotangent bundles over them.
The unit cell of $(0\le \theta_{\chi}\le 2\pi,0\le |\theta|\le \pi)$ 
is a simply connected compact and bounded domain
and biholomorphically equivalent with the Riemann sphere $\widehat{\bf C}$,
while the analytic continuation $\cos z_{1}\cos^{2}z_{2}$ is done onto 
${\bf H}$ or $\overline{\bf H}$ and the domain is non-compact. 
If we regard $(\theta_{\chi},|\theta|)$ as a torus ${\bf C}/\Gamma$,
the universal covering surface of it is biholomorphically equivalent with ${\bf C}$.
Those various domains are always Hausdorff even though we change mass parameters.
Therefore, we meet the problem to determine a stable point ( global minimum ) of a function defined on
the genus 1 Riemann surface, the variational problem of the pseudo-NG manifold.
One can propagate infinite number of closed curves on the one-dimensional complex torus
to achieve a variationally stable point.
Those curves are sometimes contractible to points, while sometimes not.
Those curves contain classical motions of a representation point on the pseudo-NG manifold as a subset
with adopting a classical Lagrangian.
At this case, $V_{eff}$ belongs to a germ of a sheaf of smooth functions ${\cal O}_{D}$
( $D$; a domain ).   
An NG-sector with no explicit symmetry breaking parameter is completely flat toward all of 
directions of its local coordinates of broken generators,
while there is a complicated geometric structure inside a pseudo-NG manifold.
From the viewpoint of "localzation",
we should mention that a wavefunction of a system, 
which is a function of a set of mass phases $\{\theta_{l}\}$, 
is localized inside such a complicated geometry,
and this fact is interesting for a fundamental problem of quantum mechanics.
This $(\theta_{\chi},|\theta|)$-dependence of $V_{eff}$ might reflect in 
thermodynamic/statistical characters of the system via a statistical weight $e^{-E/T}$,
because $|\theta|$ can give directions of excitations easier than $\theta_{\chi}$ and other
physical degrees of freedom.
Hence, an examination on collective fields under a Hamiltonian formalism is interesting for us.
If an unharmonicity of a potential of p-NG sector exists,
then it might cause a chaotic behavior of a classical motion of a representation point of the system. 
Here, we consider the case of no Majorana mass term.
It is possible to get more complicated breaking schema and p-NG manifolds 
if we consider a Majorana mass term simultaneously with a Dirac one.

\vspace{2mm}

Now, we arrive at an important/interesting matter:

\vspace{1mm}

{\bf Problem. 7}:

\vspace{1mm}

{\it The correspondences of 
the sequence of stable subspaces as points $\to$ curves $\to$ surfaces, ... ,
and various breaking schema of compact Lie groups  
demand us toward a theory which is not a simple higher-dimensional extension/generalization
of the Galois-Iwasawa theory, 
but a framework which will consider "a sequence or ramifications of several theories"
to determine positions and symmetries of stable points in pseudo-NG manifolds.}

\vspace{1mm}

( Namely, a tree of theories. ) 
Here, we consider not Lie-groups themselves but possible correspondences between
breaking schema and symmetries of "discretized" stable points.
One can consider a starting point ( before breakdown ) of Lie groups as
classical, graded, quantum, so forth.
The Iwasawa theory ( hence, the Fermat's last theorem also ) or 
a higher-dimensional Iwasawa theory may be embedded in such ramifications of mathematical theories.
In some examples discussed above, it may be the case that the simple/ordinary Galois theory
cannot handle a symmetry between stable points of an NG manifold,
while a consideration on cyclotomic points is still meaningful.  
( Similar attempt of extension of notion of Galois theory was given by Grothendieck,
namely, {\it geometric Galois actions}. )
This issue in $SU(2_{f})$ case relates with theory of tiling of a space,
and a consideration on crystallographic groups may be useful,
though an $n$-dimensional Abelian lattice isomorphic to ${\bf Z}^{n}$ seems a natural example.
In the $SU(2_{f})$ case of diagonal breaking, a unit cell of 
the domain $0\le\theta_{\chi}\le 2\pi$ and $0\le |\theta|\le \pi$ is obtained,
and which will be regarded as one of fundamental polygons in two-dimensions
( topologically classified as sphere, real projective plane, Klein bottle, and torus ).
By glueing a finite number of unit cells, we obtain a surface over ${\bf R}^{2}$,
and this can be regarded as a Riemann surface~[2,98,153].
A problem of glueing a finite number of unit cells is highly nontrivial,
and this issue has been studied/classified by notions of "origami" ( in a mathematical notion ),
Teichm\"{u}ller curves and Veech groups~[153,203].
In the cases discussed by us,
we wish to extend a domain infinitely,
and we also should take into account that a condition of smoothness
( can be examined by a fiber ( tangent ) bundle )
of $V_{eff}$ at a border of two unit cells.
This condition may limits possible forms of extensions of domains in our problem discussed above.

\vspace{2mm}

The "ramifications of theories" will formally be obtained from the following formulation as an example.
By preparing fermion ( or, matter fields, more generally ) fields $\Psi$,
mass matrix $\widehat{\cal M}^{(0)}$, 
and consider an assignment of charges of global gauge symmetry $\widehat{\cal G}$ as
\begin{eqnarray}
\Psi &\equiv& \bigl( \psi(U(1)_{A}), \psi(SU(2_{f}) \times U(1)_{A}), \psi(SU(3_{f}) \times U(1)_{A}),
\psi(SU(5_{f}) \times U(1)_{A}), \cdots )^{T},  \\
\widehat{\cal M}^{(0)} &\equiv& \left(
\begin{array}{cccccc} 
\widehat{M}^{(0)}_{U(1)_{A}} & 0 & 0 & 0 & \cdots \\ 
0 & \widehat{M}^{(0)}_{SU(2_{f})\times U(1)_{A}} & 0 & 0 & \cdots \\
0 & 0 & \widehat{M}^{(0)}_{SU(3_{f})\times U(1)_{A}} & 0 & \cdots \\
0 & 0 & 0 & \widehat{M}^{(0)}_{SU(5_{f})\times U(1)_{A}} & \cdots \\
\cdot & \cdot & \cdot & \cdot & \cdots \\
\cdot & \cdot & \cdot & \cdot & \cdots \\
\cdot & \cdot & \cdot & \cdot & \cdots 
\end{array}
\right), \\
\widehat{\cal G} &\equiv& (\hat{g}_{U(1)_{A}},
\hat{g}_{SU(2_{f})\times U(1)_{A}},\hat{g}_{SU(3_{f})\times U(1)_{A}},\hat{g}_{SU(5_{f})\times U(1)_{A}},\cdots).
\end{eqnarray}
Here, global gauge symmetries of fermi fields have been implied by their subscripts.
( For another formulation, one can consider $U(1)_{A}\bigotimes^{\infty}_{N_{f}=1}SU(N_{f})$. )
Then, one should examine the determinant 
${\rm det}(\widehat{\cal M}^{(0)}+\widehat{\cal G}\widehat{\cal M}_{dyn}\widehat{\cal G}^{-1})$
and eigenvalues of the total mass matrix. 
From this procedure, in principle, various pseudo-NG manifolds will be obtained simultaneously: 
Namely, a problem of diagonalization of a finite ( countable infinite ) dimensional matrix.
The word "various breaking schema" has been used in the sense such like
$(U(1)_{A})^{3}\times SU(3_{f})\to (U(1)_{A})^{2}\times SU(2_{f})\to (U(1)_{f})^{2}\to{\rm nothing}$,
so forth, and each step of the sequences will define a geometric object as the corresponding NG manifold.
A crucially important question is how they are (dis)connected with each other.
$V_{eff}(\theta_{\chi},\{\theta_{a}\})$ in the diagonal breaking of $U(1)_{A}\times SU(N_{f})$ is
a function of $N_{f}(N_{f}-1)+1$ variables.
From the Morse theoretical aspect, one will take first and second derivatives of those variables,
and our problem is to find a symmetry such as cyclotomic points or a Galois group,
in the above matrix $\widehat{\cal M}^{(0)}+\widehat{\cal G}\widehat{\cal M}_{dyn}\widehat{\cal G}^{-1}$.
A covering space of the matrix can be considered.

\vspace{2mm}

One can consider a more realistic situation of chiral symmetry breaking,
such as $U(2)_{R}\times U(2)_{L}\to U(2)_{R+L}\to (U(1))^{2}$.
For that case, we will consider the following Hermitian mass matrix:
\begin{eqnarray}
\widetilde{M} &=& (\eta^{\dagger}_{1},\eta^{\dagger}_{2})\Bigl( 
\widehat{M}^{(0)} + g_{L}\widehat{M}_{dyn}g^{\dagger}_{R} \Bigr)\left(
\begin{array}{c}
\xi_{1}  \\
\xi_{2}
\end{array}
\right)
+ (\xi^{\dagger}_{1},\xi^{\dagger}_{2})\Bigl( 
\widehat{M}^{(0)} + g_{R}\widehat{M}^{\dagger}_{dyn}g^{\dagger}_{L} \Bigr)\left(
\begin{array}{c}
\eta_{1}  \\
\eta_{2}
\end{array}
\right),  
\end{eqnarray}
where,
\begin{eqnarray}
\widehat{M}^{(0)} &=& \left(
\begin{array}{cc}
|M^{(0)}_{1}| & 0 \\
0 & |M^{(0)}_{2}|
\end{array}
\right),   \\
\widehat{M}_{dyn} &\equiv& \left(
\begin{array}{cc}
|M_{dyn1}|e^{i\theta^{U}_{1}} & 0 \\
0 & |M_{dyn2}|e^{i\theta^{U}_{2}}
\end{array}
\right),   \\
g_{R, L} &\equiv&
\left(
\begin{array}{cc}
\cos |\theta^{R,L}| & i\frac{\theta^{R,L}_{-}}{|\theta^{R,L}|}\sin|\theta^{R,L}| \\
i\frac{\theta^{R,L}_{+}}{|\theta^{R,L}|}\sin|\theta^{R,L}| & \cos|\theta^{R,L}|
\end{array}
\right).
\end{eqnarray}
The mass spectra becomes
\begin{eqnarray}
\lambda_{\pm} &=& \sqrt{\frac{1}{2}\bigl[ \alpha^{2}-2\beta \pm \alpha\sqrt{\alpha^{2}-4\beta} \bigr]},
\end{eqnarray}
where, the parameters in $\lambda_{\pm}$ are
\begin{eqnarray}
\alpha &\equiv& 2 ( \Re a + \Re d ),   \\
\beta &\equiv& \Re a \Re  d - |b|^{2} - |c|^{2} -2\Re (bc), 
\end{eqnarray}
and
\begin{eqnarray}
a &\equiv& |M^{(0)}_{1}| + M_{dyn1}\cos|\theta^{L}|\cos|\theta^{R}| 
+ M_{dyn2}\frac{\theta^{L}_{-}\theta^{R}_{+}}{|\theta^{L}||\theta^{R}|}\sin|\theta^{L}|\sin|\theta^{R}|,  \\
b &\equiv& -iM_{dyn1}\frac{\theta^{R}_{-}}{|\theta^{R}|}\cos|\theta^{L}|\sin|\theta^{R}|
+iM_{dyn2}\frac{\theta^{L}_{-}}{|\theta^{L}|}\sin|\theta^{L}|\cos|\theta^{R}|,   \\
c &\equiv& +iM_{dyn1}\frac{\theta^{L}_{+}}{|\theta^{L}|}\sin|\theta^{L}|\cos|\theta^{R}|
-iM_{dyn2}\frac{\theta^{R}_{+}}{|\theta^{R}|}\cos|\theta^{L}|\sin|\theta^{R}|,   \\
d &\equiv& |M^{(0)}_{2}| + M_{dyn2}\cos|\theta^{L}|\cos|\theta^{R}| 
+ M_{dyn1}\frac{\theta^{L}_{+}\theta^{R}_{-}}{|\theta^{L}||\theta^{R}|}\sin|\theta^{L}|\sin|\theta^{R}|, 
\end{eqnarray}
and
\begin{eqnarray}
\theta^{R,L}_{\pm} &\equiv& \theta^{R,L}_{1} \pm i\theta^{R,L}_{2},  \\
|\theta^{R,L}| &\equiv& (\theta^{R,L}_{1})^{2} + (\theta^{R,L}_{2})^{2}.
\end{eqnarray}
The result shows that the theory has six pseudo-NG bosons
( $\theta^{R}_{1},\theta^{R}_{2},\theta^{L}_{1},\theta^{L}_{2},\theta^{U}_{1},\theta^{U}_{2}$ )  
according to the breaking scheme.
The following two $U(1)$ symmetries ( gauge )
\begin{eqnarray}
\eta_{1,2} \to e^{i\theta^{G}_{1,2}}\eta_{1,2}, \qquad
\xi_{1,2} \to e^{i\theta^{G}_{1,2}}\xi_{1,2}
\end{eqnarray}
remain, while the chiral-symmetry types
\begin{eqnarray}
\eta_{1} \to e^{i\theta^{U}_{1}}\eta_{1}, \quad
\eta_{2} \to e^{i\theta^{U}_{2}}\eta_{2}, \quad
\xi_{1} \to e^{-i\theta^{U}_{1}}\xi_{1}, \quad
\xi_{2} \to e^{-i\theta^{U}_{2}}\xi_{2}
\end{eqnarray}
are broken.
Therefore we find that $V_{eff}$ behaves as
\begin{eqnarray}
V_{eff} &\sim& \cos|\theta^{L}|\cos|\theta^{R}|\cos\theta^{U}_{1}\cos\theta^{U}_{2},
\end{eqnarray}
and the set of stable points of the NG sector is given by
\begin{eqnarray}
& & |\theta^{L}| = (2n^{L}+1)\pi, \quad 
|\theta^{R}| = (2n^{R}+1)\pi, \quad
\theta^{U}_{1} = (2n^{U}_{1}+1)\pi, \quad 
\theta^{U}_{2} = (2n^{U}_{2}+1)\pi,   \\
& & n^{L}, n^{R}, n^{U}_{1,2} \in {\bf Z}.
\end{eqnarray}
and these hyperplanes make lines toward two directions of $|\theta^{L}|$ and $|\theta^{R}|$,
perpendicular with each other.
Due to the fact that $U(2)_{L}\times U(2)_{R}$ is a direct product,
there are couplings between $\theta^{L}_{1,2}$ and $\theta^{R}_{1,2}$ in $\lambda_{\pm}$
under a nontrivial way, while our analysis of the generalized Nambu-Goldstone theorem
works well, characteristically the same with the result of more simplified $SU(2_{f})$ case given in Sec. II A.
Thus, it is not an important matter whether the starting point of a discussion
is given by a direct product of groups or a single group.
The connectivity ( simple or not ) is also not an important matter for determination
of a periodicity of an NG sector, or a realization of our generalized NG theorem. 
A similar analysis can be done in a case of chiral group $U(N)_{R}\times U(N)_{L}$.
It is an interesting issue to apply our result of $U(N)_{R}\times U(N)_{L}$ to
a construction of non-linear sigma model.

\vspace{2mm}

Since the non-triviality of supermanifolds, 
it is also an interesting issue for us to extend our examination 
to several breaking schema of supersymmetric system
( we might need a notion of algebraic supervariety )
~[31,47,77,163,196,201,242,248].

\vspace{2mm}

Note that the widing number $n$ of the definition ( theorem 3 ) is defined by our hand, 
describes a degeneracy of the system,
thus we can set it arbitrarily, 
and it is a kind of gauge-degree of freedom ( a translation in a parameter space ).
It should be emphasized that the "cyclotomic field" in the chiral $U(1)$ case 
is defined via a variational equation:
Here we meet a ( probably very seldom ) connection between number theory and variational calculus.
A distribution of zeros defined by a gap equation in a non-Abelian case 
can not be interpreted by ordinary Iwasawa theory,
and it might stimulate us toward a generalization of the Iwasawa theory:
Our generalized Nambu-Goldstone theorem will have a quite rich mathematical structure.

\subsection{The Critical Coupling Constant}

The critical coupling with fixed $\Theta$ is obtained 
from the stationary condition $\frac{\partial V^{NJL+M}_{eff}}{\partial |m_{D}|}=0$:
\begin{eqnarray}
(G^{cr})^{-1} &=& \frac{1}{8\pi^{2}}\Bigg[ \Bigl( 1-\frac{|m_{R}|^{2}+|m_{L}|^{2}+2|m_{R}||m_{L}|\cos\Theta}{|m_{R}|^{2}-|m_{L}|^{2}}\Bigr)\Bigl(\Lambda^{2}-|m_{L}|^{2}\ln\Bigl( 1+\frac{\Lambda^{2}}{|m_{L}|^{2}} \Bigr) \Bigr)    \nonumber \\
& & + \Bigl( 1+\frac{|m_{R}|^{2}+|m_{L}|^{2}+2|m_{R}||m_{L}|\cos\Theta}{|m_{R}|^{2}-|m_{L}|^{2}}\Bigr)\Bigl(\Lambda^{2}-|m_{R}|^{2}\ln\Bigl( 1+\frac{\Lambda^{2}}{|m_{R}|^{2}} \Bigr) \Bigr) \Bigg].
\end{eqnarray}
However, we know only $G^{cr}$ at $\Theta=(2n+1)\pi$ is the correct critical coupling 
from the structure of the two-dimensional effective potential surface $V^{NJL+M}_{eff}(|m_{D}|,\Theta)$.  
Namely, only $G^{cr}$ at $\Theta=(2n+1)\pi$,
\begin{eqnarray}
(G^{cr})^{-1} &=& \frac{1}{8\pi^{2}}\Bigg[ \Bigl( 1-\frac{|m_{R}|-|m_{L}|}{|m_{R}|+|m_{L}|}\Bigr)\Bigl(\Lambda^{2}-|m_{L}|^{2}\ln\bigl( 1+\frac{\Lambda^{2}}{|m_{L}|^{2}} \bigr) \Bigr)   \nonumber \\
& & \qquad + \Bigl( 1+\frac{|m_{R}|-|m_{L}|}{|m_{R}|+|m_{L}|}\Bigr)\Bigl(\Lambda^{2}-|m_{R}|^{2}\ln\bigl( 1+\frac{\Lambda^{2}}{|m_{R}|^{2}} \bigr) \Bigr) \Bigg], 
\end{eqnarray}
is physically meaningful. 
Moreover, due to the factors coming from 
$|m_{D}|^{-1}\frac{\partial (M^{F}_{\pm})^{2}}{\partial |m_{D}|}\Big|_{|m_{D}|=0}$ in the gap equation, 
we can take the right-left symmetric condition $|m_{R}|=|m_{L}|$ in $G^{cr}$ of Eq. (130) 
only at $\Theta=(2n+1)\pi$. 
( The gap equation $\frac{\partial V_{eff}}{\partial\Theta}=0$ consists with several parameters of $V_{eff}$,
and then the gap equation defines a subset of the parameter space of $V_{eff}$.
The subset contains an unphysical part, and physical and unphysical parts may be not smoothly connected
in the parameter space in the Dirac+Majorana mass case. )
In that case, one finds
\begin{eqnarray}
(G^{cr})^{-1} &=& \frac{\Lambda^{2}}{4\pi^{2}}\Bigg[ 1-\frac{|m_{L}|^{2}}{\Lambda^{2}}\ln\Bigl( 1+\frac{\Lambda^{2}}{|m_{L}|^{2}} \Bigr) \Bigg].
\end{eqnarray}
Hence, from this expression, we find $G^{cr}$ of finite $|m_{L}|$ ( $=|m_{R}|$ ) is 
always larger than the case $|m_{L}|=0$ at a fixed $\Lambda$,
and we obtain the well-known formula $G^{cr}=4\pi^{2}/\Lambda^{2}$ at $|m_{L}|=|m_{R}|=0$~[160].

\subsection{The Collective Modes}

Now, we consider fluctuations of the collective fields. 
The effective action with taking into account the fluctuations becomes
the following form as a Taylor expansion around a point:
\begin{eqnarray}
\Gamma^{NJL+M}_{eff} &=& -\int d^{4}x \frac{|m_{D}|^{2}}{G} -\frac{i}{2}\ln{\rm Det}\Omega^{F}_{M} \nonumber \\
&\to& -\int d^{4}x  \frac{|m_{D}+\delta m_{D}|^{2}}{G} -\frac{i}{2}{\rm Tr}\ln\Omega^{F}_{M} + \frac{i}{2}{\rm Tr}\sum^{\infty}_{n=2}\frac{1}{n}\Bigl( (\Omega^{F}_{M})^{-1}\Sigma^{F}\Bigr)^{n}.
\end{eqnarray}
Here, we consider the fluctuations $\delta m_{D}$ as a small displacement of $m_{D}$:
\begin{eqnarray}
m_{D} \to m_{D} + \delta m_{D}, 
\end{eqnarray}
and the "self-energy" matrix $\Sigma^{F}$ is defined by
\begin{eqnarray}
& & \Sigma^{F} \equiv  \delta m^{\dagger}_{D}\left(
\begin{array}{cc}
0 & P_{+} \\
P_{+} & 0 
\end{array}
\right) + \delta m_{D}\left(
\begin{array}{cc}
0 & P_{-} \\
P_{-} & 0 
\end{array}
\right). 
\end{eqnarray}
We wish to expand $V^{NJL+M}_{eff}$ around the minimum $(|m_{D}|,\Theta=\pi)$ of $V^{NJL+M}_{eff}$.
Due to the phase factor at the global minimum, 
the collective field $m_{D}$ with its fluctuations $\delta|m_{D}|$ and $\delta\Theta$ 
are expressed at the global minimum $(|m_{D}|,\Theta=\pi)$ of 
the two-dimensional effective potential surface as follows: 
\begin{eqnarray}
\Theta &=& \theta_{R} + \theta_{L} -2\theta_{D} = \theta_{R} + \theta_{L} =\pi, \\
\delta\Theta &=& -2\delta\theta_{D},   \\
m_{D} &\to& (|m_{D}|+\delta|m_{D}|)e^{i\delta\theta_{D}}   
= |m_{D}| +\delta|m_{D}| -\frac{i}{2}|m_{D}|\delta\Theta.
\end{eqnarray}
Hence, 
\begin{eqnarray}
\Sigma^{F} = \Sigma^{F}_{D} + \Sigma^{F}_{\Theta},   \quad 
\Sigma^{F}_{\Theta} \equiv \frac{|m_{D}|}{2}\delta\Theta\left(
\begin{array}{cc}
0 & i\gamma_{5} \\
i\gamma_{5} & 0 
\end{array}
\right), \quad 
\Sigma^{F}_{D} \equiv \delta|m_{D}|\left(
\begin{array}{cc}
0 & 1 \\
1 & 0 
\end{array}
\right).
\end{eqnarray}
The sum of phases $\theta_{R}+\theta_{L}$ is fixed.  
In this paper, we call the collective mode associated with the $\Theta$-degree of freedom 
as "phason", and denote it as $\tilde{\Theta}$:
\begin{eqnarray}
\tilde{\Theta} &\equiv& \frac{|m_{D}|}{2}\delta\Theta.
\end{eqnarray}
Note that $\tilde{\Theta}$ is a bare quantity.
The phason corresponds to majoron at $|m_{L}|=0$, $|m_{R}|\ne 0$~[34], 
and it becomes pion at $|m_{R}|=|m_{L}|=0$~[160]. 
As discussed above, since both of the global $U(1)$ symmetries $U(1)_{V}$ and $U(1)_{A}$ are broken explicitly
at $|m_{L}|\ne 0$, $|m_{R}|\ne 0$, the phason is a p-NG boson, and it has a finite mass.

\subsection{The Decay Constant and Mass of Phason}

Before doing our estimation of the excitation mass and decay constant of phason,
we will proceed with our examination to the second-derivative, 
Hessian, of $V^{NJL+M}_{eff}$ with respect to $|m_{D}|$ and $\Theta$.
The entries of the Hessian matrix become
\begin{eqnarray}
\frac{\partial^{2} V^{NJL+M}_{eff}}{\partial|m_{D}|^{2}} &=& \frac{2}{G} - \frac{1}{8\pi^{2}}\Bigg[ \frac{\partial^{2} (M^{F}_{+})^{2}}{\partial|m_{D}|^{2}}F^{F}_{+} 
+ \Bigl( \frac{\partial (M^{F}_{+})^{2}}{\partial|m_{D}|}\Bigr)^{2}J^{F}_{+} 
+ \frac{\partial^{2} (M^{F}_{-})^{2}}{\partial|m_{D}|^{2}}F^{F}_{-} + \Bigl( \frac{\partial (M^{F}_{-})^{2}}{\partial|m_{D}|}\Bigr)^{2}J^{F}_{-}
\Bigg],\\
\frac{\partial^{2} V^{NJL+M}_{eff}}{\partial|m_{D}|\partial\Theta} &=& - \frac{1}{8\pi^{2}}\Bigg[ \frac{\partial^{2} (M^{F}_{+})^{2}}{\partial|m_{D}|\partial\Theta}F^{F}_{+} 
+ \frac{\partial (M^{F}_{+})^{2}}{\partial|m_{D}|}\frac{\partial (M^{F}_{+})^{2}}{\partial\Theta}  J^{F}_{+} 
+ \frac{\partial^{2} (M^{F}_{-})^{2}}{\partial|m_{D}|\partial\Theta}F^{F}_{-} + \frac{\partial (M^{F}_{-})^{2}}{\partial|m_{D}|}\frac{\partial (M^{F}_{-})^{2}}{\partial\Theta} J^{F}_{-}
\Bigg],   \\
\frac{\partial^{2} V^{NJL+M}_{eff}}{\partial\Theta^{2}} &=& 
- \frac{1}{8\pi^{2}}\Bigg[ \frac{\partial^{2} (M^{F}_{+})^{2}}{\partial\Theta^{2}}F^{F}_{+} 
+ \Bigl( \frac{\partial (M^{F}_{+})^{2}}{\partial\Theta}\Bigr)^{2}J^{F}_{+} 
+ \frac{\partial^{2} (M^{F}_{-})^{2}}{\partial\Theta^{2}}F^{F}_{-} + \Bigl( \frac{\partial (M^{F}_{-})^{2}}{\partial\Theta}\Bigr)^{2}J^{F}_{-}
\Bigg],
\end{eqnarray}
where,
\begin{eqnarray}
J^{F}_{\pm} &\equiv& \frac{\Lambda^{2}}{\Lambda^{2}+(M^{F}_{\pm})^{2}} -\ln\Bigl( 1 + \frac{\Lambda^{2}}{(M^{F}_{\pm})^{2}} \Bigr).
\end{eqnarray}
The derivatives appear in the Hessian become
\begin{eqnarray}
\frac{\partial^{2} (M^{F}_{\pm})^{2}}{\partial|m_{D}|^{2}} &=& \frac{1}{|m_{D}|}\frac{\partial (M^{F}_{\pm})^{2}}{\partial|m_{D}|} \pm  \frac{8|m_{D}|^{2}(|m_{R}|^{2}+|m_{L}|^{2}+2|m_{R}||m_{L}|\cos\Theta)^{2}}{[(|m_{R}|^{2}-|m_{L}|^{2})^{2}+4|m_{D}|^{2}(|m_{R}|^{2}+|m_{L}|^{2}+2|m_{R}||m_{L}|\cos\Theta)]^{3/2}} \\
\frac{\partial^{2} (M^{F}_{\pm})^{2}}{\partial|m_{D}|\partial\Theta} &=& \frac{2}{|m_{D}|}\frac{\partial (M^{F}_{\pm})^{2}}{\partial\Theta}\mp\frac{8|m_{D}|^{3}|m_{R}||m_{L}|\sin\Theta(|m_{R}|^{2}+|m_{L}|^{2}+2|m_{R}||m_{L}|\cos\Theta)}{[(|m_{R}|^{2}-|m_{L}|^{2})^{2}+4|m_{D}|^{2}(|m_{R}|^{2}+|m_{L}|^{2}+2|m_{R}||m_{L}|\cos\Theta)]^{3/2}}, \\
\frac{\partial^{2} (M^{F}_{\pm})^{2}}{\partial\Theta^{2}} &=& \cot\Theta\frac{\partial (M^{F}_{\pm})^{2}}{\partial\Theta}\pm\frac{8|m_{D}|^{4}|m_{R}|^{2}|m_{L}|^{2}\sin^{2}\Theta}{[(|m_{R}|^{2}-|m_{L}|^{2})^{2}+4|m_{D}|^{2}(|m_{R}|^{2}+|m_{L}|^{2}+2|m_{R}||m_{L}|\cos\Theta)]^{3/2}},
\end{eqnarray}
and their mass dimensions are $[{\rm mass}]^{0}$, $[{\rm mass}]^{1}$ and $[{\rm mass}]^{2}$, respectively.
$-\frac{\delta^{2}\Gamma^{NJL+M}_{eff}/\int d^{4}x}{\delta(|m_{D}|\delta\Theta/2)^{2}}=\frac{1}{2!}\frac{4}{|m_{D}|^{2}}\frac{\partial^{2}V^{NJL+M}_{eff}}{\partial\Theta^{2}}$ 
corresponds to the square of mass of phason.
The Hessian of $V^{NJL+M}_{eff}$ gives a description on the collective excitations of our model, 
and a unitary rotation in the two-dimensional space of $|m_{D}|$ and $\Theta$
corresponds to a canonical transformation of collective coordinates.
Especially, at $\Theta=(2n+1)\pi$,
\begin{eqnarray}
\frac{\partial^{2} (M^{F}_{\pm})^{2}}{\partial|m_{D}|^{2}}\Big|_{\Theta=(2n+1)\pi} &=&  2\Bigg( 1 \mp\frac{|m_{R}|-|m_{L}|}{\sqrt{(|m_{R}|+|m_{L}|)^{2}+4|m_{D}|^{2}}} \Bigg) \pm 
\frac{8|m_{D}|^{2}}{(|m_{R}|-|m_{L}|)[(|m_{R}|+|m_{L}|)^{2}+4|m_{D}|^{2}]^{3/2}},   \\ 
\frac{\partial^{2} (M^{F}_{\pm})^{2}}{\partial|m_{D}|\partial\Theta}\Big|_{\Theta=(2n+1)\pi} &=& 0, \\
\frac{\partial^{2} (M^{F}_{\pm})^{2}}{\partial\Theta^{2}}\Big|_{\Theta=(2n+1)\pi} &=& \mp\frac{2|m_{D}|^{2}|m_{R}||m_{L}|}{(|m_{R}|-|m_{L}|)\sqrt{(|m_{R}|+|m_{L}|)^{2}+4|m_{D}|^{2}}},
\end{eqnarray}
( assume $|m_{R}|>|m_{L}|$ ). Hence,
\begin{eqnarray}
\frac{\partial^{2} V^{NJL+M}_{eff}}{\partial|m_{D}|\partial\Theta}\Big|_{\Theta=(2n+1)\pi} &=& 0.
\end{eqnarray}
Because the off-diagonal element $\frac{\partial^{2} V^{NJL+M}_{eff}}{\partial|m_{D}|\partial\Theta}$ 
vanishes at $\Theta=(2n+1)\pi$ and the Hessian is diagonal, 
there is no mode-mode coupling ( decouple ) on the line $\Theta=(2n+1)\pi$ of 
the two-dimensional surface $(|m_{D}|,\Theta)$ of the potential $V^{NJL+M}_{eff}$,
at least at the description up to the second-order derivatives, 
corresponds to an RPA ( random phase approximation ),
while it does not vanish at $\Theta\ne (2n+1)\pi$, gives a mode-mode coupling. 
( If there is a mode-mode coupling, an analysis of pseudo-NG manifold becomes more complicated
than those given in the previous section. )
The eigenvalue of the Hessian will be given simply by its diagonal entries at $\Theta=(2n+1)\pi$.
Note that $\frac{\partial^{2}(M^{F}_{\pm})^{2}}{\partial\Theta^{2}}$ are singular 
at $|m_{R}|=|m_{L}|$, $\Theta=(2n+1)\pi$.
From the expression of second-derivative, 
the product of the inverse of the renormalization constant of phason field $Z^{-1}_{\Theta}$ and 
the square of phason mass $m_{\Theta}$ becomes
\begin{eqnarray}
& & Z^{-1}_{\Theta}(m_{\Theta})^{2} 
= \frac{1}{2!}\Bigl(\frac{2}{|m_{D}|}\Bigr)^{2}\frac{\partial^{2} V^{NJL+M}_{eff}}{\partial\Theta^{2}}\Big|_{\Theta=(2n+1)\pi}    \nonumber \\      
& & \qquad = -\frac{1}{2\pi^{2}}\Bigg[ \frac{|m_{R}||m_{L}|}{|m_{R}|-|m_{L}|}\frac{1}{\sqrt{(|m_{R}|+|m_{L}|)^{2}+4|m_{D}|^{2}}} \Bigg] 
\Bigl\{ (M^{F}_{+})^{2}\ln\Bigl( 1+\frac{\Lambda^{2}}{(M^{F}_{+})^{2}} \Bigr)-(M^{F}_{-})^{2}\ln\Bigl( 1+\frac{\Lambda^{2}}{(M^{F}_{-})^{2}} \Bigr) \Bigr\},  \nonumber \\
& & 
\end{eqnarray}
where, $M^{F}_{\pm}$ in this case ( namely at $\Theta=(2n+1)\pi$ ) become
\begin{eqnarray}
M^{F}_{\pm} &=& \sqrt{|m_{D}|^{2} + \frac{|m_{R}|^{2}+|m_{L}|^{2}}{2}\mp \frac{1}{2}\Bigl(|m_{R}|-|m_{L}|\Bigr)\sqrt{ \Bigl(|m_{R}|+|m_{L}|\Bigr)^{2} + 4|m_{D}|^{2} } },
\end{eqnarray}
( assume $|m_{R}|>|m_{L}|$ ).
Here, we should mention that $Z^{-1}_{\Theta}m^{2}_{\Theta}$ of (152) does not vanish  at $|m_{D}|=0$ 
with $|m_{R}|\ne 0$, $|m_{L}|\ne 0$, $|m_{R}|\ne |m_{L}|$,
while it will vanishes at $|m_{R}|=0$, $|m_{L}|\ne 0$ or $|m_{R}|\ne 0$, $|m_{L}|=0$ ( majoron ), 
$|m_{R}|=|m_{L}|=0$ ( pion ).
The limit $|m_{R}|=|m_{L}|$ of $Z^{-1}_{\Theta}m^{2}_{\Theta}$ of this expression does not give a singularity 
because the bracket of squares of $M^{F}_{\pm}$ will vanish more rapidly.

\vspace{2mm}

Phason is not a "light" particle, 
since it is a bound ( or, resonance ) state of a heavy and a light particles. 
When $\Lambda\gg |m_{R}|\gg |m_{D}|,|m_{L}| >0$ 
( this is not the same with the ( type-I ) seesaw condition $|m_{R}|\gg |m_{D}|>|m_{L}|=0$ ),
\begin{eqnarray}
Z^{-1}_{\Theta}m^{2}_{\Theta} 
&\approx& 
\frac{|m_{R}||m_{L}|}{2\pi^{2}}\frac{\Lambda^{4}}{(\Lambda^{2}+(M^{F}_{+})^{2})(\Lambda^{2}+(M^{F}_{-})^{2})}
\approx \frac{|m_{R}||m_{L}|}{2\pi^{2}}.
\end{eqnarray}
Since the condition $|m_{R}|\gg |m_{D}|$, a contribution of $|m_{D}|$ which gives cutoff dependence 
will come to the next-order of our evaluation of $Z^{-1}_{\Theta}m^{2}_{\Theta}$ given above.   
To achieve the type-I or type-II seesaw situations $\Lambda\gg |m_{R}|\gg |m_{D}|>|m_{L}|\ge 0$, 
we need a fine-tuning of $G$ because $|m_{D}|$ sharply rises to obtain ${\cal O}(\Lambda)$ 
when $G$ moves from $G_{cr}$. 
The corresponding quantity of pion in the ordinary NJL model will vanish by a self-consistent gap equation.
It is interesting for us to compare (154) with the following mass relation:
\begin{eqnarray}
m^{2}_{\Theta} &\approx& 2m_{D}(m_{R}+m_{L}) + 2m_{R}m_{L}.
\end{eqnarray}
This formula is obtained from the mass relation
$m^{2}_{pseudo}=2m_{dyn}(m^{a}_{current}+m^{b}_{current})+2m^{a}_{current}m^{b}_{current}$,
called as a PCAC ( partial conservation of axial-vector current ) relation valid both in NJL and QED 
( $m^{2}_{pseudo}$; mass of a pseudo-scalar meson, 
$m_{dyn}$; dynamical Dirac mass,
$m^{a,b}_{current}$; current mass parameters ),
shows the amount of energy of stabilization of the effective potential~[147].
Here, we have assumed that the mass relation is also valid for 
the Majorana mass parameters $m_{R}$ and $m_{L}$.
The sign of $m^{2}_{\Theta}$ reflects the (in)stability of the effective potential. 
Since $Z^{-1}_{\Theta}$ ( must be a logarithmically divergent quantity ) will give a numerical factor,
our result in (154) is almost the same with the mass relation (155).

\vspace{2mm}

We obtain $m_{\Theta}\sim 20$ GeV ( assuming $Z^{-1}_{\Theta}\sim {\cal O}(1)$ ) 
when we use $|m_{R}|=10^{11}$ GeV and $|m_{L}|=10^{2}$ eV,
and it is a little lighter than weak bosons while heavier than mesons/baryons.
The compton wave length of phason becomes 
$\lambda_{compton}\sim 10^{-2}$ fm.
A cross section between a neutrino and a nucleon is absolutely small,
thus there is no place for a high-density neutrino matter inside a star.
Phason would obtain a reaction process of the intermediate energy scale 
between electroweak to GUT breakings. 
It seems not unnatural because the radius of so-called "primeval fire ball" 
at GUT phase transition is $10^{-28}$ times smaller than those of our Universe~[202].

\vspace{2mm}

To obtain $Z^{-1}_{\Theta}$, a scalling factor of the second-derivative of $V_{eff}$, 
we evaluate vacuum polarization ( VP ) functions of our theory.
Since we have interest on the VP functions in the vicinity of a vacuum ( ground state ) 
of the theory at the one-loop level, 
we should set mass parameters/phases suitably for this purpose:
$m_{D}=m^{\dagger}_{D}$, $m_{R}=-m^{\dagger}_{R}$ and $m_{L}=-m^{\dagger}_{L}$.
The definitions of VP functions are obtained from Eqs. (133) and (139) as 
\begin{eqnarray}
\Pi^{F}_{DD}(q) &\equiv& {\rm tr}\int_{k} (\Omega^{F}_{M})^{-1}(k+q)\tau_{1}(\Omega^{F}_{M})^{-1}(k)\tau_{1},   \\
\Pi^{F}_{D\Theta}(q) &\equiv& {\rm tr}\int_{k} (\Omega^{F}_{M})^{-1}(k+q)\tau_{1}(\Omega^{F}_{M})^{-1}(k)i\gamma_{5}\otimes\tau_{1},   \\
\Pi^{F}_{\Theta D}(q) &\equiv& {\rm tr}\int_{k} (\Omega^{F}_{M})^{-1}(k+q)i\gamma_{5}\otimes\tau_{1}(\Omega^{F}_{M})^{-1}(k)\tau_{1},    \\
\Pi^{F}_{\Theta\Theta}(q) &\equiv& {\rm tr}\int_{k} (\Omega^{F}_{M})^{-1}(k+q)i\gamma_{5}\otimes\tau_{1}(\Omega^{F}_{M})^{-1}(k)i\gamma_{5}\otimes\tau_{1},
\end{eqnarray}
where, $\tau_{1}$ acts on the right-left space of the Majorana-Nambu field $\Psi_{MN}$.
After taking traces, we obtain
\begin{eqnarray}
\Pi^{F}_{IJ}(q) &=& \int_{k}\frac{A^{F}_{IJ}(k+q,k)}{D(k+q)D(k)}, 
\qquad (I,J=D,\Theta).      
\end{eqnarray}
We have an interest on $\Pi^{F}_{D\Theta}$, 
because they would give a coupling between the amplitude and phase modes, 
and also might give an imaginary part to the effective potential.
The result becomes
\begin{eqnarray}
A^{F}_{D\Theta} &=& 2(2i)^{2}|m_{R}||m_{L}|\sin\Theta  \nonumber \\ 
& & \quad \times \Bigl\{ (k+q)^{2}k^{2} +(|m_{D}|^{2}-|m_{L}|^{2})(k+q)^{2} + (|m_{D}|^{2}-|m_{R}|^{2})k^{2}  \nonumber \\ 
& & \qquad\quad + 2|m_{D}|^{2}|m_{R}||m_{L}|\cos\Theta -3|m_{D}|^{4} + |m_{R}|^{2}|m_{L}|^{2}  \Bigr\}.
\end{eqnarray}
Note that this $A^{F}_{D\Theta}$ is real.
Since both $\Pi^{F}_{D\Theta}$ and $\Pi^{F}_{\Theta D}$ will vanish at $\Theta=(2n+1)\pi$,
and again, we find there is no mode-mode coupling, 
there is no imaginary part of $V^{NJL+M}_{eff}$ at this second-order level
( and parity is also conserved ).
If we set another choice of the phases of mass parameters, 
different from CP conversing condition (74),
$A^{F}_{D\Theta}$ does not vanish, and gives a pure-imaginary matrix element
of Lagrangian of the collective fields derived from $V^{NJL+M}_{eff}$.
We cannot deny a possibility that the theory generates an imaginary part of $V^{NJL+M}_{eff}$, 
in triangle, box, ..., diagrams of our definition of the expansion of $V^{NJL+M}_{eff}$ 
before estimating them since we should be careful to apply the Furry theorem-like consideration 
in our calculations, though the $V^{NJL+M}_{eff}$ of the one-loop level (65) is real
( this is guaranteed by Hermiticity of the theory ). 
Vertex corrections of our four-fermion theory has a class of diagrams
they cannot be factorized into summations of $\Pi^{F}_{IJ}$.
They also give couplings between the amplitude and phase modes. 
If $V^{NJL+M}_{eff}$ gets an imaginary part, 
it would cause an instability of the theory 
similar to the sense of Weinberg-Wu~[228].
Because our theory is relativistic, 
a derivative of time takes the form of second-order 
inside a secular equation of the bosonic collective fields, 
we should examine energy spectra of the bosonic sector of our model
before concluding whether an imaginary part gives a remarkable effect or not.

\vspace{2mm}

The result of $A^{F}_{\Theta\Theta}$ becomes
\begin{eqnarray}
A^{F}_{\Theta\Theta}(k+q,k) &=& 
4\Bigl[(k+q)^{2}-|m_{L}|^{2}-|m_{D}|^{2}\Bigr]\Bigl[k^{2}-|m_{R}|^{2}-|m_{D}|^{2}\Bigr](k+q)\cdot k  \nonumber \\
& & -4\Bigl[ |m_{R}||m_{L}|\cos\Theta\bigl\{ [(k+q)^{2}-|m_{L}|^{2}](k^{2}-|m_{R}|^{2})+|m_{D}|^{4} \bigr\}  \nonumber \\
& & \quad +|m_{D}|^{2}\bigl\{ [(k+q)^{2}-|m_{L}|^{2}]|m_{R}|^{2}+(k^{2}-|m_{R}|^{2})|m_{L}|^{2} \bigr\}\Bigr]  \nonumber \\
& & -4|m_{D}|^{2}\Bigl[ |m_{R}|^{2}|m_{L}|^{2}\cos 2\Theta +|m_{R}||m_{L}|\cos\Theta [(k+q)^{2}+k^{2}-2|m_{D}|^{2}]  \nonumber \\
& & \quad + [(k+q)^{2}-|m_{D}|^{2}](k^{2}-|m_{D}|^{2})\Bigr]    \nonumber \\
& & + (R\leftrightarrow L). 
\end{eqnarray}
The propagator of the phason is defined as follows: 
\begin{eqnarray}
G^{-1}_{\Theta}(q) \equiv -\frac{2}{G}+\frac{i}{4}\Pi^{F}_{\Theta\Theta}(q) = Z^{-1}_{\Theta}(q^{2}-m^{2}_{\Theta}), \quad 
Z^{-1}_{\Theta} = \frac{i}{4}\frac{\partial \Pi^{F}_{\Theta\Theta}(q)}{\partial q^{2}}\Big|_{q=0}, \quad 
-Z^{-1}_{\Theta}m^{2}_{\Theta} = -\frac{2}{G}+\frac{i}{4}\Pi^{F}_{\Theta\Theta}(0).
\end{eqnarray}
We have neglected the relative angle contribution of $k$ and $q$ in the definition for $Z^{-1}_{\Theta}$. 
By a rough estimation 
( picking up several terms of the leading order for both 
the numerator and denominator of the integrand of Eq. (160) ) 
with the condition $\Lambda\gg M_{\pm}$, 
we get
\begin{eqnarray}
Z^{-1}_{\Theta} &\approx& \frac{1}{8\pi^{2}}\ln\frac{\Lambda^{2}}{2((M^{F}_{+})^{2}+(M^{F}_{-})^{2})}, \\
\frac{i}{4}\Pi^{F}_{\Theta\Theta}(0) &\approx& \frac{1}{8\pi^{2}}\Bigg[ \Lambda^{2} -2\bigl( (M^{F}_{+})^{2}+(M^{F}_{-})^{2} \bigr)\ln\frac{\Lambda^{2}}{2((M^{F}_{+})^{2}+(M^{F}_{-})^{2})} \Bigg],
\end{eqnarray}
in the vicinity of $\Theta=(2n+1)\pi$.
The coefficient factor $2$ for mass squared $(M^{F}_{+})^{2}+(M^{F}_{-})^{2}$ seems strange,
and it might have come from the case that the leading-order integration of the VP function 
which could not take into account the orders of poles/singularities due to its roughness of the approximation,
though our calculation in the approximation is correct.
$Z^{-1}_{\Theta}m^{2}_{\Theta}$ in (163) will vanish at $|m_{R}|=|m_{L}|=0$,
becomes the gap equation for $|m_{D}|$ of the ordinary NJL model.
The vertex of fermion and phason is found as follows:
\begin{eqnarray}
{\cal L}_{\Psi\Theta} &\equiv& -g_{\Psi\bar{\Psi}\Theta}\overline{\Psi_{MN}}\tilde{\Theta}^{(ren)}i\gamma_{5}\otimes\tau_{1}\Psi_{MN}, \qquad g_{\Psi\bar{\Psi}\Theta} = \sqrt{Z_{\Theta}}, \qquad \tilde{\Theta}^{(ren)} \equiv Z^{-1/2}_{\Theta}\tilde{\Theta}.
\end{eqnarray}
While, the phason decay constant is given by
\begin{eqnarray}
F^{2}_{\Theta} &\approx& 4|m_{D}|^{2}Z^{-1}_{\Theta} = \frac{|m_{D}|^{2}}{2\pi^{2}}\ln\frac{\Lambda^{2}}{2((M^{F}_{+})^{2}+(M^{F}_{-})^{2})}.
\end{eqnarray}
This result should be compared with 
$f^{2}_{\pi}\sim\frac{|m_{D}|^{2}}{4\pi^{2}}\ln\frac{\Lambda^{2}}{|m_{D}|^{2}}$ 
of the chiral limit of the ordinary NJL model obtained by the Nambu-Goldstone theorem
~[112,147,160].
Since we cannot use the Nambu-Goldstone theorem in the problem we consider here,
this is an approximate expression for $F^{2}_{\Theta}$,
though it is important because it bridges between the UV scale ( defined by the condensation scale $\Lambda$ ) 
and the IR scale ( defined by dynamical mass $|m_{D}|$ ). 
In principle, our model is independent from electroweak symmetry breaking,
though if we apply our theory to it, 
the decay constant of phason obtains the constraint from masses of charged and neutral bosons, 
$m^{2}_{W},m^{2}_{Z}\sim (g_{SU(2)_{L}}F_{\Theta}/2)^{2}$
( relations coming from the dynamical Higgs mechanism of electroweak symmetry breaking ),
and becomes $F_{\Theta}\sim 250$ GeV~[244].
Therefore, from (167),
\begin{eqnarray}
|m_{R}| &\approx& \frac{\Lambda}{2}\exp\Bigl( -\frac{\pi^{2}F^{2}_{\Theta}}{|m_{D}|^{2}} \Bigr)
\sim \frac{\sqrt{(M^{F}_{+})^{2}+(M^{F}_{-})^{2}}}{2}, 
\end{eqnarray}
and then we obtain $|m_{R}|\sim 10^{10}-10^{11}$ GeV when $\Lambda\sim 10^{16}$ GeV 
and $F_{\Theta}/|m_{D}|\sim 2$. 
In this case, seesaw mass becomes  $|m_{D}|^{2}/|m_{R}|\sim 10^{2}$ eV.

\section{Phason Mass in the Supersymmetric Nambu$-$Jona-Lasinio Model
with Left-Right Asymmetric Majorana Mass Terms}

In this section, we evaluate phason mass 
in the framework of the supersymmetric Nambu$-$Jona-Lasinio ( SNJL ) model 
with the left-right-asymmetric Majorana mass terms~[178]. 
A generalization of our analysis of the NJL-type model to the SNJL-type model is interesting,
since the quadratic divergence of $V^{NJL+M}_{eff}$ in the NJL-type model 
( which causes an inherent problem in the NJL case ) will be removed due to ${\cal N}=1$ SUSY, 
the superpartners of phason/majoron would appear ( phasino/majorino )
if we set the theory appropriately for them, 
similar to the case of axion and axino~[101],
and some cosmological results also be modified 
especially in an evaluation of reaction/decay rates of phason/phasino.
The Lagrangian of the SNJL model we consider here is defined by
\begin{eqnarray}
{\cal L}_{SNJL} &\equiv& \Bigl[ (1-\Delta^{2}\theta^{2}\bar{\theta}^{2})[ \Phi^{\dagger}_{+}\Phi_{+} +\Phi^{\dagger}_{-}\Phi_{-}] + G\Phi^{\dagger}_{+}\Phi^{\dagger}_{-}\Phi_{+}\Phi_{-} \Bigr]_{\theta\theta\bar{\theta}\bar{\theta}}    \nonumber \\
& & 
+ \Bigl[ 
\frac{m^{\dagger}_{R}}{2}\Phi_{+}\Phi_{+} + \frac{m_{L}}{2}\Phi_{-}\Phi_{-}  
\Bigr]_{\theta\theta} 
+ \Bigl[   
\frac{m_{R}}{2}\Phi^{\dagger}_{+}\Phi^{\dagger}_{+} + \frac{m^{\dagger}_{L}}{2}\Phi^{\dagger}_{-}\Phi^{\dagger}_{-}
\Bigr]_{\bar{\theta}\bar{\theta}}.
\end{eqnarray}
$\Phi_{\pm}$ are chiral superfields~[238].
The SUSY breaking mass $\Delta$ is introduced to avoid the non-renormalization theorem 
in a dynamical generation of mass:
If SNJL model keeps ${\cal N}=1$ SUSY, it cannot generate a dynamical Dirac mass~[24].
Through the method of SUSY auxiliary fields of composites
~[24,28,36,175,176,177,178], 
we get
\begin{eqnarray}
V^{SNJL}_{eff} = V^{NJL+M}_{eff} + V^{B}_{eff},    \quad 
V^{NJL+M}_{eff} = \frac{|m_{D}|^{2}}{G} + i\ln{\rm Det}\Omega^{F}_{M},  \quad
V^{B}_{eff} = - 2i\ln{\rm Det}\Omega^{B}_{M}. 
\end{eqnarray}
The matrix $\Omega^{B}_{M}$ for the sector of elementary bosons is
\begin{eqnarray}
\Omega^{B}_{M} &\equiv& \left(
\begin{array}{cc}
\Box -|m_{R}|^{2} -|m_{D}|^{2} -\Delta^{2} & m_{R}m^{\dagger}_{D} + m^{\dagger}_{L}m_{D} \\
m^{\dagger}_{R}m_{D} + m_{L}m^{\dagger}_{D} & \Box - |m_{L}|^{2} - |m_{D}|^{2} -\Delta^{2} 
\end{array}
\right).
\end{eqnarray}
Here, we take the SUSY breaking mass $\Delta$ as real.
Then we get
\begin{eqnarray}
V^{SNJL}_{eff} &=& \frac{|m_{D}|^{2}}{G} + \frac{1}{16\pi^{2}}\Bigg[ 2\Lambda^{2}\Delta^{2} + \Lambda^{4}\ln\frac{[\Lambda^{2}+(M^{B}_{+})^{2}][\Lambda^{2}+(M^{B}_{-})^{2}]}{[\Lambda^{2}+(M^{F}_{+})^{2}][\Lambda^{2}+(M^{F}_{+})^{2}]}  \nonumber \\
& & 
-(M^{B}_{+})^{4}\ln\Bigl(1+\frac{\Lambda^{2}}{(M^{B}_{+})^{2}}\Bigr)   
-(M^{B}_{-})^{4}\ln\Bigl(1+\frac{\Lambda^{2}}{(M^{B}_{-})^{2}}\Bigr)   
+(M^{F}_{+})^{4}\ln\Bigl(1+\frac{\Lambda^{2}}{(M^{F}_{+})^{2}}\Bigr)
+(M^{F}_{-})^{4}\ln\Bigl(1+\frac{\Lambda^{2}}{(M^{F}_{-})^{2}}\Bigr) 
\Bigg].   \nonumber \\
& & 
\end{eqnarray}
The mass eigenvalues of bosons become
\begin{eqnarray}
M^{B}_{\pm} &=& \sqrt{ |m_{D}|^{2} + \frac{|m_{R}|^{2}+|m_{L}|^{2}}{2} + \Delta^{2} \mp\frac{1}{2}
\sqrt{ (|m_{R}|^{2}-|m_{L}|^{2})^{2} + 4|m_{D}|^{2}(|m_{R}|^{2}+|m_{L}|^{2}+2|m_{R}||m_{L}|\cos\Theta)
}}.
\end{eqnarray}
We summarize the results of the following derivatives: 
The first derivatives, 
\begin{eqnarray}
\frac{\partial V^{SNJL}_{eff}}{\partial|m_{D}|} &=&  \frac{2|m_{D}|}{G}+\frac{1}{8\pi^{2}}\Bigg[ 
\frac{\partial(M^{B}_{+})^{2}}{\partial|m_{D}|}F^{B}_{+} + 
\frac{\partial(M^{B}_{-})^{2}}{\partial|m_{D}|}F^{B}_{-} -
\frac{\partial(M^{F}_{+})^{2}}{\partial|m_{D}|}F^{F}_{+} -
\frac{\partial(M^{F}_{-})^{2}}{\partial|m_{D}|}F^{F}_{-}
\Bigg],     \\
\frac{\partial V^{SNJL}_{eff}}{\partial \Theta} &=&  -\frac{1}{8\pi^{2}}
\Bigg[ 
 \frac{\partial (M^{F}_{+})^{2}}{\partial\Theta}F^{F}_{+}
+\frac{\partial (M^{F}_{-})^{2}}{\partial\Theta}F^{F}_{-} 
-\frac{\partial (M^{B}_{+})^{2}}{\partial\Theta}F^{B}_{+}  
-\frac{\partial (M^{B}_{-})^{2}}{\partial\Theta}F^{B}_{-} 
\Bigg]    \nonumber \\
&=& -\frac{1}{8\pi^{2}}\frac{\partial(M^{F}_{+})^{2}}{\partial\Theta}
\bigl[ 
F^{F}_{+}-F^{F}_{-}-F^{B}_{+}+F^{B}_{-}
\bigr],  
\end{eqnarray}
where, we have used 
\begin{eqnarray}
F^{B}_{\pm} &\equiv& \Lambda^{2} - (M^{B}_{\pm})^{2}\ln\Bigl( 1+\frac{\Lambda^{2}}{(M^{B}_{\pm})^{2}} \Bigr), \\
J^{B}_{\pm} &\equiv& \frac{\Lambda^{2}}{\Lambda^{2}+(M^{B}_{\pm})^{2}} -\ln\Bigl(1+\frac{\Lambda^{2}}{(M^{B}_{\pm})^{2}} \Bigr),    
\end{eqnarray}
and
\begin{eqnarray}
\frac{\partial (M^{B}_{+})^{2}}{\partial\Theta} &=& \frac{\partial(M^{F}_{+})^{2}}{\partial\Theta} = - \frac{\partial(M^{B}_{-})^{2}}{\partial\Theta} = -\frac{\partial(M^{F}_{-})^{2}}{\partial\Theta}, 
\end{eqnarray}
is satisfied.
$F^{F}_{+}=F^{B}_{+}$, $F^{F}_{-}=F^{B}_{-}$, 
$J^{F}_{+}=J^{B}_{+}$ and $J^{F}_{-}=J^{B}_{-}$ hold at $\Delta=0$.
While, the second derivatives are
\begin{eqnarray}
\frac{\partial^{2}V^{SNJL}_{eff}}{\partial|m_{D}|^{2}} &=& \frac{2}{G} - \frac{1}{8\pi^{2}}
\Bigg[  
\frac{\partial^{2}(M^{F}_{+})^{2}}{\partial|m_{D}|^{2}}\bigl( F^{F}_{+}-F^{B}_{+} \bigr) 
+\Bigl(\frac{\partial (M^{F}_{+})^{2}}{\partial|m_{D}|}\Bigr)^{2} \bigl( J^{F}_{+}-J^{B}_{+} \bigr) 
+ (+\to -)
\Bigg],   \\
\frac{\partial^{2}V^{SNJL}_{eff}}{\partial|m_{D}|\partial\Theta} &=&  
-\frac{1}{8\pi^{2}}\Bigg[
\frac{\partial^{2}(M^{F}_{+})^{2}}{\partial|m_{D}|\partial\Theta}(F^{F}_{+}-F^{B}_{+})+\frac{\partial(M^{F}_{+})^{2}}{\partial|m_{D}|}\frac{\partial(M^{F}_{+})^{2}}{\partial\Theta}(J^{F}_{+}-J^{B}_{+})
+ (+\to -)
\Bigg],
\\
\frac{\partial^{2}V^{SNJL}_{eff}}{\partial\Theta^{2}} &=& -\frac{1}{8\pi^{2}}\Bigg[ \frac{\partial^{2}(M^{F}_{+})^{2}}{\partial\Theta^{2}}\bigl[ F^{F}_{+}-F^{F}_{-}-F^{B}_{+}+F^{B}_{-} \bigr] 
+ \Bigl( \frac{\partial(M^{F}_{+})^{2}}{\partial\Theta} \Bigr)^{2}\bigl[ J^{F}_{+}+J^{F}_{-}-J^{B}_{+}-J^{B}_{-}\bigr] \Bigg].
\end{eqnarray}
From these results, 
we find $\Theta=0$ is now stable, $\frac{\partial V^{SNJL}_{eff}}{\partial\Theta}$ monotonically 
increases from $\Theta=0$ to $\Theta=\pi$, 
while $\Theta=\pi$ is unstable and tachyonic to the $\Theta$-direction,
by both the first and second derivatives given above.
Thus, the CP-conserving condition for this case becomes
\begin{eqnarray}
m_{D} &=& m^{\dagger}_{D}, \quad m_{R} = -m^{\dagger}_{R}, \quad m_{L} = -m^{\dagger}_{L},    \nonumber \\
\theta_{D} &=& 0,    \nonumber \\
\theta_{R} &=& \Bigl(2j+\frac{1}{2}\Bigr)\pi, \quad \theta_{L} = \Bigl(2l+\frac{3}{2}\Bigr)\pi, \quad {\rm or} \quad 
\theta_{R} = \Bigl(2j+\frac{3}{2}\Bigr)\pi, \quad \theta_{L} = \Bigl(2l+\frac{1}{2}\Bigr)\pi, \quad
j,l\in {\bf Z}.
\end{eqnarray}
Therefore, we get the following expression at $\Lambda\gg |m_{R}|,\Delta\gg |m_{D}|,|m_{L}|$,
\begin{eqnarray}
Z^{-1}_{\Theta}m^{2}_{\Theta} &=& - \frac{1}{2\pi^{2}}\Bigl[\frac{|m_{R}||m_{L}|}{|m_{R}|+|m_{L}|} \frac{1}{\sqrt{(|m_{R}|-|m_{L}|)^{2}+4|m_{D}|^{2}}}\Bigr]  \nonumber \\
& & \times \Bigl\{ 
(M^{F}_{+})^{2}\ln\Bigl(1+\frac{\Lambda^{2}}{(M^{F}_{+})^{2}}\Bigr)
-(M^{B}_{+})^{2}\ln\Bigl(1+\frac{\Lambda^{2}}{(M^{B}_{+})^{2}}\Bigr)   \nonumber \\
& & \qquad 
-(M^{F}_{-})^{2}\ln\Bigl(1+\frac{\Lambda^{2}}{(M^{F}_{-})^{2}}\Bigr)
+(M^{B}_{-})^{2}\ln\Bigl(1+\frac{\Lambda^{2}}{(M^{B}_{-})^{2}}\Bigr)\Bigr\}   \nonumber \\
&\approx& \frac{|m_{R}||m_{L}|}{2\pi^{2}}\cdot\frac{\Lambda^{4}[2\Lambda^{2}\Delta^{2}+\Delta^{2}(2|m_{D}|^{2}+|m_{R}|^{2}+|m_{L}|^{2}+\Delta^{2})]}{(\Lambda^{2}+(M^{F}_{+})^{2})(\Lambda^{2}+(M^{F}_{-})^{2})(\Lambda^{2}+(M^{B}_{+})^{2})(\Lambda^{2}+(M^{B}_{-})^{2})}  
\approx \frac{|m_{R}||m_{L}|}{2\pi^{2}}\cdot\frac{2\Delta^{2}}{\Lambda^{2}}.
\end{eqnarray}
This expression gives $Z^{-1}_{\Theta}m^{2}_{\Theta}=0$ at $\Delta=0$ 
( This limit is physically meaningless since $\Delta\ne 0$ must be satisfied 
to obtain a finite $|m_{D}|$ in the SNJL model ).
$Z^{-1}_{\Theta}(m_{\Theta})^{2}$ is $2\Delta^{2}/\Lambda^{2}$ times smaller than that of the NJL+Majorana case. 
The limit $|m_{R}|=|m_{L}|$ in the first expression gives $Z^{-1}_{\Theta}m^{2}_{\Theta}=0$, 
the same with the NJL+Majorana case.
If we use $|m_{R}|=10^{11}$ GeV, $|m_{L}|=10^{2}$ eV,
$\Delta=1$ TeV and $\Lambda=10^{16}$ GeV with assuming $Z^{-1}_{\Theta}\sim{\cal O}(1)$,
$m_{\Theta}\sim 2\times 10^{-3}$ eV, and phason becomes very light particle, somewhat similar to axion
( though it strongly depends on the ratio of the SUSY breaking scale $\Delta$ 
and the condensation scale $\Lambda$ )~[84].
In this case, the compton length becomes $10^{11}$ fm.

\vspace{2mm}

To evaluate VP functions of bosonic sector, again we should choose mass parameters as
$m_{D}=m^{\dagger}_{D}$, $m_{R}=-m^{\dagger}_{R}$ and $m_{L}=-m^{\dagger}_{L}$ 
in the propagator $(\Omega^{B}(k))^{-1}$ used for an expansion of $V^{SNJL}_{eff}$.
We use the following partition
\begin{eqnarray}
\Omega^{B} &\to& \Omega^{B} - \Sigma^{B}_{(1)} -\Sigma^{B}_{(2)}.
\end{eqnarray}
The definitions of $\Omega^{B}(k)$ and the fluctuation matrices of scalar sector become
\begin{eqnarray}
(\Omega^{B}(k))^{-1} &=& \frac{1}{D_{B}(k)}\left(
\begin{array}{cc}
k^{2}-|m_{L}|^{2}-|m_{D}|^{2}-\Delta^{2} & -i|m_{D}|(|m_{R}|-|m_{L}|)   \\
i|m_{D}|(|m_{R}|-|m_{L}|) & k^{2}-|m_{R}|^{2}-|m_{D}|^{2}-\Delta^{2}
\end{array}
\right),   \\
D_{B}(k) &\equiv& \bigl(s+(M^{B}_{+})^{2}\bigr)\bigl(s+(M^{B}_{-})^{2}\bigr),
\end{eqnarray}
and
\begin{eqnarray}
\Sigma^{B}_{(1)} &\equiv& \frac{|m_{D}|^{2}}{4}(\delta\Theta)^{2}\left( 
\begin{array}{cc}
1 & 0 \\
0 & 1 
\end{array}
\right),  \quad
\Sigma^{B}_{(2)} \equiv -\frac{|m_{D}|}{2}(|m_{R}|+|m_{L}|)(\delta\Theta)\left(
\begin{array}{cc}
0 & 1  \\
1 & 0 
\end{array}
\right).
\end{eqnarray}
We show the results:
\begin{eqnarray}
& & -\frac{i}{2}{\rm tr}(\Omega^{B}(k))^{-1}\Sigma^{B}_{(1)}  
= -\frac{i}{2}\frac{|m_{D}|^{2}}{4}(\delta\Theta)^{2}\frac{2}{D_{B}(k)}\Bigl[k^{2}-|m_{D}|^{2}-\frac{|m_{R}|^{2}+|m_{L}|^{2}}{2}-\Delta^{2}\Bigr],    \\
& & -\frac{i}{2}{\rm tr}\frac{1}{2}(\Omega^{B}(k))^{-1}\Sigma^{B}_{(2)}(\Omega^{B}(k+q))^{-1}\Sigma^{B}_{(2)}  \nonumber \\
& & \qquad = 
\frac{1}{D_{B}(k)D_{B}(k+q)}\frac{|m_{D}|^{2}}{4}(\delta\Theta)^{2} \nonumber \\
& & \qquad\quad \times \Bigl\{ 
(|m_{R}|^{2}+|m_{L}|^{2})\bigl[ 
(k^{2}-|m_{L}|^{2}-|m_{D}|^{2}-\Delta^{2})
((k+q)^{2}-|m_{R}|^{2}-|m_{D}|^{2}-\Delta^{2})  \nonumber \\ 
& &  \qquad \quad + 
(k^{2}-|m_{R}|^{2}-|m_{D}|^{2}-\Delta^{2})
((k+q)^{2}-|m_{L}|^{2}-|m_{D}|^{2}-\Delta^{2})   \bigr]
-2|m_{D}|^{2}(|m_{R}|^{2}-|m_{L}|^{2})^{2} 
\Bigr\}.
\end{eqnarray}
The momentum integration of the first VP function given above diverges quadratically,
while the second one logarithmically diverges.
Then we get
\begin{eqnarray}
& & -\frac{i}{2}{\rm tr}\int_{k}(\Omega^{B}(k))^{-1}\Sigma^{B}_{(1)} \approx 
\frac{1}{8\pi^{2}}(\tilde{\Theta})^{2}\Bigg(\Lambda^{2}-((M^{B}_{+})^{2}+(M^{B}_{-})^{2})\ln\frac{\Lambda^{2}}{(M^{B}_{+})^{2}+(M^{B}_{-})^{2}} \Bigg),  \\
& & -\frac{i}{2}{\rm tr}\int_{k}\frac{1}{2}(\Omega^{B}(k))^{-1}\Sigma^{B}_{(2)}(\Omega^{B}(k))^{-1}\Sigma^{B}_{(2)}  
\approx \frac{1}{8\pi^{2}}\tilde{\Theta}^{2}(|m_{R}|^{2}+|m_{L}|^{2})
\ln\frac{\Lambda^{2}}{2((M^{B}_{+})^{2}+(M^{B}_{-})^{2})}.   
\end{eqnarray}
Note that $(M^{B}_{+})^{2}+(M^{B}_{-})^{2}=2|m_{D}|^{2}+|m_{R}|^{2}+|m_{L}|^{2}+\Delta^{2}$.
Then, one finds the gap equation from the following definition of phason propagator:
\begin{eqnarray}
G^{-1}_{\Theta}(q) &\equiv& -\frac{2}{G} + \frac{i}{4}\bigl( \Pi^{F}_{\Theta\Theta}(q) - \Pi^{B}_{\Theta\Theta}(q) \bigr) 
= Z^{-1}_{\Theta}(q^{2}-m^{2}_{\Theta}),   \\
-Z^{-1}_{\Theta}m^{2}_{\Theta} &=& -\frac{2}{G} +\frac{1}{8\pi^{2}}\Bigg[ 
\bigl( (M^{B}_{+})^{2}+(M^{B}_{-})^{2} \bigr)
\ln\frac{\Lambda^{2}}{(M^{B}_{+})^{2}+(M^{B}_{-})^{2}}
-
\bigl( (M^{F}_{+})^{2}+(M^{F}_{-})^{2} \bigr)
\ln\frac{\Lambda^{2}}{(M^{F}_{+})^{2}+(M^{F}_{-})^{2}}
\Bigg].      \nonumber \\
& & 
\end{eqnarray}
Here, we have omitted minor difference between the results of the boson and fermion sectors, 
and dropped the factor 2,
from the reason of a physical consideration to construct the gap equation,
and which coincides with the previous result in literature~[178].

\vspace{2mm}

Before closing this section, let us comment on the superpartner of phason, namely "phasino."
If we consider a Coleman-Weinberg type effective potential~[39,228],
\begin{eqnarray}
V^{SNJL}_{eff(CW)} &=& \frac{m^{2}_{\Theta}}{2}\tilde{\Theta}^{2}_{c} + \frac{\lambda_{\Theta}}{4!}\tilde{\Theta}^{4}_{c} + \frac{1}{16\pi^{2}}\Bigl[m_{\Theta}^{4}\ln\frac{m^{2}_{\Theta}}{\Lambda^{2}} -m^{4}_{phasino}\ln\frac{m^{2}_{phasino}}{\Lambda^{2}}\Bigr], 
\end{eqnarray}
$m^{2}_{\Theta}\ge m^{2}_{phasino}$ should be satisfied ( $m_{phasino}$; mass of superpartner of phason ) 
for stability of the effective potential.

\section{The Schwinger-Dyson Equations in the Gauge Theory with the Majorana Mass Terms}

In this section, we will examine dynamical mass functions of a gauge model
supplemented by "current" Majorana mass terms, by the Schwinger-Dyson ( SD ) formalism.
In this model for neutrino seesaw mass, we will meet notions of 
bare, running and physical neutrino masses~[5].
We use the terminology of explicit, spontaneous, dynamical symmetry breakings as
"symmetry broken by a ( renormalized ) parameter without spontaneous mechanism",
"symmetry breaking at the tree-level, such as a Goldstone-Higgs model",
"symmetry breaking by the BCS-NJL mechanism".

\subsection{Model Building for the Type-II-like Neutrino Seesaw Mechanism}

We will consider three choices of gauge symmetries, $U(1)$, $SU(2)$ and $SU(N_{c})$ ( $N_{c}\ge 3$ ). 
It will be explained that our choice of gauge symmetries is restricted from 
the physical situations we consider here.

\vspace{2mm}

In this paper, we consider the case where two $U(1)$ ( vector and chiral ) symmetries 
are broken by both right- and left- Majorana mass terms.
If fermions couple with a $U(1)$-gauge field,
an explicit $U(1)$ symmetry breaking causes a serious trouble, 
the gauge field acquires mass through quantum radiative corrections:
In that case, in the vacuum polarization
\begin{eqnarray}
\Pi_{\mu\nu}(q) &=& -i(g_{\mu\nu}q^{2}-q_{\mu}q_{\nu})\Pi(q^{2})+ig_{\mu\nu}\Pi_{2}(q^{2}),
\end{eqnarray}
we cannot deny the possibility of non-vanishing $\Pi_{2}(q^{2})$,
and it can give a photon mass.
Due to the non-conservation of vector current $\partial_{\mu}j^{\mu}_{V}\ne 0$,
the transversal condition of a VP function $k_{\mu}\Pi_{\mu\nu}=0$ can also be violated.
In Ref.~[159], it was proven that 
a proper vertex correction of $\Pi_{\mu\nu}$ recovers the $U(1)$-gauge invariance 
$k_{\mu}\Pi_{\mu\nu}=0$ in the case of superconductivity, a spontaneous symmetry breakdown,
with a dynamical Higgs mechanism;
it does not work in an explicit symmetry breaking due to the number of physical degrees of freedom.
We should maintain gauge-invariance, at least a BRS-invariance, for keeping unitarity of the theory~[123].
A gauge theory which will permit us to make physical interpretation
must be constructed on a physical Hilbert space ( positive metric ),
$|phys\rangle\in {\cal V}_{phys}$, $Q_{BRS}|phys\rangle =0$ ( $Q_{BRS}$: BRS charge )~[123].
We should also take care about the unitarity of S-matrix, 
while at least we consider here a Lagrangian which is Hermitian. 
If we wish to consider a local $U(1)_{V}$ gauge symmetry 
from the beginning of a theory for neutrino Majorana masses,
we have to employ a Higgs potential and the Higgs mechanism or a dynamical breaking of $U(1)_{V}$
for a possible mechanism of the origin of Majorana mass terms.
If we can consider the case where mass functions $B_{R}$, $B_{L}$ and $B_{D}$ are all $SU(N_{c})$-singlet, 
$SU(N_{c})$ gauge symmetry is not broken in an SD equation,
while $U(1)_{V}$ is broken by a "possible mechanism".
We need a flavor degree of freedom to satisfy the Pauli principle 
in $SU(N_{c})$ gauge-singlet $B_{R}$ and $B_{L}$.
It seems quite difficult to achieve the hierarchy of the seesaw situation
$|B_{R}|\gg|B_{D}|\gg|B_{L}|\ge 0$ by a vector-like interaction of gauge fields.
This special situation of the seesaw mechanism of neutrino mass restricts our model building.

\vspace{2mm}

The irreducible representation of a direct product of fundamental representation
of $SU(N_{c})$ is given by 
${\bf N}_{c}\otimes\overline{\bf N}_{c}={\bf 1}\oplus ({\bf N}^{2}_{c}-{\bf 1})$, 
${\bf N}_{c}\otimes{\bf N}_{c}=\frac{\bf 1}{\bf 2}{\bf N}_{c}({\bf N}_{c}-{\bf 1})\oplus\frac{\bf 1}{\bf 2}{\bf N}_{c}({\bf N}_{c}+{\bf 1})$.
In the case of $SU(3_{c})$, ${\bf 3}\otimes {\bf 3}=\bar{\bf 3}_{a}\oplus{\bf 6}_{s}$,
and thus the color symmetry will be broken by a diquark-type fermion pair.
For the sake of simplicity of our formulation,
we wish to choose $SU(2_{c})$ gauge symmetry.
In this case, a diquark-type pair formation for 
generating Majorana-type mass functions can belong to $SU(2_{c})$-singlet
( ${\bf 2}\otimes{\bf 2}={\bf 1}\oplus{\bf 3}$ ),
and then the $SU(2_{c})$-gauge symmetry is not broken by Majorana masses.
For satisfying the Pauli principle of Majorana-type fermion bilinears,
we also introduce $SU(2_{f})_{R}\times SU(2_{f})_{L}$ chiral group, and take the flavor singlet.
Since we keep the $SU(2_{c})$ symmetry manifestly,
several relations of renormalization constants and Ward-Takahashi identities also become simple.
The situation of our model has an essential difference with 
a diquark system or a color superconductor of $SU(3_{c})$.

\vspace{2mm}

If our theory has a vector-like gauge coupling,
it does not suffer from a non-Abelian quantum anomaly,
and it is identically guaranteed in the case of $SU(2_{c})$,
though we cannot choose different gauge coupling to the right-
and left- handed fermions to keep the gauge symmetry.
While, the theory will suffer from axial anomaly. 
In cases of $SU(N_{c})$ with $N_{c}\ge 3$, Majorana mass terms will break gauge symmetries.  
A Yukawa coupling also be colored, and then an $SU(N_{c})$ gauge symmetry is broken at the tree level, 
the gauge fields become Proca, and then it is difficult ( though, not impossible in principle ) 
to use the Proca fields in a dynamics described by the SD formalism:
It is desirable to avoid a breaking of a gauge symmetry which is relevant
for a dynamical mass generation inside an SD-equation, at least at the tree-level
( for constructing a tractable model ). 
By several examinations, we recognize the fact that 
it seems difficult to make the $U(1)$-gauge symmetry as a local one by the Higgs mechanism prescription
for providing left-right asymmetric Majorana mass terms which will break the global $U(1)$-symmetry explicitly
( note that, here we are mentioning on a Majorana mass generation from a zero-mass model to a finite-mass case ),
except a GUT approach.
Since we do not need a local $U(1)$-gauge symmetry for our discussion of neutrino seesaw mechanism, 
in conclusion, we will examine an $SU(2_{c})$-gauge model with left-right asymmetric Majorana masses,
only $SU(2_{c})$ is gauged.
The bare Majorana mass parameters are introduced from outside of the theory as external parameters.
Several four-fermion interactions of an $SU(2_{c})$-gauged NJL-type model can be used 
for generating Majorana mass parameters with a large-hierarchy for seesaw machanism by the model itself,
though it means that we have to depart from the gauge principle~[224,247].

\vspace{2mm}

The symmetry of the Lagrangian becomes
\begin{eqnarray}
SU(2_{c})\times SU(2_{f})_{R} \times SU(2_{f})_{L}.
\end{eqnarray}
Both $U(1)_{V}$ and $U(1)_{A}$ are broken by Majorana masses and an axial anomaly.
The symmetry of the Lagrangian with zero-Majorana masses becomes $SU(2N_{f})$ ( we will discuss later )~[116].
We consider the generic case of broken left-right symmetry by mass terms.
We summarize the $SU(2_{c})$ gauge transformation: 
\begin{eqnarray}
\xi' &=& e^{i\omega_{A}t_{A}}\xi, \qquad \eta' = e^{i\omega_{A}t_{A}}\eta,    \\
\psi'_{MR} &=& e^{i\omega\gamma_{5}T_{A}}\psi_{MR},  \qquad 
\psi'_{ML} = e^{i\omega_{A}{\cal C}^{-1}\gamma_{5}T_{A}{\cal C}}\psi_{ML},   \\
\overline{\psi'_{MR}} &=& \overline{\psi_{MR}}e^{-i\omega_{A}\gamma_{5}{\cal C}^{-1}T_{A}{\cal C}}, \qquad
\overline{\psi'_{ML}}  =  \overline{\psi_{ML}}e^{-i\omega_{A}\gamma_{5}T_{A}},   \\
\Psi'_{MN} &=& e^{i\omega_{A}(\gamma_{5}\otimes\tau_{3}){\cal T}_{A}}\Psi_{MN},   \qquad
\overline{\Psi'_{MN}} = \overline{\Psi_{MN}}e^{i\omega_{A}(\gamma_{5}\otimes\tau_{3}){\cal T}^{*}_{A}},
\end{eqnarray}
where, the definitions of several matrices are
\begin{eqnarray}
& & t_{A} = \frac{\rho_{A}}{2}, \quad [t_{A},t_{B}] = if_{ABC}t_{C}, \quad 
T_{A} \equiv \left(
\begin{array}{cc}
t_{A} & 0 \\
0 & t^{T}_{A} 
\end{array}
\right), \qquad {\cal T}_{A} \equiv \left(
\begin{array}{cc}
T_{A} & 0 \\
0 & T^{T}_{A} 
\end{array}
\right),    \nonumber \\
& & T^{\dagger}_{A} = T_{A}, \quad 
T^{T}_{A} = T^{*}_{A} = -\gamma^{0}T^{\dagger}_{A}\gamma^{0}, 
\quad t^{T}_{A} = -\rho_{2}t_{A}\rho_{2},    \quad
{\cal C} \equiv \left(
\begin{array}{cc}
\rho_{2} & 0 \\
0 & \rho_{2}
\end{array}
\right) = {\cal C}^{-1} = {\cal C}^{\dagger}, 
\end{eqnarray}
Here, $t_{A}$ are $SU(2_{c})$ Hermitian generators.
The $SU(N_{f})$ flavor rotation is also defined in the same manner:
\begin{eqnarray}
\xi' = e^{i\omega^{R}_{i}\Upsilon_{i}}\xi, \qquad \eta' = e^{i\omega^{L}_{i}\Upsilon_{i}}\eta.
\end{eqnarray}
Here, $\Upsilon_{i}$ are generators of an $SU(N_{f})$ flavor rotation.

\vspace{2mm}

Now, we introduce the following manifestly $SU(2_{c})$-gauge invariant Lagrangian:
\begin{eqnarray}
{\cal L}_{SU(2)} &\equiv& {\cal L}_{gauge} + {\cal L}_{matter},    \\
{\cal L}_{gauge} &\equiv& -\frac{1}{4}G^{A}_{\mu\nu}G^{A\mu\nu} +
B^{A}\partial^{\mu}G^{A}_{\mu} +\frac{\xi}{2}(B^{A})^{2} - i\bar{c}^{A}\partial^{\mu}D_{\mu}c^{A},   \\
G^{A}_{\mu\nu} &\equiv& \partial_{\mu}G^{A}_{\nu} -\partial_{\nu}G^{A}_{\mu} +g^{(0)}f_{ABC}G^{B}_{\mu}G^{C}_{\nu},   \\
D_{\mu}c^{A} &\equiv& \partial_{\mu}c^{A} +g^{(0)}f_{ABC}G^{B}_{\mu}c^{C},   \\
{\cal L}_{matter} &\equiv& \frac{1}{2}\overline{\Psi_{MN}}\bigl(i\partfey\otimes \hat{1} +g^{(0)}\afey^{A}(\gamma_{5}\otimes\tau_{3}){\cal T}_{A}-{\cal M}^{(0)}_{M} \bigr)\Psi_{MN}.   
\end{eqnarray}
Here, $B^{A}$ are the Nakanishi-Lautrup $B$-fields~[158,129].
The matter fermion fields are taken as the fundamental representation of $SU(2_{c})$.
The mass matrix ${\cal M}^{(0)}_{M}$ is defined by
\begin{eqnarray}
{\cal M}^{(0)}_{M} &\equiv& \left(
\begin{array}{cc}
{\cal M}^{(0)}_{R} & {\cal M}^{(0)}_{D} \\
{\cal M}^{(0)}_{D} & {\cal M}^{(0)}_{L}
\end{array}
\right),       \\
{\cal M}^{(0)}_{R} &\equiv& [ (m^{(0)}_{R})^{\dagger}P_{+} +m^{(0)}_{R}P_{-} ]\rho_{2}\Upsilon_{2},    \\
{\cal M}^{(0)}_{L} &\equiv& [ (m^{(0)}_{L})^{\dagger}P_{+} +m^{(0)}_{L}P_{-} ]\rho_{2}\Upsilon_{2},    \\
{\cal M}^{(0)}_{D} &\equiv& (m^{(0)}_{D})^{\dagger}P_{+} + m^{(0)}_{D}P_{-}.
\end{eqnarray}
( $N_{c}=2$. )
The Majorana masses are taken to be as 
$SU(2_{c})$, $SU(2_{f})_{R}$ and $SU(2_{f})_{L}$ singlets by multiplying 
anti-symmetric matrices of color and flavor spaces $\rho_{2}$ and $\Upsilon_{2}$.
The $U(1)_{V}$ current is not conserved by the Majorana mass terms,
the $U(1)_{A}$ current is suffered from an axial anomaly,
while the currents of $SU(2_{f})_{V}$ and $SU(2_{f})_{A}$ are conserved.
In the context of neutrino physics, the $SU(2_{c})$ fields are hypothetical gauge fields 
they might relevant for dynamics of the system we consider here.
Let us discuss about symmetry breaking schema of our model.
Due to pseudo-real nature of $SU(2_{c})$, the Lagrangian has a large chiral symmetry,
i.e. $SU(2N_{f})$ at the zero-mass case ( see below )~[116].
We summarize the following breaking schema:
\begin{itemize}
\item
(1) $m^{(0)}_{D}=m^{(0)}_{R}=m^{(0)}_{L}=0$ and $m^{D}_{dyn}\ne 0$ ( dynamical Dirac mass );
$SU(4_{f}) \to Sp(4_{f})$.
\item
(2) $m^{(0)}_{D}=m^{(0)}_{R}=m^{(0)}_{L}=0$, $m^{D}_{dyn}=0$, $m^{R}_{dyn}\ne 0$, $m^{L}_{dyn}\ne 0$
with $m^{R}_{dyn}=m^{L}_{dyn}$
( both of the right and left Majorana masses are taken to be $SU(2_{f})_{R,L}$ singlet );
no symmetry breaking.
\item 
(3) $m^{(0)}_{D}=m^{(0)}_{R}=m^{(0)}_{L}=0$, $m^{D}_{dyn}=0$, $m^{R}_{dyn}\ne 0$, $m^{L}_{dyn}\ne 0$
with $m^{R}_{dyn}\ne m^{L}_{dyn}$;
$SU(4_{f}) \to SU(2_{f})_{R}\times SU(2_{f})_{L}$.
\item
(4) $m^{(0)}_{D}=m^{(0)}_{R}=m^{(0)}_{L}=0$, $m^{D}_{dyn}\ne0$, $m^{R}_{dyn}\ne 0$, $m^{L}_{dyn}\ne 0$
with $m^{R}_{dyn}\ne m^{L}_{dyn}$;
$SU(4_{f}) \to SU(2_{f})_{R+L}$.
\item
(5) $m^{(0)}_{D}=0$, $m^{(0)}_{R}\ne 0$, $m^{(0)}_{L}\ne 0$, 
$m^{D}_{dyn}\ne0$, $m^{R}_{dyn}\ne 0$, $m^{L}_{dyn}\ne 0$
with $m^{R}_{dyn}\ne m^{L}_{dyn}$;
$SU(2_{f})_{R} \times SU(2_{f})_{L} \to SU(2_{f})_{R+L}$.
\item
(6) $m^{(0)}_{D}\ne0$, $m^{(0)}_{R}\ne 0$, $m^{(0)}_{L}\ne 0$;
no dynamical symmetry breaking, no NG boson.
\end{itemize}
Here, subscript $dyn$ indicates dynamical masses, while superscript $(0)$ implies bare quantities. 
For neutrino seesaw phenomenology, 
the case (6) may be the most preferable because it does not generate an "exact" NG boson.
Usually, in the energy region of SD equation in QCD, 
we should prepare three flavors $u$, $d$ and $s$ and $SU(3_{f})_{R}\times SU(3_{f})_{L}$ chiral group,
though our main interest is on neutrino mass, and we have no insight on this
( a desirable criterion is an asymptotic freedom, $N_{f}<11N_{c}/2$ )
and thus we have determined the number of flavor from the Pauli principle, the smallest number for it.
In generic case of flavor number $N_{f}\ge 3$,
one can introduce more complicated Majorana-type condensations.
Since we should choose $SU(2_{c})$-singlet ( antisymmetric ),
a representation of a bilinear in the flavor space must be chosen from 
${\bf N}_{f}({\bf N}_{f}-{\bf 1})/{\bf 2}$, 
namely we have $N_{f}(N_{f}-1)/2$-directions ( alignment of vacuum ) of condensates.
The generic case of a trial function of a Majorana-type condensation is therefore
given in terms of a linear combination of these $N_{f}(N_{f}-1)/2$ bases.
Moreover, we can choose different directions of condensations in the flavor space
for a left and a right Majorana condensates.

\vspace{2mm}

The matter part of our Lagrangian in the Dirac-Nambu notation becomes
\begin{eqnarray}
& & {\cal L}_{matter}  \nonumber \\
& & \quad = \frac{1}{2}\overline{\Psi_{DN}}\left(
\begin{array}{cc}
i\partfey + g^{(0)}\afey^{A}t_{A} -(m^{(0)}_{D})^{\dagger}P_{+} -m^{(0)}_{D}P_{-} & -(m^{(0)}_{L})^{\dagger}P_{+}C -m^{(0)}_{R}P_{-}C   \\
-(m^{(0)}_{R})^{\dagger}P_{+}C -m^{(0)}_{L}P_{-}C & -C^{-1}\bigl( i\partfey + g^{(0)}\afey^{A}\rho_{2}t_{A}\rho_{2} \bigr)C +(m^{(0)}_{D})^{\dagger}P_{+} +m^{(0)}_{D}P_{-}
\end{array}
\right)\Psi_{DN}.   \nonumber \\
& & 
\end{eqnarray}
As we have mentioned above, a fermion model of $SU(2_{c})$-color gauge 
has a larger flavor symmetry $SU(2N_{f})$ 
due to the pseudo-real nature of fundamental representation of $SU(2_{c})$.
To show it, we will rewrite the Lagrangian in the following form
from a Majorana-like definition of $\Psi_{PR}$:
\begin{eqnarray}
{\cal L}_{matter} &=& \Psi^{\dagger}_{PR}i\sigma^{\mu}D_{\mu}\Psi_{PR}
+ \frac{1}{2}\Psi^{T}_{PR}\sigma_{2}\rho_{2}{\cal M}^{(0)}_{PR}\Psi_{PR} 
- \frac{1}{2}\Psi^{\dagger}_{PR}\sigma_{2}\rho_{2}({\cal M}^{(0)}_{PR})^{\dagger}\Psi^{*}_{PR},   
\end{eqnarray}
where,
\begin{eqnarray}
D_{\mu} &\equiv& \partial_{\mu} -ig^{(0)}A^{A}_{\mu}t_{A}, \quad
\Psi_{PR} \equiv \left(
\begin{array}{c}
\eta   \\
(i\sigma_{2})(i\rho_{2})\xi^{*}
\end{array}
\right), \quad 
{\cal M}^{(0)}_{PR} \equiv \left(
\begin{array}{cc}
m^{(0)}_{L} & -m^{(0)}_{D}  \\
m^{(0)}_{D} & m^{(0)}_{R} 
\end{array}
\right). 
\end{eqnarray}
From this form, the breaking scheme $SU(2N_{f})\to Sp(2N_{f})$ is obvious 
when the Lagrangian has/acquires only Dirac mass and/or a chiral condensate and does not have Majoranas. 
Note that, in a generic case, a fermion field will be given by a linear combination ( mixing ) 
of particle and antiparticle bases due to the pseudo-real nature of $SU(2_{c})$.

\subsection{Renormalization}

In this subsection, we summarize the renormalization property of a non-Abelian gauge theory.
We work with the mass-independent renormalization scheme of 't Hooft and Weinberg 
( or, so-called zero-mass renormalization scheme )~[222,231],
combined with the method of mass-independent homogeneous renormalization equations.
In this renormalization scheme, the renormalization constants ( and RG equations ) do not depend on
mass parameters.
Several definitions of renormalization constants in perturbative gauge theories are given as follows:
\begin{eqnarray}
\psi^{(0)} &=& \sqrt{Z_{2}(\mu,\Lambda)}\psi^{(\mu)},    \\
A(p^{2}) &=& (Z_{2}(\mu,\Lambda))^{-1}A^{(\mu)}(p^{2}),   \\
{\bf B}(p^{2}) &=& (Z_{2}(\mu,\Lambda))^{-1}{\bf B}^{(\mu)}(p^{2}),    \\
A^{(0)}_{\mu} &=& \sqrt{Z_{3}(\mu,\Lambda)}A^{(\mu)}_{\mu},   \\
g^{(0)} &=& (Z_{3}(\mu,\Lambda))^{-3/2}Z_{1}(\mu,\Lambda)g,   \\
{\bf m}^{(0)}(\Lambda) &=& ( m^{(0)}_{D}(\Lambda), m^{(0)}_{R}(\Lambda), m^{(0)}_{L}(\Lambda) )^{T}
= Z_{m}(\mu,\Lambda){\bf m}^{(\mu)},   \\
{\bf m}^{(\mu)} &=& ( m^{(\mu)}_{D}, m^{(\mu)}_{R}, m^{(\mu)}_{L} )^{T}
= {\bf B}^{(\mu)}_{current}(p^{2})/A^{(\mu)}(p^{2}),  \\
{\bf B}(p^{2}) &=& {\bf B}_{current}(p^{2}) + {\bf B}_{dyn}(p^{2}).
\end{eqnarray}
Here, ${\bf m}^{(\mu)}$ is renormalized mass, 
and ${\bf m}^{(0)}(\Lambda)$ is bare mass defined at the UV ( ultraviolet ) cutoff scale.
$A(p^{2})$ and ${\bf B}(p^{2})$ are defined by a dressed propagator such like
\begin{eqnarray}
{\cal G}^{-1}_{M}(p) &\equiv& A(p^{2})\pfey -{\bf B}(p^{2}).
\end{eqnarray}
The explicit form of ${\cal G}^{-1}_{M}(p)$ will be given in the next subsection.
These renormalization constants are flavor-independent.
Here, $Z_{m}=Z^{R}_{m}=Z^{L}_{m}=Z^{D}_{m}$ due to the zero-mass renormalization scheme, 
namely the Majorana and Dirac mass parameters share a common renormalization constant.
The solutions of RG equations of Landau gauge QCD
( corresponds to the case $\xi=0$ ) at one-loop level give~[93,147]
\begin{eqnarray}
& & A = Z_{2} = 1,   \quad
Z_{m} = (\bar{g}^{2}(\Lambda^{2})/\bar{g}^{2}(\mu^{2}))^{\gamma_{m}},   \quad 
\gamma_{m} = \frac{c}{a}.
\end{eqnarray}
Some group-theoretical factors are summarized as follows:
\begin{eqnarray}
\sum_{C,D}f_{ACD}f_{BCD} &=& C_{2}(G)\delta_{AB},  \quad 
{\rm tr}(t_{A}t_{B}) = T(R)\delta_{AB},  \quad \sum_{A}t_{A}t_{A} = C_{2}(R),    \nonumber \\ 
c &\equiv& \frac{3}{8\pi^{2}}C_{2}(R),  \quad 
a \equiv \frac{1}{24\pi^{2}}(11C_{2}(G)-4N_{f}T(R)),   \nonumber \\
C_{2}(G=SU(N_{c})) &=& N_{c},   \quad 
T(N_{c}) =\frac{1}{2}, \quad 
C_{2}(R=N_{c}) = \frac{N^{2}_{c}-1}{2N_{c}},
\end{eqnarray}
and the RG-invariant coupling constant is given by
\begin{eqnarray}
\bar{g}^{2}(p^{2}) &=& \frac{2}{a\ln(p^{2}/\Lambda^{2}_{QCD})},  \qquad
\Lambda_{QCD} = \mu e^{-\frac{1}{ag^{2}}}.   
\end{eqnarray}
Here, $a$ is the first coefficient of the QCD $\beta$-function, 
$\Lambda_{QCD}$ in $SU(3_{c})$ case ( the definition of it shows so-called "dimensional transmutation" )
is determined experimentally, $\Lambda_{QCD}\sim$ 100 MeV~[95].
Both the ladder ( non-running gauge coupling ) and improved ladder ( running case ) approximations 
which we will use in our analysis of SD equation of our model is not gauge invariant, 
though the convergence property prefers the choice of Landau gauge in ordinary QED/QCD 
( without dynamical Majorana masses )~[93,147].
We consider it is also the case in our models.
While ${\bf B}^{(\mu)}$ and ${\bf B}$ are complex,
the renormalization constants $Z_{2}$, $Z_{3}$ and $Z_{m}$ are taken to be real.
$Z_{3}=1$ is satisfied in the quenched QED, 
and the ladder approximation with Landau gauge gives $Z_{2}=1$ in QCD and $Z_{1}=Z_{2}=1$ in QED.
It is known fact that $Z_{2}=1$ can be taken also in the improved ladder approximation
~[93,138,139,147,199]. 
Hence, we should concentrate on divergences and renormalizations in fermion mass functions, $Z_{m}$.

\vspace{2mm}

The renormalization-group invariant condensation $\tilde{\phi}$ and 
its complex conjugate $\tilde{\phi}^{\dagger}$ in our model will be defined by
\begin{eqnarray}
\tilde{\phi} &\equiv& \left(
\begin{array}{c}
\phi_{D} \\
\phi_{R} \\
\phi_{L}
\end{array}
\right)
= \frac{1}{\Bigl( \ln (\mu^{2}/\Lambda^{2}_{QCD}) \Bigr)^{\gamma_{m}}}\left(
\begin{array}{c}
\langle 0 | \frac{1}{2} (\overline{{\psi}_{MR}}P_{+}\psi_{ML}+\overline{{\psi}_{ML}}P_{+}\psi_{MR})^{(\mu)} | 0 \rangle \\
\langle 0 | (\overline{{\psi}_{MR}}P_{+}\rho_{2}\Upsilon_{2}\psi_{MR})^{(\mu)} | 0 \rangle \\
\langle 0 | (\overline{{\psi}_{ML}}P_{+}\rho_{2}\Upsilon_{2}\psi_{ML})^{(\mu)} | 0 \rangle 
\end{array}
\right),    \\
\tilde{\phi}^{\dagger} &\equiv& \left(
\begin{array}{c}
\phi^{\dagger}_{D}  \\
\phi^{\dagger}_{R}  \\
\phi^{\dagger}_{L} 
\end{array}
\right) 
= \frac{1}{\Bigl( \ln (\mu^{2}/\Lambda^{2}_{QCD}) \Bigr)^{\gamma_{m}}}\left(
\begin{array}{c}
\langle 0 | \frac{1}{2} (\overline{{\psi}_{MR}}P_{-}\psi_{ML}+\overline{{\psi}_{ML}}P_{-}\psi_{MR})^{(\mu)}  | 0 \rangle  \\
\langle 0 | ( \overline{{\psi}_{MR}}P_{-}\rho_{2}\Upsilon_{2}\psi_{MR})^{(\mu)} | 0 \rangle \\
\langle 0 | ( \overline{{\psi}_{ML}}P_{-}\rho_{2}\Upsilon_{2}\psi_{ML})^{(\mu)} | 0 \rangle 
\end{array}
\right).
\end{eqnarray}
$(\bar{\psi}_{MR}P_{+}\psi_{ML} + \cdots )^{(\mu)}, \cdots$ denote renormalized composite operators.
Here, condensations are given as complex numbers in order to take into account phase factors of them.
$\rho_{2}$ and $\Upsilon_{2}$ are inserted for the Pauli principle.

\vspace{2mm}

The asymptotic behavior of the mass function is known by examinations of 
both RG equations and an operator product expansion of a fermion bilinear~[13,191]:
\begin{eqnarray}
A^{(\mu)}(p^{2}) &\to& 1,    \\
\frac{{\bf B}^{(\mu)}_{current}(p^{2})}{A^{(\mu)}(p^{2})} &\to&  \widetilde{{\bf m}}^{(ren)}\Bigl(\ln\frac{p^{2}}{\Lambda^{2}_{QCD}}\Bigr)^{-\gamma_{m}},   \\
\frac{{\bf B}^{(\mu)}_{dyn}(p^{2})}{A^{(\mu)}(p^{2})} &\to&  -\tilde{\phi}\frac{3C_{2}(R)}{4N_{c}}\frac{\bar{g}^{2}(p^{2})}{p^{2}}\Bigl( \ln\frac{p^{2}}{\Lambda^{2}_{QCD}}\Bigr)^{\gamma_{m}},  \qquad
( p^{2} \gg \Lambda^{2}_{QCD} )   \\
\tilde{\bf m}^{(ren)} &=& (\tilde{m}^{(ren)}_{D}, \tilde{m}^{(ren)}_{R}, \tilde{m}^{(ren)}_{L} )^{T},
\end{eqnarray}
( in Euclidean region ).
Here, the asymptotics of ${\bf B}^{(\mu)}_{current}$ and ${\bf B}^{(\mu)}_{dyn}$ are 
called as hard and soft mass terms, respectively.
These forms of hard and soft mass terms are essentially determined by
mass ( wavefunction ) renormalization constant, 
thus they will take the same form for Dirac and Majorana masses.
The determination of UV asymptotic behavior of ${\bf B}_{current}$ have had some discussions in the past,
and the above form is obtained by the axial current conservation at $\Lambda\to\infty$,
i.e. $\lim_{\Lambda\to\infty}m^{(0)}_{D}(Z_{m}(\mu,\Lambda))^{-1}=0$~[147].
This condition must be satisfied also in our gauge model with including Majorana masses. 
The asymptotic form of Dirac mass function given above can be used for 
a trial function for minimizing the effective potential of QCD, can describe a chiral phase transition
( an IR ( infrared ) phenomenon ), 
even though the asymptotic form valids at $p^{2}\gg\Lambda^{2}_{QCD}$~[93].
The RG-invariant ( physical ) current mass $\tilde{\bf m}^{(ren)}$ is defined by 
\begin{eqnarray}
\tilde{\bf m}^{(ren)} &=& {\bf m}^{(\mu)}\Bigl(\ln\frac{\mu^{2}}{\Lambda^{2}_{QCD}}\Bigr)^{\gamma_{m}}.   
\end{eqnarray}
Note that the RG-invariant ( physical ) current mass
$\tilde{\bf m}^{(ren)}$ are independent from the constituent part.
This fact indicates that these two parts consist a linearly independent basis set of a Hilbert space
with an appropriate definition of inner product
for expanding a solution of SD equation, i.e. an integral equation.
( The integration kernel of our SD equation which will be discussed later is not the Hilbert-Schmidt type,
since the kernel has singular points on real axis, not bounded in $L^{2}$-norm of an integration domain
of the SD equation. )
The soft mass term decreases rapidly at the UV region, 
and the divergence property of the theory is determined by the behavior of the hard mass term. 
Hence, it is established that both "symmetric" and "broken" phases are described by a theory.
The explicit symmetry breaking is defined by $\tilde{\bf m}^{(ren)}\ne {\bf 0}$.
On the other hand, the bare mass will vanish at $\Lambda\to\infty$
by the definition ${\bf m}^{(0)}(\Lambda)=Z_{m}(\mu,\Lambda){\bf m}^{(\mu)}$.
In the asymptotic free gauge theory, 
the dynamical dimension of the fermion-antifermion composite operator becomes three at 
the one-loop level of the Landau gauge $\xi=0$
( resembles to a free theory ).
The "chiral limit" is defined by $\tilde{\bf m}^{(ren)}={\bf 0}$.
The fermion mass function depends sensitively on current mass ${\bf m}^{(\mu)}$.
In our neutrino model, 
the $\Theta$-degree of freedom of $Z(N)$ dynamically arises at the case $m^{(0)}_{D}=0$.
It is known from several numerical studies on Dirac-type mass function of $SU(3_{c})$-QCD
that it will take almost constant at $p^{2}/\Lambda^{2}_{QCD}\le 1$,
while it will decrease rapidly at $p^{2}/\Lambda^{2}_{QCD}\ge 1$ in the case of zero Dirac current mass. 
The global behavior of our mass functions ${\bf B}(s)$ should
qualitatively the same with this well-known result in literature.
If a theory has an asymptotic freedom, the UV behavior of $Z_{m}$ is qualitatively the same
( the scalling exponent depends on $N_{c}$, $N_{f}$ ).
Our theory will obtain a critical coupling for dynamical mass generation
at ${\bf m}^{(0)}={\bf 0}$ as an analytic formula ( in the non-running coupling case )
and it describes a chiral phase transition at ${\bf m}^{(0)}={\bf 0}$
( precisely, $\tilde{\bf m}^{(ren)}={\bf 0}$ ), 
while dynamical mass generation at ${\bf m}^{(0)}\ne {\bf 0}$ does not give 
a phase transition since any symmetry will not be broken dynamically in that case 
while the chiral $U(1)$ symmetry is broken from the beginning.
Since the QCD running coupling has an IR divergence, 
one should employ the Higashijima-type modified running coupling~[93]:
\begin{eqnarray}
\bar{g}^{2}(p^{2}) &=& \frac{2}{a\ln((p^{2}+p^{2}_{c})/\Lambda^{2}_{QCD})}.
\end{eqnarray}
Here, the IR cutoff $p_{c}$ is a free parameter.

\subsection{The Schwinger-Dyson Equation}

In this subsection, we derive the SD-equation and examine it.
The SD equation of the improved ladder approximation in our case becomes
\begin{eqnarray}
{\cal G}_{M}(p) &=& {\cal G}^{(0)}_{M}(p) + C_{2}(R)\int_{k} \bar{g}^{2}((p-k)^{2})\Gamma^{\mu}_{A}{\cal G}_{M}(k)\Gamma^{\nu}_{A}D^{(0)}_{\mu\nu}(p-k).
\end{eqnarray}
The SD equation can also be derived from an appropriately defined effective action $\Gamma_{eff}$,
by taking the first derivative $\frac{\delta\Gamma_{eff}}{\delta{\cal G}_{M}}=0$
in an infinite-dimensional manifold.
The manifold becomes Banach, by an appropriate definition of norm of tangent and contangent vectors of it,
and the manifold is considered as a complete metric space.
Note that this variation will be done in an infinite-dimensional linear space of propagator ${\cal G}_{M}$,
not a direct variation of the space of mass functions.
This difference has come from the fact that we consider a variation in a mass ${\it matrix}$ space.
The generic form of an SD equation is written by
\begin{eqnarray} 
\widehat{\cal M}(p) &=& \widehat{\cal M}^{(0)} + \lambda\int^{a}_{b}dk\widehat{K}[p,k;\widehat{\cal M}(k)]\widehat{\cal M}(k),          
\end{eqnarray}
and an operation of a Lie group 
( one can consider it as color or flavor gauges )
to it gives
\begin{eqnarray} 
g\widehat{\cal M}(p)g^{-1} &=& g\widehat{\cal M}^{(0)}g^{-1} + \lambda\int^{a}_{b}dk\widehat{K}[p,k;g\widehat{\cal M}(k)g^{-1}]g\widehat{\cal M}(k)g^{-1},          
\end{eqnarray}
i.e., under a quite non-linear manner. 
Since ${\cal G}_{M}$ depends on a gauge, 
one should choose it before constructing an explicit form of an SD equation:
We choose the Landau gauge for the $SU(2_{c})$ interaction.
After a gauge is chosen, a norm of mass function space is appropriately determined, 
it will become a Finsler space~[33], 
and an isometry given by a Lie group of a symmetry of a theory should keep a norm.
The bare vertex $\Gamma^{\mu}_{A}$ of (235) is 
\begin{eqnarray}
\Gamma^{\mu}_{A} &\equiv& (\gamma^{\mu}\gamma_{5}\otimes\tau_{3}){\cal T}_{A}. 
\end{eqnarray}
The free gauge-boson propagator is given as follows:
\begin{eqnarray}
D^{(0)}_{\mu\nu}(k) &=& \delta_{AB}\frac{1}{k^{2}}\Bigl( g_{\mu\nu}-(1-\xi)\frac{k_{\mu}k_{\nu}}{k^{2}} \Bigr).
\end{eqnarray} 
We consider the SD equation of the Abelianized form in the neutrino model.
The propagator ${\cal G}^{(0)}_{M}(p)$ is obtained by substituting 
$m_{j}\to m^{(0)}_{j}$ ( $j=R,L,D$ ) in Eq. (47) of the definition of $\Omega^{-1}_{M}$
( $(0)$ indicates a bare quantity ).
The inverse of dressed propagator is found from Eq. (43): 
\begin{eqnarray}
({\cal G}_{M}(p))^{-1} &\equiv& \left(
\begin{array}{cc}
A(p^{2})\pfey -B^{\dagger}_{R}(p^{2})P_{+}-B_{R}(p^{2})P_{-} & -B^{\dagger}_{D}(p^{2})P_{+} -B_{D}(p^{2})P_{-}  \\
-B^{\dagger}_{D}(p^{2})P_{+} -B_{D}(p^{2})P_{-}  & A(p^{2})\pfey -B^{\dagger}_{L}(p^{2})P_{+}-B_{L}(p^{2})P_{-} 
\end{array}
\right).
\end{eqnarray}
Here, all quantities of ${\cal G}_{M}(p)$ are bare, 
and we keep mass functions of Majorana-type $B_{L}$, $B_{R}$ and Dirac-type $B_{D}$ as complex functions. 
In the neutrino model, we consider the case of all of $B_{R}$, $B_{L}$ and $B_{D}$ are color and flavor singlet,
and thus $SU(2_{c})$ gauge symmetry is conserved, in principle.
The Landau gauge is taken by setting $\xi=0$, 
and thus we can set $A(p^{2})=1$ in our ladder approximation.
This is guaranteed by the fermion wavefunction renormalization constant $Z_{2}$ given in the previous subsection.
After some manipulations, we obtain the following six coupled equations as our SD equation
( in Euclidean region ):
\begin{eqnarray}
B_{R}(p^{2}) \pm B^{\dagger}_{R}(p^{2}) &=& m^{(0)}_{R} \pm (m^{(0)}_{R})^{\dagger} 
+ C_{2}(R)\int_{k}\bar{g}^{2}((p-k)^{2})\frac{3+\xi}{(p-k)^{2}}    \nonumber \\
& & \times  \frac{[B_{R}(k^{2}) \pm B^{\dagger}_{R}(k^{2})][k^{2}+|B_{L}(k^{2})|^{2}]-(B_{D}(k^{2}))^{2}B^{\dagger}_{L}(k^{2}) \mp (B^{\dagger}_{D}(k^{2}))^{2}B_{L}(k^{2})}{D_{SD}(k^{2})},   \\
B_{L}(p^{2}) \pm B^{\dagger}_{L}(p^{2}) &=& m^{(0)}_{L} \pm (m^{(0)}_{L})^{\dagger} 
+ C_{2}(R)\int_{k}\bar{g}^{2}((p-k)^{2})\frac{3+\xi}{(p-k)^{2}}   \nonumber \\
& & \times \frac{[B_{L}(k^{2}) \pm B^{\dagger}_{L}(k^{2})][k^{2}+|B_{R}(k^{2})|^{2}]-(B_{D}(k^{2}))^{2}B^{\dagger}_{R}(k^{2}) \mp (B^{\dagger}_{D}(k^{2}))^{2}B_{R}(k^{2})}{D_{SD}(k^{2})},   \\
B_{D}(p^{2}) \pm B^{\dagger}_{D}(p^{2}) &=& m^{(0)}_{D} \pm (m^{(0)}_{D})^{\dagger} 
+ C_{2}(R)\int_{k}\bar{g}^{2}((p-k)^{2})\frac{3+\xi}{(p-k)^{2}}  \nonumber \\
& & \times \frac{[B_{D}(k^{2}) \pm B^{\dagger}_{D}(k^{2})][k^{2}+|B_{D}(k^{2})|^{2}]-B_{R}(k^{2})B_{L}(k^{2})B^{\dagger}_{D}(k^{2}) \mp B^{\dagger}_{R}(k^{2})B^{\dagger}_{L}(k^{2})B_{D}(k^{2})}{D_{SD}(k^{2})},    \\
D_{SD}(s) &=& (s+M^{2}_{+}(s))(s+M^{2}_{-}(s)).
\end{eqnarray}
Here, $M_{\pm}$ are obtained from the result of Sec. II as follows:
\begin{eqnarray}
M_{\pm}(s) &=& \Bigg( |B_{D}(s)|^{2} + \frac{|B_{R}(s)|^{2}+|B_{L}(s)|^{2}}{2}  \nonumber \\
& & \mp \frac{1}{2}\sqrt{ (|B_{R}(s)|^{2}-|B_{L}(s)|^{2})^{2} + 4|B_{D}(s)|^{2}(|B_{R}(s)|^{2}+|B_{L}(s)|^{2}+2|B_{R}(s)||B_{L}(s)|\cos\Theta) } \Bigg)^{1/2}.
\end{eqnarray}
The $U(1)$ phases of mass functions are defined as follows:
\begin{eqnarray}
B_{D}(s) = |B_{D}(s)|e^{i\theta_{D}}, \quad 
B_{R}(s) = \rho_{2}\Upsilon_{2}|B_{R}(s)|e^{i\theta_{R}}, \quad
B_{L}(s) = \rho_{2}\Upsilon_{2}|B_{L}(s)|e^{i\theta_{L}}, \quad
\Theta &=& \theta_{R}+\theta_{L}-2\theta_{D}.
\end{eqnarray}
Here, we assume these phases have no momentum dependence,
and coincide with that of bare mass parameters.
If we take different phases between a mass function and corresponding bare mass parameter,
the SD formalism will obtain a problem, an inconsistency between the real and imaginary parts of a mass function.
We will write the SD equation into a vector form as
\begin{eqnarray}
{\bf B}(s) &=& {\bf m}^{(0)} 
+ \frac{3C_{2}(R)}{16\pi^{2}}\int^{\Lambda^{2}}_{s}ds'\bar{g}^{2}(s')\frac{{\bf F}(s')}{D_{SD}(s')}
+ \frac{3C_{2}(R)\bar{g}^{2}(s)}{16\pi^{2}s}\int^{s}_{0}ds'\frac{s'{\bf F}(s')}{D_{SD}(s')}, 
\end{eqnarray}
where,
\begin{eqnarray}
F_{D}(s) &\equiv& |B_{D}(s)|\Bigl\{ s -|B_{R}(s)||B_{L}(s)|\cos\Theta + |B_{D}(s)|^{2} \Bigr\},   \\
F_{R}(s) &\equiv& |B_{R}(s)|(s+|B_{L}(s)|^{2}) - |B_{D}(s)|^{2}|B_{L}(s)|\cos\Theta,   \\
F_{L}(s) &\equiv& |B_{L}(s)|(s+|B_{R}(s)|^{2}) - |B_{D}(s)|^{2}|B_{R}(s)|\cos\Theta,   \\ 
{\bf B}(s) &\equiv& ( |B_{D}(s)|, |B_{R}(s)|, |B_{L}(s)| )^{T},  \\
{\bf F}(s) &\equiv& ( F_{D}(s), F_{R}(s), F_{L}(s) )^{T},  \\
{\bf m}^{(0)} &\equiv& ( |m^{(0)}_{D}|, |m^{(0)}_{R}|, |m^{(0)}_{L}| )^{T}.
\end{eqnarray}
The first integral which takes into account relatively high-energy region 
will give the hard mass ( current mass ) term (230),
while the second integral corresponds to the soft mass ( constituent mass ) term (231).
As we have mentioned previously, the bare mass ${\bf m}^{(0)}$ vanishes at $\Lambda\to\infty$.
To remove angular dependences inside the integrals of the above SD equation,
we have employed the following approximation~[93,147]: 
\begin{eqnarray}
& & \int \sin\theta d\theta d\phi (p-k)^{-2} \to \theta(p-k)p^{-2} + \theta(k-p)k^{-2},   \\
& & g^{2} \to \bar{g}^{2}((p-k)^{2}) \to 
\bar{g}^{2}({\rm max}(p^{2},k^{2})) = \theta(p-k)\bar{g}^{2}(p^{2}) + \theta(k-p)\bar{g}(k^{2}).
\end{eqnarray}
From (247), one finds  
\begin{eqnarray}
& & \lim_{s\to 0}\Bigg[ 
\frac{d}{ds}\Bigl( \frac{3C_{2}(R)\bar{g}^{2}(s)}{16\pi^{2}s} \Bigr) \Bigg]^{-1} 
\frac{d}{ds}{\bf B}(s) = 0,   \\
& & \lim_{s\to\Lambda^{2}}\Bigg\{ {\bf B}(s)-{\bf m}^{(0)} - \frac{3C_{2}(R)\bar{g}^{2}(s)}{16\pi^{2}s}\int^{s}_{0}\frac{s'{\bf F}(s')}{D_{SD}(s')} \Bigg\} = 0
\end{eqnarray}
as the IR and the UV boundary conditions of the differential equation, respectively.
Both the IR  and UV boundary conditions take the same forms with those of the SD equation of Dirac mass 
in QCD without Majorana mass terms.
Since the mass functions $|B_{D}(s)|,\cdots$ are much smaller than cutoff $\Lambda$ or $s$ in the UV region,
the UV asymptotic behavior of the SD equation and its solution will take the same form
with that of ${\rm QED_{4}}$ ( non-runing case in our model ) or ${\rm QCD_{4}}$ without Majorana mass terms.
Our SD equation as a coupled integral equation will be converted into the following differential equation:
\begin{eqnarray}
\frac{d}{ds}\Bigg\{\Bigl[ \frac{d}{ds}\Bigl( \frac{3C_{2}(R)\bar{g}^{2}(s)}{16\pi^{2}s} \Bigr) \Bigr]^{-1}\frac{d}{ds}{\bf B}(s) \Bigg\} 
&=& \frac{s{\bf F}(s)}{D_{SD}(s)} = \frac{s}{D_{SD}(s)}\Xi(s){\bf B}(s).
\end{eqnarray}
The definition of $\Xi$ is
\begin{eqnarray}
\Xi(s) &\equiv& \left(
\begin{array}{ccc}
s + |B_{D}(s)|^{2} -|B_{R}(s)||B_{L}(s)|\cos\Theta & 0 & 0 \\
0 & s + |B_{L}(s)|^{2} & -|B_{D}(s)|^{2}\cos\Theta  \\
0 & -|B_{D}(s)|^{2}\cos\Theta & s+|B_{R}(s)|^{2} 
\end{array}
\right).
\end{eqnarray}
The global behavior of ${\bf B}(s)$ is determined by the first derivative $\frac{d}{ds}{\bf B}(s)$
and it will decreases monotonically at $s\to\infty$ in ordinary QCD.
In our SD equation, it is non-trivial due to the existence of $\Theta$.

\vspace{2mm}

First, let us consider the SD equation with an appropriate linearization.
The SD equation is Fuchsian, has four regular singular points $s=0,M_{+},M_{-},\infty$,
and after a linearlization, it becomes a Heun equation ( see Appendix ), can be solved analytically~[91,92].
To obtain our SD equation as a matrix Heun equation, we will linearize it
by substituting all of mass functions of $\Xi$ by their value at the origin:
${\bf m}_{dyn}=(m^{D}_{dyn},m^{R}_{dyn},m^{L}_{dyn})^{T} \equiv {\bf B}(s=0)$. 
This kind of linearization is also frequently used in ordinary QED/QCD, 
and it is justified at a "critical" region $0<|B|^{2}/\Lambda^{2}\ll\infty$ and $s\to\Lambda$.
Moreover, $\lim_{s\to\infty}{\bf B}(s)\to {\rm const.}$ ( in fact, vanish ) 
if $s=\infty$ is a regular singular point. 
( The Heun equation is important for theory of integrable nonlinear wave equations, 
and appears in the scalar-vector sector of a spinor-spinor Bethe-Salpeter equation~[92]. 
Both the Heun equation and the Gauss hypergeometric differential equation are defined on the Riemann sphere, 
and the Heun equation has 192 solutions ( local ), 
similar to the case that the Gauss equation has Kummer's 24 local solutions~[73,133,220]. )
SD equations of QED or QCD with non-running couplings, and without Majorana mass terms, 
are converted into the Gauss hypergeometric differential equation of three singular points,
and our Heun-type SD equation ( in the non-running coupling case ) 
also becomes the Gauss hypergeometric differential equation
when we consider some limiting cases of mass functions/parameters
( for example, $|B_{R}(s)|=|B_{L}(s)|$ ).
From the usual procedure, we get all of the scalling behaviors
of $|B_{D}(s)|$, $|B_{R}(s)|$ and $|B_{L}(s)|$ in (258) are the same,
and the exponents are found to be ( in the non-running coupling case )
\begin{eqnarray}
l &=& \frac{1}{2}\Bigg(1\pm \sqrt{1-\frac{3C_{2}(R)g^{2}}{4\pi^{2}}}\Bigg),
\end{eqnarray}
if we neglect the off-diagonal elements of $\Xi$ in our matrix Heun-type equation.
We can obtain analytic expressions of exponents for the case of non-vanishing off-diagonal elements 
of $\Xi$ by solving a $2\times2$ matrix eigenvalue problem as the indicial equation of the problem,
when we assume the UV behaviors $|B_{R}(s)|\sim|B_{L}(s)|\sim s^{l,l\pm 1,l\pm 2,\cdots}$ at $s\to\infty$,
but the correction will become small.
Hence the contribution of the off-diagonal element $|B_{D}|^{2}\cos\Theta$ 
inside the SD equation will vanish at $s\to\infty$.
Moreover, from the discussion of renomalized mass function given previously,
$|B_{R}|$ and $|B_{L}|$ should take the same exponent at the UV asymptotic region.  
This off-diagonal elements clearly shows the coupling between $|B_{R}|$ and $|B_{L}|$.
Thus the coupling tends to zero at $s\to\infty$.
The parameter $q$ in Appendix C has no role for the character of singularity in the generic Heun equation,
and thus the mass functions of the matrix elements of $\Xi$ 
do not contribute to the SD equation of the UV asymptotic region.

\vspace{2mm}

From the well-known asymptotic behavior of mass functions in QCD, 
we can discuss the characteristic aspect of our matrix SD equation.
In our treatment, current and constituent parts of a mass function share the same phase,
and it seems the only choice for us in the SD formalism. 
Moreover, as mentioned above, only the choice to take the same phase for bare mass parameters
and mass functions ( self-energies ) can keep consistency between the real and imaginary parts of
a mass function inside the SD equation. 
We have taken the convention of negative sign of $\tilde{\phi}$.
The current mass part becomes dominant than the constituent part inside the mass function at large $s$.
Hence, when we consider our SD equation without the linearization, 
there are several cases of the behavior of ${\bf B}(s)$:
Its first derivative is
\begin{eqnarray}
\frac{d{\bf B}(s)}{ds} &=& \frac{3C_{2}(R)}{16\pi^{2}}\Bigl(
\frac{1}{s}\frac{d\bar{g}^{2}(s)}{ds}-\frac{\bar{g}^{2}(s)}{s^{2}} \Bigr)
\int^{s}_{0}s'ds'\frac{{\bf F(s')}}{D_{SD}(s)}.
\end{eqnarray}
${\bf B}(s)$ decreases monotonically at $s\to\infty$ when the matrix elements of $\Xi$ is always positive,
while if, for example, $s+|B_{D}|^{2}-|B_{R}||B_{L}|\cos\Theta$ in (259) 
has the region where it becomes negative
( especially in the IR region ),
the mass function $|B_{D}(s)|$ can have regions of $s$ where it decreases/incleases, 
implies an instability of vacuum of the theory similar to the NJL-type model case we have discussed previously,
and can have a node, can cross the $s$-axis before it will vanish at $s\to\infty$.
The character of matrix elements of $\Xi$ for the SD equation are determined by 
relative sizes between $|B_{D}|$, $|B_{R}|$, $|B_{L}|$ and the phase factor $\Theta$.
To find the value of $\Theta$ for vacua of the theory,
we should examine its effective potential:
In our speculation, a stationary point in $\Theta$-coordinates is non-degenerate critical
from the result of our NJL-type model, namely we should find a specific VEV of $\Theta$. 
The behavior of mass function at the IR region, namely a mass function 
at "relatively" long distance, has strong connection with the stability 
of the effective potantial of the theory ( and also confinement~[3,64,93] ),
though the asymptotic behavior of mass functions given in the previous subsection 
can be used for discussion of the structure of the effective potential due to confinement of fermions~[93].  
An interesting fact of our case is that the $\Theta$-degree of freedom can
qualitatively affect the global behavior of the solution of our matrix SD equation.

\vspace{2mm}

Since ${\bf B}(s)$ is a three-dimensional real vector, we can represent it as
\begin{eqnarray}
{\bf B}(s)^{T} &=& (|B_{D}(s)|,|B_{R}(s)|,|B_{L}(s)|) = r(s)( \cos\theta, \sin\theta\cos\phi, \sin\theta\sin\phi ),    \\
r(s) &\equiv& \sqrt{|B_{D}(s)|^{2}+|B_{R}(s)|^{2}+|B_{L}(s)|^{2}}.
\end{eqnarray}
Then, our SD equation becomes 
\begin{eqnarray}
& & \frac{d}{ds}\Bigg\{\Bigl[ \frac{d}{ds}\Bigl( \frac{3C_{2}(R)\bar{g}^{2}(s)}{16\pi^{2}s} \Bigr) \Bigr]^{-1}\frac{d}{ds}r(s) \Bigg\} 
= \frac{r(s)s}{D_{SD}(s)}\bigl[ s+r^{2}(s)\Phi(\theta,\phi)\bigr],     \\
& & \Phi(\theta,\phi) = \cos^{4}\theta + 2\sin^{4}\theta\sin^{2}\phi\cos^{2}\phi -3\cos^{2}\theta\sin^{2}\theta\sin\phi\cos\phi\cos\Theta.
\end{eqnarray}
Here, we have taken the approximation of the complete neglection 
on momentum $s$ dependence of $\theta$ and $\phi$.
This is a rough approximation because it fixes a relative ratio of $|B_{D}|$, $|B_{R}|$, and $|B_{L}|$ 
for whole region of $0\le s < \infty$, 
though we mainly have our interest on the UV asymptotic behavior of the mass functions 
( which determines critical couplings of mass generations ),
it should work well at the asymptotic region.
The seesaw condition $|B_{R}|^{2}\gg|B_{D}|^{2}\gg|B_{L}|^{2}$ is obtained at $\phi\sim 0$, $\theta\sim\pi/2$.
By this prescription with the linearization $r(s)\to r(0)$ for the right-hand side of bracket of (264), 
again our SD equation, as a Heun equation, will give the same exponents (260) for $r(s)$.
$\phi=\pi/4$ of the left-right symmetric case gives a Gauss hypergeometric equation.
In the vicinity of critical region, 
we obtain the following well-known asymptotic behaviors 
from the UV boundary condition at ${\bf m}^{(0)}=0$ limit with using the non-running coupling
( the Maskawa-Nakajima solution~[142] ):
\begin{eqnarray}
r(x) &\propto& x^{-\frac{1}{2}\Bigl(1\pm\sqrt{1-\frac{3C_{2}(R)g^{2}}{4\pi^{2}}}\Bigr)},  
\qquad {\rm for} \quad 1 \ge \frac{3C_{2}(R)g^{2}}{4\pi^{2}},  \\
&\propto& x^{-1/2}\cos\Bigl( \frac{1}{2}\sqrt{\frac{3C_{2}(R)g^{2}}{4\pi^{2}}-1}\ln x \Bigr), \qquad
\qquad {\rm for} \quad 1 \le \frac{3C_{2}(R)g^{2}}{4\pi^{2}}.     
\end{eqnarray}
Hence, the critical coupling satisfies the well-known result $3C_{2}(R)g^{2}/4\pi^{2}=1$~[93,147]. 
Here, we consider the case where the coupling constant is close to the critical value.
By the UV boundary condition with this situation,
and with the zero bare-mass limit $|m^{(0)}_{D}|=|m^{(0)}_{R}|=|m^{(0)}_{L}|=0$, gives
( non-running coupling case )
\begin{eqnarray}
M_{dyn} &=& \Lambda \exp\Bigg[ -\frac{(2n+1)\pi}{2\sqrt{\frac{3}{4\pi^{2}}C_{2}(R)g^{2}-1}} \Bigg],  \quad (n=0,1,2\cdots ),  \\
M_{dyn} &=& \sqrt{ (m^{D}_{dyn})^{2} + (m^{R}_{dyn})^{2} + (m^{L}_{dyn})^{2} }, \\
(m^{D}_{dyn},m^{R}_{dyn},m^{L}_{dyn}) &=& (|B_{D}(0)|,|B_{R}(0)|,|B_{L}(0)|).
\end{eqnarray}
By utilizing the vector notation of mass function ${\bf B}$, one finds
\begin{eqnarray}
\Phi(\theta,\phi) &=& \frac{1}{r^{4}(s)}\Bigl( |B_{D}(s)|^{4} +2|B_{R}(s)|^{2}|B_{L}(s)|^{2} - 3|B_{D}(s)|^{2}|B_{R}(s)||B_{L}(s)|\cos\Theta  \Bigr).
\end{eqnarray}
For example, when we consider $|B_{R}|\sim 10^{11}$ GeV, $|B_{D}|\sim 10^{2}$ GeV and $|B_{D}|\sim 10^{2}$ eV, 
then we find 
$|B_{D}|^{4} \sim 
|B_{R}|^{2}|B_{L}|^{2} \sim 
|B_{D}|^{2}|B_{R}||B_{L}| \sim 10^{8}$ $({\rm GeV})^{4}$.
In this case, they will take the same magnitude in $\Phi(\theta,\phi)$.
Thus, under the seesaw condition with the three-dimensional vector approximation 
for mass function, the effect of $\Theta$ in the SD equation is not negligible.

\vspace{2mm}

Our SD equation of several limiting cases can be solved more explicitly.
Let us examine the SD equation of the Dirac mass part.
We list the following cases:
\begin{eqnarray}
\frac{sF_{D}(s)}{D_{SD}(s)} 
&\to& \frac{s|B_{D}(s)|}{s+|B_{D}(s)|^{2}}, \qquad ( case (a); \,\, |B_{R}|= |B_{L}|= 0 ),   \\
&\to& \frac{s|B_{D}(s)|}{s+|B_{D}(s)|^{2}+|B_{R}(s)|^{2}}, \qquad 
( case (b); \,\, |B_{R}| = |B_{L}|, \, \Theta=(2n+1)\pi ),   \\
&\to& \frac{s|B_{D}(s)|(s+|B_{D}(s)|^{2})}{\Bigl(s+\Bigl(\frac{|B_{R}(s)|}{2}-\sqrt{\frac{|B_{R}(s)|^{2}}{4}+|B_{D}(s)|^{2}}\Bigr)^{2}\Bigr)\Bigl(s+\Bigl(\frac{|B_{R}(s)|}{2}+\sqrt{\frac{|B_{R}(s)|^{2}}{4}+|B_{D}(s)|^{2}}\Bigr)^{2}\Bigr)},   \nonumber \\ 
& & \qquad\qquad\qquad\qquad\qquad\qquad\qquad\qquad\qquad\qquad\qquad ( case (c); \,\, |B_{R}|\ne 0, \, |B_{L}|= 0 ).  
\end{eqnarray}
The case (c) with $|B_{R}(s)|^{2}\gg |B_{D}(s)|^{2}$ gives the type-I seesaw condition~[152].
We will set $|B_{R}(s)|=m_{R}={\rm const}.$ for solving the SD equation 
( becomes a hypergeometric differential equation by this linearization ) of case (b).
It is impossible to obtain a single function of analytical solution 
for the linearized SD equation in the whole region of momentum space $0\le s < \infty$ 
( no global solution, though we can use analytic continuations )~[2,220]. 
Since the hypergeometric differential equation has $s=0,1,\infty$ as regular singular points
from its definition, we obtain the following convergent serieses 
( expansions of solutions inside several convergent radii defined around $s=0,1,\infty$ ) 
from the 24 Kummer local solutions of hypergeometric functions $_{2}F_{1}=F(\alpha,\beta,\gamma,x)$ 
as special solutions ( up to a multiplicative constant ) of the linearized SD equation~[73,220]:
\begin{eqnarray}
& & F(\alpha,\beta,\gamma,x), \quad {\rm and} 
\quad  \frac{x^{1-\gamma}F(\alpha-\gamma+1,\beta-\gamma+1,2-\gamma,x)}{\Gamma(2-\gamma)} \quad {\rm for} \quad  |x| < 1,  \\ 
& & x^{-\alpha}F(\alpha,1+\alpha-\gamma,1+\alpha-\beta,1/x) \quad {\rm and}
\quad  x^{-\beta}F(\beta,1+\beta-\gamma,1+\beta-\alpha,1/x) \quad {\rm for} \quad |x| > 1,  \\
& & \frac{F(\alpha,\beta,\alpha+\beta-\gamma+1,1-x)}{\Gamma(\alpha+\beta-\gamma+1)} \quad {\rm and} \quad
\frac{(1-x)^{\gamma-\alpha-\beta}F(\gamma-\beta,\gamma-\alpha,\gamma-\alpha-\beta+1,1-x)}{\Gamma(\gamma-\alpha-\beta+1)} \quad {\rm for} \quad |1-x| < 1, 
\end{eqnarray}
where,
\begin{eqnarray}
x \equiv -\frac{s}{|m_{R}|^{2}+(m^{D}_{dyn})^{2}}, \qquad
\alpha \equiv \frac{1+\sqrt{1-\frac{3C_{2}(R)g^{2}}{4\pi^{2}}}}{2},
\qquad \beta \equiv  \frac{1-\sqrt{1-\frac{3C_{2}(R)g^{2}}{4\pi^{2}}}}{2}, 
\qquad \gamma \equiv 2. 
\end{eqnarray}
Here, we use gamma functions to remove singularities at several points of auguments 
in the hypergeometric functions.
The IR boundary condition is satisfied by $F(\alpha,\beta,\gamma,x)$,
while $|B_{D}(s)|$ of the UV region is given by a linear combination of
$x^{-\alpha}F(\alpha,1+\alpha-\gamma,1+\alpha-\beta,1/x)$ and 
$x^{-\beta}F(\beta,1+\beta-\gamma,1+\beta-\alpha,1/x)$.
Obviously, under the approximations mentioned above, the critical coupling becomes 
the same with (266) and (267).

\vspace{2mm}

In summary, a numerical value of $\Theta$ can affect the structure and solution of our SD equation, 
and then it reflects "global" structure of the effective potential of the theory,
the effect is not negligible even in the case of seesaw situation of a huge mass hierarchy. 
While, various parts of our SD equation show the same characters/behaviors with those of QED/QCD
even though our model includes Majorana masses.
At least in non-running coupling case, $\Theta$ does not affect the critical coupling of
dynamical mass generation. In the case of NJL-type model, $\Theta$ affects essentially
on structure of effective potential, gap equation and critical coupling. 
An evaluation of phason mass and decay constant demands us a hard and complicated calculation 
in our gauge model, while a numerical calculation is not meaningful due to a huge hierarchy of 
the explicit symmetry breaking mass parameters $\tilde{\bf m}^{(ren)}$. 
If a mass prediction is the purpose, one should consider a renormalization group analysis 
usually employed in the SM/MSSM
~[12,28,29,30,36].
We should mention that, here we do not prove that the seesaw situation takes place
in our SD equation with taking a large hierarchy of bare mass parameter,
and discuss several characteristic features of the equation.
Our SD equation has a non-linear vector radial Schr\"{o}dinger operator at the non-running coupling case.
It is an interesting subject to examine the operator by differential Galois theory 
to understand our SD equation more deeply
~[140,162,194,195,208].
Connes and Kreimer found/established Hopf-algebra structures in
perturbative expansions of several renormalizable field theories
~[10,41,42,43,121,122]. 
Needless to say, quantum field theory is constructed by Feynman-Dyson perturbative expansion,
renormalization, and renormaization-group prescription,
and thus the Connes-Kreimer theory may radically modify our understanding and interpretations
on quantum field theory.
It is an interesting issue for us to understand our SD equation 
by the Connes-Kreimer theory.
Our SD equation is a differential equation and a variational equation.
The SD equation is defined locally, 
and thus the method to know informations on the global structure of the problem
( from local to global ),
beyond a simple analysis of the effective potential, 
is an interesting issue.
In the $SU(3_{c})$-QCD of diagonal breaking of the chiral group $SU(2)_{R}\times SU(2)_{L}$
can also be formulated analytically, 
though the result of the SD equation becomes a complicated one
for analysing a mathematical manner.

\section{Interactions, Reactions and Thermodynamics of Phason: Physical Implications}

\subsection{Two-body Phason Decay}

The two-body pion decay $\pi^{-}\to l^{-} + \bar{\nu}_{\mu}$ 
is the best laboratory to examine the $\mu e$-universality of weak interaction~[61].
This reaction is used to determine $f_{\pi}=93$ MeV.
Since phason $\tilde{\Theta}$ is a "pseudo-scalar meson"ic boson field,
we can evaluate the two-body phason decay $\tilde{\Theta}\to \varphi_{1} +\bar{\varphi}_{2}$ 
by a similar way of pion weak decay, with assuming a hypothetical V-A type interaction.
Since $\tilde{\Theta}$ is a neutral boson,
it does not have a decay like $\tilde{\Theta}^{\dagger}\to\bar{\varphi}_{1}+\varphi_{2}$
which could have a role in lepton/baryon-number non-conservation.
By assuming $m_{1}\ne 0$, $m_{2}=0$,
and temperature-dependence as $|m_{D}(T)|=|m_{D}(T=0)|\sqrt{1-T^{2}/T^{2}_{c}}$
( $T_{c}$; critical temperature ), one finds the decay width
\begin{eqnarray}
\Gamma(\tilde{\Theta}\to\varphi_{1}+\bar{\varphi}_{2}) = \frac{G^{2}_{p}F^{2}_{\Theta}}{4\pi}\frac{(m^{2}_{\Theta}-m^{2}_{1})^{2}m^{2}_{1}}{m^{3}_{\Theta}}
\end{eqnarray}
will vanish at $T\to T_{c}$ and $|m_{D}(T=0)|=0$ in our NJL-type model.
Here, $G_{p}$ is the Fermi constant of this hypothetical interaction.
We wish to mention that, quite recently, an implication of a rare pion decay ( a two-body type )
to dark matter was given in literature~[104].
The two-body phason decay could also obtain such an implication in universe.

\subsection{Tunneling, Instanton, and Topological Nature of the Theory}

We have found that $\Theta=(2n+1)\pi$ ( $n\in{\bf Z}$ ) give infinitely degenerate vacua of the NJL-type model, 
and $V^{NJL+M}_{eff}$ is almost proportional to $\cos\Theta$.
Hence, there is a tunneling probability $P_{t}=\exp(-S_{E})$ between two vacua. 
Here, the effective action $S_{E}$ for describing dynamics in the direction $\Theta$ 
in a field-theoretical framework can be written as follows:
\begin{eqnarray}
Z(\beta) &=& \int{\cal D}\tilde{\Theta} e^{-S_{E}},   \\
S_{E} &=& {\cal V}_{3D}\int^{\beta}_{0}d\tau
\Bigl( \frac{1}{2}\bigl(\frac{|m_{D}|}{2}\bigr)^{2}\bigl(\frac{d\Theta}{d\tau}\bigr)^{2} + \frac{\Delta V_{eff}}{2}\cos\Theta \Bigr),  \\
\Delta V_{eff} &\equiv& V_{eff}(\Theta=0)-V_{eff}(\Theta=\pi) \sim {\cal O}(|m_{D}|^{2}|m_{R}||m_{L}|), 
\end{eqnarray}
where,
\begin{eqnarray}
|m_{D}|^{2} \sim F^{2}_{\Theta}, \quad \Delta_{eff} \sim F^{2}_{\Theta}m^{2}_{\Theta}.
\end{eqnarray}
Here, ${\cal V}_{3D}$ is the three dimensional volume of the system.
For example, we obtain 
\begin{eqnarray}
\Delta V_{eff} \sim 10^{4}({\rm GeV})^{4}
\end{eqnarray}
by using $|m_{D}|\sim 1{\rm GeV}$, $|m_{L}|\sim 1{\rm eV}$ and $|m_{R}|\sim 10^{13}{\rm GeV}$. 
We will consider quantum mechanics to describe the system.
The system has kink ( instanton, namely a classical solution which makes action functional as a finite quantity )
solution~[38].
We can evaluate a tunneling effect of instanton/anti-instanton to the system,
and it will give a perturbation to eigenvalues. 
For an estimation of it, we can use the virial theorem, with assuming
the harmonic potential in the vicinity of the bottom of a valley by specifying the order of $\Theta$.
The system is translation invariant in time direction, 
and thus it has a zero-mode which will be removed by a Faddeev-Popov determinant 
in the path-integration of a partition function.
Moreover, the system is perfectly periodic, 
and thus a wavefunction of the potential can take a Bloch form obtained 
( for example ) 
by solving a canonically quantized Schr\"{o}dinger equation of the Hamiltonian in $S_{E}$.
Hence, the potential of $\Theta$ resembles to the $\theta$-vacua of ${\rm QCD}_{4}$.
We find the canonical commutation relation $[\Theta,P_{\Theta}]=i$ for quantization, 
and the uncertainty relation $\Delta P_{\Theta}\Delta\Theta\ge 1/2$ also be satisfied. 
We wish to emphasize that this is an uncertainty relation of fluctuation of 
phase of mass parameters and its canonical momentum.
We can consider a squeezed state for $\Theta$ via a harmonic approximation 
for describing the vicinity of the bottom of a valley. 
If an external field like $-C_{ext}\frac{\Theta}{2\pi}$
is added to the system (281) and becomes stronger than a threshold, 
a representation point of the system ( a wave packet ) will move toward the $\Theta$-direction, 
passes through places where CP is broken, 
and the dynamics seems like that of a density wave in (1+1)-dimensions. 
( Note that CP is obviously broken at $\Theta\ne\pi$ at the classical Lagrangian level. )

\vspace{2mm}

To examine thermodynamics of the one-dimensional system defined above,
we assume the temperature-dependence of the dynamical Dirac mass
as $|m_{D}(T)|\sim |m_{D}(T=0)|\sqrt{1-T^{2}/T^{2}_{c}}$.
When temperature goes down from $T\sim T_{c}$ to $T\ll T_{c}$,
the amplitude of modulation of $V_{eff}$ develops,
and a decoherence between quatum states of valleys can take place.
For example, in the case ${\cal O}(|m_{D}(T=0)|)\sim T_{c}\sim{\cal O}(1 {\rm MeV})$,
a vanishing of Dirac mass by thermal effect exists inside a neutron star 
( $T_{c}$ of superfluidity in a neutron star becomes $\sim {\cal O}(1 {\rm MeV})$ )~[173].
A thermal energy of our model will be obtained by $\Theta$-direction, 
a thermodynamic system like a neutron star would have a statistical density matrix 
which includes CP-violating states ( a CP-violation by thermalization ).
This effect cannot be observed in the one-loop calculation of 
our model under the Matsubara formalism because each of the mass spectra of 
our model at this approximation level has two-fold degeneracy.
On the contrary, the Lagrangian obviously breaks CP-invariance at $\Theta\ne 0$,
and this fact indicates that the effect of the CP-violation cannot be described by a mean-field approximation. 
To clarify this issue, we need a further examination.
A thermal spectrum will also be obtained in a curved spacetime or a uniform acceleration
~[52,71,88,223].  
Hence, a uniformly accelerated observer would detect a CP-violation effect 
by the Hawking-Unruh effect
~[52,71,88,174,216,223].
Hence, if the evolution of Universe had an era of a huge accelerated expansion,
a CP violation by the acceleration might occur.
Namely, a large-scale dynamics of the Universe itself causes a CP-violating effects.
In that case, different observer of different Rindler coordinates will detect 
different result of CP violation from observations of a single system.
If our theory obtains a Lorentz-symmetry violating perturbation,
such as gravitational effect, then CPT theorem cannot be utilized by us, 
and the perturbation would lift the degeneracies of mass spectra and CP and T violation might occur.
CP and T are conserved at the vacua ( ground state ) of our Lorentz symmetric NJL-type model 
even though it has Majorana mass terms in the left-right asymmetric manner.
While, if Lorentz symmetry is explicitly broken in a Dirac equation 
with having only Dirac-type mass, the model can give a CPT violation
~[79,119,120].
In this paper, we mainly consider neutrino seesaw mass, 
though a similar situation might take place in mixon/cpon of a CKM/PMNS-type matrices.
If the CKM/PMNS matrices are physical, 
they give p-NG modes and if they have their origins in dynamical symmetry breakings, 
then they may have decay constants, they would contribute to cosmological processes such as leptogenesis.
Hence, it may be possible that entries of the CKM/PMNS matrices vary by environment or a uniform acceleration
in our Universe.
Moreover, the entries of the CKM/PMNS matrices may have kink solutions.

\vspace{2mm}

By concerning the fact that a scalar field is dimensionless in (1+1)-dimensions,
we introduce the following action for describing dynamics of phason beyond a single-valley 
( i.e., a harmonic potential ) approximation:
\begin{eqnarray}
S_{p} &=& \int dx_{0}dx_{1} \Bigl[ \frac{F^{2}_{\Theta}}{2}(\partial_{\mu}\Theta)^{2} -\frac{m^{2}_{\Theta}}{2}\cos \Theta \Bigr].
\end{eqnarray}
We have redefined $m^{2}_{\Theta}$ from mass dimension.
This action is a relativistic version of a commensurate charge density wave,
and it will have a soliton solution as an excited state of phason
~[51,136,165,198,249].
By using a soliton solution, we obtain the energy of a soliton as
\begin{eqnarray}
E_{soliton} &\sim& \sqrt{F_{\Theta}m_{\Theta}},
\end{eqnarray}
namely, 
energy of a soliton can have the same order of magnitude of phason mass. 
By putting $m_{\Theta}$ and $F_{\Theta}$ of (154), (167) 
and $|m_{D}(T)|=|m_{D}(0)|\sqrt{1-T^{2}/T^{2}_{c}}$,
one finds,
\begin{eqnarray}
E_{solition} &\sim& \Bigg( \frac{2}{\pi^{2}}|m_{R}||m_{L}||m_{D}(0)|^{2} \Bigg)^{1/4}\Bigg[1-\frac{T^{2}}{T^{2}_{c}} \Bigg]^{1/4}.
\end{eqnarray}
Since $T_{c}$ and $|m_{D}(T=0)|$ take the same order of magnitude,
the Dirac mass term will vanish before a soliton is thermally excited
when $\Lambda\gg T_{c}$.
Therefore, tunneling and thermal effects are more important 
for a transport of energy between two nearest valleys
than a solitonic excitation.

\vspace{2mm}

It is interesting for us to extend our analysis on the $U(1)$ periodic potential to 
the $SU(N_{f})$ case discussed in the previous section.
In the $SU(2)$ cases of diagonal chiral mass breakings discussed in the previous sections, 
we have
\begin{eqnarray}
{\cal L} &=& 
\frac{1}{2}\Bigl( F^{2}_{\theta_{\chi}}(\partial_{\mu}\theta_{\chi})^{2}
+F^{2}_{|\theta|}\bigl\{(\partial_{\mu}\theta_{1})^{2}+(\partial_{\mu}\theta_{2})^{2}\bigr\} \Bigr)   
-|V|\cos\frac{\theta_{\chi}}{F_{\theta_{\chi}}}\cos^{2}\frac{\sqrt{\theta^{2}_{1}+\theta^{2}_{2}}}{F_{|\theta|}},
\end{eqnarray}
or
\begin{eqnarray}
{\cal L} &=& 
\frac{1}{2}\Bigl( 
F^{2}_{\theta^{L}}(
(\partial_{\mu}\theta^{L}_{1})^{2} + (\partial_{\mu}\theta^{L}_{2})^{2} 
)
+
F^{2}_{\theta^{R}}(
(\partial_{\mu}\theta^{R}_{1})^{2} + (\partial_{\mu}\theta^{R}_{2})^{2} 
)
+
F^{2}_{U1}(\partial_{\mu}\theta^{U}_{1})^{2}
+ 
F^{2}_{U2}(\partial_{\mu}\theta^{U}_{2})^{2} \Bigr)  \nonumber \\
& & -|V|\cos\frac{|\theta^{L}|}{F_{\theta^{L}}}
\cos\frac{|\theta^{R}|}{F_{\theta^{R}}}
\cos\frac{\theta^{U}_{1}}{F_{U1}}
\cos\frac{\theta^{U}_{2}}{F_{U2}}.
\end{eqnarray}
In those Lagrangians, we have made a trigonometric-function-approximation of the mass terms,
they show non-perturbative effects.
More generic case might be given by
\begin{eqnarray}
{\cal L} &=& \frac{1}{2}F^{2}\partial_{\mu}\Xi\partial_{\mu}\Xi^{\dagger}
- |V|\prod^{N}_{l=1}\cos\frac{\theta_{l}}{F},   \\
\Xi &\equiv& \exp\Bigl(i\sum^{N}_{l=1}\frac{\theta_{l}T_{l}}{F}\Bigr).
\end{eqnarray}
Here, we have used the notion of non-linear realization,
and we should mention that the definition of $\Xi$ has an invariance under a homeomorphism of 
the Lie algebra ( if it exists ), 
while the mass term is defined under a fixed algebra.
Thus, strictly speaking, our definition of the generic Lagrangian has a mathematical "inconsistency". 
A Hamiltonian with a potential $g(1+x^{2}/2!+x^{4}/4!)$, terminated at the fourth-order,
is a model of nonlinear lattice vibration, can be solved by a mean-field approximation.
If there is an unharmonic term for phases in their mass terms, 
then it might be possible that a chaotic behavior is caused in the dynamical systems.
In fact, our phason system can have a similar structure with the quantum kicked rotor~[37]:
\begin{eqnarray}
H &=& \frac{p^{2}}{2} -g\Bigl(\sum^{N}_{n=1}f(t-n)\Bigr)\cos\theta.
\end{eqnarray}
Here, the function $f(t-n)$ defines the shape of kick-force, and $t$ is time,
$p$ is a canonical momentum of $\theta$.
This is an important physical implication since our generalized Nambu-Goldstone theorem has some effects
they can erase histories of Universe, chiral phase transition of QCD, so on:
Axion or pion potentials can cause chaotic behaviors.
In the quantum kicked rotor, quantum localization and classical chaos compete with each other,
and it is a model which will give a similar physics with an Anderson localization in (1+1)-dimensions.
Namely, axion/pion/majoron/phason potentials can show an Anderson localization
which has been investigated in condensed matter physics.
An examination of relaxation from an excited state is an iteresting issue for us.
There might be a combined effect between the CP-violation and a chaotic motion around a bottom of valley.
Since our model gives a kind of Sine-Gordon equation, a solution of it may gives a soliton.
Therefore, we arrive at the notion of GNG soliton ( generalized-Nambu-Goldstone soliton ).
It is interesting for us to extend such a theory of GNG soliton to a non-Abelian case.

\subsection{Effective Action of the Phason-Photon System}

In case of pion, two-photon pion decay has a deep relation with the axial anomaly.
In Ref.~[190], an effect of temperature to $\pi^{0}\to 2\gamma$ was examined.
If phason $\tilde{\Theta}$ couples with electromagnetic field under the similar way with the case of axion,
the Lagrangian of the phason-photon system can be given as follows
~[19,59,60,75,108,236]: 
\begin{eqnarray}
{\cal L}_{\Theta -em} &\equiv& 
{\cal L}_{\Theta(K)} + {\cal L}_{\Theta{M}} + {\cal L}_{em} + {\cal L}_{\Theta -em},  \\ 
{\cal U} &\equiv& \exp\bigl[i\tilde{\Theta}/F_{\Theta}\bigr],    \\
{\cal L}_{\Theta(K)} &\equiv& F^{2}_{\Theta}\partial_{\mu}{\cal U}^{-1}\partial^{\mu}{\cal U},    \\
{\cal L}_{\Theta(M)} &\equiv& -F^{2}_{\Theta}\frac{m^{2}_{\Theta}}{2}\cos \frac{\tilde{\Theta}}{F_{\Theta}} = -F^{2}_{\Theta}\frac{m^{2}_{\Theta}}{4}\bigl( {\cal U} + {\cal U}^{-1} \bigr),    \\
{\cal L}_{\Theta-em} &\equiv&  \kappa\tilde{\Theta}F_{\mu\nu}\tilde{F}^{\mu\nu},   \\
{\cal L}_{em} &\equiv& -\frac{1}{4}(F_{\mu\nu})^{2} -\frac{1}{2\xi}(\partial_{\mu}A^{\mu})^{2}.
\end{eqnarray}
We will obtain the Raffelt-Stodolsky-type Lagrangian if we expand the potential of (296) 
up to the fourth-order of $\tilde{\Theta}$.
In our case, phason is massive boson.
Integrating out fluctuation of phason field by employing a harmonic potential approximation 
gives a four-body photon coupling 
$\kappa^{2}\int d^{4}y F_{\mu\nu}(x)\Tilde{F}_{\mu\nu}(x)G_{\Theta}(x-y)F_{\mu\nu}(y)\Tilde{F}_{\mu\nu}(y)$.
The interaction $\tilde{\Theta} F_{\mu\nu}\tilde{F}_{\mu\nu}$ 
( $\tilde{\Theta}$ is a pseudoscalar ) of the Lagrangian (297) 
causes a two-photon phason decay $\tilde{\Theta}\to 2\gamma$.
Note that this interaction gives gauge-invariant results in invariant scattering amplitudes.
The coupling $\kappa$ becomes
\begin{eqnarray}
\kappa &\equiv& \frac{e^{2}}{16\pi^{2}}F^{-1}_{\Theta}.
\end{eqnarray}
If we assume the temperature dependence of Dirac mass as $|m_{D}(T)|=|m_{D}(T=0)|\sqrt{1-T^{2}/T^{2}_{c}}$, 
we have
\begin{eqnarray}
F^{2}_{\Theta} &\approx& \frac{|m_{D}(T=0)|^{2}}{2\pi^{2}}\Bigl( 1 -\frac{T^{2}}{T^{2}_{c}} \Bigr)\ln\frac{\Lambda^{2}}{2|m_{R}|^{2}}. 
\end{eqnarray}
The result of Pisarsky et al. for pion is $f_{\pi}(T)\sim (1-T^{2}/(12f^{2}_{\pi}))f_{\pi}$
~[190], 
and it could be interpreted as the expansion of our formula at $T\ll T_{c}$.
Since we have obtained $F_{\Theta}$ by utilizing the Nambu-Goldstone theorem,
our $F_{\Theta}$ is estimated quite roughly,
though this formula tells us that a finite-temperature effect suppresses $F_{\Theta}$.
Similar effect of temperature can arise from a uniform acceleration 
or a curved spacetime, by the Hawking-Unruh effect
~[52,71,88,223].
Hence the decay width of $\tilde{\Theta}\to 2\gamma$ becomes
\begin{eqnarray}
\Gamma(\tilde{\Theta}\to 2\gamma) 
&\propto& \Bigl( \frac{e^{2}}{4\pi} \Bigr)^{2}\frac{m^{3}_{\Theta}}{F^{2}_{\Theta}}    \nonumber \\
&\propto& \Bigl( \frac{e^{2}}{4\pi} \Bigr)^{2}m^{3}_{\Theta}\Bigl\{ \frac{|m_{D}(T=0)|^{2}}{2\pi^{2}}\Bigl( 1-\frac{T^{2}}{T^{2}_{c}} \Bigr)\ln\frac{\Lambda^{2}}{2|m_{R}|}  \Bigr\}^{-1}.
\end{eqnarray}
Since we know that $m_{\Theta}$ does not vanish at $|m_{D}|=0$ with
$|m_{R}|\ne |m_{L}|$, $|m_{R}|\ne 0$, $|m_{L}|\ne 0$, 
we find that $\Gamma(\tilde{\Theta}\to 2\gamma)$ diverges 
at $T\to T_{c}$ or at $|m_{D}(T=0)|\to 0$.
An interesting subject is to know how to combine the phason effective action with
the Euler-Heisenberg effective Lagrangian of photon system of QED
under a consistent manner ( approximation level, gauge invariance, so forth )
~[90].

\section{The Mathematical Aspects of the Generalized Nambu-Goldstone Theorem}

( A reference: Mathematical Dictionary, 3rd. ed., Iwanami, Tokyo, 1985.
Caution: There are a lot of literatures/papers/works as our references 
and we cannot put all of them into the end part of this paper. )

\vspace{2mm}

The purpose of this section is to show the total structure of the generalized Nambu-Goldstone ( GNG ) theorem.
We wish to axiomatize several results to enlarge our perspective as much as possible.
We intend to put our GNG theorem into larger mathematical system/structure.
We will see that the GNG theorem contains a large part of modern mathematics and physics:
The GNG theorem gives in fact a view point to summarize them under a unified manner. 
It is also inside our perspective to enlarge our GNG theorem beyond "physical" situations.
Especially, cosets such as $G_{2}\backslash G_{1}$ 
( for example, $\Gamma\backslash G$ where $\Gamma$ is a discrete subgroup of $G$ )
and $G_{2}\backslash G_{1}/G_{3}$ 
( for example, $\Gamma\backslash G/H$ where $H$ is a closed subgroup of $G$ )
are interesting for us,
while we should keep the fact in our mind that 
sometimes a breaking scheme we will meet in a physical problem 
does not give a coset and it has no quotient topology:
The essential part of the GNG theorem must not depend on any special character/aspect of a homogeneous space.
Our theory of the GNG theorem must not be restricted on a coset or a homogeneous space $G/H$
( for example, any breaking scheme described by $SU(2)\to U(1)$ cannot become a homogeneous space ),
though a coset is mathematically easier for us to handle since $H$ is a closed subgroup of $G$, 
the breaking scheme $\pi:G\to G/H$ is just a canonical homomorphism,
and $G/H$ keeps the Hausdorff separable nature.
The theorem of Borel-Harish-Chandra-Mostow-Tamagawa states that the condition of compactness of 
$\Gamma\backslash G$ is that $G_{\mathbb{Q}}$ or $G_{\mathbb{Z}}$ are semisimple.
The arithmeticity theorem ( Margulis, 1974 ) is interesting for us:
In that theorem, it is proved that an irreducible lattice $\Gamma$ is an arithmetic subgroup
when ${\rm rank}_{R}G\ge 2$.
The central theorem of discrete subgroup of a semisimple Lie group
( called as a I-type topological group in the classification of
von Neumann algebra ):
If $G$ is a non-compact semisimple Lie group with ${\rm rank}_{\bf R}G\ge 2$,
then all of irreducible lattices of $G$ are arithmetic.
( A. Borel and Harish-Chandra,
{\it Arithmetic Subgroups of Algebraic Groups},
Margulis. )
The definition of {\it lattice}:
A discrete subgroup $\Gamma$ is called as a lattice when the volume form is finite, 
${\rm vol}(G/\Gamma)<\infty$.
For a variational calculas, 
it should be noted that any connected simple Lie group $G$ is decomposed as
$G=KS$, where $K$ is compact simple and $S$ is noncompact simple components of $G$.
One should also consider on "representations by homogeneous space",
especially from a perspective of ( Pontrjagin-Tannaka ) duality. 
An important example is provided from $U(1)\to{\rm nothing}$: 
In that case, by taking $G=\mathbb{R}$, $H=\mathbb{Z}$, 
then $S^{1}$ is given by a homogeneous space expression such that $G/H=\mathbb{R}/\mathbb{Z}\simeq S^{1}$.
A large part of our investigation of this section 
is schematically given in the following relations of groups:
\begin{eqnarray}
\begin{array}{ccccc}
G & \longrightarrow & G/H & \longrightarrow & \Gamma\backslash G/H  \\
  & & \Bigg\downarrow & & \\
  & & G_{\mathbb{C}}/H_{\mathbb{C}} & \longrightarrow & \Gamma'\backslash G_{\mathbb{C}}/H_{\mathbb{C}}.
\end{array}
\end{eqnarray}
Here $\Gamma$ ( $\Gamma'$ ) is a discrete subgroup of $G$ ( $G_{\mathbb{C}}$ ),
and it may capture a "periodicity" of an effective potential of a theory.
A very ideal situation is given by the situation when $\Gamma\backslash G/H$ is a Clifford-Klein form~[114].
In that case, $\Gamma$ acts on $G/H$ properly discontinuously and fixed point free,
and then $\Gamma\backslash G/H$ becomes a manifold, has no singularity:
A variational calculus of a function over it is safely employed.  
We should mention on a complexification of a Lie group in our context.
The following relation is a known fact and useful for us to consider the diagram given above: 
\begin{eqnarray}
\begin{array}{ccc}
G & \subset & G_{\mathbb{C}}  \\
\cup & & \cup  \\
H & \subset & H_{\mathbb{C}}.
\end{array}
\end{eqnarray}
This is satisfied in a case of homogeneous space of reductive type~[114].
For example, we have several choices for a domain in a $U(1)$ case discussed in the previous section, 
i.e.,
$\mathbb{C}$, $\widehat{\mathbb{C}}$ or $\bf{H}$ by taking into account the uniformization theorem
of Riemann surfaces.
In the case of $\bf{H}$, the analyticity of the Lie group at the origin may be lost
because it does not contain the real axis $\mathbb{R}$, 
though the algebraic property of Lie algebra/group are unchanged.
Thus, the domain of complexification of $U(1)$ group can be taken on any Riemann surface of genus $g\ge 0$.

\vspace{2mm}

Note that the periodicty of $V_{eff}$ of the $U(1)$ case given in the previous section 
can be described by two ways:
(1) Use a congruence subgroup of $SL(2,\mathbb{C})$ with taking the domain $z\in {\bf H}\equiv \Im z>0$,
(2) use a torus $\bf{T}$ with a fundamental period $(1,\tau)$, 
and make a mass function to contain the parameter $\tau\in{\bf H}$ while $z\in\mathbb{C}$. 
The second example gives an Abelian variety $(\tau_{j},z_{j})\in \mathfrak{S}_{g}\times\mathbb{C}^{g}$
with $g=1$.

\vspace{2mm}

Let us turn to a discussion on topology of a group.
In the mathematical side, "continuous" and "discrete" in topology is different,
while physicists know by their experiences that the difference is not essential
in the physics described by the ordinary NG theorem.
For example, cardinality numbers of $\mathbb{N}$ and $\mathbb{R}$ are different.
On the contrary, a phonon can live on a discrete lattice of a solid,
and a lattice Hamiltonian can be approximated by a continuous model
in a low-energy limit. 
Therefore,

\vspace{2mm}

{\bf Theorem:}
{\it physics of the ordinary NG theorem is inert with topology "discrete" or "continuous"},
while,
{\it physical predictions derived from the viewpoint of our GNG theorem distinguish
"discrete" and "continuous"}.

\vspace{2mm}

In fact, it seems the case that the nature of a topological space 
where a quantum field is defined is physically 
not essential for the NG theorem and one can give a field defined on various topological spaces,
but the ( quantum ) field theory constructed by the field 
must give a continuous model at the low-energy limit for the (G)NG theorem.

\subsection{Main Theorem.1}

{\bf Theorem.I:}
If $G_{2}\backslash G_{1}/G_{3}$ is compact and $G_{2}$ is finite, 
$f(g)$ of the integral $\int_{G_{2}\backslash G_{1}/G_{3}} f(g)dg$ 
always has at least a minimum and a maximum in the integration domain.
Then, if $f(g)$ has only one minimum inside $G_{2}\backslash G_{1}/G_{3}$, 
the symmetry between minima of $f(g)$ on $G_{1}$ or on $G_{1}/G_{3}$ 
will be {\it generated} by the action of $G_{2}$ on $f(g)$.

\vspace{2mm}

{\bf Theorem.II:}
Under the above setting, $V_{eff}={\rm const}$ gives a set of level sets,
and they divide the base manifold of $\Gamma\backslash G$.

\vspace{2mm}

For example, when $G_{2}$ is a fundamental group, 
then the symmetry between minima also aquires its representation in the same fundamental group.
The following fact, a fundamental fact of class field theory~[252,255], 
should be emphasized ( and, very useful ) 
to consider Galois representations they will appear in our GNG theorem: 
\begin{eqnarray}
{\rm a\,\, finite\,\, group\,\, scheme\,\, over}\, K\,\, {\rm is\,\, equivalent\,\, with}\,\, 
G_{K} = {\rm Gal}(\overline{K}/K), \quad ({\rm char} K=0),
\end{eqnarray}
where ${\rm Gal}(\overline{K}/K)$ is an absolute Galois group.
Recall the known facts:
(1) A finite group of order $n=p$ ( prime ) uniquely exists and it is a cyclic group.
(2) A finite group of order $n=p^{2}$ is an Abelian group.
(3) A finite group of order $n=pq$, $p\ne 1\mod q$ ( $p,q$; primes ) is a cyclic group,
while the case $p=1 \mod q$ gives a cyclic or a non-Abelian group.   
For example, 
if a Lie group ( or, an algebraic group ) $G$ contains a discrete subgroup $\Gamma$ 
implicitly, explicitly, or "generated" dynamically,
the quotient $\Gamma\backslash G$ ( possibly a universal covering ) gives an automorphic representation.
$\Gamma$ gives a Galois representation if $\Gamma$ is essentially defined 
over a field $K$ with ${\rm char}K=0$.
Thus, representations $V(g)$ ( $g\in\Gamma$, $g\in\Gamma/G$, ... ) of 
both continuous and discrete groups always contain Galois group actions.
When we prepare a Lie group by an exponential mapping from a Lie algebra,
then the automorphic representation realizes on a set of local coordinate system
( a coordinate ring ) and any function defined on the system. 
Therefore, we propose the following scheme in the case of Lie groups:
\begin{eqnarray}
& & {\rm Lie\,\, algebras \,\, and \,\, coordinate \,\, rings} \to {\rm Lie \,\, groups \,\, and \,\, their \,\, representations}  \nonumber \\
& & \qquad \to {\rm automorphic \,\, representations} \to {\rm Galois \,\, representations}.
\end{eqnarray}

\vspace{2mm}

We will see the following statement:

\vspace{2mm}

{\bf Theorem.III:}
{\it The mechanism of our GNG theorem, from continuous to discrete,
works well in various one-dimensional groups. This fact does not imply that our theory/theorem
is restricted to an Abelian class field theory, but contains also a non-Abelian class field theory,
as shown in the double-periodic potential of $SU(2)$-case.}

\subsection{Lie Algebras/Lie Spaces}

A Lie algebra is generically written as follows~[??]:
\begin{eqnarray}
\widehat{\mathfrak{g}} &=& \bigl( \mathfrak{g}\otimes_{A_{1}}{\cal F}_{1}\bigr) \oplus {\cal F}_{2}c \oplus {\cal F}_{3}d.
\end{eqnarray}
To include a Kac-Moody algebra, we put the center of algebra by $c$, 
and $d$ acts as $[d,t^{n}\otimes\mathfrak{g}]=nt^{n}\otimes\mathfrak{g}$ 
in the case of affine Lie algebra of one variable. 
In the case of the algebra over a complex manifold $X$, 
$A_{1}=\mathbb{C}$, ${\cal F}_{1,2}\in {\cal O}_{X}$ ( locally flat free sheaf over $X$ ),
and it will define a stalk ${\cal F}_{p}$ at $p\in X$ ( local ring ).
In the case of affine Lie algebra, the "formal" power series field does not belong to a scheme
but an ind-scheme.
$X=M\times S$ will be introduced if one consider a deformation of complex structures,
where $M$ is the base manifold, $S$ is a parameter space for deformations,
and the analytic family is obtained via the map $\pi:X \to S$.
Such a deformation, if it exists ( no obstruction ), 
gives a possible effect in the mass function ${\cal M}\in \Gamma(M\times S)$
because a mass function contains such a deformation parameter implicitly/explicitly. 
A usual treatment on a mass function in a calculation of symmetry breaking 
is employed under a fixed complex structure,
while a deformation of a Lie group ( and a deformation of a Galois group/representation also ) 
gives a deformation of ${\cal M}$. 
A consideration of deformation ( in a very generic sense ) is important 
because it can find a theory larger than that of the starting point of a deformation, 
usually contains it, 
and sometimes a deformation clarifies the structure of a theory.
Of course, one can "localize" the Lie algebra in the sense of gauge theory in physics, for example,
\begin{eqnarray}
\widehat{\mathfrak{g}}(x^{\mu}) &=& \Bigl( \mathbb{C}[[t(x^{\mu})]]\otimes\mathfrak{g} \Bigr) \oplus \mathbb{C}c(x^{\mu}).
\end{eqnarray}
Here, $x^{\mu}$ ( $\mu=1,\cdots,D$, $D\in\mathbb{N}$ ) are spacetime coordinates.

\vspace{2mm}

Beside the Hilbert 5th problem, 
one can consider various rings/fields for constructing a Lie algebra $\mathfrak{g}\otimes_{A}M$
under an algebraic manner:
For example, local fields $\mathbb{R}$, $\mathbb{C}$, global fields, 
or a power series ring/field over a ring.
To construct a Lie group from a Lie algebra via an exponential mapping,
one may need an algebra defined over a locally compact topological field, namely $\mathbb{R}$ or $\mathbb{C}$, 
for the sake of analyticity and convergence property. 
( For example, the exponential function over $\mathbb{Q}_{p}$ does not always converge in the whole domain 
of $\mathbb{Q}_{p}$. )
The theorem of Ostrowski states that there are only four example
( $\mathbb{R}$, $\mathbb{C}$, the rational function field over $\mathbb{F}_{q}$ ( $q$; prime ), 
and the p-adic number field $\mathbb{Q}_{p}$ ) 
they are locally compact topological fields of paracompact type ( not a discrete topology ). 
While, there are various examples of general linear groups they are employed in number theory
( for example, a group defined over an adele ring )
and we do not restrict ourselves on a specific type of Lie groups 
because we wish to enlarge our consideration on GNG theorem as much as possible. 
In cases where $M$ takes its value on a complex projective space,
a complex Grassmann space and a complex Abelian variety, 
we can consider the Kodaira-Spencer-type deformations of complex structures of them 
because all of them are Riemannian symmetric, Hermitian symmetric, K\"{a}hler and compact.  
Moreover, algebraic groups ( thus, Lie spaces also ) and their breaking schemes are considered
in this paper.

\subsection{Lie Groups, Algebraic Groups and their Representations}

\subsubsection{Mass Matries}

Since we extend our theory of the GNG theorem to include not only Lie groups
but also topological groups 
( a topological space with the condition that
$G\times G\to G$, $(x,y)\mapsto xy$, $G\to G^{-1}$, $x\mapsto x^{-1}$ are continuous ),
especially of locally compact type, 
the criterion of a realization of broken symmetry should be modified:
The ordinary NG theorem handles only Lie groups,
while we should extend groups in our GNG theorem due to the result given in the previous sections. 
We also have our interest on how the usual Lie-group-type topology
( locally Eucleadian ) will/can take place ( generate )
from more generic topological group, via our GNG theorem.
We should mention the fact that a completion $\widehat{G}$ of its corresponding
topological group $G$ does not always exists. 
The criterion of a realization of broken symmetry is simply expressed by a mass matrix in our GNG theorem.

\vspace{2mm}

{\bf The Criterion of Breakdown of a Symmetry:}

\vspace{2mm}

Let $\rho_{0}:G\to GL_{n}(R_{0})$ be a representation of a general linear group,
and let $R_{0}\in R$ be a ( commutative ) ring.
A completion or a localization of $R_{0}$ is achieved by using its maximal ideal if one needs it.  
Let $H$ be a subgroup of $G$, gloup elements are denoted by $h\in H$, $g\in G$,
and we assume the beginning of the vacuum state $|0\rangle$ of a system ${\cal L}(G,H,\cdots,R_{0},R,\cdots)$ 
( a Lagrangian ) is invariant ( singlet ) 
under any operation of the groups.
For the definition of a broken symmetry, 
we do not have to specify the topology of a group, its represenation, and a vector space.
In a case of a matrix represenation of a topological group,
usually we assume the existence of a continuous Abelian structure in its corresponding vector space. 
It is also possible to assume that a space of mass matrices has a definition of 
distance under a specified sense, namely, metrizable
( this is not the same with the question that a group $G$, which a mass matrix is its representation,
is metrizable or not ).
For the criterion whether the breakdown of a symmetry is spontaneous or not,
we use the usual definition:

\vspace{2mm}

{\bf Definition:}
{\it If the symmetry expressed by a group-operation is kept in the beginning of a theory,
while it is broken by a mechanism inside the theory, 
then the breakdown of the symmetry spontaneously takes place.}

\vspace{2mm}

Thus, we consider 
(1) the beginning of a theory is a ( well-defined ) representation of a symmetry of a group, or not,
(2) the symmetry is broken in a specific, well-defined sense, in the spontaneous manner, or not. 
Hereafter, we assume both the case.
More generally, a "symmetry" in physics is given as a homomorphism
\begin{eqnarray}
\varrho: G \longrightarrow {\rm Aut} S,
\end{eqnarray}
namely, an automorphism of a set $S$.
One can choose a total space of equivalence classes of finite-dimensional irreducible unitary representations
for a consideration/construction on representation of a compact group, 
to remove a redundancy: 
If a mass matrix belongs to a representation of a compact group, 
we only need 
its total space of equivalence classes of finite-dimensional irreducible unitary representations,
while we have to prepare
(i) the total space of equivalence classes of finite-dimensional irreducible representations,
(ii) the total space of equivalence classes of finite-dimensional irreducible unitary representations
and
(iii) the total space of equivalence classes of irreducible unitary representations,
for obtaining the total space of representations of the mass matrix spaces of a non-compact group.

\vspace{2mm}

To go beyond the defintion of a broken symmetry by Lie algebra,
we introduce the following definition of a broken symmetry by the notion of groups.
Let $G_{1}$ be a group, and let $G_{2}$ be its subgroup.
Let $\widetilde{\mathfrak{M}}_{0}$ be a mass function generated spontaneusly by a system,
it is defined on a point which is contained in a space of mass matrices $\widetilde{\mathfrak{M}}$,
and the relation ( for example, the distance ) between two points of the space $\widetilde{\mathfrak{M}}$
is defined under a specific manner ( a topological space is the usual situation ).
There are the following cases under an adjoint action of $g_{2}\in G_{2}$, 
\begin{eqnarray}
{\rm Ad}(g_{2})\widetilde{\mathfrak{M}}_{0} \equiv g_{2}\widetilde{\mathfrak{M}}_{0}(g_{2})^{-1} 
\begin{cases}
= \widetilde{\mathfrak{M}}_{0}, \qquad ({\rm invariant}),   \\
\ne \widetilde{\mathfrak{M}}_{0}, \quad {\rm but} \,\, \in \widetilde{\mathfrak{M}}, \qquad ({\rm belongs}),   \\
\notin \widetilde{\mathfrak{M}}, \qquad ({\rm broken}), 
\end{cases} 
\end{eqnarray}
( $(g_{2})^{-1}$ is an appropriately defined inverse operation of $g_{2}$ inside the group $G_{2}$ ).
This is a classification by $G_{2}$-orbits.
( A $G$-orbit decomposition has a deep relation with the Nagata-Mumford theorem,
especially, when a set of $G$-orbits are defined over an affine scheme $X={\rm Spec}R$.
Each closed $G$-orbit corresponds to an equivalence class used to define a moduli of the affine scheme. )  
The first and second cases are the usual notion of "broken", 
and they are related with a centralizer and a normalizer, respectively.
Note that our definition includes not only Lie groups or topological groups,
but also discrete and finite groups.
From this aspect, a definition of topological space of mass matrices 
( topology of the space of mass matrices )
does not give an important condition.
We should emphasize the fact that a mass matrix can have several topology simultaneously.
Thus, one can consider the following quantum field for spontaneous broken symmetries,
\begin{eqnarray}
\Psi_{a,\mathfrak{a},\alpha},
\end{eqnarray}
where, $a$, $\mathfrak{a}$ and $\alpha$ denote gauge-charge, supercharge, and Galois charge, respectively.

\vspace{2mm}

One of the motivations of our definition of a broken symmetry is to
take into account a deformation of a Lie/general-linear group defined 
over a scheme $X$ over $k$. 
It is formally obtained by utilizing a fiber product for algebraic family of schemes:
\begin{eqnarray}
GL_{n}(X_{y}), \quad X_{y} = X_{0}\times_{Y}{\rm Spec}k(y).
\end{eqnarray} 
Here, $k(y)$ denotes a residue field at $y$.
A construction of mass matrix via a gloup element of $GL_{n}(X_{y})$ contains those deformations.
Then, the interesting problem is whether there is a family, 
and then how such a deformation affects to the broken symmetry.
In that case, the group element is defined over $X_{y}$, i.e., $g(X_{y})$.

\vspace{2mm}

In the ordinary NG theorem, broken or symmetric is judged by an examination of 
algebra ( in fact, a commutator ) between the basis set ( generators ) of a Lie algebra 
and an operator ( fields, VEVs, mass functions ),
because it implicitly assumes an existence of a Lie group over $\mathbb{A}^{n}_{\mathbb{R}}$
in a theory/model.
This fact implies that the ordinary NG theorem only considers the vicinity of the origin
of Lie group by utilizing its tangent space,
and thus the geometric aspect of the ordinary NG theorem is understandable
by the method of Maurer-Cartan form ( the dual of a Lie algebra ). 
Hence, the essential part of the ordinary NG theorem may be understood by an examination/extension
of the following exact sequence:
\begin{eqnarray}
1 \longrightarrow \mathfrak{g} = T_{0}G \longrightarrow TG \longrightarrow G \longrightarrow 1.
\end{eqnarray}
We will emphasize the fact that we do not need 
an exponential mapping from a Lie algebra to the corresponding Lie group
to construct a mass matrix, in principle, 
and the periodicity we discuss in this paper must arise without this procedure.
If there is a bijective morphism between a Lie algebra and a Lie group,
then the usual criterion of broken symmetry acts well.
It seems the case that there is no obstruction to make a physical/mathematical situation 
considered in the NG theorem, 
from Lie groups to topological groups, 
and sometimes they are obtained by a completion of inverse limit,
an $I$-adic filteration, or Banach/Hilbert-Riemann/Finsler, so on.
Even in the case of a Lie group/algebra, one takes a Banach ring $A$ in $A\otimes{\rm Lie(G)}$.
Hence, in our GNG theorem, we have topology or groups distinguished such as
"discrete", "continuous", "compact" or "locally compact."

\vspace{2mm}

{\bf The Definition of Stable Subspaces and the GNG Manifold:}

The definition of a derivation $D$ is defined by~[251,253]
\begin{eqnarray} 
D: R_{0} \to M_{0}, \quad D(xy) = xD(y) + D(x)y, \quad \forall x,y \in R_{0} \subset R,
\end{eqnarray}
where $R_{0}$ is a ring over a field $k$ assumed as a subring of $R$, 
$M_{0}$ is an $R_{0}$-module.
Let $f$ be a function defined on $R_{0}$, 
let $R_{0}$ be a ring defined on a space of broken symmetry,
and assume a derivation $D$ of a broken symmetry exists. 
Then the NG manifold of the ordinary NG theorem is defined by the gap equation
for "phase degrees of freedom", 
\begin{eqnarray}
D_{j}f(x\in R_{0}\subset R) = 0, \qquad \forall D_{j}, \quad j\in\mathbb{N} < \infty.
\end{eqnarray} 
However, this differential equation does not crucially depend on a local coordinate system
of a Lie group, and can be extended to $D$ defined over a ring or a field.
In our GNG theorem, this equation gives us only a set of stable points
they locate along with the NG manifold.
Thus,

\vspace{2mm}

{\bf Axiom:}
The condition of realization of our GNG theorem is stated as follows:
$D_{j}f(x)=0$ ( $\forall j$ ) gives a true subset $x\in S_{0} \subsetneqq R_{0}\subset R$ 
with a finite codimension.

\vspace{2mm}

A consideration on conjugacy classes is important for us since we do not restrict ourselves to
the logic $f:{\rm Lie}G\to G$ for our GNG theorem.
From the definition of the notion of "conjugate",
\begin{eqnarray}
H_{2} &=& gH_{1}g^{-1}, \quad \exists g\in G, \quad H_{1},H_{2} \subseteq G,
\end{eqnarray}
( the case $H_{1}=H_{2}$ corresponds to the case of the normalizer of $G$ ),
one notice $\mathfrak{M}\in H_{1}\oplus H_{2}$ has a symmetric case as a special example,
under operation of $g$.
Then, it is possible to exist an equivalence class of sets of 
stable points of our GNG theorem defined to be
\begin{eqnarray}
D_{j,p}V_{eff}(\mathfrak{M}_{p}) = 0, \qquad D_{j,p} \in D({\rm conj}), \qquad 
\mathfrak{M}_{p} \in \mathfrak{M}^{\rm conj},
\end{eqnarray}
( $p$; a point of a conjugacy class, "${\rm conj}$" implies a conjugacy class. )
which reflects the symmetry of the NG manifold of the problem.

\vspace{2mm}

In our GNG theorem, we will extend our theory to consider not only a Lie group,
but also a compact group, a topological group, a pro-finite group, a finite group, so forth.
It is a very important ( and famous ) fact that any compact group $F$ will be obtained
by a projective limit of a compact Lie group $\{G_{n}\}$: 
\begin{eqnarray}
F = \lim_{\leftarrow n}G_{n},
\end{eqnarray}
and it is possible that a realization of a discrete group in a theory/model
given by a "superposition" ( or, a cooperative effect ) of a set of breaking schema $G_{n}$. 
One of other possibilities is a direct product topological group:
\begin{eqnarray}
G &=& \prod_{\alpha\in A}G_{\alpha}, 
\end{eqnarray}
where, $G$ is defined on a direct product topology of 
a set ( family ) of topological groups $\{G_{\alpha}\}_{\alpha\in A}$.
A projective limit group ( projective system of topological groups ),
\begin{eqnarray}
G = \lim_{\leftarrow \alpha}G_{\alpha}, \qquad &{\rm or}& \qquad 
G \simeq \lim_{\leftarrow\alpha}G/H_{\alpha} 
\end{eqnarray}
also provides another possibility for us.
In those cases, we can find various breaking schema and realizations of symmetries 
in physical systems, some examples might not be recognized until now.
One can consider a breakdown ( breakdowns ) of symmetry ( symmetries ) of the group ( those groups )
of a specific way ( several ways ).  
In the ordinary NG theorem, the notions of both $continuous$ and $differentiable$ 
are essentially important.
In our theory of the GNG theorem, 
we can utlize the fact that a regular ( left/right ) representation of 
a locally compact topological group $G$ with a topology of uniform convergence
becomes continuous. 
If $G$ is unimodular ( = compact topological group ), 
then there is an invariant measure ( both right and left ), 
and an integration ( Lebesgue ) which would be required in a theory/model becomes easier to handle
( and also can employ several approximation techniques such as the steepest decent, so on ).

\vspace{2mm}

Let us consider the case where a mass matrix $\widetilde{\mathfrak{M}}$ 
is defined over a scheme $X={\rm Spec}R$ ( $R$; a ring ) and an algebraic group $G$ acts on $X$.
Then, if $X$ contains its subring as a set of invariants of $G$,
the following diagram is found~[253],
\begin{eqnarray}
\begin{array}{ccc}
X = {\rm Spec}R & \to & Y = {\rm Spec}{R}^{G}   \\
\downarrow & & \downarrow \\
\widetilde{\mathfrak{M}}(X) & \to & \widetilde{\mathfrak{M}}(X)^{G},
\end{array}
\end{eqnarray}
and it can give a moduli for both $X$ and $\widetilde{\mathfrak{M}}$$(X)$.

\vspace{2mm}

In the case of a product of a Lie group $G\subseteq SL(n,A)$ ( $A$, a ring or a field ) and 
a Galois group of Abelian extension,
\begin{eqnarray}
\mathfrak{G} &=& G\otimes {\rm Gal}(L/K),
\end{eqnarray}
the Galois part defines a center of $\mathfrak{G}$. 
The result of the breaking scheme $SU(N)\times U(1)\to SU(N)$ 
with an explicit symmetry breaking parameter for $U(1)$
discussed previously provides an example of this case.

\vspace{2mm}

{\bf A Mass Matrix on an Abelian Variety:}

\vspace{2mm}

An interesting application of theta function ( which appears as a character of affine Lie algebra )
is found in our mass matrix.
For example, 
\begin{eqnarray}
{\cal M} &\equiv& {\rm diag}(\Theta_{j_{1},m_{1}}(\tau_{1},z_{1}),\cdots,\Theta_{j_{n},m_{n}}(\tau_{n},z_{n})), \\
\Theta_{j_{l},m_{l}}(\tau_{l},z_{l}) &=& \sum_{k\in \frac{j_{l}}{2m_{l}}+\mathbb{Z}}\exp\Bigl[2\pi im_{l}\bigl( k^{2}\tau_{l}+kz_{l} \bigr)  \Bigr].
\end{eqnarray}
On the other hand, the linear combination the holomorphic and anti-holomorphic theta functions gives
the following expression on the real axis $(\tau,z)\in\mathbb{R}^{1}\times\mathbb{R}^{1}$ :
\begin{eqnarray}
\Theta_{j_{l},m_{l}}(\tau_{l},z_{l}) + \bigl( \Theta_{j_{l},m_{l}}(\tau_{l},z_{l})\bigr)^{*}
= 2\sum_{k}\cos(2\pi m_{l}kx_{l}).
\end{eqnarray} 
While, the character of a representation of dihedral group $\rho:G\to D_{2n}$
is
\begin{eqnarray}
\chi_{\beta} &=& q^{\beta} + q^{-\beta}, \quad q \equiv e^{2\pi i/\beta}, \quad \beta \in \mathbb{N}.
\end{eqnarray}
Thus, when a dynamical mass of a model is expressed by the theta function and its anti-holomorphic counterpart,
then the model gives an "approximated" represenation of 
dihedral group at $\mathbb{R}^{1}\ni x=1/\beta m_{l}k$,
because $V_{eff}\sim \cos\theta$. 
If each $\Theta_{j_{l},m_{l}}$ is defined over a Jacobian variety $J(R)$, 
then the each pair $(J(R),\Theta_{j_{l},m_{l}}=0)$ 
uniquely defines a closed Riemann surface from the Torelli theorem~[254]. 
Thus, the mass matrix defined on a set of Jacobian varieties $J(R_{l})=\mathbb{C}^{g}/\Omega_{l}$
expressed by a maximal torus defines a multiple of closed Riemann surfaces.
An Abelian variety 
( complex torus, the number of moduli $m=\dim H^{1}(\mathbb{C}^{n}/\Lambda,\Theta)=n^{2}$ ) 
is a variety possible to be embedded into $\mathbb{C}P^{n}$ as a complex subvariety, 
and a theta function is a holomorphic section of a complex line bundle of an Abelian variety.
The special case will happen when 
\begin{eqnarray}
{\cal M} \equiv {\rm diag}\bigl( \Theta_{j,m}(\tau,z),\cdots,\Theta_{j,m}(\tau,z)\bigr),
\end{eqnarray}
namely, the all diagonal elements share the same theta function of the same value.
Let $R$ be the Riemann surface corresponds to $\Theta_{j.m}$.
Then the mass matrix defines
\begin{eqnarray}
R^{(n)} \equiv R^{n}/\mathfrak{S}_{n}, \quad R^{n} \equiv \underbrace{R\times \cdots \times R}_{n}.
\end{eqnarray}
Here, $\mathfrak{S}_{n}$ is a symmetric group.
The multiple Riemann surface $R^{(n)}$ is a variety
of compact, complex, non-singular, complete and projective:
The mass function has a deep connection with the moduli problem of Riemann surfaces.

\vspace{2mm}

Beside the analysis of mass functions on an Abelian variety, 
there is an interesting interpretation on the Schwinger-Dyson equation of dynamical masses 
as a Picard-Fuchs equation~[254].
It is a well-known fact that the Gauss-Manin connection defined on an Abelian variety
gives 2$g$-dimensional vector-type first-order differential equation,
and it is converted into a 2$g$th-order Picard-Fuchs equation.
Such a Picard-Fuchs equation has a strong connection with a moduli problem of Riemann surfaces.
Since the Gauss hypergeometric equation or the Heun equation are special types of Picard-Fuchsian,
they and their solutions ( i.e, mass functions ) will obtain an interpretation from the moduli theory.

\vspace{2mm}

Matrix elements of a representation of a general linear group form a graded ring 
if they are obtained by an exponential mapping from a Lie algebra.
The essence of deformation theory and the Kac-Moody theory 
( an affine Lie algebra is contained as a special case of a Kodaira-Spencer deformation of complex structures
when one consider a complex manifold, especially a curve, as the base space of the theory~[??],
and thus the ordinary Lie algebra is a subalgebra of an affine Lie algebra,
while a case of several complex variables will give an affine Lie algebra of several variables )
are interpreted by sequences of graded rings
\begin{eqnarray}
& & {\rm Exp}: \mathfrak{g} \to G, \qquad X \in \mathfrak{g}, \quad g \in G,  \\
& & e^{X} = 1 + X + \frac{X^{2}}{2!} + \cdots   
= 1 + S^{1} + S^{2} + \cdots 
\to 1 + P^{1} + P^{2} + \cdots,
\end{eqnarray}
namely, they provide a change of orders of homogeneous polynomials ( ideals ) such as
\begin{eqnarray}
S^{j} &\to& C_{j1}P^{1}+ C_{j2}P^{2}+ \cdots.
\end{eqnarray} 
Thus, such a change has a strong relation with natural actions of the multiplication group $\mathbb{G}_{m}$
to the homogeneous elements of polynomials.
One should notice that the action of $\mathbb{G}_{m}={\rm Spec}k[x,x^{-1}]$ 
to a homogeneous ideal contains a translation over a lattice represented by a Galois group. 
A graded ring has an action of $\mathbb{G}_{m}$, 
and a representation of $\mathbb{G}_{m}$ is isomorphic with the a graded ring
$R=\oplus_{m\in\mathbb{Z}}R_{m}$, $R_{m}R_{n}\subset R_{m+n}$.
By taking a set of representations of $\mathbb{G}_{m}$, 
one can find a moduli of the graded ring contained in the exponential mapping. 
Those mathmatical structures are contained in mass matrices
given by our method of constructions.

\vspace{2mm}

In fact, a Chevalley group which is a finite group of Lie type is given by taking a Chevalley basis
of Lie algebra $\mathfrak{g}$, taking a $\mathbb{Z}$-linear combination $\mathfrak{g}_{\mathbb{Z}}$ of the basis,
and then taking an exponential map. 
A case of group over an arbitrary field $k$ is obtained by a tensor product 
$\mathfrak{g}_{k}= k\otimes_{\mathbb{Z}} \mathfrak{g}_{\mathbb{Z}}$.
The prescription discussed in the above is a general case of this.
Let $G$ be a connected reductive algebraic group scheme with a Frobenius $F:G\to G$
( an $\mathbb{F}_{q}$-rational structure ).
Then one obtains a finite group $G^{F}$: 
This is a special case of the procedure to obtain a finite group in our GNG mechanism.
Beside this, one can also consider a diagonal breaking of a Chevalley group
( namely, a theory or a Lagrangian has a symmetry of Chevalley group from the beginning )
because it has a Cartan subgroup.
As we have discussed in previous section, a diagonal breaking of a non-Abelian unitary group is more involved.
For example, the Frobenius map is defined as $F(\{x_{ij}\})=\{(x_{ij})^{q}\}$, $\{x_{ij}\}=g\in G$,
namely, the Chevalley groups or Deligne-Lusztig theory 
( repesentation theory of finite groups of Lie-type,
see, 
C. W. Curtis,
{\it Representations of Finite Groups of Lie Type},
Bull. Am. Math. Soc. {\bf 1}, 721 (1979),
R. W. Carter,
{\it A Survey of the Work of George Lusztig}, Nagoya J. Math. {\bf 182}, 1 (2006). ) considers
a Frobenius map of all of the elements of a group.
On the other hand,

\vspace{2mm}

{\bf Theorem}:
{\it Let $G$ be an algebraic group ( scheme ), and let us consider a breaking diagram $G\to H$ with $X=G-H$.
If our GNG mechanism gives a set of discrete equilibria in the topological space $X$,
and they have a symmetry of Frobenius map, then the subspace aquires a representation
of Deligne-Lusztig theory.}

\vspace{2mm}

We mention some facts: If $H$ is a closed normal subgroup of $G$ defined in the theorem above,
and if $H$ is connected, $G^{F}/H^{F}\simeq (G/H)^{F}$ holds.
Especially, when $G=GL_{2}$, then
\begin{eqnarray}
G^{F}/B^{F} &=& \mathbb{P}^{1}(\mathbb{F}_{q}),
\end{eqnarray}
( $B$, a Borel subgroup ), and it realizes in the compact-type ( i.e., genus = 0 ) mass function of 
the $U(1)$ case of our GNG theorem. 
Of course, our complexification $G=SL(2,\mathbb{C})$ contains a Frobenius map of complex conjugations
$F:\{x_{ij}\}\to\{x^{*}_{ij}\}$. Then $G^{F}=SL(2,\mathbb{R})$.
Since we consider a symmetry breaking of a group $G$, 
we should examine carefully "a ( partial ) discrete Lie group" arised from our GNG mechanism,
a comparison with Chevalley groups.

\vspace{2mm}

Let us assume the case that the set of equilibria of an NG manifold corresponds to
a Chevalley group scheme.
After one has a set of "winding numbers" 
( or periods, especially, in integers or primes or powers of primes ) 
of our effective potential,
one obtains a familiy of finite groups of Lie types $G(q)$ such as $GL_{n}(q)$. 
This is very interesting from the viewpoint of "the ramifications of theories",
mentioned in the previous section.

\subsubsection{Determinants}

A fermionic determinant which may appear in a path integration 
has several aspects.
It has a similarity with zeta-functions ( zeta-regularization ), 
an invariant function relates with moduli theory,
given by a section of line bundle ( or, an invertible sheaf ) defined on a certain topological space,
so on.
For example, $\det(\pfey-{\cal M})$ gives the residue ring
\begin{eqnarray}
k[X]/\det\bigl(\pfey-{\cal M}(X)\bigr) \simeq k(\lambda_{j}),
\end{eqnarray}
where, $k$ is a field, and $\lambda_{j}$ imply the eigenvalues of the determinant $\det(\pfey-{\cal M})$.
This is the algebraic meaning of the dispersion relations.
Hence $V_{eff}$ is a superposition of infinite-number of extension $k(\lambda_{i})$.
A fermionic determinant is generically written in the following form:
\begin{eqnarray}
\det\Bigl( 1- u(p)\bigl( \mathfrak{Y}+{\rm Ad}(g)\mathfrak{X} \bigr) \Bigr),
\end{eqnarray}
where, $u(p)$ is a parameter depends on a momentum $p$,
$\mathfrak{Y}$ and $\mathfrak{X}$ belong to a Lie algebra/group under matrix forms,
and ${\rm Ad}(g)\mathfrak{X}$ gives a set of adjoint orbits:

\vspace{2mm}

{\bf Theorem:}
{\it The periodicity discussed in this paper corresponds to the adjoint orbits.}
$\mathfrak{Y}+{\rm Ad}(g)\mathfrak{X}={\rm const}$ 
{\it gives a level set and then the adjoint orbit divides} $V_{eff}$ 
{\it into ( generally uncountable number of ) pieces with a topology ( discrete, continuous ). 
A special case of the adjoint orbits contains a set of stable points as a subset of NG manifold,
possiblly becomes discrete.}

\vspace{2mm}

One should notice that the definition of the adjoint orbits in the above statement is
slightly different from the usual definition.

\vspace{2mm}

If the submanifold given by an NG-mode is isomorphic with $S^{1}$ ( for example, the $U(1)$ case ),
then the set of adjoint orbits will be divided into three subsets:
(a) isomorphic with $S^{1}$,
(b) not isomorphic with $S^{1}$ but given by a set of closed curves,
(c) given by a set of points.
And, we can say all of those cases reflect the topological nature of $S^{1}$.

\vspace{2mm}

Let us consider a $U(1)$ case.
Since $g$ contains $U(1)\ni e^{i\theta}\to S^{1}$, 
the fundamental group $\pi_{1}(S^{1}) \simeq\mathbb{Z}$ acts on the domain $\theta\in\mathbb{R}$,
and after converting it into a mod $n$ representation $\mathbb{Z}/n\mathbb{Z}$ 
( it is an Artinian ring, namely, the Krull dimension is 0 ),
we get a Frobenius morphism:  
\begin{eqnarray} 
\theta \to \theta + 2m\pi, \quad m \in \mathbb{Z}/n\mathbb{Z}.
\end{eqnarray}
The determinant of our GNG theorem keeps the symmetry of Frobenius exactly:
It is a representation of invariant of the Frobenius, defines a functor
$\varrho: f(g\in U(1))\to f(g\in U(1))^{Frob}$.
Moreover, one can introduce an $l$-adic representation via the mod $n$ representation.
Thus, a Tate-type module and a Galois action exists inside the determinant. 
These "symmetry" and the Galois action exist inside a path integration,
must be excluded by the notion of "moduli space" ( of our mass matrix ) taken by a quotient. 
For obtaining a more general case, recall the fact that 
a finitely generated Abelian group $\cal{A}$ has the following decomposition
by considering the fundamental theorem on Abelian groups and the Chinese remainder theorem:
\begin{eqnarray}
{\cal A} &\simeq& \mathbb{Z}^{\oplus s}\bigoplus \mathbb{Z}/n\mathbb{Z},  \\
\mathbb{Z}/n\mathbb{Z} &\simeq& \mathbb{Z}/p^{e_{1}}_{1}\mathbb{Z}\times \cdots\times \mathbb{Z}/p^{e_{r}}_{r}\mathbb{Z}, \quad
n = p^{e_{1}}_{1}\times\cdots\times p^{e_{r}}_{r},
\end{eqnarray} 
where, $p_{1},\cdots,p_{r}$ are prime numbers.
Thus, the Frobenius-like symmetry of the determinant 
belongs to the torsion part of the decomposition of an Abelian group.
It is a suggestive fact for us that a solvable group 
is almost an Abelian group given by a stack ( Aufbau ) of Abelian groups. 
Since our determinant can be replicated with keeping a certain symmetry, eigenvalues and energy,
the determinant aquires a representation of a ( almost ) solvable group. 
The eigenvalues of the determinant contain the set of stable points they reflect those symmetries.
Because the fermionic determinant $\det(x-M)$ is factorized as $(x-\lambda_{1})\cdots(x-\lambda_{n})$,
the determinant has a correspondence with a point on an affine space $k^{n}$ via the Hilbert Nullstellensatz,
and it defines a subspace of the polynomial ring $k[x]/\det(x-M)$, 
a localization of $k[x]$ by $\det(x-M)$.

\vspace{2mm}

Since the mass matrix belongs to a group $G$,
the periodicity of the determinant of the mass matrix will be described by
\begin{eqnarray}
& & G \ni \widetilde{\cal M} \to \det\widetilde{\cal M} \simeq \mathbb{G}_{m}, \\
& & \chi \ni \det\Bigl( 1- u \bigl( \mathfrak{Y}+{\rm Ad}(g)\mathfrak{X} \bigr) \Bigr) \in {\cal M}(\Gamma).
\end{eqnarray} 
$\chi$ is the space of character of $G$.
This implies a deep connection with the Galois representation of an automorphic form on $\Gamma\backslash G$.

\vspace{2mm}

{\bf Theorem:}
{\it Our determinant gives a functor} $\det$ {\it from a category of Lie-algebra/group-valued matrices
to a category of Abelian group which contains a set of characters ( Galois representations )
and authomorphic forms.}

\vspace{2mm}

This statement will be generalized into the cases of commutative and solvable group schemes.

\vspace{2mm}

Next, we consider the case of coset $G/H$.
In that case, the determinant takes the form of Chern-Weil invariant polynomial,
defined locally inside the space of group manifold,
\begin{eqnarray}
\det\Bigl( 1 + \frac{\Omega}{2\pi i} \Bigr), \quad \Omega \propto \mathfrak{Y}+{\rm Ad}(g)\mathfrak{X}, \qquad
{\rm Ad}(h)\Omega = \Omega, \qquad \forall h \in H.
\end{eqnarray}
Since any complex $p$-form has a Hodge decomposition 
\begin{eqnarray}
H^{p}(X)^{\mathbb{C}} &=& \bigoplus_{r+s=p}H^{r,s}(X), 
\end{eqnarray}
the 2-form $\Omega$ also acquires the Hodge decomposition after a complexification, 
i.e, $\Omega_{\mathbb{C}}=H^{2,0}\oplus H^{1,1}\oplus H^{0,2}$,
and then a period mapping and a period domain of Griffiths 
will be obtained from the decomposition employed for our determinant.
Because the Hodge decompostion has the weight 2, it relates with a moduli of surface.

\subsubsection{Path Integrals}

A path integral of dynamical symmetry breaking is generically written as follows:
\begin{eqnarray}
{\cal Z} \sim \int {\cal DM}{\cal DM^{\dagger}}e^{-i{\rm Tr}\int d^{d}x\frac{{\cal MM}^{\dagger}}{2\lambda}}
{\rm Det}(\pfey -{\cal M}-{\cal M}^{\dagger}).
\end{eqnarray}
When ${\cal M}$ belongs to a matrix space defined over $\mathbb{C}$,
then it has a specific orientation, relates with cobordism.
It gives a functor $\pi:G\to \mathbb{C}^{1}$, a character ( class function ) of a group $G$, 
or it belongs to a space of set of charaters:

\vspace{2mm}

{\bf Axiom:}
{\it A path integral of a dynamical symmetry breaking is defined as a functor from a category of groups
to a category of fields/rings/schemes.}

\vspace{2mm}

Thus it will obtain an expansion by a set of characters via the Peter-Weyl theorem.
We should emphasize the fact that the Peter-Weyl theorem exists
not only for a compact Lie group, but for any compact group of a very large class,
with a left-right regular representation:
In those cases, any finite dimensional representation must be unitalizable for the theorem,
and a certain unitary equivalence in a class of finite-dimensional representations must be satisfied.
Then a matrix representation of $G$ is chosen "$\mod$ equivalence class",
which provides an orthonormal basis set $\pi(g)$ ( $g\in G$ ),
and any function $f(g)$ defined on $G$ is uniformly approximated 
by a linear combination of the basis set $\pi(g)$, a discrete sum due to the Pontrjagin duality.
Especially, a class function is expanded by the set of characters $(\chi_{\pi_{1}},\cdots\chi_{\pi_{n}})$
such that 
\begin{eqnarray}
f(g) = \sum^{n}_{i=1}a_{i}\chi_{\pi_{i}}(g).
\end{eqnarray}
$V_{eff}$ or $\Gamma_{eff}$ have such kind of expansions 
according to their symmetries/invariances.
The Weyl character formula of Lie groups is a special case of the Peter-Weyl theorem,
while a Weyl-Kac character formula of Kac-Moody ( affine Lie ) algebra 
( i.e., a character of Kac-Moody group constructed algebraically from the corresponding Kac-Moody algebra )
are given by theta functions~[114].
Hence, $V_{eff}$ as an automorphic representation may be interpreted by the following diagram:
\begin{eqnarray}
\begin{array}{c}
V_{eff}(\Gamma\backslash G) \sim {\rm characters}  \\ 
\downarrow  \\
{\rm theta \,\, functions} \,\, ({\rm modular \,\, transformation, \,\, complex \,\, tori}) \\ 
\uparrow  \\
{\rm infinite-dimentional \,\, Lie \,\, algebra \,\, (group)}.  
\end{array}
\end{eqnarray}
The Pontrjagin duality $G\times G^{\vee}\to T$ is obtained from the map $\chi:G\to T$ 
( $G$; a locally compact Abelian group, $G^{\vee}$; the dual of $G$, $\chi$; a character of $G$, $T$; a torus ).
Such a dual theory is not yet established in a case of locally compact non-Abelian group,
while there is a duality in a case of compact non-Abelian group ( Tannaka ). 
Therefore, 
\begin{eqnarray}
\pi(f) = \int_{G}f(g)\pi(g)dg &\leftrightarrow& f(g) = \sum_{\lambda \in G^{\vee}}{\rm tr}\hat{f}_{\lambda}(\pi)\pi_{\lambda}(g). 
\end{eqnarray}
If $f$ is a class function, then one needs only a set of characters in the expansion of $f(g)$.
For the case of $f$ defined as an automorphic representation, 
the basis set $\pi(g)$ should be constructed by invariants of $\Gamma$:
Namely, an automorphic represenation may take a form of expansion by a set of automorphic forms.
After a preparation of Hilbert space by the Petersson innerproduct,
one can obtain an integration kernel
\begin{eqnarray}
K_{\Gamma\backslash G}(g,h) &=& \sum_{k}f_{k}(g)\bar{f}_{k}(h). 
\end{eqnarray}
It is possible that an automorphic representation contains a lot of automorphic forms
in such an expansion,
and therefore their corresponding Galois represenations
( various weights, levels, ... ). 
For example, a function as an automorphic form have the following expansion:
\begin{eqnarray}
f(\Gamma\backslash G) = \sum\sum\sum a_{w,N}\beta_{l}(e^{2\pi iz})^{l}, \quad (w;{\rm weight},\,\,N;{\rm level}). 
\end{eqnarray}
Here, we argue that our effective potential $V_{eff}$ on $\Gamma\backslash G$
is an example of $f(\Gamma\backslash G)$.  
In the context of our GNG theorem, it is interesting to introduce the method of 
integral equation in scattering theory ( beside an examination of convergence property of the kernel ):
\begin{eqnarray}
{\cal M}_{\Gamma\backslash G}(g) &=& {\cal M}^{(0)}_{\Gamma\backslash G}(g)
+ \lambda \int K_{\Gamma\backslash G}(g,h){\cal M}_{\Gamma\backslash G}(h)dh.
\end{eqnarray}   
If our GNG mechanism is correct for the realization of the CKM/PMNS matrices,
then they have a scattering picture in their "imaginary" domains:
After analytic continuations or complexifications of local coordinate systems,
they ( dynamics of CKM/PMNS matrices ) aquire scatterings inside the imaginary regions,
similar with that of instanton.
This might give us a new {\it interpretation} on the history of the Universe
( and it will make us to meet the notion on real and imaginary parts of the Universe ). 
Moreover, a cusp form is obtained from a regular differential form of a modular curve ( a Riemann surface ),
under an algebro-geometric manner.
If this fact can be generalized into higher-dimensional varieties/manifolds,
then our automorphic CKM/PMNS matrices may have their "modular varieties",
and this implies that there is a set of moduli for the CKM matrix. 
Note the following diagram is induced by the Pontrjagin duality which relates with our context:
\begin{eqnarray}
\begin{array}{ccc}
G &\stackrel{\chi}{\longrightarrow}& T  \\
\downarrow &  & \downarrow \\
C(G) &\longrightarrow& C(T)  \\
\cup &  & \cup \\
C(G)^{cf} &\longrightarrow& C(T)^{sym},
\end{array}
\end{eqnarray} 
where, $C(G)$, $C(T)$ are spaces of continuous functions on $G$ and $T$, respectively,
while $C(G)^{cf}$ and $C(T)^{sym}$ denotes spaces of class functions on $G$
and spaces of continuous symmetric functions on $T$, respectively.  
Our Chern-polynomial-like determinant of a breaking scheme expressed by a maximal torus,
based on a finite-dimensional representation of a unitary group, 
is interpreted by this diagram
( for example, a breaking scheme may change a Chern invariant, a class function,  
into a non-invariant but a continuous function ).

\vspace{2mm}

A quantum field theoretical calculation, which is defined by a systematic expansion
and renormalization ( Hopf algebra ), 
reflects the algebraic/geometric/number-theoretic structure of characters
of a group $G$, which might be handled by the theory of character sheaves of G. Lusztig
( G. Lusztig, {\it Character Sheaves}, 
I, Adv. Math., {\bf 56}, 193 (1985), 
II, Adv. Math., {\bf 57}, 226 (1985),
III, Adv. Math., {\bf 57}, 266 (1985),
IV, Adv. Math., {\bf 59}, 1 (1986),
V, Adv. Math., {\bf 61}, 103 (1986),
the theory of character sheaves of Lusztig can be used for any group over a field of arbitrary characteristic ).
An action functional of field theory ( physics ) defined over a finite field
is given by an expansion of characters via
the Deligne-Lusztig theory which is a method to give characters of a group of finite fields, 
and then the theory explicitly has a symmetry such as a Frobenius 
( P. Deligne and G. Lusztig, 
{\it Representations of Reductive Groups over Finite Fields},
Ann. Math. {\bf 103}, 103 (1976) ).
Characters of $G^{F}$ ( $G$; an algebraic group, $F$; a Frobenius map ) 
of the Deligne-Lusztig theory is defined by an $l$-adic cohomology
( a Lefschetz number, used also in the Weil zeta function ),
looks like a similar form with the Weyl character formula
( the Deligne-Lusztig theory also uses a set of characters of maximal torus, namely
$\theta:T^{F}\to\mathbb{C}$, $T\subset G$ ),
and it relates with counting numbers of rational points of a variety over a finite field. 
They also satisfy an orthogonal relation.
By substituting the $l$-adic cohomology to an intersection cohomology with perverse sheaves
( $G$-equivariant sheaves, or especially, character sheaves ), the Deligne-Lusztig theory
relates with the Kazhdan-Lusztig theory.
Since cohomology can be defined over various manifolds/varieties/fields, 
the method to define characters by cohomology has a large possibility to apply our GNG theorem. 
The Kazhdan-Lusztig theory gives a representation of Iwahori-Hecke algebra coming from
an induced representation ${\rm Ind}^{G^{F}}_{B^{F}}(1)$.
As we have mentioned, one might obtain a family of Chevalley groups in the expansion of
a path integral or an effective action "in terms of Frobenius $\mathbb{F}_{q}$-rational structures" 
in our GNG mechanism. Thus, such an expansion gives a series/family of Iwahori-Hecke algebras. 
From the GNG theorem of non-Abelian cases, we have an interest on to extend the theory of Lusztig
to other number-theoretical group actions than Frobenius, 
and it might give us a set of new zeta functions.

\subsubsection{Effective Potentials}

A distance between two points defined on the potential $V_{eff}$ and the base ( topological ) space $X$ is
usually different. A fiber product of them is given by
\begin{eqnarray}
X\otimes V_{eff}.
\end{eqnarray}
In our case, the relation $V_{eff}={\rm const}.$ gives a set of points
( or, curves, surfaces, ... ) they have a symmetry,
which may be represented by a Galois group,
inside a topological space such as a ${\rm Spec}A$ or a ${\rm Proj}S$.
This fact is quite similar to the situation of a map multiplication by $N$ in an elliptic curve $E$
over $\mathbb{Q}$ ( finite, \'{e}tale ), i.e. $[N]:E\to E$.
It is a well-known fact that $E$ over $\mathbb{C}$ defines an algebraic equation over $\mathbb{Q}$,
$E[N]\simeq \mathbb{Z}/N\mathbb{Z}\oplus\mathbb{Z}/N\mathbb{Z}$,
and the Galois group and its representation becomes 
$G_{\mathbb{Q}}:{\rm Gal}(\overline{\mathbb{Q}}/\mathbb{Q})\to GL_{2}(\mathbb{Z}/N\mathbb{Z})$.
After a reduction mod $p$, $E$ aquires a geometric Frobenius
$(x,y)\to (x^{p},y^{p})$.
By a Tate module $T_{p}(E)\equiv \lim_{\leftarrow}E[p^{n}]$, 
the representation $\rho_{p}:G_{\mathbb{Q}}\to GL_{2}(\mathbb{Z}_{p})$ is obtained.
Then the mod $N$ cyclotomic character and the $l$-adic cyclotomic character,
\begin{eqnarray}
\overline{\chi}_{N}: G_{\mathbb{Q}} \to (\mathbb{Z}/N\mathbb{Z})^{\times}, \quad 
\chi_{l}: G_{\mathbb{Q}} \to \mathbb{Z}^{\times}_{l},
\end{eqnarray}
are obtained. 
An automorphic form will be systematically constructed by $a_{l}={\rm Tr}\rho_{p}({\rm Frob}_{l})$
( ${\rm Frob}_{l}$; a Frobenius conjugacy class at $l$ ).
These situation of elliptic curves must also be realized in an effective potential with a double periodicity
( the case $SU(2_{f})$ we have shown in the previous section is an example of it ).
More generally,
let $X$ be a scheme, let $G$ be a finite flat commutative group scheme.
Then a map of multiplication by $N$, 
$[N]: G \to G$, gives a group $(\mathbb{Z}/N\mathbb{Z})\oplus\cdots\oplus(\mathbb{Z}/N\mathbb{Z})$
( here, $G$ should contain an \'{e}tale morphism for our context )
in our theory. 
If one considers such a group scheme with a connected geometric fiber,
then $G$ is an Abelian scheme ( tori, elliptic curves ).
Thus, we conclude that a realization of a class of morphisms $[N]: G \to G$
gives Abelian schemes in an effective potential
( one should notice the fact that any Abelian variety is projective ).
Moreover, ${\rm Ker}([N]: G \to G)$ is isomorphic with 
\begin{eqnarray}
{\mathcal Hom}(X,\mu_{N}) &=& {\mathcal Hom}({\mathcal Hom}(G,\mathbb{G}_{m}),\mu_{N}),
\end{eqnarray}
where, $\mu_{N}$ is the Abelian scheme of $N$-th root of unity. 
The theory of moduli of vector bundles on an algebraic/Abelian group scheme may be useful for us.
Such a theory could be constructed by a decomposition of a vector bundle into a direct sum of line bundles.

\vspace{2mm}

Let us consider a $V_{eff}$ in the words of differential ring and the index theorem:
\begin{eqnarray}
& & K_{F} \ni V_{eff}(\theta_{1}, \cdots, \theta_{n}) \in \mathbb{C}[[\theta_{1},\cdots,\theta_{n}]],  \\
& & \forall \frac{\partial}{\partial\theta_{i}}V_{eff}(\theta_{1}, \cdots, \theta_{n})=0 \in 
K^{\partial}_{F} \equiv \{c\in K_{F}| \partial(c)=0 \},
\end{eqnarray}
where $K_{F}$ is a differential ring with an appropriately defined derivation
which satisfies the Leibniz rule, and $K^{\partial}_{F}$ denotes a constant ring
( $K^{\partial}_{F}$ is not an ideal of $K_{F}$ ).

\vspace{2mm}

{\bf Theorem:}
{\it The gap equation belongs to the set of constant ring} $K^{\partial}_{F}$.

\vspace{2mm}

( More abstruct effective potential can be introduced by a completion of local ring 
$\widehat{A}=\lim_{\leftarrow n}A/\mathfrak{m}^{n}$. 
A differential/derivative on $K$ is possibly non-Abelian,
though here we only consider an Abelian case, especially over $\mathbb{R}$.
A connection, i.e., a differential can be defined under an algebraic manner~[251] ) 
Then, the gap equation is written in the following form:
\begin{eqnarray}
D\hat{V}_{eff}=0, \quad  
D \equiv {\rm diag}\Bigl(\frac{\partial}{\partial\theta_{1}},\cdots,\frac{\partial}{\partial\theta_{n}}\Bigr), \quad
\hat{V}_{eff} \equiv (V_{eff},\cdots,V_{eff})^{T},
\end{eqnarray}
Therefore, it determines a zero mode of $\widehat{V}_{eff}$ with respect to $D$.
The derivative $D$ defines a connection/differential $\nabla$ 
which acts on $V_{eff}$ as a section $\in\Gamma(X,{\mathcal E})$, 
( $X$, a base space, ${\cal E}$ a fiber bundle ) and it defines a local constant sheaf on $X$ as a subset $Y$. 
By utilizing direct image functors, 
one will meet with a cohomology group for $X$, $Y$ and $V_{eff}$.
Thus the set of stable points/subvarieties/subschemes is given by the kernel of $D$.
By the following definitions,
\begin{eqnarray}
D: \Gamma(M,E) \to \Gamma(M,F), \quad D^{\dagger}:\Gamma(M,F) \to \Gamma(M,E), \quad
s \in \Gamma(M,E), \quad s' \in \Gamma(M,F), \quad 
\langle s', Ds\rangle_{F} = \langle D^{\dagger}s',s\rangle_{E},
\end{eqnarray}
where, $M$ is a base space, $E,F$ fiber bundles on $M$, and $\Gamma$ implies a set of sections. 
The kernel of $D$ and its adjoint are defined by
\begin{eqnarray}
{\rm ker}D \equiv \{s\in\Gamma(M,E)|Ds=0\}, \quad 
{\rm ker}D^{\dagger} \equiv \{s\in\Gamma(M,E)|D^{\dagger}s=0\}.
\end{eqnarray}
Then the index we consider is given by the following definition:
\begin{eqnarray}
{\rm ind}D \equiv {\rm dim} {\rm ker}D - {\rm dim} {\rm ker}D^{\dagger}.
\end{eqnarray}
In the very basic theory of index theorem, a compact manifold with no boundary is assumed,
while in our case, the base space can become compact while it has a boundary in usual,
especially in an automorphic representation.
Thus, we refer the result of Atiyah which discusses a spectral flow with a boundary condition~[250].
Note that the differential operator $\partial/\partial\theta$ is elliptic, 
will become Fredholm on a compact space:
We assume our case on a compact space of $\Gamma\backslash G$.
In the result of Atiyah, the first-order differential operator $D=\partial_{\theta}+i\partial_{t}-it$ 
defined on $M=S^{1}$ is considered, and the boundary condition of eigenfunctions 
( it becomes a theta function ) of $D$ on $S^{1}$ gives a spectral flow.
An interesting fact is that $t$ is regarded as a $U(1)$ gauge potential.
The result of Atiyah must be contained in our $V_{eff}$ of the $U(1)$ case, 
and it is generalized on a torus $T=S^{1}\times\cdots\times S^{1}$ as follows:
We introduce the following theta function,
\begin{eqnarray}
\Theta(\tau_{j},\theta_{j}) \equiv \prod^{g}_{j=1}\exp\Bigl[ -\frac{(l_{j}+\tau_{j})^{2}}{2} -il_{j}\theta_{j} \Bigr].
\end{eqnarray}
Then the differential operator which has the theta function as its eigenfunction will be found 
in the following form:
\begin{eqnarray}
D^{\dagger} &\equiv& {\rm diag}(\partial_{\theta_{1}}+i\partial_{\tau_{1}}-i\tau_{1}, \cdots, \partial_{\theta_{g}}+i\partial_{\tau_{g}}-i\tau_{g}).
\end{eqnarray}
Therefore, by the same logic of Atiyah, 
we conclude that

\vspace{2mm}

{\bf Theorem:}
{\it The set} $K^{\partial}_{F}\ni V_{eff}$ {\it defined over a torus has generically a spectral flow.}

\vspace{2mm}

More precisely, a $V_{eff}$ of an automorphic form 
is embedded into a set which contains a spectral flow ( a subset ).
Since the theta function contains a set of moduli $\tau_{j}$ ( $j=1,\cdots g$ ) 
of a complex structure of torus 
( after a complexification $\theta_{j}\in\mathbb{C}$, 
and $\tau_{j}$ may be regarded as an $U(1)\times\cdots\times U(1)$-gauge potential ), 
this phenomenon also relates with a Kodaira-Spencer deformation of a complex manifold 
which belongs to a family of given torus.

\vspace{2mm}

Let us consider the case where a set of discrete vacua realizes 
in the model of explicit+dynamical symmetry breaking of our mechanism.
Then from the following projection of an affine space to a field,
\begin{eqnarray}
\pi : K^{\partial}_{F} \to \mathbb{C}^{n}, 
\end{eqnarray}
one obtains a finite group scheme over $\mathbb{C}$, i.e., 
it gives a representation of an absolute Galois group.
Moreover, 
\begin{eqnarray}
K_{F} \ni V_{eff} \in {\rm automorphic \,\, representation},  \\
K^{\partial}_{F} \ni {\rm gap \,\, equation} \in {\rm automorphic \,\, representation}.
\end{eqnarray}
Hence the automorphic representation/forms give a subset of $K_{F}$,
the theory gives a kind of restriction of $K_{F}$ to the subspace.
Similarly, the gap equation picks up an element of automorphic representation.

\vspace{2mm}

{\bf Theorem:}
$\pi$ {\it is a functor from a category of constant ring to a category of absolute Galois group 
with an effective potential} $(V_{eff})^{\mathbb{C}}$.

\vspace{2mm}

Hence, the theory is summarized as follows:
\begin{eqnarray}
\varrho: K_{F} \to K^{\Gamma\backslash G}_{F},   \quad
\varrho_{g}: K^{\partial}_{F} \to K^{\partial,\Gamma\backslash G}_{F},  \quad
K^{\partial}_{F} \subset K_{F}, \quad 
K^{\partial, \Gamma\backslash G}_{F} \subset K^{\Gamma\backslash G}_{F}.
\end{eqnarray}

\vspace{2mm}

Now, a very interesting fact is that an automorphic form, especially a cusp form,
may be contained in an effective potential which is defined over an elliptic curve:
There is a connection between mathematics of the Fermat's last theorem
( modular forms, Galois representations,... )
and the differential ring realized in an effective potential.
Moreover, those effective potentials will be expressed as a section of ( homogeneous ) fiber bundle,
or a section of line bundle on manifold/variety/scheme, 
and then one gets a geometric construction on those mathematics for various aspects
( differential rings, Shimura-Taniyama-Weil theorem, zeta/L-functions, connections and curvatures, 
de Rham cohomology, the Chern-Weil theory, the Borel-Weil theory, moduli spaces, 
deformation theory, index theorem with a boundary of $G$ or of $\Gamma\backslash G$, 
algebras of characters such as fusion algebra, the Verlinde formula, so forth ),
namely there is a deep connection between elements of Lie groups, number theory and geometry
in our effective potential.
If an effective potential is defined on a curve ( or, curves ),
and if a curve contains a moduli, 
then the curve may be deformed by the moduli,
and the effective potential chooses a point or a subset of the moduli space.
In the Borel-Weil theory, a finite-dimensional representation of $U(N)$ is realized 
as a total space of holomorphic sections of a holomorphic line bundle
( invertible sheaf ).
Thus, a moduli theory of a line bundle over a curve can be applied if the holomorphic line bundle
is defined over a curve as a base manifold.
We wish to emphasize the fact that the mathematical structures discussed here
is very generic both in the NG and GNG theorems.

\vspace{2mm}

The Lefschetz fixed point formula gives a number of fixed points of a map $f:X\to X$:
\begin{eqnarray}
\sum^{dim X}_{i=0}(-1)^{i}{\rm tr}\bigl( f^{*}:H^{i}(X,\mathbb{Z})\to H^{i}(X,\mathbb{Z}) \bigr).
\end{eqnarray}
This formula is applied to our GNG theorem when we consider a set of equilibria ( a subspace of the NG manifold ), 
defined by $\frac{\partial V_{eff}}{\partial \theta_{a}}=0$, which is invariant under a map $f$.
For a close discussion with number theory, we consider a Frobenius map in the $U(1)$ case,
while we should consider a permutation group of higher-dimensional ( not a zero dimensional object )
in a diagonal breaking of non-Abelian gauge/flavor symmetry.
This may imply that the permutation represents a counting of "rational varieties",
including a case of a set of rational points.

\subsubsection{Fiber Bundles and the Chern-Weil Theory}

Since there are several invariances ( given by $H\subset G$ ) in our GNG theorem of a homogeneous space, 
the Chern-Weil theory is interested by the following diagram:
\begin{eqnarray}
& & {\rm Homogeneous \,\, fiber \,\, bundles \,\, on \,\, G/H, \,\, \Gamma\backslash G/H} \to {\rm Chern-Weil \,\, theory}  \nonumber  \\
& & \qquad \to {\rm \,\, cohomology, \,\, deformation \,\, theory \,\, and \,\, moduli \,\, theory}.
\end{eqnarray}
Especially, a Chern-Weil theory on an automorphic represenation is important.

\vspace{2mm}

When we take a fundamental representation in a fermion determinant of a theory, 
\begin{eqnarray}
\det\Bigl( \pfey -\widehat{M}_{0} -( g\widehat{M}_{dyn}g^{-1} + {\rm h.c.} ) \Bigr)
\end{eqnarray}
gives a Chern-Weil theory with a restriction of the mass matrices to a maximal torus:
In a represenation of $U(n)$, for example, 
a character is expressed by a symmetric function derived by a maximal torus ( Weyl ).
After a complexification of the mass matrix $\widehat{M}_{dyn}$, 
we obtain what we want for 
our examination of a symmetry breaking such as the case of $SU(2)$ previously discussed.
By introducing a complex analytic family in our Chern-Weil-"like" determinant,
the infinitesimal deformation of complex structure of Kodaira and Spencer in the determinant is obtained.
It is important to examine the first Chern class in the determinant
because it corresponds to the obstruction of a deformation.
To consider them systematically, we use the method of fiber bundles.
If one wants to discuss them on a homogeneous space $G/H$,
a homogeneous fiber bundle and its Kodaira-Spencer-type deformation theory are needed. 
A homogeneous fiber bundle is a bundle associated with a principal bundle $(G,G/H,H,\pi)$,
such as $G\times_{H}F\to G/H$ where $F$ is a topological space.

\vspace{2mm}

The tangent space $T(G/H)$ of $G/H$ is a homogeneous fiber bundle,
\begin{eqnarray}
T(G/H) \simeq G\times_{{\rm Ad}(H)}\mathfrak{g}/\mathfrak{h},
\end{eqnarray} 
and it has a roll in a theory of moduli space of connections,
and also in a complex analytic family of the Kodaira-Spencer deformation theory.
If a breaking scheme gives a homogeneous space of reductive type,
it always has a Riemannian mentric:
The Riemannian metric of $G/H$ is defined by a $C^{\infty}$-class section of the fiber bundle
$T(G/H)\times T(G/H)\to G/H$.
Generally, a $p$-form on $G/H$ ( contains a Maurer-Cartan form ) is a section of
\begin{eqnarray}
G\times_{H}\Bigl( \wedge^{p}\bigl(\mathfrak{g}/\mathfrak{h}\bigr)^{\vee}\Bigr),
\end{eqnarray}
and by a complexification of this map, we can introduce a Hodge decomposition to the $p$-form
and possiblly a period mapping, and the de Rham cohomology contains them.
If $G$ is compact and $G/H$ is a symmetric space, 
then the de Rham cohomology of it has the following isomorphism:
\begin{eqnarray}
H^{p}(G/H,\mathbb{R}) \simeq \Bigl(\wedge^{p}\bigl(\mathfrak{g}/\mathfrak{h}\bigr)^{\vee}\Bigr)^{H},
\end{eqnarray}
and it also has a Hodge decomposition after a complexification of it.
Especially, a case of Hermitian symmetric space of K\"{a}hlerian type, 
a projective variety ( a compact K\"{a}hlerian manifold can be embedded into $\mathbb{P}^{n}$ ), 
a flag variety and a generalized flag variety ( they are compact K\"{a}hler, and algebraic ),
and a complex Grassmannian manifold ( a compact K\"{a}hlerian manifold ) 
are interesting for the context of the Hodge theory and period maps.
The isomorphism 
\begin{eqnarray}
\bigl( \mathfrak{g}/\mathfrak{h} \bigr)^{G}\otimes_{\mathbb{R}}\mathbb{C} \simeq 
\bigl( \mathfrak{g}_{\mathbb{C}}/\mathfrak{h}_{\mathbb{C}} \bigr)^{G_{\mathbb{C}}}
\end{eqnarray}
will be employed, and several Hodge decompositions and filtrations will be obtained.
For a Clifford-Klein form $\Gamma\backslash G/H$, the Hirzebruch proportionality theorem is useful~[114], 
which allows us to estimate the Chern classes of $\Gamma\backslash G/H$ from those of $G/H$.

\vspace{2mm}

Now, let us turn to the result of $SU(2)$ discussed in the previous section. 
The following determinant is the problem:
\begin{eqnarray}
\det\Bigl[  e^{-i\theta_{3}\tau_{3}}\Bigl( \pfey -\widehat{M}_{0} -( \hat{g}\widehat{M}_{dyn}\hat{g}^{-1} + {\rm h.c.})\Bigr)e^{i\theta_{3}\tau_{3}} \Bigr].
\end{eqnarray}
Then, $[\tau_{3},\widehat{M}_{dyn}]=0$ holds,
while $g(\tau_{1},\tau_{2})\ne g(\tau_{3})g(\tau_{1},\tau_{2})g(\tau_{3})^{-1}$, 
and the determinant has a Chern-invariant-like invariance at $\widehat{M}_{dyn}=0$
while it disappears at $\widehat{M}_{dyn}$.
This implies that the eigenvalues of the determinant at a stable minimum will not change, 
while there are couplings between the modes $\tau_{3}$ and $(\tau_{1},\tau_{2})$,
because the breaking scheme $SU(2)\to U(1)_{3}$ does not give a homogeneous space,
no quotient topology.
In that case, it is impossible to gauge ( so-called "local gauge" ) the $\tau_{3}$-direction
at $\widehat{M}_{dyn}\ne 0$, while it is possible at $\widehat{M}_{dyn}=0$.
Therefore, the $\tau_{3}$ mode can be absorbed by a Higgs mechanism,
and after that, there is no coupling between $\tau_{3}$ and $(\tau_{1},\tau_{2})$ modes. 
It should be mentioned that $(\theta_{1},\theta_{2})=(0,0)$ ( or, $\hat{g}=1$ ) is a special point
because the mass matrix $\widetilde{M}=\widehat{M}_{0}+(\hat{g}\widehat{M}_{dyn}\hat{g}^{-1}+{\rm h.c.})$ 
does not break $\tau_{3}$ even at $\widehat{M}_{dyn}\ne 0$.
Since 
\begin{eqnarray}
X = \pfey - \widehat{M}_{0} - \hat{g}\widehat{M}_{dyn}\hat{g}^{-1} = c_{1}\tau_{1}+c_{2}\tau_{2}+c_{3}\tau_{3} 
\in \mathfrak{su}(2,\mathbb{R}),
\end{eqnarray}
the determinant will be expressed as
\begin{eqnarray}
\det\Bigl[  \hat{g}^{-1}\Bigl( \pfey -\widehat{M}_{0} -\hat{g}"\widehat{M}_{dyn}(\hat{g}")^{-1}\Bigr)\hat{g} \Bigr] 
= \det \hat{g}X\hat{g}^{-1} = \det X.
\end{eqnarray}
Namely, the determinant $\det X={\rm const}.$ defines an adjoint orbit 
( a submanifold of the Lie ${\it algebra}$ $\mathfrak{su}(2,\mathbb{R})$ ), 
\begin{eqnarray}
G/G(X), \quad G(X) \equiv \{ g\in G,\, X\in\mathfrak{g} | {\rm Ad}(g)X = X \}, 
\end{eqnarray}
and the periodicity corresponds to the fact that the adjoint orbit 
gives a level set of an effective potential and a base manifold $M$ usually discussed in the Morse theory.
After an extension of fundamental domain of the group,
the adjoint orbits close inside the domain/manifold of the group and the closed orbits are compact.
The adjoint orbit relates with a topological nature of $M$ summarized into a critical ( singular ) point
of the effective potential.

\vspace{2mm}

It is well-known fact that a variation of an invariant polynomial of $2j$-form $P$
under the map,
\begin{eqnarray}
\varphi : (\omega,\Omega) \to (\tilde{\omega},\widetilde{\Omega})
\end{eqnarray} 
( $\omega,\tilde{\omega}$: connections, $\Omega,\widetilde{\Omega}$: curvatures )
is given by the following famous formula:
\begin{eqnarray}
& & P(\widetilde{\Omega}) - P(\Omega) = d\int^{1}_{0}Q(\eta,\Omega_{t})dt, \\ 
& & \eta = \tilde{\omega} - \omega, \\
& & \omega_{t} = \omega + t\eta,  \\
& & \Omega_{t} = d\omega_{t} -\omega_{t}\wedge\omega_{t},
\end{eqnarray}
where, $Q$ is a $2j-1$ form.
Needless to say, the formula is defined locally since it contains curvatures explicitly, as an exact form.
This Chern-Weil theory can be applied to any manifold $M$ as the base space 
defined by a local coordinate system of a Lie group.
What we are interested in here is to know how such a variation affects to,
our Chern-like polynomial with several extra "symmetries", 
such as a Galois action ( for example, Frobenius ) arised from the mechanism of our GNG theorem, 
which contains automorphic forms.

\vspace{2mm}

Now, we wish to axiomatize our result.

\vspace{2mm}

{\bf Axiom}:{\it A fermionic determiant and a Chern-Weil invariant of a certain Lie group are contained in
a larger class of determinants derived from a space of a matrix of 2-form, 
especially which is defined over $\mathbb{C}$. 
A determinant of the class has a Hodge decomposition in the 2-form, deformations inside a family,
contains a set of moduli parameters,
and by performing several restrictions on the determinant such as a condition expressed by an adjoint orbit,
a fermionic determinant of quantum field theory will be obtained.}

\subsubsection{Deformation Theory and Families}

A homogeneous space can have a Riemannian metric simultaneously with a complex structure,
a simplectic structure, so forth.
Thus, it is interesting problem for us to find a theory which can describe deformations of them
in the GNG theorem of a homogeneous space.
For the purpose, the most convenient method is provided from the Kodaira-Spencer theory.
The Hodge theory and a period domain is also important for us.
We summarize our interest on the following scheme:
\begin{eqnarray}
\begin{array}{ccccccc}
{\rm Compact \,\, K\ddot{a}hler \,\, manifold} &\longrightarrow& {\rm Hodge \,\, structure} &\longrightarrow&  {\rm period \,\, domain} 
&\to& {\rm automorphic \,\, forms}   \\
\downarrow & & & & & & \\
{\rm Siegel \,\, upperspace} &\longleftarrow& {\rm Shimura \,\, variety}. & & & &    
\end{array}
\end{eqnarray}
It should be mentioned that the theorem of Hironaka states that an infinitesimal deformation 
( such as a Kodaira-Spencer deformation )
of a compact K\"{a}hler manifold gives again a K\"{a}hler manifold, 
while an arbitrary deformation can give a non-K\"{a}hler manifold
( H. Hironaka, {\it An Example of a Non-K\"{a}hlerian Complex-analytic Deformation
of K\"{a}hlerian Complex Structures}, Ann. Math. {\bf 75}, 190-208 (1962) ).
Let ${\rm Sieg}(2g,\{z_{i}\})$ be the space of Siegel modular form.
Then a function $F(\{z_{i}\})\in {\rm Sieg}(2g,\{z_{i}\})$ can have an expansion 
with respect to the Peter-Weyl theorem,
and this implies that $F(\{z_{i}\})$ aquires an expression given by higher-dimensional theta functions.
Any theta function, defined on a projective space, has deformation/moduli parameters from its definition.
A mass matrix defined over a Siegel upperhalf space will aquire those mathematics,
and brings it into an effective potential.

\vspace{2mm}

An infinitesimal deformation of complex structures of a compact K\"{a}hler manifold
or a complex Lie group are given by the following Kodaira-Spencer mappings~[115,254].
An analytic family of compact complex manifolds is defined by 
a proper smooth morphism
\begin{eqnarray}
\pi: {\cal V} \to {\cal W}, \quad \rho_{0}: T_{0}{\cal W} \to H^{1}(V_{0},\Theta),
\end{eqnarray}
where, ${\cal W}$ is the parameter space for a deformation,
$V_{0}\equiv \pi^{-1}(0=\omega_{i}\in{\cal W})$.
While, an analytic family of principal $G$-bundle is given by
\begin{eqnarray}
\pi: {\cal P} \to {\cal M\times S}, \quad \rho_{0}: T_{s_{0}}{\cal S} \to H^{1}({\cal M},{\rm Ad}(P_{s_{0}})), \quad
P_{s_{0}} \equiv {\cal P}|_{\cal M}\times\{s_{0}\},
\end{eqnarray}
where a Lie group $G$ is defined over a complex manifold ${\cal M}$,
${\cal S}$ is the space of parameters for a deformation ( an analytic space ).
The definition ${\rm Ad}(P_{s_{0}})\equiv P_{s_{0}}\times_{Ad}\mathfrak{g}$ implies that
the curvature 2-form $\Omega$ of $G$, which is an element of $P\times_{Ad}\mathfrak{g}$,
can also be extended to a form parametrized in the space ${\cal S}$ and be affected by the deformation.  
We need an extension of this theory to a homogeneous space $G/H$,
and then apply it to the fermionic determinant.

\vspace{2mm}

For our problem, one can consider the following method.
Let $G/H$ be a homogeneous space, and set a Lie algebra according to it by the decomposition 
$\mathfrak{g}=\mathfrak{h}+\mathfrak{m}$.
Then an exponential mapping will be introduced in the vicinity of the origin of the Lie group as
\begin{eqnarray}
\exp\bigl( \theta_{s}\otimes\mathfrak{h} + \theta_{b}\otimes\mathfrak{m} \bigr) &\to&
\exp\bigl( f_{s}(\theta,t)\otimes\mathfrak{h} + f_{b}(\theta,t)\otimes\mathfrak{m} \bigr). 
\end{eqnarray}
Here, $t$ implies a set of deformation parameters.
Then, after the procedure we have discussed in the previous section,
we obtain an effective potential and a set of gap equations.
While a Kodaira-Spencer map gives a set of tangent vectors at the origin of a parameter space
after projecting a total space to the space.
Especially this procedure works well in a case of Riemannian symmetric space
$(G,H,\sigma,g)$ ( here $g$ denotes metric ) and a Riemannian symmetric Lie algebra
$(\mathfrak{g},\sigma,g)$:
In that case, especially in a Hermitian symmetric space, 
the orthogonal complement $\mathfrak{m}$ of an example takes its value
on a Siegel upperhalf space ( or, isomorphic with it ), 
and after taking a polarization of Abelian variety, 
and if the space of deformation parameters is devided into two parts,
namely for $\mathfrak{h}$ and for $\mathfrak{m}$,
then a moduli space ( and also an analytic family of deformation ) will be obtained. 
In that case, the space of orthogonal complement at the origin itself gives an analytic family
( do not confuse! ). 
Hence a mass matrix constructed by the exponential functions contains ( a certain type of ) 
deformation parameters.
Recall the definition of a $G$-invariant Riemannian metric on a homogeneous space $G/H$:
\begin{eqnarray}
& & \langle {\rm Ad}(h)X, {\rm Ad}(h)Y \rangle = \langle X, Y \rangle, \quad h \in H, \quad X,Y \in \mathfrak{m}, \quad 
{\rm Ad}(h)X \in \mathfrak{m},  \nonumber \\
& & \mathfrak{m} \equiv \{ \langle X,Y \rangle =0, X\in\mathfrak{g}, \forall Y\in \mathfrak{h}, \mathfrak{g}=\mathfrak{h}\oplus\mathfrak{m} \}.  
\end{eqnarray}
Here the first equation is the condition for the metric.
It is well-known fact that any Riemannian symmetric Lie algebra has the following decomposition:
\begin{eqnarray}
(\mathfrak{g},\sigma,g) &=& 
(\mathfrak{g}_{a},\sigma_{a},g_{a})\bigoplus(\mathfrak{g}_{c},\sigma_{c},g_{c})\bigoplus(\mathfrak{g}_{n},\sigma_{n},g_{n}),
\end{eqnarray}
where, $a$, $c$ and $n$ impliy the commutative, compact and non-compact parts of the total algebra, respectively.
This form is convenient because it is linearly decomposed 
in terms of all elements of the Riemannian symmetric Lie algebra:
For example, the K\"{u}nneth formula of cohomology can be applied into them,
and then one can estimate a global aspect of the algebra, especially via a universal enveloping algebra.
Since a Kac-Moody algebra $\widehat{\mathfrak{g}}$, especially an affine Lie algebra, 
can be regarded as a special case of infinitesimal-deformation of complex structure
in the context of conformal field theory, namely,
\begin{eqnarray}
\begin{array}{ccccc}
{\rm Lie \,\, algebras} & & & &  \\
\downarrow & & & & \\
{\rm complex \,\, analytic \,\, families \,\, of \,\, Lie \,\, algebras} & \supset & {\rm Kac-Moody \,\, (affine \,\, Lie) \,\, algebras} & \leftrightarrow & {\rm Virasoro \,\, algebra}, 
\end{array}
\end{eqnarray}
then the following algebra becomes interesting for our GNG theorem:
\begin{eqnarray}
& & \widehat{\mathfrak{g}} \equiv \bigl( \widehat{\mathfrak{g}}_{1}/\widehat{\mathfrak{r}}_{1} \bigr)\bigoplus \cdots \bigoplus \bigl( \widehat{\mathfrak{g}}_{n}/\widehat{\mathfrak{r}}_{n} \bigr),  \\
& & \widehat{\mathfrak{g}}_{l} \quad {\rm for} \quad (\mathfrak{g}_{l},\sigma_{l},g_{l}), \quad l = 1, \cdots, n,   \\
& & \widehat{\mathfrak{r}}_{1} \quad ( {\rm ideal \,\, of} \,\, \widehat{\mathfrak{g}}_{l} ),
\end{eqnarray}
and in that case the total $\widehat{\mathfrak{g}}$-module $V$ will have the following decomposition:
\begin{eqnarray}
V &=& V_{1}\oplus \cdots \oplus V_{n},
\end{eqnarray}
and the Riemannian metric also be obtained under a systematic manner.

\vspace{2mm}

Let us consider the case of a mass function defined over a Riemann surface.
In principle, a Riemann surface on which a mass function is defined can take all numbers of genus $g\ge 0$.
The obstruction $H^{2}(R,\Theta)=0$ at $g\ge 2$, 
while the number of moduli becomes $\dim_{\mathbb{C}}H^{1}(R,\Theta)=3g-3$.
The moduli space of Riemann surfaces is given by the following holomorphic map:
\begin{eqnarray}
{\rm Mod}_{g}\backslash T_{g} &\to& Sp(2g,\mathbb{Z})\backslash \mathfrak S_{g}, \\
\mathfrak{S}_{g} &=& Sp(2g,\mathbb{R})/U(n).
\end{eqnarray}
These considerations on Teichm\"{u}ller spaces are useful not only for the $U(1)$ case discussed above,
but also for a mass function which is defined over a multiple of Riemann surfaces as stated above, 
in terms of theta functions. 
It is well-known fact that 
the dual of tangent vector $T_{p}(T_{g})$ at $p$ obtained from the tangent bundle $T(T_{g})$ of 
Teichm\"{u}ller space $T_{g}$ is a holomorphic cusp form of $\Gamma$ of $\bf{H}$. 
When a mass function as a section of line bundle ( invertible sheaf ) 
is defined over a Riemann sphere $\mathbb{P}^{1}_{k}$ ( $k$: a field ), 
then ${\rm Aut}\mathbb{P}^{1}_{k}=PGL(1,k)$ acts on the mass function.

\subsubsection{Variational Calculus, Singularity and the Morse Theory}

In general, an infinite-dimensional Lie group $G$ consists of
an infinite-dimensional vector space defined locally,
with an appropriately defined norm such as Banach, Hilbert, Fr\'{e}chet, so forth.
For example, let $G\equiv C^{\infty}(M,\mathbb{R})$ with $M$ a finite-dimensional manifold.
In that case, $\mathfrak{g}=T_{0}G=T_{0}C^{\infty}(M,\mathbb{R})\simeq C^{\infty}(M,\mathbb{R})$.
It is well-known fact ( a theorem of J. Eells ) that 
a variational calculus is described by an infinite dimensional Lie group with a Finsler metric.

\vspace{2mm}

The essence of the Morse theory will be stated as follows
( J. W. Milnor, {\it Singular Points of Complex Hypersurfaces},
Ann. Math. Stud., Princeton University Press, 1969 ):
Let $M$ be a differentiable manifold of $\dim M=m$,
and let $f$ be a function on $M$.
The topological nature of $M$ has a deep connection with $f$.
Especially it will be summarized in a charater of $f$ at a critical point.
A decomposition of $M$ by utilizing the level set of $f$ contains a lot of informations of
the topological nature of $M$, and $M$ at the critical point of $f$ 
is homotopy equivalent with a CW complex.
The critical point of $f$ relates with the homology of $M$,
the Euler characteristic of $M$ can be estimated from it
( a Morse index and a number of critical points of $f$ satisfies a certain relation of a Betti number of $M$ ).
These aspects are also contained in our $V_{eff}$.
For example, the CKM matrix might contains such a mathematical aspect 
if our theory of it is correct.
Of course, a critical point has a strong relation with a singular point in the sense of algebraic geometry.

\vspace{2mm}

{\bf Theorem:}
{\it A critical point of an effective potential of our GNG theorem gives a singularity.}

\vspace{2mm}

In algebraic geometry, a singular point of an argebraic variety ( curve ) in $\mathbb{A}^{2}$ is defined by 
$f(a,b)=\frac{\partial f}{\partial x}(a,b)=\frac{\partial f}{\partial y}(a,b)=0$ 
in the following expansion:
\begin{eqnarray}
f(x,y) &=& f(a,b) + \frac{\partial f}{\partial x}(a,b)+\frac{\partial f}{\partial y}(a,b) + \cdots.
\end{eqnarray}
Thus, our effective potential $V_{eff}(x,y)$, for example in the $U(1)$ case, 
at a critical point becomes a singular point after a suitable normalization $V_{eff}(x=a,y=b)=0$.

\vspace{2mm}

It is known fact that the set of singular points of a two-dimensional normal analytic space $X$
is discrete.
The definition of geometric genus $p_{g}(X,x)$ of a singular point $x\in X$
will be obtained by ${\rm rank}R^{1}\pi_{*}{\cal O}_{\tilde{X}}$
( rank of the direct image functor ) at $x$ ( thus, it is defined locally ),
where ${\cal O}_{\tilde{X}}$ is the structure sheaf of the resolution $\pi:\widetilde{X}\to X$.

\vspace{2mm}

In the Morse theory, a function $f$ over $\mathbb{R}^{n}$ in the vicinity of
a non-degenerate critical point $p$ will be written down as
\begin{eqnarray}
f(x) &=& f(p) +x^{2}_{1} + \cdots + x^{2}_{j} -x^{2}_{j+1} - \cdots -x^{2}_{n},
\end{eqnarray}
and this quadratic form defines a $GL_{n}(\mathbb{R})$-orbit, 
an element of the set $GL_{n}(\mathbb{R})/O(j,n-j)$.
In a complexification of a Lie group, the following Hermitian form should be considered,
will be obtained from $f$ under the manner to keep the total potential a real form
( thus, a non-holomorphic ):
\begin{eqnarray}
f(z,\bar{z})^{\mathbb{C}} &=& f(p)^{\mathbb{C}} +z_{1}\bar{z}_{1} + \cdots + z_{j}\bar{z}_{j} -z_{j+1}\bar{z}_{j+1} - \cdots -z_{n}\bar{z}_{n}.
\end{eqnarray}
This form belongs to the set $GL_{n}(\mathbb{C})/U(j,n-j)$.
Of course, an effective potenial $V_{eff}$ of our GNG theorem also takes those functional forms.
It should be mentioned that those potentials given above are defined only by the local coordinate systems
of an explicit+dynamical symmetry breaking ( namely, $x_{l}$ or $z_{l}$ ). 
On the other hand, the total space of a Lie group which is contained in a theory/model has more coordinates, 
those potentials must be embedded into a larger space defined over the following base spaces 
$\mathbb{R}^{N}$ or $\mathbb{C}^{N}\otimes\overline{\mathbb{C}}^{N}$ 
( $N\ge n$ ). 
When a mass function is devided into a holomorphic and an anti-holomorphic parts,
then the holomorphic part is a section of a vector bundle ( locally free sheaf ) defined 
on $X \subset \mathbb{C}^{n}$,
and $X$ may be embedded into a topological space $Y \subset \mathbb{C}^{N}$
( if it is projective ).
These considerations presented here will be generalized as follows:
Let $G$ and $H$ be Lie/algebraic/finite groups 
( one can consider a compact topological group 
${\rm Gal}(L/K)\simeq \lim_{\leftarrow_{\lambda}}{\rm Gal}(L_{\lambda}/K)$ )
defined over schemes $X$ and $Y$, respectively.
If a breaking $G\to H$ is considered, then an NG scheme ( manifold/variety ) is realized in the subscheme $X-Y$.  
One may go back to a physically relevant situation via a change of base space
such as schemes $\to$ varieties $\to$ manifolds.
If both $(X,\mathfrak{m}_{X})$ and $(Y,\mathfrak{m}_{Y})$ are N\"{o}therian regular local rings,
their Krull dimensions $\dim{X}$ and $\dim{Y}$ are the same with dimensions of those of
Zariski tangent spaces $\dim_{X/\mathfrak{m}_{X}}$ and $\dim_{Y/\mathfrak{m}_{Y}}$, 
coincide with coordinate rings of them: 
Thus $\dim{X}=\dim{Y}+\dim(X-Y)$.
One can employ a completion $\widehat{X}=\lim_{\leftarrow_{n}}X/\mathfrak{m}^{n}_{X}$ 
to convert $X$ to a topological ring to aquire the Hausdorff nature, 
and the completion of $\mathfrak{m}_{X}$-adic topology does not change the dimension.

\vspace{2mm}

We consider a very generic situation.
Let $X$ be a scheme over an algebraically closed field $k$,
and let $Y$ be an irreducible closed subscheme of $X$ defined by a sheaf of ideal ${\cal I}$. 
For our GNG theorem, we set the NG base space as $X-Y$
( $V_{eff}$ can give a variety/manifold over $X-Y$,
and it can have a singularity, though it is always removed by setting $V_{eff}\ne 0$ ),
while the behavior of an effective potential as a section on the symmetric subscheme $Y$ 
defines a constant sheaf ( since it does not depend on $Y$ ),
$\Omega_{Y/k}$ must be locally free ( vector bundle ), 
and must have the following short exact sequence:
\begin{eqnarray}
0 \to {\cal I}/{\cal I}^{2} \to \Omega_{X/k}\otimes {\cal O}_{Y} \to \Omega_{Y/k} \to 0.
\end{eqnarray} 
It is known fact that there is a surjective homomorphism if Y is a proper closed subset of X: 
\begin{eqnarray}
{\mathcal Cl} X \to {\mathcal Cl}(X-Y),
\end{eqnarray}
( ${\mathcal Cl}$; a divisor class group )
and the following exact sequence exists if $Y$ is an irreducible subset with codimension 1:
\begin{eqnarray}
\mathbb{Z} \to {\mathcal Cl}X \to {\mathcal Cl}(X-Y) \to 1.
\end{eqnarray}
For example, the following relation exists 
if $Y$ is an irreducible curve of degree $n$ in $\mathbb{P}^{2}_{k}$:
\begin{eqnarray}
{\mathcal Cl}(\mathbb{P}^{2}_{k}-Y) = \mathbb{Z}/n\mathbb{Z} \simeq {\rm Gal}(\mathbb{F}_{q^{n}}/\mathbb{F}_{q}).
\end{eqnarray}

\vspace{2mm}

The definition of variation vector field along with $\phi$ ( will be explained soon ), 
a set of $C^{\infty}$-class section $\Gamma(\phi^{-1}TN)$, 
is defined by the following diagram:
\begin{eqnarray}
\begin{array}{ccc}
M & \stackrel{\phi}{\longrightarrow} & N \\
\psi\Bigg\uparrow & & \Bigg\uparrow\psi' \\
\phi^{-1}TN & \longrightarrow & TN,
\end{array}
\end{eqnarray}
where, $\phi$ is $C^{\infty}$-class, $M$, $N$ manifolds,
and $\phi^{-1}TN$ is an induced bundle on $M$.
These arrows directed from down to up are projections of those bundles.
Hence, under a variation of $\phi$, it will give a tangent space $T_{\phi}C^{\infty}(M,N)$
as an infinite-dimensional Lie group.
A combination with a Kodaira-Spencer deformation of complex structure of a principal bundle 
$(P,X,\omega,G)$ of a complex Lie group is given by
\begin{eqnarray}
\begin{array}{ccccccc}
P & \stackrel{\omega}{\longrightarrow} & X\times S & \stackrel{{\rm restriction}\,\, r}{\longrightarrow} & S & & \\
& & \uparrow &  & \uparrow & & \\ 
& & r^{-1}T_{s_{0}}S & \longrightarrow & T_{s_{0}}S & \stackrel{\rho}{\longrightarrow} & H^{1}(X,{\rm Ad}(P_{s_{0}})) \\
& & \downarrow & & & & \\
& & \Gamma(r^{-1}T_{s_{0}}S), & & & &  
\end{array}
\end{eqnarray}
where, $\Gamma(r^{-1}T_{s_{0}}S)$ is a set of variation vector fields.
The Kodaira-Spencer deformation can be applied 
if the base manifold of $G$-bundle or a homogeneous bundle are locally compact.
( It may be extended algebraically via a proper morphism. )
In our GNG theorem, an automorphic representation $L^{2}(\Gamma\backslash G)$ is obtained
in an effective potential, and it may be expanded by a set of cusp forms, for example.
Then the set of cusp forms, eigenfunctions of a Hecke operator, will enter into the
diagram given above. 
A geodesic of a finite periodicity is called as a closed geodesic,
and it relates with a fundamental group of a Riemannian manifold.
A classical dynamics over a manifold ( or a topological space ) of our GNG theorem
without any chaotic behavior can give, but not always, a closed geodesic on an NG manifold. 
A variational calculas given by a holomorphic ( harmonic ) mapping of two compact K\"{a}hler manifolds
consists of a set of homotopic paths, and there is a work on relation between such a set of homotopic paths
and Teichm\"{u}ller theory ( S. A. Wolpert ).

\vspace{2mm}

One can assumes that a fermionic determinant ( a real form ) is factorized such that
\begin{eqnarray}
\frac{1}{2}|f|^{2} = \frac{1}{2}f^{*}f = \det (\pfey-{\mathcal M}),  \quad f:M \to N,  
\end{eqnarray} 
where, $(M,g)$ and $(N,h)$ are Riemannian manifolds.
Then the so-called "energy density" in theory of variational calculus is obtained
via a definition on Finsler metric ( it can always be taken if the base space is paracompact ) 
over a tangent space:
\begin{eqnarray}
\int \det(\pfey-{\cal M})dx = \frac{1}{2}\int_{M} |f|^{2}dx,   \quad
E(f) = \frac{1}{2}\int_{M} |df|^{2}dx,   \quad  
df_{x}: T_{x}M \to T_{f(x)}N. 
\end{eqnarray}
From the first variation of $E(f)$, an Euler-Lagrange equation will be derived,
and it has a Laplacian defined on a local coordinate system $\{x\}\in M$.
Since the first variation of $E(f)$ is the second variation in our theory of this expression,
the Hessian of the fermionic determinant must contain a Laplace operator under the definition given above.
After a complexification of Lie groups,
several breaking schema will have the structure of K\"{a}hler manifolds.
As mentioned above, 
a holomorphic mapping $\phi:M\to N$ between two K\"{a}hler manifolds $M$ and $N$ becomes harmonic, 
and $\phi$ satisfies a Laplace equation $\triangle\phi=\lambda\phi$ 
derived from the action ${\cal E}[\phi]=\int||d\phi||dv$.
Since a Casimir element 
${\cal C}_{G}:\{\mathfrak{g}\otimes\mathfrak{g}\simeq T_{0}G\otimes T_{0}G|{\rm Ad}{\cal C}_{G}={\cal C}_{G}\}$ 
is a Laplacian ( $\triangle\in$ the space of second-order derivatives ) of a Lie group,
it will have the same roll for a harmonic mapping between Lie groups defined over K\"{a}hler manifolds,
with the energy functional ${\cal E}[\psi]=\int_{G}\langle\psi, {\cal C}_{G}\psi\rangle dg$.
$\langle\,\, ,\,\,\rangle$ denotes an appropriately defined inner product.

\vspace{2mm}

On the other hand, a generic zeta function of a Lie/algebraic group $G$ is given by
\begin{eqnarray}
\zeta_{G} \sim \det\Bigl( {\cal C}_{G}-s + {\rm const}. \Bigr),
\end{eqnarray}
and after solving the eigenvalue equation ${\cal C}_{G}\psi_{a}=\lambda_{a}\psi_{a}$,
we get
\begin{eqnarray}
{\cal E}[\psi] &=& \int_{G}\langle\psi, {\cal C}_{G}\psi\rangle dg = {\rm Tr}{\cal C}_{G} = \sum_{a}\lambda_{a}.
\end{eqnarray}
Especially, the definition of Selberg zeta function 
( a zeta function for congruence subgroups of the upper half plane )
on $\Gamma\backslash {\bm H}=\Gamma\backslash SL(2,\mathbb{R})/SO(2)$,
where $\Gamma$ consists with the fundamental group $\pi_{1}$,
are interesting from the context of our GNG theorem:
$\zeta^{Selberg}_{\Gamma}=\prod_{p\in Prim(\Gamma)}(1-N(p)^{-s})$,
where $Prim(\Gamma)$ implies a set of conjugacy classes prime with each other.
In fact, the fermionic determinant discussed above generically has such a Selberg-zeta-like structure,
and in addition, an L-function structure.
An $L$-function of a finite group is constructed from the eigenvalues of the following determinant:
\begin{eqnarray}
\det\bigl( X-\varrho \bigr),
\end{eqnarray}
where, $\rho$ is a representation of a conjugacy class $\varphi$ of a finite group $G$.

\vspace{2mm}

Another interesting example for our study is the Dirichlet-Dedekind-Epstein zeta function of a quadratic form.
This zeta function relates with the $SU(2)$ ( or, $U(2)$ ) case discussed in the previous section
since the exponential mapping of them gives a clear example of a quadratic form,
defined over a two-dimensional surface.
The definition of the zeta function of positive-definite quadratic form is
\begin{eqnarray}
\zeta_{Q}(s,M) &\equiv& \sum_{x\in M,x\ne 0}Q(x)^{-s} = \sum^{\infty}_{n=1}\frac{a(n)}{n^{s}}, \qquad \Re s > m/2.
\end{eqnarray}
Here, $Q$ is a positive-definite quadratic form with $m$ the number of indeterminates/variables,
$M$ is a lattice defined on an $m$-dimensional vector space $V$ over $\mathbb{R}$,
and $a(n)$ the number of points $x\in M$ with the condition $\mathbb{N}\ni n=Q(x)$.
The series absolutely converges at $\Re s>m/2$. 
Let us introduce the dual basis $(x^{*}_{1},\cdots,x^{*}_{m})$ with the condition
$Q(x_{i},x^{*}_{j})=\delta_{ij}$, $Q(x,y)\equiv (Q(x+y)-Q(x)-Q(y))/2$.
Then a theta function is obtained by the following definition:
\begin{eqnarray}
\vartheta_{Q}(u,M) &\equiv& \sum_{x\in M}\exp\bigl[ -\pi u Q(x) \bigr], \qquad \Re u >0.
\end{eqnarray}
In is known fact in literature that those zeta and theta functions satisfy certain functional equations.
Here, we find that those zeta and theta functions are embedded into the mathematical structure of 
the effective potential of our GNG theorem of the $SU(2)$ or $U(2)$ flavor case.

\subsubsection{Connections and Moduli Spaces of Lie Groups}

We consider a homogeneous space with the fiber bundle of a Lie group,
\begin{eqnarray}
(G,G/H,\pi,H),
\end{eqnarray}
where, $G$ the total space, $G/H$ the base space, $\pi:G\to G/H$ the projection and $H$ the structure group.
The Maurer-Cartan form ( namely, the dual of the tangent space of the Lie group ) is given as
\begin{eqnarray}
\omega_{\mathfrak{g}} &=& \omega_{\mathfrak{h}}+\omega_{\mathfrak{m}},
\end{eqnarray}
and the connection of the principal $G$-bundle is defined by $\omega_{\mathfrak{h}}$.
Then one can find a "gauge-transformation" group for the homogeneous space,
but here we wish to consider the following suitable for our purpose:
Generically, for a $G$-bundle $P$, one has
\begin{eqnarray}
& & \omega \in \mathfrak{C}(P) \simeq \phi^{1}(P\times_{Ad}\mathfrak{g}),  \\
& & \mathfrak{R} \in \phi^{2}(P\times_{Ad}\mathfrak{g}),  \\
& & \mathfrak{G}(P) = \Gamma(P\times_{Ad}G),  \\
& & {\rm Lie}(\mathfrak{G}) = \Gamma(P\times_{Ad}\mathfrak{g}),  \\
& & {\rm Exp}: \Gamma(P\times_{Ad}\mathfrak{g}) \to \mathfrak{G}(P), 
\end{eqnarray}
where, $\omega$, $\mathfrak{R}$, $\mathfrak{G}$, are a connection, a curvature, and the gauge group,
respectively. 
$\Gamma$ denotes a total set of differentiable sections.
Then the gauge transformation is given as follows:
\begin{eqnarray}
\mathfrak{G}(P): \mathfrak{C}(P) \to \mathfrak{C}(P), \quad \varphi: \{\omega \} \to \{\omega\}.
\end{eqnarray}
We wish to emphasize the fact that one can always has a "gauge-transformation" group $\mathfrak{G}$
of a Lie group,
if one prepares a principal bundle and the space of its connections ( differentials on a set of sections ):
It is also the case in our GNG theorem.
Then the moduli space of the connection and the tangent space of it ( at $\omega_{0}$ ) are defined by
\begin{eqnarray}
\mathfrak{M}_{\mathfrak{C}} &=& \mathfrak{C}(P)/\mathfrak{G}(P),   \\
T_{\omega_{0}}(\mathfrak{M}_{\mathfrak{C}}) &\to& H^{1} 
\equiv T_{\omega_{0}}\mathfrak{C}(P)/T_{\omega_{0}}\mathfrak{G}(\omega). 
\end{eqnarray} 
( $\mathfrak{G}(\omega)$; a $\mathfrak{G}$-orbit. )
Since we consider a usual Lie group in here, 
the dimension of $\mathfrak{M}_{\mathfrak{C}}$ coincides with that of the tangent space:
${\rm dim}\mathfrak{M}_{\mathfrak{C}}={\rm dim}H^{1}$.
In the case of a broken symmetry,
the set of gauge transformations contains gauge orbits of both "broken" and "symmetric",
classified by a set of irreducible representations of the "gauge-transformation" group,
and the action functional of the theory lost the $\mathfrak{G}$-invariance partially/totally.
However, the definition of connection does not ( or, do not have to ) change, 
the same for both "broken" and "symmetric" cases.
Thus, the number of moduli parameters $\in \mathfrak{M}_{\mathfrak{C}}=\mathfrak{C}/\mathfrak{G}$ 
of a symmetry-breakdown case is the same with that of no symmetry-breaking case, 
though the equivalence/flatness along with gauge-orbits of the action functional 
on the parameter space are partially/completely lost.   
The set of stable points of an effective potential gives 
a subset of the space of gauge transformation and the moduli space.
Since a Riemannian/Hermitian metric of a symmetric space is given by a Killing form,
a deformation of it is obtained by the method of Yang-Mills functional discussed in a moduli theory.

\subsection{Automorphic Forms and Automorphic Representations}

( References; J. S. Milne, {\it Modular Functions and Modular Forms},
{\it Introduction to Shimura Varieties}, can be obtained from his website.
G. van der Geer,
{\it Siegel Modular Forms}, arXiv:math/0605346. )

\vspace{2mm}

An automorphic form 
( such as a modular form, a Hilbert modular form, a Siegel modular form, so forth ) 
is an example of semi-invariants. 
The definition of a semi-invariant $f$ of a weight $\chi$ is given as follows:
\begin{eqnarray}
f(x) \in R, \quad X = {\rm Spec}R, \quad f(gx) = \chi(g)f(x), \quad \forall g\in G, \quad \forall x\in X.
\end{eqnarray}
For example, theta functions over the torus $\mathbb{C}/\Lambda$ ( $\Lambda$: a lattice group ) give,
\begin{eqnarray}
f(z+\lambda) &=& e_{\lambda}(z)f(z), \quad (\forall z\in\mathbb{C},\,\, \lambda\in\Lambda), \quad
e_{\lambda+\lambda'}(z) = e_{\lambda}(z+\lambda')e_{\lambda'}(z).
\end{eqnarray} 
The canonical example of it is
\begin{eqnarray}
\Theta(\tau,z+m+n\tau) &=& e^{-i\pi(2nz+n^{2}\tau)}\Theta(\tau,z).
\end{eqnarray}
In a classical theory of cusp forms, 
it is well-known fact that any cusp form can be given by a linear combination of 
Poincar\'{e} serieses they are invariants under a certain group action.
Namely, one can consider a very generic invariant function such as
\begin{eqnarray}
F(x_{1},\cdots,x_{n})^{G} \equiv \sum_{g\in G}f(g\cdot x_{1},\cdots, g\cdot x_{n}),
\end{eqnarray}
and modular forms in various dimensions over various spaces also be given under this manner.
In general, an automorphic form $f$ is defined over a coset $\Gamma\backslash G$ 
( or, $\Gamma\backslash G/H$ ).
Therefore, if 
\begin{eqnarray}
f(\Gamma\backslash G) \in L^{2}(\Gamma\backslash G)
\end{eqnarray}
( namely, a Hilbert space with an appropriately defined inner product ) is the case, 
then a set of $f$ gives a unitary represenation.

\vspace{2mm}

The space of automorphic forms 
( module of a Hecke algebra )
as a vector space over $\mathbb{C}$ is given as follows:
\begin{eqnarray}
{\cal M}(\Gamma) &=& \bigoplus^{+\infty}_{k=0}{\cal M}_{k}(\Gamma),  \\
\bigoplus^{+\infty}_{k=0}{\cal M}_{k}(\Gamma)^{\mathbb{C}} &=& \mathbb{C}[G_{2},G_{3}],  \\
\bigoplus^{+\infty}_{k=0}{\cal M}_{k}(\Gamma)^{\mathbb{Z}} &=& \mathbb{Z}[G_{2},G_{3}], 
\end{eqnarray}
( a graded commutative ring ), where $k$ denotes the weight of the space of automorphic forms of $\Gamma$,
and $G_{2}$, $G_{3}$ are the Eisenstein serieses.
The algebra of automorphic forms is 
\begin{eqnarray}
{\cal M}_{k}(\Gamma){\cal M}_{l}(\Gamma) \to {\cal M}_{k+l}(\Gamma), \quad
{\cal M}_{k}(\Gamma)\cap{\cal M}_{l}(\Gamma) = \{0\} \,\, ({\rm if}\,\,k\ne l),
\end{eqnarray}
and it is finitely generated, and ${\rm dim}{\cal M}_{k}<\infty$.
The system of generators of the algbera is given by $g_{2}$ and $g_{3}$ of
the Weierstrass form $y^{2}=4x^{3}-g_{2}x-g_{3}$ of the elliptic curve.
For example,
\begin{eqnarray}
{\cal M}_{2}(\Gamma(1)) = \mathbb{C}g_{2}, \quad 
{\cal M}_{3}(\Gamma(1)) = \mathbb{C}g_{3}, \quad 
{\cal M}_{4}(\Gamma(1)) = \mathbb{C}g^{2}_{2}, \quad 
{\cal M}_{5}(\Gamma(1)) = \mathbb{C}g_{2}g_{3}, \quad 
\cdots
\end{eqnarray}  
( $\Gamma(1)$ the full modular group ),
and thus they are given by Eisenstein serieses.
The space of cusp form 
\begin{eqnarray}
{\cal M}^{0}(\Gamma) &=& \bigoplus^{+\infty}_{k=0}{\cal M}^{0}_{k}(\Gamma)
\end{eqnarray}
is an ideal of ${\cal M}(\Gamma)$.
Therefore, if an effective potential is expanded in terms of cusp forms,
then the potential is expressed as an element of algebra of Eisenstein serieses,
expanded by them.
The space of cusp forms has an inner product ( the Petersson inner product ) 
in the upper half plane $\bf{H}$, it gives a Hilbert space, 
and then the set of cusp forms can expand a function 
( belongs to an automorphic representation ) over $\bf{H}$.
For a modular form of $\Gamma=SL_{2}(\mathbb{Z})$ of real-analytic, 
the Maass wave form is also useful.

\vspace{2mm}

It is clear that an effective potential with a specific period
will obtain an expansion given by a set of automorphic function $f\in {\cal M}_{k}$ 
( $\forall k\in\mathbb{Z}_{\ge 0}$ ).
A cusp form $f$ over $K$ ( ${\rm char}K=0$ ) is uniquely expanded such as 
$f=\sum^{\infty}_{m=1}a_{m}(f)q^{m}$ ( $q=e^{2\pi iz}$ ) with a specific period defined by $\Gamma$.
This implies that the expansion has a coherency between waves $e^{2\pi imz}$ 
which is not destroyed in the infinite-order series.
While the modularity will get the connection between a cusp form 
over $\mathbb{Q}$ and a Galois represenation of an elliptic curve.
Due to its clear periodicity, this coherency must be held in our $V_{eff}$ 
or a fiber which gives $V_{eff}$ as its section.
Thus, we wish to call this phenomenon as {\it modular Galois coherence}. 
$V_{eff}$ can have ( not always ) a modulation along the coordinate of periodicity,
while a regular cusp form must be a constant inside a compact region.
Therefore, the modulation of $V_{eff}$ should be generated 
via a combination of holomorphic and anti-holomorphic cusp forms.
Moreover, this mathematical structure should be kept ( reflects ) in the dual of $G$. 
Here, we wish to show a diagram for those relations for a consideration on our GNG theorem:
\begin{eqnarray}
{\rm Peter-Weyl \,\, Theorem \,/\, Pontrjagin-Tannaka \,\, duality} \to
{\rm automorphic \,\, representation} \to
{\rm automorphic \,\, forms}.
\end{eqnarray}

\vspace{2mm}

Let $X_{0}(N)$ be a modular curve, 
and let $S(N)\equiv \Gamma(X_{0}(N),\Omega^{1}_{X_{0}(N)/\mathbb{Q}})$
be a space of regular differential forms, a finite-dimensional $\mathbb{Q}$-linear space.
It is known in literature that the space of modular forms of level $N$ is equal to $S(N)$~[252].
While our holomorphic mass function, especially of $U(1)$ case, 
\begin{eqnarray}
{\cal M}(z,\tau) &\propto& e^{if(z,\tau)}, \quad z \in \mathbb{C}, \quad \tau \in {\bf H}, 
\end{eqnarray}
belongs to a space of automorphic forms and thus,

\vspace{2mm}

{\bf Theorem:}
$S(N)$ {\it as a finite-dimensional} $\mathbb{Q}$-{\it linear space gives a basis set 
to express a mass function with level} $N$.

\vspace{2mm}

If we generalize this, then we obtain the following expression:
\begin{eqnarray}
{\cal M}(z,\tau) \in d \Bigl( \bigoplus_{N} X_{0}(N) \Bigr). 
\end{eqnarray}
This formal expression may give us an insight on our string-like action of mass function
discussed in the previous section.
We show the following logical sequence:
\begin{eqnarray}
& & {\rm Modular \,\, curves} \to {\rm cusp \,\, forms} \to {\rm mass \,\, functions} \to {\rm string}  \nonumber \\ 
& & \qquad \to {\rm conformal \,\, field \,\, theory} \to {\rm Kac-Moody \,\, algebras}.
\end{eqnarray}

\vspace{2mm}

An interesting interpretation to the breaking scheme $SU(4_{f})\to Sp(4_{f})$ is provided from
a coset representation of Siegel upperhalf space.
In that case,
\begin{eqnarray}
Sp(4_{f},\mathbb{Z})\backslash Sp(4_{f},\mathbb{R})/U(2_{f})
\end{eqnarray}
will be considered, and it gives ${\bf H}^{2}$.
The effective potential and mass matrix can be given 
as sections of the coset $Sp(4_{f},\mathbb{Z})\backslash Sp(4_{f},\mathbb{R})$,
belong to the automorphic representation. 
Moreover, one knows the following scheme by a complexification:
\begin{eqnarray}
SU(2)\times U(1) = U(2) \to GL(2,\mathbb{C}).
\end{eqnarray}
On the other side, 
\begin{eqnarray}
GL(4,\mathbb{R})/GL(2,\mathbb{C})
\end{eqnarray}
gives all complex structures they can be introduced in $\mathbb{R}^{4}$.
This fact may have an interesting application to the electroweak symmetry breaking...

\subsection{Galois Representations}

The following schematical generalization of the ( a lot of ) works related with the Fermat's last theorem
is desirable for our ultimate purpose:
\begin{eqnarray}
\begin{array}{ccc}
{\rm modular\,\, variety} & \to & {\rm (projective, Abelian) variety\,\, with\,\, finite \,\,groups} \\
({\rm motive}) & & \\
\downarrow & & \downarrow  \\
{\rm automorphic\,\, form/representation, \,\, perverse \,\, sheaves} & \leftrightarrow & {\rm Galois\,\, representation} \\
( {\rm as \,\, a \,\, differential \,\, form \,\, of \,\, modular \,\, variety} ) & & ( {\rm a \,\, finite\,\, group\,\, scheme,\,\, motivic \,\, Galois} ) \\
\downarrow & & \downarrow \\
{\rm Hecke}-{\rm type \,\, algebra} & \leftrightarrow & {\rm Galois \,\, deformation} \\
( {\rm defined \,\, on \,\,} H\backslash G/H ) & & 
\end{array}
\end{eqnarray}
The fundamental problem of class field theory is the question on
whether a finite group scheme always has its corresponding Galois represenation or not.
From the work of Artin and Tate, we know the following logical correspondence:
\begin{eqnarray} 
{\rm finite}\, {\rm group} \to {\rm Galois}\, {\rm represenation} \to
{\rm Galois}\, {\rm cohomology} \to {\rm class}\, {\rm field}\, {\rm theory}.
\end{eqnarray}
In the context of our GNG theorem, the fundametal group $\pi_{1}$ of a base space
or a ( compact ) Lie group is the origin of the number-theoretical aspect of the theorem:
A geometric Galois action naturally be generated in the mechanism of our GNG theorem.
By the definitions of congruence subgroups, we know
\begin{eqnarray}
\Gamma_{0}(N)/\Gamma_{1}(N) \simeq 
S_{0}(N)/S_{1}(N) \simeq 
(\mathbb{Z}/N\mathbb{Z})^{\times} \simeq {\rm Aut}(\mathbb{Z}/N\mathbb{Z}) \simeq  
{\rm Gal}(\mathbb{Q}(\zeta_{N})/\mathbb{Q}).
\end{eqnarray}
The following isomorphisms are known in literature:
\begin{eqnarray}
\mathbb{Z}/N\mathbb{Z} \simeq {\rm End}(\mathbb{Z}/N\mathbb{Z}) \simeq {\rm Gal}(\mathbb{F}_{q^{N}}/\mathbb{F}_{q}) \simeq 
H^{2}({\rm Gal}(K/k),C_{K}).
\end{eqnarray}
Here, $C_{K}$ denotes an idele class group.
As discussed above, those Galois represenations realize in the effective potential of the $U(1)$ case.

\vspace{2mm}

Moreover, one can introduce the following automorphic representations on $\Gamma\backslash G$ 
to symmetry breakings of physics:
\begin{eqnarray}
\Gamma = SL_{2}(K) &\subset& SL_{2}(\mathbb{A}_{K}) = G,   \\
\Gamma = GL_{n}(K) &\subset& GL_{n}(\mathbb{A}_{K}) = G,   \\
\Gamma = \mathbb{Z}^{n} &\subset& \mathbb{R}^{n} = G,   \\
\Gamma = K^{n} &\subset& \mathbb{A}^{n}_{K} = G.
\end{eqnarray}
Here, $\mathbb{A}_{K}$ is the adele ring, 
and it can become a locally compact topological group after a completion.
Then the group of $\mathbb{A}_{K}$ is continuous,
and one can introduce an auxiliary field which is defined over the ring "globally",
it may be embedded into a fermionic/bosonic determinants, 
gives an effective action in terms of them, and then a gap equation of gap function
defined over $\mathbb{A}_{K}$.
For example, $\Gamma$ can be chosen as a fundamental group of polyhedron.
Then an interesting question is that there is a Galois representation in each of them.
It might be useful to employ the Grothendieck-Galois-Teichm\"{u}ller theory on those questions.

\subsubsection{Arithmetic Varieties and Galois-Riemannian Geometry}

In the case of N\"{o}therian regular local ring, 
a Riemannian metric may be given as follows:
\begin{eqnarray}
\hat{g}_{Y}\Bigl(\frac{\partial}{\partial X},\frac{\partial}{\partial X}\Bigr) 
\simeq \hat{g}_{Y}\Bigl( (\mathfrak{m}_{x}/\mathfrak{m}^{2}_{x})^{\vee},(\mathfrak{m}_{x}/\mathfrak{m}^{2}_{x})^{\vee} \Bigr).
\end{eqnarray}
This definition may give us a road toward a very generic construction of a Riemannian metric.
For example, Grassmann coordinates of a supermanifold may be interpreted by the notion
of Galois-Riemannian geometry.
One may introduce the following ring:
\begin{eqnarray}
R &=& \mathbb{C}[x_{1},\cdots,x_{j}] + \mathbb{F}_{q}[x_{j+1},\cdots,x_{n}],
\end{eqnarray}
or, for example, as a Lie algebra 
\begin{eqnarray}
\mathfrak{g} &=& \mathfrak{h}\otimes_{\mathbb{R}}\mathbb{C} 
+ \mathfrak{m}\otimes_{\mathbb{F}_{q}}\overline{\mathbb{F}_{q}}.
\end{eqnarray}
Note that the affine line and a Riemann sphere over $\mathbb{F}_{q}$ 
are defined by the global field $\mathbb{F}_{q}[x]$,
\begin{eqnarray}
\mathbb{A}^{1}_{\mathbb{F}_{q}} \simeq {\rm Specm}\mathbb{F}_{q}[x], 
\quad \mathbb{P}^{1}_{\mathbb{F}_{q}} \equiv \mathbb{A}^{1}_{\mathbb{F}_{q}}\cup\{\infty\}.
\end{eqnarray}
Let $C$ be a curve defined over $\mathbb{F}_{q}$.
The Frobenius map acts on the divisor class group of $C\otimes_{\mathbb{F}_{q}}\overline{\mathbb{F}_{q}}$,
and the characteristic polynomial of the matrix representation of the Frobenius map 
on to the divisor class group becomes the congruence zeta function.
Thus, our theory will aquire a relation with the Weil conjecture, 
class field theory of global fields 
( algebraic function field $\mathbb{F}_{q}[x]$ ), and 
algebraic field of finite extension 
( via the well-known correspondence between divisor class group and ideal class group of Dedekind domain ).
The function field can take arbitrary genus, and our mass matrix at the set of stable points in 
the $U(1)$ case defines a vector bundle ( locally free sheaf ) over the function field.
Thus, beside the question on an invariance of an effective potential,
a holomorphic mass function on a ( abstract ) Riemann surface should belong to a birational/biholomorphic 
equivalence class since a genus must be an invariant.
This birational equivalence does not distinguish whether the breaking scheme is spontaneous or explicitly. 
Hence we have arrived at the notion "chiral/anti-chiral birational transformation", 
with the emphasis on a physical interpretation.  
An intersting application of our Galois-Riemannian geometry is supersymmetry.
In that case, $\mathbb{F}_{2}$ may be chosen in the ring/algebra shown above.

\vspace{2mm}

For the context of our GNG theorem, arithmetic varieties and so-called Arakelov geometry
( A. Moriwaki, {\it Arakelov Geometry}, Iwanami, Tokyo, 2008 ) is interesting.
In an arithmetic variety, a finitely generated free $\mathbb{Z}$-module $R$ is defined
( it is integral over $\mathbb{Z}$ ),
and a finitely generated Hermitian $R$-module is considered.  
Then, by the following ring homomorphisms,
\begin{eqnarray}
R \longrightarrow K \stackrel{\sigma}{\longrightarrow} \mathbb{C}
\end{eqnarray}
( $K$ is the total quotient ring of $R$ ),
one can define an inner product such as $h_{\sigma}(x\otimes 1, y\otimes 1)$.
For example, a metric on $R$ is given by
\begin{eqnarray}
h^{can}_{\sigma} &=& \sigma(x)\overline{\sigma(y)}.
\end{eqnarray}
Then, one obtains the connection between the metric and a trace of number theory: 
\begin{eqnarray}
{\rm Tr}_{K/\mathbb{Q}}(\omega_{i},\omega_{j}) &=& \sum^{n}_{k=1}\sigma_{k}(\omega_{i})\sigma_{k}(\omega_{j}),  \\
\langle\omega_{i},\omega_{j}\rangle_{h^{can}} &=& \sum^{n}_{k=1}\sigma_{k}(\omega_{i})\overline{\sigma_{k}}(\omega_{j}), 
\end{eqnarray}
where, the set $(\omega_{1},\cdots,\omega_{n})$ gives the $\mathbb{Z}$-free basis of $R$,
and consider the case $K(\mathbb{C})\equiv(\sigma_{1},\cdots,\sigma_{n})$.
The volume function on $R$ becomes
\begin{eqnarray}
{\rm vol}(R,\langle\, ,\, \rangle_{h^{can}}) &=& \sqrt{\det\bigl({\rm Tr}_{K/\mathbb{Q}}(\omega_{i},\omega_{j})\bigr)}.
\end{eqnarray}
The framework of arithmetic variety can be systematically extended into the Lie group theory.
In that case, theory of arithmetic varieties handle $V/\Lambda$, 
while the arithemtic Lie group may consider the arithmeticity of $\Gamma\backslash G$,
automorphic forms and Kac-Moody groups.
The summation of arithmetic Chern classes should contains a correspondence with our Chern-like polynomial.

\subsection{Aspects of Dynamical Systems}

This subsection will consider several aspects of dynamical systems in our GNG theorem.
( S. Smale, {\it Differential Dynamical Systems}, Bull. Am. Math. Soc. {\bf 73}, 747 (1967),
J.-P. Eckmann and D. Ruelle,
{\it Ergodic Theory of Chaos and Strange Attractors},
Rev. Mod. Phys. {\bf 57}, 617 (1985),
A. Katok, 
{\it Fifty Years of Entropy in Dynamics:1958-2007},
J. Mod. Dyna. {\bf 1}, 545 (2007),
Handbook of Dynamical Systems ( B. Hasselblatt and A. Katok ed., Elsevier ), especially,
B. Hasselblatt and A. Katok, {\it Principal Structures},
M. Pollicott, {\it Periodic Orbits and Zeta Functions},
J. Franks and M. Misiurewicz, {\it Topological Methods in Dynamics},
R. Feres and A. Katok, {\it Ergodic Theory and Dynamics of G-Spaces},
D. Kleinbock, N. Shah and A. Starkov, 
{\it Dynamics of Subgroup Actions on Homogeneous Spaces of Lie Groups and Applications
to Number Theory},
E. Glasner and B. Weiss,
{\it On the Interplay between Measurable and Topological Dynamics}. )

\vspace{2mm}

The definition of a period $n$ of a map $\varphi$ is  
\begin{eqnarray}
\varphi^{n}x = x, \quad {\rm or} \quad \widehat{\varphi}^{n}|x\rangle = |x\rangle,
\end{eqnarray}
where $x$ denotes a point/object,
and correspond to a problem to obtain an $n$-th root of unity.
( Hence a dynamical zeta function is a summation of a set of $n$-th root of unity with respect to $n$. 
Therefore, in the case of $U(1)$ defined over a circle $0\le\theta<2\pi$, 
the problem is to devide the interval by positive integer $\mathbb{Z}_{+}$, 1 to infinity. )
We emphasize here this is an eigenvalue equation.
Then, from the context of dynamical systems, 
a Dirac equation can be regarded as a period 2 operator of a Klein-Gordon equation:
A period defined on a ( elliptic ) operator space,
and an additional mass term or a gauge field of Dirac equation explicitly break the period.
We can introduce the notion of periodic and anti-periodic $G$-orbits, namely,
\begin{eqnarray}
(\varphi)^{n}x &=& \pm x,
\end{eqnarray}
and we should understand the weak mixing 
\begin{eqnarray}
(\varphi)^{n}x &=& e^{i\alpha}x
\end{eqnarray}
by this viewpoint.
Let $X$ and $Y$ topological spaces, and consider a morphism $\varphi: X\to Y$ with $X\cap Y\ne \emptyset$.
Then, if elements of those sets ( possibly $X=Y$ ) give
\begin{eqnarray}
\varphi: x_{0} \to y_{0} \ne x_{0}, \quad \exists x_{0}\in X, \exists y_{0} \in Y,
\end{eqnarray}
we call $x_{0}$ is not invariant under $\varphi$, close to the notion of broken symmetry.
Therefore, we can call
\begin{eqnarray}
\varphi^{n}: x_{0} \to y_{0}\ne \varphi^{n}(x_{0}),  \\
\varphi^{n}: x_{0} \to y_{0} = \varphi^{n}(x_{0}),
\end{eqnarray}
as $n$-periodically broken and $n$-periodically symetric, respectively.
A formalism closer to physical situation is given by adjoint action of period $n$,
\begin{eqnarray}
{\cal M}_{1} = {\rm Ad}(\varphi){\cal M}_{0}, \quad \cdots, \quad 
{\cal M}_{n} = {\rm Ad}(\varphi){\cal M}_{n-1} = {\cal M}_{0}.
\end{eqnarray}
Here, one can regards ${\cal M}_{j}$ as mass functions.
Of particular importance in here is these periods are defined locally, and they reflect the local structure
of an effective potential.
Note that the complexification of a $U(1)$ group, $e^{iz}$ ( $z\in\mathbb{C}$ )
can have a periodic orbit when $\Im z=0$ and $\Re z$ rational, 
while it has no period at $\Im z\ne 0$.

\vspace{2mm}

Let us consider a Morse-Smale dynamics in our GNG theorem.
( Such a dynamics is, in fact, not the same with collective modes of the (G)NG theorem,
though the consideration/examination by dynamical system theory is useful for us to understand (G)NG theorem. )
In this dynamics, a gradient flow will be considered:
\begin{eqnarray}
\frac{d\theta(t)}{dt} &=& -{\rm grad}V_{eff}(\theta).
\end{eqnarray}
Note that this equation has {\it no} kinetic part. 
The attractor of this dynamics is given by a set of stationary points of $V_{eff}$, 
topology of it will be distinguished between the ordinary NG and our GNG cases.
In the case of ordinary NG theorem, the gradient vanishes along the $U(1)$ circle of the wine bottle,
and there is no dynamics along this.
On the other hand, our GNG theorem lifts the degeneracy of vacua along the $U(1)$ degree of freedom,
and gradient flow ( and thus, the Morse-Smale dynamics ) will be defined everwhere.
Let us introduce the notion of stractural stability:
Let X be a set with a topology and $X$ has an equivalence relation.
Then a stracturally stable element is included in the equivalence class,
and it remains inside the class under a perturbation, 
the equivalence relation is kept under the perturbation.
Since an NG potential of $U(1)$ case has non-hyperbolic equilibria along the $U(1)$ circle,
it is structurally unstable, while the GNG case has only hyperbolic equilibria:
An explicit symmetry breaking perturbation causes NG $\to$ GNG,
and converts a stracturally unstable NG potential into a stable GNG potential,
with changing "nature" of sigularities of them, 
and the GNG potential shows an arithmetic nature of the Lie group.  
The periodic effective potential $V_{eff}$ belongs to automorphic representation,
\begin{eqnarray}
-{\rm grad} V_{eff}(\gamma\cdot\theta) &=& -{\rm grad} V_{eff}(\theta), \quad \forall \gamma \in \Gamma.
\end{eqnarray}
For example, let us consider a Morse-Smale dynamics with a gradient flow of the ordinary 
Ginzburg-Landau ( GL )-type potential which is $\mathbb{Z}_{2}$-equivaliant ( $z\to -z$ ):
\begin{eqnarray}
\frac{d^{2}z}{dt^{2}} &=& z -\lambda |z|^{2}z.
\end{eqnarray}
Here, we consider a second-order derivative which make the theory closer to 
a physical situation ( though still there is no kinetic term ).
A variation of $\lambda$ gives a bifurcation.
The Galois map $z\to -z$ restricts a solution of period 1.
After a linearization, one gets
\begin{eqnarray}
\frac{dz}{dt} &=& y, \quad
\frac{dy}{dt} = \bigl(1-\lambda|z_{0}|^{2}\bigr)z.
\end{eqnarray}
The eigenvalue of the Jacobian 
( the definition; ${\rm Mat}(\partial f_{i}/\partial x_{j})$ obtained from a linearization of $\dot{x}=f(x)$,
and it is used for a classification of types of bifurcations ) 
of this equation becomes $\pm\sqrt{1-\lambda|z_{0}|^{2}}$,
and they will vanish at an equilibrium defined by $0=z-\lambda|z|^{2}z$ ( the gap equation ).
Namely, the GL equation of the ordinary NG situation has two zero-eigenvalues of its Jacobian 
along the $U(1)$ NG mode, intrinsically implies/corresponds to a massless NG mode, 
thus non-hyperbolic equilibria.
While the GL equation of the case of $\Lambda<0$ ( symmetric, no symmetry breaking )
has a pair of positive and negative eigenvalues and thus the unique equilibriam $(z,y)=(0,0)$ 
is hyperbolic and structurally stable.  
Thus, we summarize our result as the

\vspace{2mm}

${\bf Theorem}$:
{\it The analysis of NG space by a Morse-Smale gradient flow shows a stractural instability
coming from the non-hyperbolicity of massless NG particle, 
and after the NG particle aquires a mass in our GNG theorem,
it recovers the structural stability.}

\vspace{2mm}

We summarize our consideration as a scheme:
\begin{equation}
\begin{array}{ccccc}
{\rm symmetric} & = & {\rm a \, hyperbolic \, equilibrium} & = & {\rm structurally \, stable} \\
\downarrow & & & & \\
{\rm NG} & = & {\rm "always" \, non-hyperbolic \, equilibria} & = & {\rm structurally \, unstable} \\
\downarrow & & & & \\
{\rm GNG} & =& {\rm recover \, the \, hyperbolicity} & = & {\rm structurally \, stable}
\end{array}
\end{equation}

\vspace{2mm}

Therefore, if the KM matrix is generated via our GNG mechanism, 
especially the degrees of "mixons" show a hyperbolicity, structural stability.
One can say our GNG theorem demands an extension of the catastroph/bifurcation theory to complex manifolds/domains.

\vspace{2mm}

We also have another theorem for our $U(1)$ case.

\vspace{2mm}

{\bf Theorem}:
{\it The Morse-Smale dynamics of gradient flow of the effective potential of the $U(1)$ case of our GNG theorem devides 
a ( finite/infinite ) countable number of basins of attraction, and the total of them gives the total probability.}

\vspace{2mm}

While, we need a careful examination for a non-Abelian Lie group case of our GNG theorem.

\vspace{2mm}

A Morse function has the following quadratic form in the vicinity of a critical point $p$ with 
the signature $(m,n)$, $m+n=N$
\begin{eqnarray} 
f(x) &=& f(p) - x^{2}_{1} - x^{2}_{2} - \cdots - x^{2}_{m} + x^{2}_{m+1} + \cdots + x^{2}_{N}.
\end{eqnarray}
This takes the following form with a Riemannian metric $g_{\mu\nu}$,
\begin{eqnarray}
f(x) &=& f(p) + g_{\mu\nu}x^{\mu}x^{\nu}, 
\end{eqnarray}
while one can consider a Ricci flow for a deformation of $g_{\mu\nu}$,
\begin{eqnarray}
\frac{\partial}{\partial t}g_{\mu\nu}(t) &=& -2R_{\mu\nu}.
\end{eqnarray}
The quadratic form is also used for studying a Diophantine apploximation for geomery of numbers,
especially in dynamical system theory.
( The Diophantine problem; define a set of finite integers they will vanish the quadratic form $f(x)-f(p)$. )
For our GNG situation, one can assume the signature is an invariant under a Ricci flow deformation,
and the sigunatures between a pure NG ( no explicit breaking ) and a GNG stuations are different
around their vacua.
Moreover, the signature reflects the number of explicit symmetry breaking parameters of a model
( how many local coordinates $\{\theta_{a}\}$, $0\le a \le N$ are broken explicitly,
and which direction is flat in $V_{eff}$ ), 
one has an interest on what quantity is invariant under 
a deformation of a critical point of an NG manifold via a Ricci flow. 
( G. Perelman showed that a Ricci flow is a gradient flow, 
and it can also be interpreted as a scalling by a (1+1)-dimensional quantum field theory. 
Terminology of the dynamics of $g_{\mu\nu}$; breathers $\to$ periodic orbits, solitons $\to$ fixed points.
In an effective potential of spontaneous symmetry breaking shows a quantitative 
but not qualitative difference under a renormalization.
Thus, a topological entropy is interesting from the aspect of scalling.  )

\vspace{2mm}

A formulation of a gauge theory as a dynamical system is given as follows.
Let $X$ be a topological space, let $x$ be a point of $X$, i.e. $x\in X$.
Let ${\cal L}(x)$ be a Lagrangian under our consideration.
\begin{eqnarray}
{\cal L}: X \to \mathbb{R}^{1}, \\
g\cdot{\cal L}(x) = {\cal L}'(x), \quad \exists g \in G.
\end{eqnarray}
Here, we consider the gauge ( or, flavor, Galois, so on ) transformation 
$g$ of a group $G$ as a map which is defined on a specific point $x$.
Then, one arrives at the following equation via a generalization,
\begin{eqnarray}
\frac{d}{dt}{\cal L}^{\alpha} &=& F(g,{\cal L}^{\alpha},t).
\end{eqnarray}
Here, $\alpha$ paramerizes a family of the Lagrangian.
For example, the case $g\cdot{\cal L}={\cal L}'={\cal L}+\Delta{\cal L}$ gives 
a linearization of the above first-order differential equation, namely $\Delta{\cal L}=(g-1){\cal L}$.
The case ${\cal L}$ is invariant under the operation of $g$ gives $\Delta{\cal L}=0$,
and which corresponds to $F(g,{\cal L},t)=0$. 
Of course, this implies that the invariance is an equilibrium point on $\mathbb{R}^{1}$. 
The symmetric Lagrangian and the ordinary NG case define a space of equilibria 
\begin{eqnarray}
F(g_{i},{\cal L}^{\alpha},t) = 0,
\end{eqnarray} 
while a non-invariant Lagrangian and explicit(+dynamical) symmetry breaking give a flow
\begin{eqnarray}
F(g_{i},{\cal L}^{\alpha},t) \ne 0.
\end{eqnarray} 
Especially in the $U(1)$ case, we need only one parameter and it is brounded,
thus the flow of ${\cal L}$ generated by the $U(1)$ must be periodic:
This is a dynamics of the Dashen theorem in the $U(1)$ case,
while we need more careful examination for a non-Abelian Lie group, as shown in the previous section.
For example, $z$-direction is flat while $(x,y)$-space is nontrivial 
in the case $SU(2)\to U(1)$ disscused above.
In the case where $g$ is a discrete group, the flow gives a set of discrete equilibria.
When ${\cal L}$ depends on a cutoff or coupling constant, then several spaces,
each of them is possibly an infinite-dimensional topological space, emerge vertically to 
$\mathbb{R}^{1}$.
Our arguement here implies that the CKM matrix or CPT transformation also give flows of a set of Lagrangians,
provides a new method to depict a phase diagram of a theory.
Moreover, one can puts a stochastic part of the flow equation, then one obtains
\begin{eqnarray}
a \,Ito \, stochastic \, differential \, equation \, of \, the \, Lagrangian \, flow \, in \, our \, GNG \, theorem.
\end{eqnarray}
We wish to emphasize the fact that the flow of a group ( "symmetry" at an equilibrium ) operations 
is recognized via our GNG mechanism,
while the ordinary NG theorem does not have such a flow.

\vspace{2mm}

Let us consider the following Siegel formula:
\begin{eqnarray}
(\mu(\Omega_{n}))^{-1}\int_{\Omega_{n}}\hat{f}(\Delta)d\mu(\Delta) &=& \int_{\mathbb{R}^{n}}fd\lambda.
\end{eqnarray}
Here, $\Omega_{n} \equiv SL(n,\mathbb{R})/SL(n,\mathbb{Z})$, $\Delta\in\Omega_{n}$ is a lattice,
$\mu$ is a Haar measure defined on $\Omega_{n}$, $\lambda$ is a Lebesgue measure of $\mathbb{R}^{n}$, and
\begin{eqnarray}
f &\in& C(\mathbb{R}^{n}), \\
\hat{f}(\Delta) &\equiv& \sum_{v\in\Delta\backslash\{0\}}f(v).
\end{eqnarray}
Here $C(\mathbb{R}^{n})$ is a support. 
This Siegel formula is useful as an integral transformation for an auxiliary field method
and a variational calculas.
A normalized Haar measure on $G/\Gamma$ is equivalent with 
the uniquely defined ergodic measure of maximal entropy for a geodesic flow.
( G. Knieper, {\it The Uniqueness of the Maximal Measure for Geodesic Flows on Symmetric Spaces of Higher Rank},
Israel J. Math. {\bf 149}, 171 (2005). )
In a path integral quantization ${\cal Z}\sim\int{\cal D}\phi_{G}\exp[-\int f(\phi_{G})]$,
an ergodicity will be examined by ${\cal D}\phi_{G}$, while the integration weight
will choose the largest contribution for ${\cal Z}$ which corresponds to the classical mechanical subspace
and then a set of quantum fluctuations will be taken into account.
Note that a non-ergodic $G$-invariant measure can be decomposed into a sum of ergodic components.
The following famous theorems are useful for us to consider a homogeneous dynamics
which would arise in our GNG theorem:
(1) When $G$ is a noncompact simple Lie group with a finite center, $\Gamma$ a lattice of $G$
and $H$ a noncompact closed subgroup of $G$, 
then a left action of any $H$ to $G/\Gamma$ is ergodic.
(2) The Howe-Moore ergodicity theorem:
Let $G$ be a semisimple Lie group with a finite center,
let $X$ be an irreducible $G$-space ( such as $G/\Gamma$ ) which has a finite $G$-invariant measure.
Then $H$ acts to $X$ ergodically if $H$ is a noncompact closed subgroup of $G$. 
For our "cyclotomic" effective potential of the $U(1)$ case, the following expression of average
by the Sinai-Ruelle-Bowen ( SRB ) measure $\mu$ is intersting:
\begin{eqnarray}
\frac{1}{n}\sum^{n-1}_{j=0}\varphi(f^{j}x) &\to& \int \varphi d\mu, 
\end{eqnarray}
where, $\varphi: M \to \mathbb{R}^{1}$ is a continuous map, 
$\mu$ is an ergodic SRB measure when all of the Lyapunov exponents are non-zero,
and $x\in V$ where $V$ is a set of positive-definite Lebesgue measure.
We have the following,

\vspace{2mm}

{\bf Theorem}:
{\it The measure and the integration weight of the path integral gives the functor,}
\begin{eqnarray}
{\cal L} \to \mathbb{R}^{1} \, ({\rm or}\,\mathbb{C}^{1}\otimes\bar{\mathbb{C}}^{1}) \to \mathbb{F}_{p},
\end{eqnarray}
{\it where, ${\cal L}$ is a Lagrangian,
and the periodicity and ergodicity ( both in the sense of dynamical system ) are coming from
the measure, while the finite field is given from both the measure and weight via a variational calculus.}

\vspace{2mm}

For considering dynamical systems in our GNG theorem, we should decompose our examination into several steps:
(1) dynamics of mass functions related with gauge/chiral/flavor degrees of freedom, 
(2) dynamics on the hypersurface defined by $V_{eff}$,
(3) a Hamiltonian dynamics of classical mechanics,
(4) a quantum dynamics.
For example, if $T$ denotes an action of a group corresponds to a symmetry of a system, then,
\begin{eqnarray}
T: {\cal H}(x) \to {\cal H}(T\cdot x) = {\cal H}(x).
\end{eqnarray}
One can consider $T$ as a gauge symmetry, and quantization of gauge field is subtle, though,
\begin{eqnarray}
\langle phys|{\cal H}(T\cdot x)|phys \rangle &=& \langle phys|{\cal H}(x)|phys\rangle
\end{eqnarray}
should be satisfied in a exact calculation ( no gauge-parameter dependence ).
Namely,

\vspace{2mm}

{\bf Theorem}:
{\it A continuous symmetry of a system, whether it is broken spontaneously or not, 
gives infinite-number of period 1 orbits.
This theorem cannot be applied into our GNG case, an explicit+dynamical symmetry breaking.}

\vspace{2mm}

The notion of "mixing" in dynamical systems is oposite of "isometric."
The definition of topologically weak mixing of one-parameter subgroup $g_{t}\in G$ is the follows:
\begin{eqnarray}
& & F(g_{t}x) = e^{i\alpha t}F(x), \quad (\alpha>0), \nonumber \\
& & g_{t}: \Lambda \to \Lambda,  \quad
F: \Lambda \to \mathbb{C}.
\end{eqnarray}
If we consider a unitary representation $\rho(G)$,
it is reformulated as follows:
\begin{eqnarray}
\rho(g_{t})f = e^{i\lambda t}f, \quad \forall f \in L^{2}(X,\mu), \quad \exists \lambda \in \mathbb{R}^{1}, \quad \forall t \in \mathbb{R}^{1}.
\end{eqnarray}
The stabilizer $\rho(g)f=f$ is a special case of it.
The transformation of the mass matrix in the $U(1)$ case discussed above,
\begin{eqnarray}
\phi_{t}: & & |M|e^{i\alpha} \to |M|e^{i(\alpha+\beta)}
\end{eqnarray}
is an example of topologically weak mixing.
This is interesting from the viewpoint of vortex and confinement algebra
( except a single-value condition of the vortex ).

\vspace{2mm}

Let us consider a homogeneous flow $(G/\Gamma,g_{\mathbb{R}})$.
The definition of partial hyperbolicity is the following:
Let $G=SL(n,\mathbb{R})$ and $g_{\mathbb{R}}$ as a subgroup of $G$.
Then, let us consider the torus
\begin{eqnarray}
g_{t} &=& {\rm diag}(e^{\omega_{1}t},\cdots,e^{\omega_{n}t}), \quad \{\omega_{j}\}\in \mathbb{R}^{n}, \quad
\sum^{n}_{j=1}\omega_{j} = 0 \quad ( {\rm namely}, \,\, {\rm det}g_{t}=1 ).
\end{eqnarray}
If the torus has at least two non-zero $\omega_{j}$, the flow of $g_{t}$ is hyperbolic
( = the eigenvalues distribute outside the unit circle ).
Note that this torus gives a form of non-Abelian vortex:
Thus,

\vspace{2mm}

{\bf Theorem}:
{\it A complexification of non-Abelian vortex-type solution of mass function of $SU(N)$ is hyperbolic.}

\vspace{2mm}

In a diagonal symmetry-breaking scheme discussed above, a torus subgroup
${\rm diag}(e^{a_{1}t},\cdots,e^{a_{N}t})$ is hyperbolic, 
contains a stable, unstable, and possiblly a center manifolds.
When a flow is partially hyperbolic, then its entropy is positive.
It is well-known fact that a partially hyperbolic flow shows quite sometimes a chaotic behavior.

\vspace{2mm}

Note that a linearlized dynamical system can be solved formally by a Laplace transformation:
\begin{eqnarray}
\dot{x} &=& Ax,   \\
x(t) &=& e^{tA}x(0), \\
e^{tA} &=& {\cal L}^{-1}\Bigl( (s1-A)^{-1} \Bigr).
\end{eqnarray}
Here, $x(t)\in \mathbb{R}^{n}$ is a vector, $A$ is a matrix, assume $t\ge 0$, and ${\cal L}^{-1}$
implies an inverse-Laplace transform.
Hence one may obtain an expression of exponential mapping from a Lie algebra to a Lie group
by solving a characteristic polynomial of $A$ ( a resolvent ),
and this method can be used ( at least formally ) for constructing a mass matrix in our GNG situation
discussed in the previous sections.
A method of spectral theory and also perturbation theory of resolvent $(s1-A)^{-1}$ 
of Tosio Kato can be utilized.
The stability/hyperbolicity in the sense of dynamical system also be converted into 
the Laplace transform ( the Laplace transform will give a classification of
stable, unstable and center manifolds in the linearized dynamics $\dot{x}=Ax$ ).
Thus, a homogeneous flow, ergodicity, and its number-theoretical implications also can be understood
via a Laplace transform.

\vspace{2mm}

Both the measure-theoretic ( Kolmogorov-Sinai, 1959 )
and topological entropies give the exponential rate of growth of $n$-orbits
( rate of growth of area/volume of the map $f:M\to M$ of a Riemannian manifold $M$ ).  
It is known fact that it gives a positive entropy if and only if $g_{\mathbb{R}}$ is partially hyperbolic.
In a case of homogeneous flow $(G/\Gamma,g_{\mathbb{R}})$, 
the entropy is given by the following Pesin-Ruelle-type formula:
\begin{eqnarray}
h(g_{\mathbb{R}}) \le \sum_{|\lambda_{i}|>1}m_{i}\log|\lambda_{i}|,
\end{eqnarray}
where $(\lambda_{1},\cdots,\lambda_{n})$ are the set of eigenvalues of ${\rm Ad}(g_{\mathbb{R}})$
( Lyapunov exponents ),
and $(m_{1},\cdots,m_{n})$ are their multiplicities.
Since our mass matrices take the form of this set of eigenvalues,

\vspace{2mm}

{\bf Theorem}:
{\it A mass matrix of our GNG theorem has a positive entropy.}

\vspace{2mm}

Moreover, since we consider a topological dynamics in our mass function,
and it has a positive topological entropy,
then it is Li-Yorke chaotic.
Recall our discussion on the Morse-Smale dynamics of the gradient flow given above.
Then we arrive at the following theorem.

\vspace{2mm}

{\bf Theorem}:
{\it In the Morse-Smale dynamics of gradient flow of} $V_{eff}$,
{\it the Lyapunov exponents are given by the Hessian of} $V_{eff}$,
{\it namely, the Lyapunov exponents and topological/measure-theoretical entropy are given by
mass matrix of NG bosons of our GNG theorem.}

\vspace{2mm}

A. Katok et al. examined a deformation ( family ) of Anosov flow $\{\phi\}$ and showes differentiability
of its topological entropy $h_{top}(\phi)=\lim_{T\to\infty}\frac{1}{T}\log P(T)$ 
( $P(T)$; the number of closed trajectories of prime periods ). 
( A. Katok, G. Knieper, M. Pollicott and H. Weiss,
{\it Differentiability of Entropy for Anosov and Geodesic Flows},
Bull. Am. Math. Soc. {\bf 22}, 285 (1990), 
A. Katok, G. Knieper and H. Weiss,
{\it Formulas for the Derivative and Critical Points of Topological Entropy for Anosov and Geodesic Flows},
Commun. Math. Phys. {\bf 138}, 19 (1991). )
For a geodesic flow of negatively curved manifold ( an example of Anosov flow ), 
they considered a set of Riemannian metrics and their perturbation/deformation. 
Since several formulas of topological entropy can be related with the Selberg trace formula
via a set of characters of a Lie group, we find the following conjecture/problem:

\vspace{2mm}

{\bf Conjecture/Problem}:
{\it Establish the following scheme in our GNG theorem},
\begin{eqnarray}
{\rm Selberg \, trace \, formula} \leftrightarrow {\rm dynamical \, zeta}
\leftrightarrow {\rm topological \, entropy} \leftrightarrow 
{\rm Ricci \, flow \, of \, a \, negatively \, curved \, manifold}.
\end{eqnarray}

\vspace{2mm}

Another characterization of the dynamics in our mass function is "distal."
The definition of distal is
\begin{eqnarray}
\inf_{n\in{\mathbb{Z}}}d(f^{n}x,f^{n}x') > 0, \quad
(x,x'\in X),
\end{eqnarray}
where $d$ implies a distance defined appropriately ( Banach, Riemann-Hilbert, so forth ).
This "distal" is completely realized in our $U(1)$ dynamics of dynamical mass.
While in a non-Abelian dynamics of our mass functions, 
we wish to consider not only a one-parameter subgroup,
but also several transformations/maps they may be noncommutative 
( then we will generalize theory of dynamical systems ), 
we should generalize the notion of distal for a characterization of a non-Abelian dynamical mass functions
( and also $V_{eff}$ ).
An ergodicity of a measure-theoretic dynamics corresponds to 
a topological transitivity of a topological dynamics.
While, an ergodic dynamics is measure distal if and only if it admits a Furstenberg tower
( R. J. Zimmer, 1976 ).
Since the $U(1)$ dynamics of mass function contains ergodic elements
( a rotation of unit circle $R:S^{1}\to S^{1}$, $R\cdot x = x+a$, $a\in{\mathbb{R}}^{1}$ 
is ergodic if $a$ is irrational ),
the theorem of Zimmer should be applied. 
Similarly, a translation over the group ${\mathbb{Z}}_{p}$ of a $p$-adic integer is ergodic
if the translation $\lambda$ cannot be divided by $p$. 
This theorem will be applied for our GNG theorem when we consider a $p$-adic ( Lie ) group.

\vspace{2mm}

The dynamical zeta function is a tool to give a global character of a map $f$ acts on a space $X$,
by periodic orbits as a set of local quantity, and has a deep relation
with both measure-theoretic and topological entropies.
The definition of Artin-Mazur zeta function for map $f$ is
\begin{eqnarray}
\zeta_{f}(z) &=& \exp\Bigg[ \sum^{\infty}_{n=1}\frac{z^{n}}{n}N_{f}(n) \Bigg].
\end{eqnarray}
Here, $f:M\to M$, $N_{f}(n)$ is the numer of points of period $n$, $f^{n}x=x$ ( $n\ge 1$ ).
The dynamical zeta of Ruelle is a slight modification/generalization of the Artin-Mazur zeta,
and the Weil zeta function considers number of periods a Frobenius map on a finite field.
For example, a period of $f$ acts on a mass function ${\cal M}$ is an invariant 
( and thus a dynamical zeta for $f$ is also an invariant ) 
under a Ricci flow in the vicinity of a critical point of $V_{eff}$.
From the definition of dynamical zeta, one may find a situation where 
one should consider a dynamical zeta of an abstract operator.
Also, one can consider an eigenvalue equation $H\phi=\epsilon\phi$ as a dynamics,
can introduce the notion of hyperbolicity.
In the Tomita-Takesaki theory, it is shown that
a one-parameter subgroup of an automorphism ( Tomita-Takesaki modular automorphism )
of $C^{*}$-algebra of statistical mechanics
corresponds to the Kubo-Martin-Schwinger condition.  
A theorem of Connes:
The Tomita-Takesaki automorphism belongs to the center of ${\rm Out}N={\rm Aut}N/{\rm Inn}N$.
An application of both the Connes theorem and dynamics theory to a gauge theory is interesting for our context.

\vspace{2mm}

In our context, we meet the case where a map $f$ itself is a manifold defined over $G$ or $G/\Gamma$,
thus it is meaningful to consider a family of the maps,
\begin{eqnarray}
\{ f_{1}, \cdots, f_{N}\} \in {\cal M}\times{\cal S},
\end{eqnarray}
where, ${\cal M}$ and ${\cal S}$ are a manifold and a parameter space, respectively.
Then, we handle a deformation of $f_{j}$.
For example, a homogeneous flow aquires a deformation.
Our interest is whether a measure-theoretic and a topological entropies, and also a zeta function, 
are invariant under a deformation of $f_{j}$.
Another, one can consider a dynamics of Kac-Moody group which may partially contain deformation effects 
of complex structures of the corresponding Lie group.
In this case, 
we have particular interest on a classification of 
types of dynamics a set of eigenvalues arised from a Kac-Moody group.

\vspace{2mm}

For a summary, we have our interest on the following scheme:
\begin{equation}
\begin{array}{ccc}
{\rm Pontrjagin, \, Tannaka \, dualities} & \to & {\rm homogeneous \, dynamics \, on} \, G/\Gamma  \\
{\rm spectral \, properties, \, unitary \, representations} & & \\
\downarrow & & \\
{\rm automorphic \, form \, on } \, G/\Gamma & &
\end{array} 
\end{equation}

\subsection{Toward a Generalization of the Generalized Nambu-Goldstone Theorem}

Here, we intend to generalize the definition of automorphic representation.
The purpose of the generalization is to give our GNG theorem from a very generic group-theoretical viewpoint.
As we have observed, the situations where mass matrices, determinants and effective potentials
belong to automorphic representations of topological/Lie groups have the key-roll in our GNG theorem. 
Let $G$ be a topological group ( for example, $\mathbb{R}$ ), 
and let $G^{\vee}$ be its dual group 
( for example, a torus $\mathbb{T}\simeq \mathbb{R}/\mathbb{Z}$ ).
Let us assume the situation that $G^{\vee}$ ( such as a character group of a locally compact Abelian group ) 
is isomorphic with $H\backslash G$, where $H$ ( for example, a lattice group $\Gamma$ ) is a subgroup of $G$. 
Then, one can consider groups defined over those groups,
\begin{eqnarray}
\mathfrak{G}(G), &\quad& \mathfrak{G}(H), \\
GL_{n}(G), &\quad& GL_{n}(H).
\end{eqnarray}
Under the setting, we consider the following quotients: 
\begin{eqnarray}
\mathfrak{G}(H)\backslash \mathfrak{G}(G), \qquad
GL_{n}(H)\backslash GL_{n}(G).
\end{eqnarray}
If there is a case in which a construction of those quotients is possible,
then a generalized automorhic representation becomes,
\begin{eqnarray}
F(\mathfrak{G}(H)\backslash \mathfrak{G}(G)), \qquad 
F(GL_{n}(H)\backslash GL_{n}(G)),
\end{eqnarray}
i.e., functions on the quotients.
Constructons of homogeneous fiber bundles also can be introduced,
\begin{eqnarray}
\mathfrak{G}(G)\times_{\mathfrak{G}(H)}\mathfrak{F}  \to \mathfrak{G}(H)\backslash \mathfrak{G}(G).
\end{eqnarray}
In general, a discrete subgroup $\Gamma$ of a separated topological group $G$ is closed,
and then the quotient topology of $G/\Gamma$ is separated. 
From the manner of the construction, 
$\mathfrak{G}(H)\backslash \mathfrak{G}(G)$ can become a moduli space if it is Hausdorff.
( If $\mathfrak{G}(G)$ is a topological group and if $\mathfrak{G}(H)$ is its closed subgroup,
then the quotient is Hausdorff. ) 
Under a specification of those groups $G$, $H$, ..., 
principal bundle, connection 1-forms and curvature 2-forms,
representations such as Chern-Weil theory or Borel-Weil theory, 
and deformations of them also be introduced.
By considering a tangent space over $F(\mathfrak{G}(H)\backslash \mathfrak{G}(G))$,
one may meet with an infinite-dimensional Lie groups.

\vspace{2mm}

Let $S$ be a scheme, and $G$ be a $g$-dimensional torus.
Let $X={\mathcal Hom}(G,\mathbb{G}_{m})\simeq\mathbb{Z}^{g}$ be an \'{e}tale sheaf,
a character group $G^{\vee}$.
In this case, $\mathfrak{G}(G^{\vee})\backslash\mathfrak{G}(S)$ gives an automorphic representation.
More generally, one can consider a method of category theory to enlarge possible applications
of our GNG theorem.
Let ${\cal C}\to{\cal C}^{\circ}$ be a dual category, 
and let us introduce the functor from the category ${\cal C}$ to a quotient category ${\cal C/C'}$.
Let us assume the following isomorphism 
\begin{eqnarray}
{\rm Ob}({\cal C}^{\circ}) &\simeq& {\rm Ob}({\cal C})/{\rm Ob}({\cal C}')
\end{eqnarray}
exists. 
Then by groups defined over those categories ( so-called group-objects ), 
we consider the quotient
\begin{eqnarray}
\mathfrak{G}({\cal G(C)})/\mathfrak{G}({\cal H(C')}), \qquad 
{\cal G(C)} \in {\rm Ob}({\cal C}), \quad 
{\cal H(C')} \in {\rm Ob}({\cal C}'). 
\end{eqnarray} 
If such a quotient exists and especially it becomes Hausdorff,
then it will be called as an automorphic representation over the categories.
Those consideration can be applied into topological/Lie/algebraic groups and group schemes.
$\mathfrak{G}({\cal G(C)})/\mathfrak{G}({\cal H(C')})$ can become a moduli of groups belong to 
certain categories, if it has a topological nature such as the Hausdorff, so forth.

\subsection{The Principle of Broken Symmetries}

It is clear from our discussion given above, 
the ordinary Nambu-Goldstone theorem is a very special case of the category of broken symmetries 
in physics and mathematics.
Here, {\it we must carefully distinguish} the essential differences between 
"symmetry breakdown" and "broken symmetry":
In the exact sense, very strictly, "symmetry breakdown" implies the ordinary NG theorem,
while the word "broken symmetry" simply means that 
there is a symmetry and it is broken in a ( physical ) system.
The ordinary discussions of "symmetry breakdowns" are 
included in the category of those of "broken symmetries",
while we have to know/find ( physically, mathematically ) a symmetry before 
we discuss it is whether broken or not.
One of important examples is supersymmetry:
Needless to say, one cannot consider a breakdown of SUSY if he/she does not know it.
For example, the differences of notions between 
"the SUSY is a broken symmetry", 
"it is under a spontaneous symmetry breakdown", and
"it is explicitly broken "
must be distinguished.

\vspace{2mm}

{\it If one finds/puts/sets/recognize a symmetry in physical/mathematical nature, 
then one can consider/examine it is broken or not, under several methods and steps of discussions,
both by physics and mathematics.
The ordinary Nambu-Goldstone theorem of spontaneous symmetry breaking 
belongs to the special case of the category of those considerations/methodologies.
We have established the mechanism of our generalized Nambu-Goldstone theorem 
as one of examples of those methodologies, by a slight modification of ordinary NG theorem.}

\vspace{2mm}

On the other hand, we have mentioned on topological nature of an NG space.
A physical quantity which should/can be regarded so as to defined over a continuous metric space
in low-energy limit can be given over a discrete topological space in high-energy region.
Moreover, an explicit symmetry breaking parameter can have a quantum fluctuation
in the corresponding underlying theory: 
An explicit symmetry breaking parameter can become a localization in space of a field.
Hence, several conditions of our GNG theorem are rather "relative",
and this implies that {\it an arithmetic variety, an algebraic variety and a complex manifold
arises "relatively" in the mechanism of our GNG theorem.}
The "relative" relations between those varieties/manifold 
are given over ring homomorphisms between each other,
while they have physical implications.

\vspace{2mm}

A more detailed classification of broken symmetry could be done, now in progress.

\subsection{Perspective}

In this subsection, we consider various extensions, applications and related problems of our GNG theorem.

\vspace{2mm}

Needless to say, in the Nambu-Goldstone theorem of spontaneous symmetry breakdown,
a set of degenerate vacua arises associated with a "transition" from
a Wigner phase to a Nambu-Goldstone phase. 
In this paper, 
we have considered phenomena of dynamical mass generation with parameters they
break global continuous symmetries explicitly.
Due to an explicit symmetry breaking parameter from the beginning of a theory/model,
the Nambu-Goldstone theorem is modified ( cannot be applied under a naive manner )
since the effective potential is not a class function
( for example, a characteristic function is a special example of class function ) 
of a Lie group action which corresponds to a Nambu-Goldstone mode,
and we have found various interesting physical/mathematical phenomena.
A phase degree of freedom of mass parameters in a dynamical model we have considered in this paper
cannot be absorbed by any field redefinition 
due to explicit symmetry breaking parameters of the models.
It has been shown that, such a phase is interpreted as a pseudo Nambu-Goldstone mode.
In the case of U(1) Lie group, 
the effective potential of the model aquires a periodic modulation.
From the gap equation $\frac{\partial V_{eff}}{\partial\theta}=0$ with the periodicity, 
we can consider an application of Galois theory to the effective potential,
and it is generalized into a coset $\Gamma\backslash G$ of Lie/algebraic group.

\vspace{2mm}

Because of the importance of the Nambu-Goldstone theorem in
condensed matter, nuclei, particles ( in fact, whole region of physics ),
even though the main motivation of this paper has been come from
neutrino mass and CP-violation ( breaking of time-reversal invariance ),
we believe the results are quite universal and general:
Our results will obtain various applications/insights/predictions in physical phenomena.

\vspace{2mm}

(1) A statistical physics is defined by a statistics of particles,
and it is described by a permutation group, and symmetric group, namely a finite group.
Then if a finite group of statistical physics is defined over a field $K$ of ${\rm char}K=0$,
then it is a representation of absolute Galois group.
Therefore, the fundamental problem of statistical physics may be understandable
by a theory of Lie+Galois groups.
This fact aquires a remarkable insight when one considers dimensions of spacetime:
In the case of two-dimensional system, any permutation of two particles is given by a rotation inside a plane.
A clockwise and counterclockwise rotations cannot be distinguished 
because there is no orientation in the two-dimensional world, as the discussion of anyon.
While, two permutations, rotations inside a surface embedded into 3D space, 
are distinguished by a guy lives in the three-dimensional Euclidean world. 
This implies that a permutation described by a Galois group ( for exmaple, given in a cyclotomic field ), 
aquires a new insight if the Galois action is embedded into a higher-dimensional system.
Thus, we have an interest on how to "generate" a statistics via our GNG mechanism.

\vspace{2mm}

(2) A Fourier analysis over a locally compact field $k$ gives a representation of "harmonic oscillators",
a solvable model,
and thus we can consider a ( second ) quantization or a Wigner-Weyl quantization by utilizing the model.
This implies that the set of matrix elements of a represenation $\rho: G\to GL_{n}(k)$
gives a basis set for constructing a quantum theory via the Peter-Weyl theorem,
also be considered in our GNG theorem and an effective potential as an automorphic representation.
Here, one should notice that $GL_{n}(k)$ contains various interesting objects such as 
a way to construct a homogeneous space or a Riemannian symmetric space,
certain classes of congruence subgroups, so forth.
An interesting question related with our GNG theorem is that how we can quantize 
( for example, by deformation quantization ) such an automorphic represenation obtained dynamically
by our GNG mechanism.

\vspace{2mm}

(3) Since a two-dimensional lattice group could have a morphism into a certain Galois group,
it is interesting for us to interpret a generic procedure of renormalization group
by a limit ( inverse limit ) of Galois group.

\vspace{2mm}

(4) If there is a non-Abelian symmetry between stable points on an NG manifold,
then it might be expressed by symmetry operations of the Rubik's cube.

\vspace{2mm}

(5) Methods of constructions of non-linear sigma models over several
topological spaces ( discrete, continuous, metric, ... ).
Our generalized Nambu-Goldstone theorem needs to find a relation between those models.
( While, the ordinary NG theorem argues that the theorem is inert with
differences of topology. 
Thus, we will consider both "universalities over" and "characteristics of" a topological space
for finding a suitable way to construct our GNG theorem. )
This problem has an essential relation with zeta/L-functions
and string theory.

\vspace{2mm}

(6) A role of finite/infinite-dimensional
( also, unitary/non-unitary, compact/non-compact ) representations of groups
in non-linear sigma models.
This problem should clarifies a relation between represenations and deformations
of them in a (un)harmonic mapping ( thus/then, (un)holomorphic, (non-)Kaehlerian,
namely a problem of variational calculus
( for example, several works of M. Kreck, S. Stolz, so on ) ),
and it also has a deep relation with string theory.
In supersymmetry
( both the SUSY of the textbook of "Wess and Bagger", 
and supermanifolds/supergeometry of random matrix theory ),
there are several examples of compact/non-compact types of Lie groups,
and a consideration on differences of compact/non-compact types of Lie groups
seems important for studying phenomenological applications of symmetry breakings, 
supergravity
( Ref: N. Seiberg, {\it Modifying the Sum over Topological Sectors
and Constraints on Supergravity}, arXiv:1005.0002 ),
and also on a relation with the Langlands reciprocity of number theory
( since the Langlands' theory is a conjecture of a correspondence between
infinite-dimensional irreducible representations of global fields 
and finite-dimensional representations of Galois groups obtained from them ).

\vspace{2mm}

(7) As we have discussed in previous subsections, 
it is well-known fact that the affine Lie algebra of conformal field theory
is a deformation of complex structure of a curve, i.e., a Riemann surface,
and a mathematically rigolous theory on conformal field theory is constructed
on (semi)stable curves and those moduli.
( Ref: A. Tsuchiya, K. Ueno and Y. Yamada,
{\it Conformal Field Theory on Universal Family of Stable Curves with Gauge Symmetries},
Adv. Stud. Pure Math. {\bf 19}, (1989). )
While, such a deformation of complex structure of Riemann surfaces can be 
described by a quasi-conformal mapping,
and the author wrote a paper on a relation between deformation quantization and 
quasi-conformal mapping of a complex structure of Riemann surfaces.
( Ref: T. Ohsaku, arXiv:math-ph/0610032. )

\vspace{2mm}

Let us consider possible applications/extensions/generalization of them.
For example, a mathematical generalization of deformation quantization can be set 
toward "a higher-dimensional case", where there is "a set of Planck constants",
namely we will have several deformation parameters,
\begin{eqnarray}
\hbar_{j}\in \mathbb{R}^{1}, \qquad j\in \mathbb{N}.
\end{eqnarray}
Under the setting, one can consider a simplectic manifold with a deformation quantization
introduced into the direction of "simplectic structure". 
Then, after a complexification of the variables $\hbar_{j}\in \mathbb{C}^{1}$, 
we obtain a deformation quantization theory of several complex variables:
This is a generalization of a power series of deformation quantization such that 
\begin{eqnarray}
\mathbb{C}[[\hbar]] \to \mathbb{C}[[\hbar_{1},\cdots,\hbar_{n}]].
\end{eqnarray}
We can employ an $\hbar_{j}$-adic completion with a discrete topology to them.
One can refer several theorems of Kiyoshi Oka to find a complex analytic space for several variables
$\hbar_{j}$ ( $j=1,\cdots,n$ ), and more generally one can utilize the method of sheaves.  
A fundamental and impressive introduction on algebraic aspects of deformation quantization
from those point of views,
see: M. Kashiwara and P. Schapira,
{\it Deformation Quantization Modules I,
Finiteness and Duality},
arXiv:0802.1245. 
We may consider that those viewpoint clearly shows or, implies, that a naive Moyal product definition 
of deformation quantization is a special case of the algebra of it: 
Namely, there is a deformation of Moyal-product itself 
( "Deformation of Mayal-Weyl Deformations" in different directions ).
See also, M. Kontsevich,
{\it Deformation Quantization of Algebraic Varieties},
arXiv:math/0106006.
In this paper, Kontsevich considers an affine Poisson variety,
and glueing them as the usual prescription of algebraic geometry,
and then he arrives at a projective space, and also "quantum coherent sheaves." 
C. Fronsdal and M. Kontsevich,
{\it Quantization on Curves},
arXiv:math-ph/0507021,
may also be useful to consider a case of (quasi-)conformal theory. 
In this work, they consider the case of algebraic variety with singularity.
From their work, 
we have recognized the fact that the deformation quantization of ( several, complex ) Planck constant(s)
should be contained in / overlapped with a Kac-Moody theory ( of several variables )
and the Kodaira-Spencer infinitesimal deformation theory.

\vspace{2mm}

As we have observed in the previous discussions of this paper,
a mass function ( or, a Higgs VEV, or an auxiliary fields ) is a Riemann surface
( and then a supersymmetric auxiliary field may be a super-Riemann surface ),
or one can say it defines a Riemann surface.
( Ref: Y. V. Fyodorov, Y. Wei and M. R. Zirnbauer, 
{\it Hyperbolic Hubbard-Stratonovich Transformation made Rigorous},
J. Math. Phys. {\bf 49}, 053507 (2008), arXiv:0801.4960 )
While we have discussed deformation theory 
( in the sense of both Kodaira-Spencer and Teichmueller-type ) 
of generic dynamical mass functions.
Thus, our theory gives an insight on relations between
deformation theory of complex structures,
Kac-Moody theory ( both algebras and groups,
Ref:
Victor Kac,
{\it Infinite-dimensional Lie Algebras},
Shrawan Kumar, 
{\it Kac-Moody Groups, their Flag Varieties and Representation Theory},
Birkhaeuser, Basel-Berlin-Boston, 2002 ),
conformal field theory 
( including superconformal algebras and so-called "conformal superalgebra" given by Victor Kac ),
and deformation quantization.
Toward a geometric understanding for those issues releted with Kac-Moody theory,
applications of methods of quiver varieties of Nakajima 
( quiver variety representation of the universal enveloping algebra in terms of Hall algebra ) 
might be important/helpful for us.
( Ref: 
H. Nakajima,
{\it Quiver Varieties and Finite Dimensional Representations of Quantum Affine Algebras},
arXiv:math/9912158,
Victor Ginzburg, {\it Lectures on Nakajima's Quiver Varieties},
arXiv:0905.0686. ) 
Super-Riemann surfaces and super-Teichm\"{u}ller theory are well known in literature. 
( Ref: L. Crane and J. M. Rabin,
{\it Super Riemann Surfaces: Uniformization and Teichm\"{u}ller Theory},
Commun. Math. Phys. {\bf 113}, 601 (1988). ) 
A very interesting problem of those relations for our context 
might be found in topological and number-theoretical natures of those theories
( for example, via theta functions and modular forms ).
Especially, an investigation on arithmeticity and modularity is important.
Theory of Kodaira-Spencer-type infinitesimal deformation of a supermanifold can be constructed in principle,
in general,
by introducing an analytic family $\rho: {\cal W}\to{\cal V}$ ( ${\cal W}$ and ${\cal V}$; supermanifolds )
with taking a principal super-bundle,
and it may contain a super-Kac-Moody ( BKM ) algbera.
For this purpose, some parts of the work of O. Goertsches,
{\it Riemannian Supergeometry}, Math. Z. {\bf 260}, 557 (2008)
may be useful, especially its algebraic discussions ( sheaves, so on ).
See also, 
M. V. Movshev and A. Schwarz,
{\it Supersymmetric Deformations of Maximally Supersymmetric Gauge  Theories. 1},
arXiv:0910.0620.
For a connection with the global analysis of our GNG theorem in a supersymmetric case,
the papers of Edward Witten,
{\it Supersymmetry and Morse Theory}, 
J. Diff. Geom. {\bf 17}, 661 (1982),
{\it Constraints on Supersymmetry Breaking},
Nucl. Phys. {\bf B202}, 253 (1982),
may be useful for us.

\vspace{2mm}

(8) In our periodic potential of the generalized Nambu-Goldstone theorem, 
by changing "angle" of our view,
the amplitude of modulation of potential changes, 
and an appropriate choice of the angle
gives us a flat potential, namely, the periodicity "disappears."
This phenomenon has a similarity with procedures of resolution of singularity
in algebraic geometry.
The stable subspace of local coordinate system of Lie group 
( the subspace given by a set of gap equations ) defines a singularity 
in the sense of algebraic geometry, 
and the "disappearance" of periodicity implies a qualitatve change 
of the singularity.
For example, a singular point $X$ $\to$ a singularity $Y$ defined by a curve, ..., like that.
It seems the case that there are changes of cardinalities between two sets $X$ and $Y$ of singularities, 
in general.  
To understand those behavior of singularity of effective potential
via theory of resolution of singularity $\varrho:\widetilde{X}\to X$ is an important issue for us.
Recently, H. Hironaka gived his study on theoy of resolution of singularity
over a finite characteristic field $K$, ${\rm char}K\ne 0$
( Ref: H. Hironaka,
{\it A Program for Resolution of Singularities, in all Characteristics $p>0$ and in all Dimensions},
Summer School on Resolution of Singularities, 12-30 June 2006, 
The Abdus Salam International Center for Theoretical Physics ), 
and his theory might have an important application in our GNG theory.

\vspace{2mm}

(9) In this paper, we have recognized that a sigma model can have
several topologies ( discrete, metric, continuous, I-adic, ... ) of target spaces 
( probably we can emphasize it as a "target space topology" ),
from the context of our generalized Nambu-Goldstone ( GNG ) theorem,
though we have not yet examined them systematically.
Arithmeticity and automorphic representation/forms are emphasized in the GNG theorem.

\vspace{2mm}

While, a lot of sigma models on a complex projective space $\mathbb{C}{\bf P}^{n}$ are studied,
by introducing the constraint of quantum fields.
It is intersting for us to generalize it to cases of Grassmannian varieties, ( generalized ) flag varieties,
Abelian and Jacobian varieties, Calabi-Yau manifold, and those of supersymmetric counterparts.
Since an Abelian variety or a Grasmannian variety ( affine and projective varieties also ) has 
those moduli expressed by cosets, the target spaces should be deformed continuously.
If a model of random matrix theory contains a quantity which is defined over 
an algebraic variety or a complex manifold,
those variety/manifold can be deformed under variations with respect to moduli,
and we obtain different theories continuously under the condition of the absence of obstruction:
Thus, we have arrived at the notion "cohomological supersymmetric sigma models and random matrix theory".
If a quantum correction to a target space exists in the nature of a theory/model,
it might give a notion on "quantum moduli". 
( Beside this issue, a Ricci flow of target space obtained from
a renormalization group prescription of a sigma model,
given firstly by Daniel Friedan, also provides a description of a kind of deformation
depends on energy scale. )
There are various works on sigma models with target-space supersymmetry,
from random matrix theory to AdS/CFT. 
( Ref: 
T. Creutzig, Thesis {\it Branes in Supergroups}, Dept Phys. 2009-Hamburg-Germany
arXiv:0908.1816,
C. Candu, V. Mitev, T. Quella, H. Saleur and V. Schomerus, 
{\it The Sigma Model on Complex Projective Superspaces}, arXiv:0908.0878, 
C. Candu, T. Creuzig, V. Mitev and V. Schomerus, 
{\it Cohomological Reduction of Sigma Models}, arXiv:1001.1344,
M. R. Zirnbauer, 
{\it Riemannian Symmetric Superspaces and Their Origin in Random-matrix Theory}, 
J. Math. Phys., vol.37, 4986 (1996), 
{\it Symmetry Classes}, arXiv:1001.0722. )

\vspace{2mm}

On the other hand, the Sato-Tate conjecture of number theory
( it has a deep relation with the Shimura-Taniyama-Weil, or Serre theorem/conjecture, 
or the Fermat's last theorem )
was proved in the last year.
( Ref:
M. Harris, N. Shepherd-Barron and R. Taylor, 
{\it A family of Calabi-Yau varieties and potential automorphy}, 
available from the homepage of Richard Taylor. )
In that work, Hodge theory and period mappings are employed on 
a family of Calabi-Yau varieties.
I think those mathematical structure should realize ( be contained ) in a certain class
of sigma models defined over complex ( algebraic, via the Serre-GAGA principle ) 
manifolds, especially Calabi-Yaus.
Moreover, an algebra of Calabi-Yau can be treated under algebraic/axiomatic manner
( Ref: V. Ginzburg, {\it Calabi-Yau Algebras}, arXiv:math/0612139,
a similar but incomplete attempt; T. Ohsaku, arXiv:math-ph/0606057 ),
and it can be applied into a target space construction of a sigma model.
( Note that the attempt of the paper of Victor Ginzburg itself is to give a deformation quantization
on a Calabi-Yau-type complex manifold. )
Beside this issue, a quantization of our generalized NG space can also be performed
by the theory of the paper of Victor Ginzburg 
since it is constructed generally, beyond a Calabi-Yau manifold.

\vspace{2mm}

Mirror symmetry 
( a notion coming from symmetries of Betti numbers, Hodge diamonds of complex manifolds )
is a duality between two $N=2$ superconformal theories,
and it is obtained by a quantization of a sigma model ( i.e., a conformal field theory ) 
with choosing a compact Calabi-Yau manifold as a target space
( Ref: A. Kapustin and D. Orlov,
{\it Lectures on Mirror Symmetry, Derived Categories, and D-Branes},
arXiv:math/0308173 ).
There are several papers in literature they study relations between mirror symmetry of
Calabi-Yau supermanifolds.
( Ref: M. Aganagic and C. Vafa, {\it Mirror Symmetry and Supermanifolds}, hep-th/0403192 )
Maxim Kontsevich shows the condition/criterion on a mirror pair
by words of Fukaya category, so on.
( Ref: M. Kontsevich, 
{\it Homological Algebra of Mirror Symmetry}, arXiv:alg-geom/9411018,
{\it Deformation Quantization of Poisson Manifolds},
Lett. Math. Phys. {\bf 66} 157 (2003),
P. Bressler and Y. Soibelman,
{\it Mirror Symmetry and Deformation Quantization},
arXiv:hep-th/0202128.
Lagrangian Floer Homology of Fukaya-Oh-Ohta-Ono also gives some insight on mirror symmetry. )
From our interest on relation between our GNG theorem and the proof of Sato-Tate conjecture mentione above,
K. Hori and C. Vafa,
{\it Mirror Symmetry}, hep-th/0002222, contains a lot of important discussions.

\vspace{2mm}

Recently, some papers on an emergence of Calabi-Yau manifold 
from a statistical-mechanical model has been published
( Ref: A. Okounkov, N. Reshetikhin and C. Vafa,
{\it Quamtum Calabi-Yau and Classical Crystals},
arXiv:hep-th/0309208.
H. Ooguri and M. Yamazaki, 
{\it Emergent Calabi-Yau Geometry},
arXiv:0902.3996 ),
and the mechanism of emergence is physically close to a random surface
and mathematically close to ramdom matrix theory, integrable systems and the Gromov-Witten theory.
( Ref: A. Okounkov, 
{\it The Uses of Random Partitions},
arXiv:math-ph/0309015, 
{\it Noncommutative Geometry of Random Surfaces},
arXiv:0907.2322,
A relation between random partions and the Seiberg-Witten theory; 
N. A. Nekrasov and A. Okounkov, 
{\it Seiberg-Witten Theory and Random Partitions},
arXiv:hep-th/0306238. )

\vspace{2mm}

Therefore, it is interesting for us to find a mechanism of ( dynamical ) emergence of Calabi-Yau supermanifold
( for example, a generalization of theory of Okounkov, with emphasis on randomness )
which has a relation with a mirror pair in superconformal theory.
In the theory of Okounkov-Reshetikhin-Vafa, a mathematical similarity 
between partition function and a determination equation of a complex Calabi-Yau manifold is utilized, 
while a mathematical similarity between a 3D quantum gravity partition function ( its modular invariance )
and a modular form
( which has a crutial importance in number theory )
is investigated in the paper of Maloney and Witten, 
{\it Quantum Gravity Partition Functions in Three Dimensions},
arXiv:0712.0155.
We wish to emphasize the fact that the modular invariance of partition function
implies that it is an automorphic representation.
S. Stolz and P. Teichner also study mathematical structures
with emphases on relations between so-called topological modular forms,
(co)bordisms, partition functions of string theory, conformal field theory
and supersymmetric field theory.  
( Ref: S. Stolz,
{\it Supersymmetric Euclidean Field Theories and Generalized Cohomology}, his lecture note. )
One of our interests on works of Kreck, Stolz and Teichner is to reveal number theoretical aspects
of them ( total convergence of their works toward number theoretical frameworks ),
and realize those mathemaical structure of theory via our GNG mechanism.
How to find the Monster and Moonshine in the works of Stolz et al. also a very interesting problem for us.

\vspace{2mm}

It is well-known fact that statistics of spaces of zero points of the Riemann zeta 
belongs to the Gaussian unitary ensemble ( GUE ) of random matrix theory,
and thus it is interesting for us to find zeta/L-functions inside various theories/models mentioned here,
by utilizing similarities and relations between them.
This issue will naturally relate/result with the question ( ultimalte understanding of ) 
why the distribution of Riemann zeta arises in GUE of random matrix theory.
Moreover, a dynamical generation of a zeta function ( via our GNG theorem ) might give us
a controlled interpretation on various zeta functions.

\vspace{2mm}

In this paper, we have discussed on Hodge structures in our GNG theorem.
Hodge/deformation theoretical aspects of mirror symmetry also has been studied.
( Ref: Y. Soibelman, 
{\it Quamtum Tori, Mirror Symmetry and Deformation Theory},
arXiv:math/0011162, 
L. Katzarkov, M. Kontsevich and T. Pantev, 
{\it Hodge Theoretic Aspects of Mirror Symmetry},
arXiv:0806.0107,
Under the context to understand an aspect of Hodge theory in quantum field theory,
an application of the results of Kontsevich-Soibelman, 
{\it Cohomological Hall Algebra, Exponential Hodge Structure and Motivic Donaldson-Thomas Invariants},
arXiv:1006.2706, 
may give us several insights into our GNG theorem, and also large-N, sigma models, mirror symmetry, so forth.
This paper of Kontsevich-Soibelman includes a discussion on Chern-Simons theory
( thus, it can be utilized to analyze a topological field theory, 
"possibly" topological insulators, so forth,
and thus the work may give us a way toward "Hodge theory and moduli theory of topological insulators",
namely the question "how to deform continuously a topological insulator", ... ),
while their work on rapid decay cohomology might be generalized into more general functional forms,
or, for a Feynman functional of situations of supersymmetric random matrix theory.

\vspace{2mm}

Several papers given in literature show that
Mirror symmetry has a certain importance in geometric Langlands dual(s)/correspondence(s) ( conjecture ),
a number-theoretical issue.
( Ref: 
P. B. Gothen,
{\it The Topology of Higgs Bundle Moduli Spaces},
Thesis, The Mathematical Institute,
University of Warwick, Coventry, ( Aug. 1995 ), 
available by internet.
T. Hausel and M. Thaddeus, 
{\it Mirror Symmetry, Langlands Duality, and the Hitchin System}, 
arXiv:math/0205236; 
E. Frenkel, 
{\it Lectures on the Langlands Program and Conformal Field Theory}, 
arXiv:hep-th/0512172;
A. Kapustin,
{\it A Note on Quantum Geometric Langlands Duality, Gauge Theory,
and Quantization of the Moduli Space of Flat Connections},
arXiv:0811.3264, 
K. C. Schlesinger, 
{\it A Physics Perspective on Geometric Langlands Duality}, 
arXiv:0911.4586;
M. A. A. de Cataldo, T. Hausel and L. Migliorini, 
{\it Topology of Hitchin Systems and Hodge Theory of Character Varieties},
arXiv:1004.1420,
J. Teschner, 
{\it Quantization of the Hitchin Moduli Spaces, Liouville Theory, 
and the Geometric Langlands Correspondence. I},
arXiv:1005.2846. )
For example, it seems the case that only the case Langlands correspondence 
of $SL(2,{\bf Z})$ congluence subgroup is considered in physicists:
How about $Sp(N,{\bf Z})$ case or Siegel upper half space ( several $S$-dualities )?

\vspace{2mm}

It seems the case that the sine-Gordon type Lagrangian derived from our GNG theorem
given in the previous section
has a similarity with so-called Liouville theory, or "tachyon condensation", of string theory. 
Thus, the sine-Gordon-type effective sigma model may have its mirror pair via a duality,
possibly the Fateev-Zamolodchikov-Zamolodchikov duality ( see the paper of J. Teschner given above ).
Moreover, our GNG theorem gives a "quartet" sine-Gordon model in the case of $SU(2)$:
In this case, the total configuration space/manifold of the model contains four "Liouville sections"
defined by a set of infinite number of Liouville theories.
Then we may have our interest on to know how a mathematically characteristic aspect 
in the sine-Gordon model as an integrable system,
such as the infinite-dimensional Grassmannian manifold/variety 
of the Korteweg-de Vries equation, will be introduced/contained in our GNG theorem. 
( Ref: for example, G. Niccoli and J. Teschner,
{\it The Sine-Gordon Model Revisited. 1}, arXiv:0910.3173. )
It is also another interesting problem for us to consider how the differntial Galois theory
should apply to those integrable models and ultimately ( namely, implicitly/explicitly ) to our GNG theorem.
( Ref: M. van der Put and M. F. Singer,
{\it Differential Galois Theory}, available from internet. )
In the works mentioned above, a Higgs bundle ( a bundle of Higgs field defined over a curve ) 
of so-called Hitchin system is considered,
several Hodge structures and moduli spaces 
( locally free sheaves on curves, mostly (semi)stable cases in the sense of Mumford )
have been studied,
and there is a mathematical similality with our GNG theorem.

\vspace{2mm}

In our GNG theorem, Higgs fields given in terms of fermion composites 
of bilinear-type will be defined as follows:
\begin{eqnarray}
\langle \psi\bar{\psi} \rangle &=& \Phi^{S} + \Phi^{V}_{\mu}\gamma^{\mu} 
+ \Phi^{T}_{\mu\nu}\sigma^{\mu\nu} + \Phi^{A}_{\mu}\gamma^{\mu}\gamma_{5} + \Phi^{P}(i\gamma_{5}), \\
\langle \psi\psi^{T} \rangle &=& \Bigl( \Delta^{S} + \Delta^{V}_{\mu}\gamma^{\mu} 
+ \Delta^{T}_{\mu\nu}\sigma^{\mu\nu} + \Delta^{A}_{\mu}\gamma^{\mu}\gamma_{5} + \Delta^{P}(i\gamma_{5})
\Bigr)\gamma_{5}C. 
\end{eqnarray}
( Ref: for example, T. Ohsaku, 
{\it BCS and Generalized BCS Superconductivity in Relativistic Quantum Field Theory I and II},
Phys. Rev. {\bf B65}, 024512 (2002), {\bf B66}, 054518 (2002), 
{\it Relativistic Model of Two-band Superconductivity in (2+1)-Dimensions},
Int. J. Mod. Phys. {\bf B18}, 1771 (2004),
{\it Dynamical Chiral Symmetry Breaking and Superconductivity in the Supersymmetric Nambu$-$Jona-Lasinio Model at finite Temperature and Density},
Phys. Lett. {\bf B634}, 285-294 (2006),
T. Ohsaku,
{\it Dynamical Chiral Symmetry Breaking, Color Superconductivity, and Bose-Einstein Condensation in an $SU(N_{c})\times U(N_{f})_{L}\times U(N_{f})_{R}$-invariant Supersymmetric Nambu$-$Jona-Lasinio Model at finite Temperature and Density},
Nucl. Phys. {\bf B803}, 299-322 (2008).
 )
Here, $C$ is a charge conjugation matrix.
Those Higgs fields can be constructed by taking the Nambu-Gor'kov notation
$\Psi\equiv (\psi,C\bar{\psi}^{T})^{T}$, and the notation is often taken in 
supersymmetric approach of random matrix theory.
Gauge or flavor symmetries also be introduced into those composite Higgs fields,
with a restriction to satisfy the Pauli principle:
In that case, the composite Higgs fields take their values on the corresponding Lie (super)algebras, 
with adjoint actions of Lie (super)groups.
Note that all of components of the expansions of Higgs fields can be given by sections of sheaves.
Then, we obtain principal bundle ( or associate bundle ) 
by taking a product of a base space and these Higgs fields, in cases of trivial bundle,
and it can be generalized.
Thus, our Higgs fields contains the definition of Hitchin Higgs bundle as a special case,
namely, by taking the base space $X$ as a curve, and it is defined as a ( holomorphic ) section
( 0th-order cohomology such like $H^{0}(X,\Theta_{X})$, 
more precise definition, see the literature mentioned above ).
More precisely, we prepare three components for the total Higgs field, 
\begin{eqnarray}
{\cal M} &=& |M^{(0)}| + M_{dyn} + M^{\dagger}_{dyn}  
\end{eqnarray}
where, $|M^{(0)}|$, $M_{dyn}$ and $M^{\dagger}_{dyn}$ are real, holomorphic and anti-holomorphic parts,
respectively. 
( For a case of supersymmetric model, see:
T. Ohsaku, 
{\it Generalized Seesaw Mechanism of Neutrino and Bose-Einstein Condensation
in the Modified O'Raifeartaigh Model}, 
arXiv:0811.0617. )
A topological and number-theoretical character of the GNG theorem arise from 
an inconsistency inside the total Higgs ${\cal M}$.
Our Higgs fields have Lorentz indices and they have transformation laws 
under charge conjugation, time-reversal and parity. 
Hence, those transformation laws can be implemented into the ( a generalized ) Hitchin Higgs bundle
( for an algebraic construction, 
we need parity-even/odd locally free ( or, invertible ) sheaves, so forth ).
Therefore, it is important for us to consider/examine how the Hitchin Higgs bundle theory connects 
with usual fermion pairing models ( how they overlap, or (dis)connected, with each other ),
and our GNG theorem.
Furthermore, "parity-odd/even Langlands", "charge-conjugation Langlands",
and "time-reversal Langlands" might be constructed for the {\it geometric} Langlands theory.
( How about a possible connection between those "Langlands" and S/T-duality?
It might be possible to study a theory/model which has a "parity-odd/even",
or "time-reversal broken" coupling constant with such a special "symmetry." )
From the method of construction of Hitchin Higgs bundle, it can be applied to a case of
homogeneous space, and a Higgs bundle for $SL(2,{\bf R})/SO(2)={\bf H}$ ( the upper half space )
can be obtained. 
Therefore, for example, the theory of Hitchin can be used into a method of auxiliary fields
in supersymmetric random matrix approach ( Fyodorov-Wei-Zirnbauer, Spencer-Zirnbauer )
or a SUSY sigma model arised from a symmetry breakdown,
and probably we will arrive at the notion "supersymmetric Hitchin Higgs bundle and its moduli space".
Furthermore, the Hitchin Higgs bundle can also be applied into 
theory of "Abelian and Non-Abelian bosonizations in (1+1)-dimensions", 
namely in fermion systems and spin systems,
and then they might aquire a geometric Langlands.
( Ref: too many literatures. For example,  
E. Witten, {\it Non-Abelian Bosonization in Two Dimensions}, 
Commun. Math. Phys. {\bf 92}, 455 (1984). )
From our context, 
an interesting issue is to find the mathematical structure in a condensed matter situation
( such as a ( disordered ) quantum wire, so forth ),
examined by using a bosonization prescription.
Since a Hitchin Higgs bundle can be defined in a bosonic field defined over a curve
with a (semi)stable condition, for example, we could find the mathematical structure
inside the MSSM 
( namely, "Minimal Supersymmetric Standard Model Higgs bundle and a geometric Langlands" ).  
Our GNG theorem can derive those mathematical/physical contents under a systematic manner.

\vspace{2mm}

Therefore, we have an interest on to find a realization and a role of a number theoretical conjecture
in a sigma model, Calabi-Yau supermanifolds, so forth, from the context of our GNG theorem.
Then, we hope we can arrive at a point where number theory, 
mirror symmetry, supersymmetry/supermanifolds, random matrix theory,
AdS/CFT and our GNG theorem meet with each other.
For another different viewpoint for us, note the fact that 
the space $AdS_{5}$ is defined by a G-orbit decomposition, a coset $GL(6,{\bf R})/O(4,2)$, 
and thus it is also an interesting problem for us to make a theory which spontaneously emerges the coset,
like the case of Calabi-Yaus, topological insulators, or supermanifolds of random matrices.
( The paper of Fyodorov-Wei-Zirnbauer mentioned above 
considers an $O(p,q)$-invariant Hubbard-Stratonovich transformation,
and this method itself might be extended into $GL(p+q,{\bf R})/O(p,q)$.
The bosonic part of supersymmetric auxiliary field of paper of Disertori-Spencer-Zirnbauer mentioned above
is defined over ${\bf H}=SL(2,{\bf R})/SO(2)$. 
Note that finite-dimensional irreducible representions of  
$SU(1,1)$, $SU(2)$, $SL(2,{\bf R})$ and $SL(2,{\bf C})$ 
are "equivalent" in the sense of representation theory.
Thus, a congruence subgroup $SL(2,{\bf Z})$, which has a special role in number theory, 
is contained in the model of Disertori et al. 
Those consideration can be used in a SUSY sigma model, 
and thus it relates with our GNG theorem in a SUSY case. )
In a possible realization of mirror symmetry in our GNG theorem,
a theory under the GNG situation should contain mirror doublet,
or, we should intend to construct a model which spontaneously/dynamically generates a mirror doublet.

\vspace{2mm}

For a possible context in condensed matter physics, 
so-called ${\bf Z}/2{\bf Z}$ topological insulator is interesting for us.
In this physics, there are strong and weak topological insulators,
and the Anderson localization of an electron in a disorder/random potential
can take place in the latter case, as a localization-delocalization transition.
( Ref; An example of supersymmetric sigma model description on 
a localization/delocalization transition:
M. Disertori, T. Spencer and M. R. Zirnbauer,
{\it Quasi-diffusion in a 3D Supersymmetric Hyperbolic Sigma Model},
arXiv:0901.1652. )
It is well-known fact that a topological term in a non-linear sigma model,
which chooses a sector of a specific winding, 
causes a non-pertubative effect ( studied by Haldane, Pruisken, Fendley, Furusaki et al., ... ).
Furusaki et al. showed that a sigma model with ${\bf Z}/2{\bf Z}$ topological term 
gives a transition from a trivial to a non-trivial sectors of the system.
Those topological insulators are believed to be realized in a graphene with a random potantial,
and the random potential is regarded as ripples of the 2D system
( which is also realized in a Wigner crystal ).
While, recently, T. Hartman, W. Song and A. Strominger
examined a dynamical fermion model ( which has a kind of Fermi sea ) near 
a Kerr blackhole, and they derived an effective theory from it:
It is given as ripples in the Fermi surface.
( Ref: T. Hartman, W. Song and A. Strominger,
{\it The Kerr-Fermi Sea}, arXiv:0912.4265. )
Thus, it might be the case that the effective theory of the system of blackhole 
is given as a non-linear sigma model of a random potential. 
It is interesting for us to interpret those works and their relations by
the supersymmetric random matrix approach or a SUSY sigma model,
and interprete by the (G)NG theorem. 
In that approach, considerations/classifications 
on symmetries of randomness and model Lagrangians will be important.

\vspace{2mm}

A possible relation/application of 
the theory of C. T. McMullen on "Braid Groups and Hodge Theory"
( published at 5 May, 2009 ) to those sigma models and generalized NG theorem
also an interesting subject.
The Hodge theory derives a deformation of complex structure and period mapping, 
while braid groups are found in several classes of solvable models.
An appliation of p-adic Hodge theory of Faltings 
( might be relate with p-adic string/sigma-model ) 
and techniques of \'{e}tale cohomology also an intersting issue from number theory, 
complex ( analytic ) manifolds and symmetry breaking.
An application of \'{e}tale cohomology, and also Galois cohomology = class field theory, 
to a target space might derive various interesting mathematical structures, including number theory,
in sigma models.

\vspace{2mm}

Final comment/opinion of myself on (9) is in order:
In this paper, 
we have employed several Lie-group/algebra techniques to reveal the mathematical structure
of our generalized Nambu-Goldstone theorem, especially a complexification 
( "analytic" continuation ) of Lie group.
This is a trick, and then our attempt goes beyond the "Grenze" between physical and unphysical regions
( "ein grosses Aufheben", a notion/Idee of G. W. F. Hegel ),
though the prescription gives us quite rich results, 
and we still have not yet summarized all of them in here.
Therefore, we wish to examine how our results relate ( has relations, or no relation ) with other theories, 
for example, the recent work of Edward Witten
given in {\it Analytic Continuation in Chern-Simons Theory}, arXiv:1001.2933.
Moreover, the supersymmetric approach to random matrices
( A typical example: 
M. R. Zirnbauer,
{\it The Supersymmetry Method of Random Matrix Theory},
arXiv:math-ph/0404057,
and "beyond",
P. Littelmann, H.-J. Sommers and M. R. Zirnbauer,
{\it Superbosonization of Invariant Random Matrix Ensembles}, arXiv:0707.2929 )
can also be interpreted as an "analytic" continuation from
real spaces to supermanifolds.
What we should know ( my ultimate purpose ) 
is a common ( and "essential", in various senses ) feature of 
those theoretical/mathematical frameworks...

\vspace{2mm}

(10) Recently, Nikita A. Nekrasov and Samson L. Shatashvili
extenstively have studied on to prove/show the fact that 
there is a bijective correspondence ( map )
between supersymmetric vacua of 2D ${\cal N}=4$ SUSY gauge theory with a massive matter field
softly broken into ${\cal N}=2$ and eigenstates of integrable spin chains:
In this attractive picture, a bijection between the superpotential of the gauge theory
and so-called Yang-Yang function of the spin system has been established.
( Ref: 
N. A. Nekrasov and S. L. Shatashvili,
{\it Supersymmetric Vacua and Bethe Ansatz},
arXiv:0901.4744,
{\it Quantum Integrability and Supersymmetric Vacua},
arXiv:0901.4748,
{\it Quantization of Integrable Systems and Four Dimensional Gauge Theories},
arXiv:0908.4052,
N. Nekrasov and E. Witten,
{\it The Omega Deformation, Branes, Integrability, and Liouville Theory},
arXiv:1002.0888. )
For example, it is interesting to find isomorphisms/automorphisms in the bijection,
since it may give us a way toward more general ( generalized ) theory 
for the Nekrasov-Shatashvili theorem.
How about Hecke algebra and braid groups, and how to generalize it to two- and three-dimensional spin systems? 
Especially our GNG theorem has a similarity with the treatment of 
mass matrices of matter of the SUSY gauge theory of Nekrasov and Shatashvili.
It is interesting for us to know how we can map a "super" spin chain to such a ( SUSY ) gauge model,
if possible.

\vspace{2mm}

(11) If we consider our GNG theorem or a sigma model 
over a solvmanifold ( or a Heisenberg manifold ) as a target
( and solvable Lie group, it is intersting also from the construction
of Borel subgroup ( Borel-Weil theory of representation ) or Heisenberg group ), 
what will be found?
( Ref: L. Auslander, 
{\it An Exposition of the Structure of Solvmanifolds.
Part 1: Algebraic Theory},
Bul. Am. Math. Soc., {\bf 79}, 227 (1973),
H. Tilgner, 
{\it A Class of Solvable Lie Groups and Their Relation to the Canonical Formalism},
Ann. Inst. Henri Poincar\'{e}, {\bf 13}, 103 (1970),
S. Semmes,
{\it An Introduction to Heisenberg Groups in Analysis and Geometry},
Notice of the AMS, vol. 50, 640 (2003). )
The Klein bottle is an example of a compact solvmanifold.
This problem is interesting from algebraic/number-theoretical/arithmetic point of view in our GNG theorem.
For example, can we construct a solvable Chern-Simmons theory?
( Ref: 
S. Bloch and H. Esnault,
{\it Algebraic Chern-Simons Theory}, 
arXiv:alg-geom/9602002. )
Those attempts will relate with a unification/embedding of topological-field-theorerical
perspective into our GNG theorem,
and an extension of theory of sigma models.

\vspace{2mm}

(12) Throughout the previous sections, and in several plans/questions given above, 
we consider mainly regular cases of cosets $G/H$ or $\Gamma\backslash G$.
In a regular case ( no singular point, no fixed point in $\Gamma\backslash G$ ), 
conformal field theory,
deformation theory of complex structures and deformation quantization works well.
On the contrary, a moduli theory based on Kodaira-Spencer-type infinitesimal deformation 
which assumes an analytic family $\pi: M\to S$ ( $S$; a parameter space ), 
is usually applied to a compact K\"{a}hlerian manifold,
and thus it can not be applied "naively" into a case of singular variety
( of course, a generalization of Kodaira-Spencer theory was considered in the paper of
Tsuchiya-Ueno-Yamada mentined above ). 
However, a deformation quantization defined over a singular variety exists in literature
( the work of Fronsdal and Kontsevich ).
Thus, we conclude their relations are not simple inclusion/generalization, 
but they may have overlaps in "the space of theories."
Therefore, we should know how they relate with each other
to understand physical methods such as conformal field theory, 
and also theory of resolution of singularity and deformation quantization, 
and ultimately our GNG theorem more deeply.
Especially a cohomological aspect is important:
In algebraic geometry, cohomology ( under Zariski topology ) 
of affine and projective schemes become very simple by vanishing theorems ( see the book of Hartshorne ). 
On the other hand, conformal field theory can be defined over a curve with a singularity
( ordinary double point ).
It is very interesting for us to investigate a relation between theory 
of deformation quantization on sigular curves of Fronsdal-Kontsevich, 
conformal field theory, and quantization of D-branes, so forth.
To go beyond the vanishing theorems under the Zariski topology of algebraic varieties mentioned above,
especially to introduce notions of universal coverings or homotopic loops,
we need to employ \'{e}tale cohomology of Grothendieck, to introduce techniques of algebraic topology. 
( The purpose of Grothendieck is to give a ( \'{e}tale ) fundamental group to an algebraic scheme.
In our context of GNG theorem defined over a field, 
a fundamental group has a crutial importance since it corresponds to 
( or, a generalization of ) the Galois group. 
Note that the theory of \'{e}tale cohomology is constructed on a new topology, 
Grothendieck topology ( or, {\it site} ). )
Namely, we can schematically write as follows:
\begin{eqnarray}
{\rm Zariski\,\, cohomology} \leftrightarrow 
{\rm etale\,\, cohomology} \leftrightarrow
{\rm Galois\,\, cohomology}.
\end{eqnarray}
Furthermore, if we construct a sigma model via a cohomological treatment of target space,
then we will meet with Zariski cohomological, \'{e}tale cohomological and Galois cohomological 
sigma models.

\vspace{2mm}

In our context of GNG theorem, a topological consideration is not enough to understand 
a GNG space since it considers a global aspect of a topological space in general.
While a manifold given from the GNG theorem has its characteristic aspect in a local quantity,
given at a point of a manifolds ( for example, a critical point ),
though the Morse theory states that a topological nature of a base space is summarized
into a critical point of a function defined over the base.
An \'{e}tale morphism defined by a morphism between tangent spaces of two varieties,
and thus it is locally defined.
( A tangent cone is used for a singular point of a variety. )
Therefore, a question how to apply method of \'{e}tale cohomology to a local analysis of a GNG space
is important for us. 
One of a possible way to analyze the GNG space is to decompose it in terms of level sets,
given by effective potential.
Then, for example in the case of $U(1)$, the GNG space is a set of points closed and open curves.
A point or a curve can be analyzed by a usual manner of Riemann-Roch, cohomology, so forth.
Thus, a GNG space consists with those elements.
Another possible way to analyze the GNG space is to investigate a ( algebraic ) generalization of
Morse theory, especially around a critical point.

\vspace{2mm}

(13) Connes-Kreimer Hopf Algebra of Quantum Field Theory
and the Generalized Nambu-Goldstone Theorem.

\vspace{2mm}

The fundamental framework of quantum field theory
consists with field quantization by path integration,
perturbative expansion, regularization and renormalization, 
and renormalization group prescription.
In the works of A. Connes and D. Kreimer
( Ref:
A. Connes and D. Kreimer,
{\it Renormalization in Quantum Field Theory and the Riemann-Hilbert Problem I:
The Hopf Algebra Structure of Graphs and the Main Theorem}
Commun. Math. Phys. {\bf 210}, 249 (2000),
{\it Renormalization in Quantum Field Theory and the Riemann-Hilbert Problem II:
The $\beta$-Function, Diffeomorphisms and the Renormalization Group},
Commun. Math. Phys. {\bf 216}, 215 (2001) ), 
it was shown that a perturbative expansion of series of Feynman diagrams and
a procedure of renormalization/renormalization-group 
can be understood in terms of Hopf algebra and a Birkoff decomposition.  
Works of Connes-Kreimer, Marcolli
( Ref: 
A. Connes and M. Marcolli,
{\it From Physics to Number Theory via Noncommutivative Geometry. I and II},
arXiv:math/0404128, hep-th/0411114 ), 
seem that they mainly handle/analyze "renormalizable models", 
while we have our interest on the relation between "renormalizable" and "non-renormalizable" models,
especialy a relation between gauge theories and four-fermi-like models.
This may be achieved, because a non-renormalizable four-fermion model 
can be regarded as a specific limit of ( one-boson exchange-type Yukawa model ) a renormalizable model,
and can examine a non-renormalizable theory from corresponding renormalizable one.

\vspace{2mm}

Needless to say, the framework of Schwinger-Dyson theory of a gauge model 
is very important for studying our generalized Nambu-Goldstone theorem, as shown in No.28.
Thus, we speculate that the Connes-Kreimer theory can provide us a method 
for generalization/extension of the Nambu-Goldstone theorem, 
more deep understanding on spontaneous/dynamical symmetry breaking,
or, more generally, phase transitions and critical phenomena.
In fact, Dirk Kreimer also mentioned on possible applications of
his theory to understand critical phenomena and universality classes.
( D. Kreimer, "Dyson-Schwinger Equations: From Hopf Algebras to Number Theory",
arXiv:hep-th/0609004. )

\vspace{2mm}

In a NL$\sigma$M, which describes collective modes associated with a phase transition, 
an effect of renormalization will modify the metric of target space, 
and it can be described as a Ricci flow generated by a Wilsonian renormalization group equation
( an attempt of Connes-Kreimer-type Hopf albegra analysis in Wilsonian RG,
see T. Krajewski and P. Martinetti, arXiv:0806.4309 ),
especially in (1+1)-dimensional spacetime.
There is an equivalence between bosons and fermions in (1+1)-dimension. 
(1+1)-Gross-Neveu, a four-fermion model which shows the asymptotic freedom, 
an exactly solvable model by several methods ( for example, Bethe ansatz ),
and an important model for considering various phenomena in condensed matter physics,
will provide us an interesting subject:
A toy model for examining a relation between BPHZ, phase-space expansion,
critical exponents, Bethe ansatz method, and the Connes-Kreimer theory.
It is interesting for us to understand a symmetry and its breakdown in those theories.

\vspace{2mm}

A recent paper of S. Bloch and D. Kreimer studies on mixed Hodge structure 
( Deligne, used in family of a complex manifold, number theoretical aspects of a period domain )
in Feynman path integrals.
( Ref: S. Bloch and D. Kreimer,
{\it Mixed Hodge Structures and Renormalization in Physics},
arXiv:0804.4399,
{\it Feynman Amplitudes and Landau Singularities for 1-Loop Graphs},
arXiv:1007.0338. )
It is also interesting for our GNG theorem that how the works of Bloch-Kreimer, Kontsevich-Soibelman
relate with each other, and what kind of insight their results give us.

\vspace{2mm}

(14) A Neural Network Model and Hopf Algebra:

\vspace{2mm}

It is a famous fact that the structure of neuron network of a brain has a self similarity, 
namely a kind of fractal structure.
In fact, some neural network models are constructed by 
a graph theoretical technique combined with numerical recipes.
Thus, we propose a scientific ( not a psychological ) investigation on brain, 
its mechanism of memory storages and destructions of memories, 
disorders, and a possibly schizophrenic situation in a brain, by a Hopf algebra.
For this purpose, probably we need a huge amount of computational resources
( we need more powerful resources than a lattice field simulation,
i.e., a brain is more complicated than QCD ),
and our numerical simulation of brain by generating a Hopf algebra
( similar to a complex dynamics of fractals, a sequence of mappings )
becomes similar with those of gravity.
Moreover, this attempt on analysis of brain "dynamics" 
by Hopf algebra might give us an insite into a dynamical model 
( i.e., Lagrangian/Hamiltonian for a brain ). 
This is an "axiomatic" or "constructive" method for a field theory of brain, 
beyond a neural-network simulation.
It might be a possible quantization of brain by introducing quantum algebra.
Moreover, as mentioned above, a structure of neurons itself is fractal,
while renormalization-group viewpoint can be understood ( or, proved ) 
as showing a fractal structure from a prescription of a zoom-in/out ( Suzuki Masuo ), 
thus I speculate that we can comnsider a motivic Galois theory ( and thus, number theory ) 
for understanding a neuron system ( graph/geometrical ) of brain.

\vspace{2mm}

Comment: The author wrote a paper on quantum brain dynamics ( QBD ) with an emphasis on symmetry breaking.
In that paper, it was shown that the QEB Hamiltonian can be cast into a Heisenberg spin system.
On the other hand, as we have mentioned above, 
Nekrasov and Shatashvili proved that several spin chains ( $XXX$, $XXZ$, $XYZ$ )
have bijective correspondences with ${\cal N}=2$ SUSY gauge theories.
Thus, we have found the following correspondence, 
\begin{eqnarray}
{\rm SUSY\,\, Gauge\,\, Theories} \leftrightarrow 
{\rm XXX,\,\, XXZ, \,\, XYZ} \leftrightarrow 
{\rm Quantum \,\, Brain \,\, Dynamics}. 
\end{eqnarray}
It is interesting for us to combine the model of QBD with the Hopf algebra approach.
Furthermore, a supersymmetric version of QBD ( namely, SQBD ) can also be considered.

\vspace{2mm}

(15) On quark confinement and dynamical mechanism of CKM/PMNS matrices of flavor mixings, {\it Again}.

\vspace{2mm}

In our GNG mechnism, we obtain a potential of phase degree of freedom, 
and the phase becomes a complex field after its complexification.
Thus, if we consider the mechanism in an $SU(N)$-QCD, 
we obtain the theory of massive $SU(N)$-QCD coupled with a complex boson field of a Higgs-type potential.
Futhermore, if we introduce a trick to make the complex boson field coupled with a gauge field,
then we obtain another $SU(N')$ QCD with a Ginzburg-Landau-type model.
In our speculation, the Higgs VEV might vanish under a naive treatment of these models
similar to our GNG mechanism on the origin of CKM/PMNS matrices.
Though if the Higgs-type bosonic field aquires a VEV, it is very interesting:
We have an interest on to apply this step-wise mechanims into quark confinement.
Our arguement is that the quark confinement and CKM/PMNS matrices have essentially the same dynamical origin.
My plan given in here is not so strange:
For example, the Polyakov model for confinement in (2+1)-QED utilized a dual theory
of Maxwell field, and it is a scalar with a periodic potential.
( Ref: A. M. Polyakov, 
{\it Quark Confinement and Topology of Gauge Groups},
Nucl. Phys. {\bf B120}, 429 (1977).
Of course, the confinement mechanism of Polyakov and our naive speculation shown in here is different,
though it is interesting for us to find a possible relation between them.
See also the paper of Hori and Vafa, 
{\it Mirror Symmetry}, arXiv:hep-th/0002222.
Therefore, our GNG mechanism might have a certain relation with confinement phenomenon,
and possibly mirror symmetry and (1+1)-D SUSY gauge theory.

\vspace{2mm}

(16) Some relations with the Kuramoto model: A viewpoint from symmetry breaking and (G)NG theorm.

\vspace{2mm}

We make a brief comment on the Kuramoto model, 
with an emphasis on relations with phenomena of broken symmetries.
( Y. Kuramto, Lecture Notes in Physics, ${\bf 39}$, H. Araki ed. ( Springer, New York, 1975 ) p.420,
S. H. Strogatz,
{\it From Kuramoto to Crawford:
Exploring the Onset of Synchronization in Populations of Coupled Oscillators},
Physica {\bf D143}, 1 (2000),
M. A. Lohe,
{\it Non-Abelian Kuramoto Models and Synchronization},
J. Math. Phys. A. Math. Theor. {\bf 42}, 395101 (2009).
See also, a "Hamiltonian system" with couplings; 
H. Morita and K. Kaneko,
{\it Collective Oscillation in a Hamiltonian System},
Phys. Rev. Lett. {\bf 96}, 050602 (2006). )
The Kuramoto model gives a topological dynamics ( usually a continuous topology is assumed ),
and it is well-known fact than a lot of theories will be reduced to the Kuramoto model by the phase reduction.
The Kuramto model is defined as follows:
\begin{eqnarray}
\frac{d}{dt}\theta_{a} &=& \omega_{a} + \kappa{\rm Tr}_{b}C_{ab}\sin(\theta_{a}-\theta_{b}), \quad
0 \le a,b \le N.
\end{eqnarray}
Here, $\kappa$ is a coupling constant.
Namely, the set of the equations gives a flow in a topological space.
The phase reduction can be interpreted as a procedure to subtract a non-hyperbolic part of a theory
by removing hyperbolic part 
( namely, describe a system only by $e^{i\alpha t}$ of 
${\rm diag}(e^{\omega_{1}t},e^{i\alpha t},e^{-\omega_{2}t})$, $\omega_{1,2}, \alpha\in{\bf R}$ ).
Based on the philosophy of Kuramoto model, 
a lot of works consider $\theta_{a}$ and $\omega_{a}$ as real, 
but of course we can consider them as complex, and then it gives a complex dynamics.
( Our GNG theorem contains aspects of complex dynamics. )
Here, $\omega_{a}$ is a constant, and this condition can be interpreted as the system obtains energy
constantly from outside to keep $\omega_{a}$ constants. 
$a,b$ denote sites of a lattice/network, and the form of sine function removes a self interaction automatically.
$C_{ab}$ is adjacency matrix which defines a network structure of the coupled oscillators,
and one can consider, for example, a stochastic geometry for the matrix:
A stochasticity can be introduced into the model via a characteristic function of a matrix form
$\int \exp[i{\rm tr}X]f(X)$.
Of course, one can define an entropy of the network $\{ C_{ab}\}$.
The Kuramoto model has been used intensively for studying various types of complex systems and networks.
The crucially important feature of the model is that it will show a synchronization
by varying $\kappa$:
The model is consisted with a specific number of coupled phase oscillators, 
and the oscillators show a coinsidence of phases of them with a threshold of coupling constant.
Since the phase degree of freedom of a phase oscillator can be interpreted as an NG mode,
and thus the phase transition can be interpreted as a kind of symmetry breakdown.
Note that each oscillator describes a unit circle with a specific frequency before a synchronization,
like a motion along the bottom of the wine bottle. 
After a perfect synchronization of the oscillators,
they share the common frequency with common phase:
The onset of synchronization is expressed by the Kuramoto ${\it ansatz}$:
\begin{eqnarray}
re^{i\psi} &=& \sum^{N}_{a=1}e^{i\theta_{a}}.
\end{eqnarray} 
Namely, a very specific, a linear combination of equal weight of all elements.
$r$ and $\psi$ are meanfield amplitude and phase, respectively.
An interesting interpretation of the meanfield may obtain from a comparison with the condition
of Poincar\'{e} recurrence:
\begin{eqnarray}
\mu(f^{s_{j}}B\cap B) > 0, \quad {\rm for} \quad \exists s_{j},
\end{eqnarray}
where, $\mu$ is a measure, $\{s_{j}\}$ give a sequence, satisfy
\begin{eqnarray}
\lim_{N\to\infty}\frac{1}{N}\sum^{N}_{j=1}e^{i\alpha s_{j}} &=& 0, \quad \forall \alpha \in (0,2\pi).
\end{eqnarray}
Since the left hand side of the definition of meanfield is assumed that it is always finite, 
we can say the meanfield assumption of the Kuramoto model satisfies the Poincar\'{e} recurrence
condition, gives a special example of sequence.
Another important fact of the Kuramoto model we argue is that it has essentially a hierarchical structure:
In fact an oscillator can be regarded as a set if synchronized subspace of the system,
and then the set of subspace show a synchronization via some interactions between them.
Namely, a system will be devided into a set of subsets after a rescalling, 
and then each subset will be divided into a set of subsets of the subset, ... ( continue ). 
One can conside an NG mode for giving a phase oscillator ${\it before}$ a synchronization,
though, the "conserved" ( classical mechanical ) system of spontaneous symmetry breaking 
does not give such an "oscillation" along a unit circle
( simply gives a point which locates on the unit circle, and the choice of a point 
is completely equivalent, the same probability if there is no external field ).
The "phase description" of a ( physical ) system has a similar philosophy with that of NL$\sigma$M.
If the $U(1)$ NG mode is obtained by our GNG mechanism of explicit+dynamical symmetry breaking,
then the phase oscillation has two possible states, 
(1) a circle motion, (2) a fluctuation around a stationary point 
( $\omega_{a}$ is not a frequancy of circle motion in this case ). 
While, there are several interesting possible motions of each oscillator for a non-Abelian Lie group case.
If one constructs a Kuramoto-type model defined on a homogeneous space $G/\Gamma$,
then the "expectation value" of the model should be given in a set of automorphic forms.

\vspace{2mm}

We consider a group-theoretical formulation of the Kuramoto model:
\begin{eqnarray}
& & -i(g_{a})^{-1}\frac{d}{dt}g_{a} = H_{a} + {\cal F}\bigl[(g_{a})^{-1}C_{ab}g_{b} - (g_{b})^{-1}D_{ba}g_{a}\bigr], \\
& & g_{a} \in G_{1}, g_{b} \in G_{2}, \cdots, 
\end{eqnarray}
where, $G_{j}$ ( $j=1,\cdots,N$ ) imply ( Lie ) groups and they do not have to belong to the same group,
whether compact or non-compact types, though the philosophy of ordinary Kuramoto model
is based on a set of limit cycle, thus compact.
( Mathematically, the notion of limit cycle is almost equivalent with a compact Lie group. )
The second term of the differential equation will 
remove self-interactions of group elements under $C_{ab}=D_{ba}$ ( symmeric adjacency matrix )
and a function ${\cal F}(0)=0$,
and then the model aquires some symmetries such as an invariant under a simultaneous gauge rotation
of every lattice points:
$H_{a}$, $C_{ab}$ and $D_{ab}$ act as explicit symmetry breaking parameters in the model.
A gauge field is introduced by 
\begin{eqnarray}
{\cal A} &\equiv& g^{-1}dg,
\end{eqnarray}
and a winding number of the gauge field may be estimated by the usual manner.
The ordinary Kuramoto model will be obtained by $G_{1}=G_{2}=\cdots=G_{N}=U(1)$,
take a real matrix for ${\rm diag}(H_{a})$,
substituting
\begin{eqnarray}
{\cal F} \to \frac{\kappa}{N\sqrt{-1}}\sum^{N}_{b=1}\Bigl( (g_{a})^{-1}C_{ab}g_{b} - (g_{b})^{-1}D_{ba}g_{a} \Bigr),  
\end{eqnarray}
and with a specific orientation, in the sense of surface, in each site:
Namely, the ordinary Kuramoto model consists with a pair of sheets 
( imagine!, a bilayer consisted by a chiral and an anti-chiral sheets!,
and one should notice that they do not have to be "a mirror pair" ),
and one of them is given by a set of chiral rotations while another is given by anti-chiral rotations,
and the interaction term of the Kuramoto model is defined by combinatorial sum of interactions between sites of 
two ( chiral and anti-chiral ) sheets.
If we consider an amplitude to each site, then the Kuramoto model
will become a dynamical system of Riemann surfaces with interactions between chiral and anti-chiral surfaces
( interacting Riemann-surface system ).
In that extension of the Kuramoto model, one considers a deformation of complex structure
( which may give a deformation of the Kuramoto model ), 
conformal mappings and uniformization of surfaces with genus $g\ge 0$. 
More generally, the following dynamical equation of group elements is considered for our context,
\begin{eqnarray}
\frac{d}{dt}g_{i} &=& F(g_{j},t)
= {\rm const.} + C^{(1)}_{j}g_{j}(t) + C^{(2)}_{jk}g_{j}(t)g_{k}(t) + C^{(3)}_{jkl}g_{j}(t)g_{k}(t)g_{l}(t) + \cdots,
\end{eqnarray}
since our GNG theorem gives an effective potential or an effective Lagrangian as 
polynomials or series of group elements.
The matrix $C^{(2)}_{jk}$ is interpreted as an adjacency matrix, 
while there are superpositions to the adjacency matrix given by $C^{(3)}_{jkl}$, $C^{(4)}_{jklm}$, ..., 
namely the dynamical equation contains a notion beyond the ordinary/canonical network theory.

\vspace{2mm}

Now, we arrive at a generalization of the Hilbert's 16th problem:

\vspace{2mm}

{\bf Problem}:
Find and determine the symmetries and numbers of limit-cycle type solutions of 
ordinary differential equations of Lie groups.

\vspace{2mm}

The importance of the Kuramoto model comes from the fact that 
quite a lot of phenomena of complex systems can be described by a set of limit cycles.
There are several models of economics given by a ( set of ) limit cycle(s), and its dynamics.
The Kuramoto model is derived from a phase reduction of corresponding underlying theory,
and the phase degree of freedom can be provided by the circle of winebottle of the NG theorem.
Thus, it is interesting for us whether one can find an underlying theory of 
the limit cycle mode of economics, and also a mechanism of spontaneous/explicit symmetry breaking in economics.
We would like to emphasize that we find the fact that the Kuramoto model gives us the notion
of interaction between group elements.
From this fact, the model indicates a road toward a lattice/graph/network model of ( Lie ) groups.

\vspace{2mm}

(17) Similar to the case of Kuramoto model, it is a well-known fact that a reduction of Boltzmann equation 
gives a Navier-Stokes equation.
From our context of GNG theorem,
it is interesting for us to consider a $G$-Boltzmann equation ( $G$, a ( Lie ) group ) with 
a distribution function $f$,
\begin{eqnarray}
f &=& f\bigl( g, \frac{dg}{dt}, t\bigr), \quad g \in G, \quad t \in {\bf R}^{1}, \\
\frac{df}{dt} &=& \Bigl(\frac{\partial}{\partial t}+\frac{\delta}{\delta g}\frac{\partial g}{\partial t} \Bigr)f = Q(f,f), \\
Q(f,f) &\equiv& {\rm Tr}_{\sigma}{\rm Tr}K\bigl(\frac{dg}{dt}-\frac{dg_{*}}{dt},\sigma\bigr)\Bigl\{ 
f\bigl( g', \frac{dg'}{dt}, t\bigr)f\bigl( g'_{*}, \frac{dg'_{*}}{dt}, t\bigr) -
f\bigl( g, \frac{dg}{dt}, t\bigr)f\bigl( g_{*}, \frac{dg_{*}}{dt}, t\bigr)
\Bigr\},
\end{eqnarray} 
and its reduction to the corresponding $G$-Navier-Stokes theory.
Note that the "Boltzmann collision term" gives a transition between group elements of two specific points
to other two points ( or, one can also consider a transition from a pair of two group elements to
other pair ).
Our insight is the $G$-Boltzmann equation may have a similar nature with a ( non-Abelian ) Kuramoto-type model.

\vspace{2mm}

(18) There is a famous method to map a network into a Bose gas, and the Bose gas shows a BEC,
and the BEC can interpreted by a network theory.
( G. Bianconi and A.-L. Borab\'{a}si,
{\it Bose-Einstein Condensation in Complex Networks},
Phys. Rev. Lett. {\bf 86}, 5632 (2001). )
Therefore, in our context,
\begin{eqnarray}
{\rm network} \to {\rm Bose \, gas} \to {\rm BEC} \to {\rm BCS-NJL}
\end{eqnarray}
is an interesting scheme.
Namely, we are interested on a network-theoretical interpretation of a macroscopic quantum effect,
or a description of network by a bosonization of interacting Fermi system.

\vspace{2mm}

(19) Information geometry is an interesting subject for us.
( S. Amari,
{\it Information Geometry on Hierarchy of Probability Distributions},
IEEE Trans. Info. Theor. {\bf 47}, 1701 (2001). 
See, for various discussions include a graph-theoretical approach;
C. E. Shannon,
{\it A Mathematical Theory of Communication},
Bell System Tech. J. {\bf 27}, 379 (1948). )
It provides a method to evaluate an entropy and Riemannian metric of 
a set of statistical probability {\it density} functions.
The Riemannian metric gives a parametrization of a set/family of specific type 
( for example, Gauss, Poisson, ... ) of probability density functions.
Let $\Theta$ be an $n$-dimensional parameter space, 
let $p_{\theta}$ be a probability density function at $\theta\in\Theta$ with 
the normalization $\int_{\Omega} p_{\theta}(x)dx=1$.
Then the Riemannian metric, entropy and distance of the manifold are defined as follows:
\begin{eqnarray}
g_{\mu\nu}(\theta) &\equiv& \frac{\partial^{2}}{\partial\theta^{\mu}\partial\theta^{\nu}}S(\theta),  \\
S(\theta) &\equiv& -\int_{\Omega}p_{\theta}(x)\log p_{\theta}(x)dx,  \\
ds^{2} &=& g_{\mu\nu}(\theta)d\theta^{\mu}d\theta^{\nu}.
\end{eqnarray}
Note that the entropy is a potential ( K\"{a}hler potential after a complexification ),
and the Riemannian metric is a Hessian of the entropy.
An interesting issue is to apply the method of information geometry to
interacting Bose/Fermi systems: 
The distribution functions of those systems are parametrized by several order parameters
or model parameters of Lagrangians, 
thus a distribution function changes under a phase transition or a bifurcation.
This method can handle a quantum phase transition, of course,
then a quantum phase transition may generate a flow of a Riemannian metric.
Thus, we can examine our GNG situation and its number-theoretical structure,
and singularities of a theory by a dynamics of information geometry.
An interesting subject is to understand a relation between such a flow of phase transition and a Ricci flow.
Since the method of auxiliary fields gives a "Gaussian",
one can estimate an entropy and a Riemannian metric for an auxiliary field ( thus, dynamical mass ). 
This is a kind of emergence of a geometry via a bosonization of interacting fermi system
and the prescription of information geometry.
Since a breaking scheme of a symmetry reflects to a matrix form of Gaussian, 
the information geometry might provide 
a systematic interpretation/classification of spontaneous/dynamical/explicit symmetry breakdowns.

\vspace{2mm}

(20) S. Smirnov considers a planar graph and an embedding of it into a Gauss plane,
then he discusses a discrete Cauchy-Riemann equation.
( S. Smirnov,
{\it Discrete Complex Analysis and Probability},
arXiv:1009.6077. )
Both a planar and a non-planar graphs can be embedded into Riemann surfaces with genus $g\ge 0$.
It is interesting to generalize the analysis of S. Smirnov to cases of arbitrary genus Riemann surface,
From our context of this paper, mass functions of our GNG theorem,
it is also important to generalize the one-dimensional discrete complex analysis of S. Smirnov to
higher dimensional complex manifolds.
If it is possible, then one could find a Kodaira-Spencer deformation theory of discrete complex manifold
and discrete complex Lie group.

\vspace{2mm}

(21) The analogue of Riemann hypotheis was proved in cases of finite fields or algebraic varieties of all dimensions.
Thus, we can imagine if continuations, "finite field$\to {\bf C}$, or ${\bf R}$ or a point ( constant )", 
are established in theory of zeta function, the hypothesis might be proved.
Our GNG theory might give a hint on it via a modification of singularity with the vanishing limit
of an explicit symmetry breaking parameter.

\section{Concluding Remarks}

We will summarize our results:
\begin{itemize}
\item
In our NJL-type model for neutrino seesaw mechanism, 
a pseudo Nambu-Goldstone boson arises after the dynamical generation of the Dirac mass,
and the phase ( we call its quantum fluctuation as "phason" in this paper ) 
is self-consistently chosen as $\Theta=(2n+1)\pi$ as infinitely countable degenerate vacua of $V_{eff}$.
We have examined the mechanism of ( dynamical ) generation of infinitely countable degenerate vacua, 
summarized into a theorem by mathematical language.
The non-trivial structures of pseudo-Nambu-Goldstone manifolds given by explicit$+$dynamical symmetry breakings
have been examined. 
We emphasize that the non-trivial geometry of a pseudo-NG manifold 
( manifold of pseudo-NG bosons given over an NG sector )
cannot be found in ordinary Nambu-Goldstone theorem. 
The nontrivial structure of the NG sector cannot be understood 
by a naive topological/gauge-theoretical/Lie-group analysis.
\item
The p-NG boson has a non-vanishing mass in its excitation spectrum which corresponds to 
the observable quantity, the mass-gap of collective excitation.
The phason mode becomes a majoron at $|m_{L}|=0$ 
( for example, in the case of type-I seesaw condition of neutrino ),
and gives a pion at $|m_{R}|=|m_{L}|=0$ in our model.
Evaluations of phason mass in NJL and SNJL have been presented,
and the NJL gives the mass as little lighter than weak bosons, while that of SNJL becomes very light,
almost the same order of axion mass today accepted by people widely.
\item
The $SU(2_{c})\times SU(N_{f})_{R}\times SU(N_{f})_{L}$ model for neutrino seesaw mass
has been constructed, and the SD equation for current and dynamical masses has been examined.
The linearized version of the SD equation will take the form of a matrix Heun equation under
the non-running coupling case.
The effect of mass phase $\Theta$ appears in the SD equation 
and can become crucial under the seesaw mechanism situation.  
\item
We have considered the two-body/two-photon phason decay rates,
which might have some contributions to astrophysical/cosmological processes.
Possible physical implications coming from the phase degree $\Theta$ have been discussed
with the emphasis on CP-violation. 
\item
To intend to generalize our GNG theorem mathematically as much as possible,
we have described various mathematical aspects/structures of our GNG theorem
by employing various (very well-)known mathematical methods/theorems. 
We believe it is an important Arbeit to establish the basis of our GNG theorem.
\end{itemize}

\vspace{2mm}

We argue that some parts of mathematical structure of our GNG theorem can be applied
both chiral and flavor symmetry breakings and confinement of $SU(N)$ gauge symmetry:
We consider that both of these phenomena share essentially the same mathematical structure,
and one can consider each problem by starting from this common structure with adding
a characteristic aspect of it.
For example, our mass matrix of diagonal symmetry breaking has similar form
with the non-Abelian vortex widely used in theory of confinement of non-Abelian gauge theory.
Let us consider a matrix of the form $M={\rm diag}(\lambda_{1},\cdots,\lambda_{N})$.
In fact, a chiral transformation on $M$ gives a chiral anomaly,
while a gauge transformation on $M$ gives a vortex of 't Hooft.
It might be the case that a gauge fixing causes a corresponding vortex solution, 
and then it shows a confinement.
For example, G. 't Hooft ( Nucl. Phys. {\bf B138}, 1 (1978) ) 
considers a system of $SU(N)$ gauge fields associated with a Higgs field,
and introduces the Higgs field of the following torus form:
\begin{eqnarray}
H(\theta) &\equiv& {\rm diag}(e^{-i\theta/N},\cdots,e^{-i\theta/N},e^{i\theta(1-N^{-1})}).
\end{eqnarray}
Then he argues/assumes that the vortex solution gives a stationary point of a functional of the gauge fields,
and this implies an existence if a set of discrete vacua, 
with the cyclotomic Galois nature we have discussed in the previous section.
It seems the case that the discussion of 't Hooft can be obtained by both elementary and composite Higgs fields.
( 't Hooft also discussed a {\it quantum algebra}, 
written in his paper as $A(C)\phi(x_{0})=\phi(x_{0})A(C)\exp(2\pi in/N)$ and its hierarchy,
though he did not recognize it. )
Since $N$ is an integer, the vortex-type Higgs is periodic, not ergodic.
One can consider a complexification of the Higgs by our prescription considered in this paper.
A non-Abelian monopole also has been considered by the same mathematical structure until various people. 
Hence,

\vspace{2mm}

{\bf Theorem}:
The $SU(N)$ confinement Higgs field of the 't Hooft Ansatz is periodic and not ergodic. 
After the complexification $\theta\to z\in\mathbb{C}^{1}$,
it shows a hyperbolicity, and its topological entropy is positive.

\vspace{2mm}

Moreover,

\vspace{2mm}

{\bf Theorem}:
A Frobenius-like substitution in our GNG theorem of such a vortex gives a permutation 
between two vortices.

\vspace{2mm}

The Dashen theorem 
( R. Dashen, Phys. Rev. {\bf D3}, 1879 (1971) )
corresponds to the $U(1)$ case we discussed,
while in the non-Abelian cases we have examined do not show the situation of Dashen "theorem":
For example, there is no unique vacuum, infinite number of degeneracy in the case of $SU(2)$
since the manifold is devided into the $z$ and $(x,y)$ directions.
The several results of Creutz
( M. Creutz, Phys. Rev. {\bf D52}, 2951 (1995), Phys. Rev. Lett. {\bf 92}, 201601 (2004) ) 
are known as "tilted potentials", though they are not recognized as fundamental theorems.
An idea of a generalization of the NG theorem was first given in the work of R. F. Streater,
{\it Generalized Goldstone Theorem}, Phys. Rev. Lett. {\bf 15}, 475 (1965),
and he discussed a 2-flavor model of explicit symmetry breaking, obtained massive NG bosons.
( See also, J. Nuyts, Phys. Rev. Lett. {\bf 26}, 1604 (1971). )

\vspace{2mm}

Until now, the equivalence, the NG theorem = NL$\sigma$M/ChPT, is believed by people widely.
There are a lot of discussions on spontaneous CP violation, mainly by the method of
chiral perturbation theory ( ChPT ). 
( For example, T. D. Lee, Phys. Rep. {\bf 9C}, 143 (1974).
The works to extend the idea of Lee to technicolor models are,
E. Eichten, K. Lane and J. Preskill, Phys. Rev. Lett. {\bf 45}, 225 (1980),
T. Inagaki, Nucl. Phys. B ( Proc. Suppl. ), {\bf 37A}, 197 (1994). 
In those works, they introduce four-fermion interactions they break flavor symmetries explicitly,
and then they result CP violations by a coupling of flavor and chiral degrees of freedom.
Namely, they derived dynamical generations of the KM matrix. )
Our results given by the BCS-NJL mechanism does not give a spontaneous CP violation,
at least in the vacuum ( vacua ).
From the viewpoint of CP violation, the main difference is that our BCS-NJL method
examines the Bogoliubov quasiparticle spectra, while the method of ChPT discusses
meson mass spectra by introducing a mass matrix to a NL$\sigma$M, 
and seeking a VEV of complex number.
Sometimes someone introduces a set of bare mass parapeters of complex numbers into 
two or three flavor NL$\sigma$M of by his/her hand, 
and then the models can violate CP symmetry but not spontaneously. 
In several works of ChPT, a mass term of mesons is given as $-F^{2}_{\pi}\sum_{j}\mu^{2}_{j}\cos\theta_{j}$,
namely a sum of cosine functions. 
While in our examination, we put several angles they cannot be absorbed by any field redefinition,
and then obtain the non-Abelian vortex-type ( like a confinement of QCD ) dynamical mass.
The angles of our matrix couple with each other, 
and thus we have otained the product form $\prod\cos\theta_{j}$ 
or $\cos(\sum \theta^{2}_{j})$ by our BCS-NJL method.

\vspace{2mm}

Now, we will mention on several aspects of the future of this work.
It is summarized into the following categories:
(1) Mathematical physics and mathematics contained in our generalized NG theorem,
(2) on applications of early Universe and evolution of Universe,
leptons/quarks and pseudo-NG bosons,
(3) CP and flavor violations, 
(4) beyond the (MS)SM,
(5) on gauge theoretical ( differential geometrical ) aspects.

\vspace{2mm}

In this paper, we have assumed the order parameter $m_{D}$ of the NJL/SNJL-type model as spacetime-independent.
If we investigate the relation between spacetime-dependent $m_{D}(x)$ and the seesaw mass spectra,
we should employ a "Bogoliubov-de Gennes" ( BdG )-type equation often used in theory of superconductivity.
This is also the case in a curved spacetime.
One can obtain spacetime-dependences of $m_{D}$ and the angle $\Theta$ 
( it might be possible that $\Theta$ forms a vortex ) in the BdG theory,
and possible exciations around the bottom of a valley,
a time-dependent collective motion of $\Theta$,
it may contain a CP-violation effect, can be described. 
We only consider the (3+1)-dimensional case, though our method and also the dispersion relations 
we have obtained in this paper can be applied/used 
to study similar situations in other spacetime dimensions.

\vspace{2mm}

For a three-flavor model, we can introduce mass matrices as
${\rm diag}(m^{e}_{R},m^{\mu}_{R},m^{\tau}_{R})$, ${\rm diag}(m^{e}_{L},m^{\mu}_{L},m^{\tau}_{L})$.
If we wish to consider a $U(3)$-flavor symmetry in dynamical Dirac mass terms,
we can utilize
$g_{1}(\bar{\psi}\psi)^{2} + g_{2}(\bar{\psi}\frac{\lambda_{3}}{2}\psi)^{2} + g_{3}(\bar{\psi}\frac{\lambda_{8}}{2}\psi)^{2}$.
We denote $\lambda_{3}/2$ and $\lambda_{8}/2$ as the Cartan subalgebra.
This interaction can generate the Gell-Mann$-$Okubo type mass relation
usually used in the classical three-flavor quark model. 
An implementation of the Froggatt-Nielsen machanism of mass hierarchy
of a flavon field~[32,67] is also an interesting direction of an extension 
of our models.
This could reduce model parameters to get a mass hierarchy or a mass matrix of a theory.
For example, we can consider the following Lagrangian for 
a dynamical Froggatt-Nielsen mechanism of neutrino sector:
\begin{eqnarray}
{\cal L} &=& \bar{\psi}i\partfey\psi + G\Bigg[ \Bigl(\bar{\psi}\frac{1+\tau_{3}}{2}\psi\Bigr)^{2} + \Bigl(\bar{\psi}i\gamma_{5}\frac{1+\tau_{3}}{2}\psi\Bigr)^{2} \Bigg] +  \epsilon\bar{\psi}\frac{\tau_{1}+i\tau_{2}}{2}\psi + \epsilon^{\dagger}\bar{\psi}\frac{\tau_{1}-i\tau_{2}}{2}\psi   \nonumber \\
& & -\frac{1}{2}\psi^{T}(m^{\dagger}_{R}CP_{+}+m_{L}CP_{-})\psi 
-\frac{1}{2}\bar{\psi}(m^{\dagger}_{L}CP_{+}+m_{R}CP_{-})\bar{\psi}^{T},
\end{eqnarray}
where, $\psi\equiv (\psi_{1},\psi_{2})$ is a two-flavor Dirac field.
By using $\tau_{3}$ Pauli matrix ( here $\tau_{j}$ ( $j=1,2,3$ ) act on the two-dimensional flavor space ), 
we give an NJL interaction only to the first flavor $\psi_{1}$, 
while $\epsilon$ gives a cross term between two flavors.
We can replace the NJL interaction to an $SU(2)$ gauge interaction similar to our model discussed in Sec. IV.
The Majorana mass parameters $m_{R},m_{L}$ are universal for the two flavors.
After a dynamical generation of Dirac mass takes place in the first flavor, 
we integrate out it, and then we will get a Dirac mass term of the second flavor.
This model has no massless majoron, could obtain a mass hierarchy between the first and second flavors.

\vspace{2mm}

Next, we present an approach to CKM/PMNS matrices based on an NL$\sigma$M
( non-linear sigma model )
~[19,59,60,75,108,236].
When CKM/PMNS matrices are regarded as physical degrees of freedom, 
it is possible to describe them by an NL$\sigma$M Lagrangian,
due to a similarity of the mathematical structures of NL$\sigma$M Lagrangians 
with flavon or little Higgs models. 
We concentrate on the PMNS matrix.
The diagonalization of PMNS matrix is achieved by 
\begin{eqnarray}
\widehat{M}_{neu} &=& U^{*}_{PMNS}\lambda_{neu} U^{\dagger}_{PMNS}.
\end{eqnarray}
Here, $\lambda_{neu}$ is the matrix of mass eigenvalues.
While, the p-NG field $\Sigma$ of non-linear realization parametrized by broken generators 
( an exponential mapping of a Lie group ) will be transformed according to 
the transformation law of $\widehat{M}_{neu}$ as follows:
\begin{eqnarray}
\Sigma &\equiv& e^{i\Pi_{a}/F}, \quad
\Sigma \to \Sigma' = U_{PMNS}\Sigma U^{T}_{PMNS}.
\end{eqnarray}
Note that $\Pi_{a}$ are matrices of a flavor space.
Therefore, the NL$\sigma$M Lagrangian will be determined as 
\begin{eqnarray}
{\cal L}_{neu} &\equiv& \frac{F^{2}}{2}\Bigl[ \partial_{\mu}\Sigma\partial^{\mu}\Sigma^{\dagger} -\kappa_{neu}{\rm tr}\bigl( \widehat{M}_{neu}\Sigma + \Sigma^{\dagger}\widehat{M}^{\dagger}_{neu} \bigr)\Bigr].
\end{eqnarray}
( $\kappa_{neu}$; a constant. )
From a minimization of a potential given by the mass term, 
ultimately, we can obtain VEVs of mixons ( mixing angles ).
For obtaining a realistic mixing matrix, one should choose $\widehat{M}_{neu}$ suitably for the purpose. 
Our consideration on our generalized Nambu-Goldstone theorem implies that
the CKM/PMNS matrices are special examples of realization of our mechanism of pseudo-NG boson:
Namely, one can find several "realizations" 
( phenomenology of our generalized NG theorem )
they can cause physical effects
implicitly/explicitly contained in the CKM/PMNS matrices. 
As we have stated in the Introduction of this paper, flavor fraction of mass eigenstates 
in neutrino oscillation must be the same for neutrinos and anti-neutrinos, if CPT is conserved.
A CPT/Lorentz violating effect can rather easily be incorporated into our non-linear sigma model
approach of flavor oscillation.

\vspace{2mm}

Let us consider a possible experiment which can prove fluctuations of mixing angles are physical. 
If it is the case, then there should exist thermal excitations. 
Oscillation of a very strong and thermal beam of neutrinos would show an effect of 
such kind of thermal/collective excitations. 
( We wish to emphasize that our viewpoint argues that 
neutrino oscillation experiments can find a key for understanding interactions
in a flavor space of the neutrino/lepton sector. )
Similarly, a very strong gravitational field ( curvature of spacetime ) can give a similar effect.
Recently, a paper on a possible modification of neutrino oscillation 
by gravity from a context of CPT violation was published in literature~[209].

\vspace{2mm}

Another interesting subject for the next step of our investigation
is to construct theory of noncommutative harmonic superanalysis
and its application to a (p)NG supermanifold, 
supersymmetry and Morse theory
~[23,144,242], 
generalizations/extensions of the MSSM and the SNJL
( namely, a geometrical understanding of the MSSM/SNJL ), so forth.
The Kobayashi-Maskawa sector may take a (SUSY) NL$\sigma$M,
and gauge symmetry for a flavor space may take $SU(N_{F})$,
it is global, cannot be given by a complete description of local gauge interactions
because the symmetry should be broken explicitly.
Thus, if we try to generate such a mass matrix, 
we need an NJL-type four fermion interaction like the top-condensation model.  
Explicit symmetry breaking parameters/matrices may determine the total structure and mixing character 
of a mass matrix as "boundary conditions" of it.  
In this approach, dynamics of the Kobayashi-Maskawa sector 
is similar to that of phason which has been studied in this paper.
A mathematical definition of path integral should be done on a Finsler manifold~[33].
Thus, measure of path-integration, symmetry of a system, 
and a definition for a ( supersymmetric ) NL$\sigma$M could be examined by Finsler (super)geometry.
This issue relates not only pseudo-NG (super)manifold, but also random matrix theory~[248], 
and method of SUSY auxiliary fields.
For example, we can consider the following example: 
\begin{eqnarray}
|\Psi\rangle = \hat{g}_{super}|\Phi\rangle, \quad
\hat{g}_{super} \equiv \exp[iG_{super}], \quad
G_{super} \equiv \left(
\begin{array}{cc}
G_{bb} & G_{bf}  \\
G_{fb} & G_{ff}
\end{array}
\right),   \quad
|\Phi\rangle \sim |\Phi_{b}\rangle\otimes |\Phi_{f}\rangle, 
\end{eqnarray}
where $|\Phi_{b}\rangle$ ( $|\Phi_{f}\rangle$ ) is a linear combination of $n_{b}$-boson permanents
( a linear combination of $n_{f}$-fermion determinants ),
$G_{bf}$ and $G_{fb}$ are broken generators.
A supermatrix contains a set of explicit supersymmetry breaking parameters,
and it will lift degeneracies of eigenvalues of bosons and fermions. 
Then, a dynamical effect generates a VEV toward the broken direction.
From an examination of a supermatrix and a superdeterminant given in terms of $\hat{g}_{super}$,
one would obtain some interesting (super)manifolds/varieties of a pseudo-NG sector. 
Since ${\cal N}=1$ SUSY is handled by a set of explicit supersymmetry breaking parameters in MSSM,
an application of our mechanism to the MSSM is interesting from a particle-phenomenological point of view.

\vspace{2mm}

As mentioned in the previous section, 
an explicit symmetry breaking parameter gives an inconsitency 
with a ( part of ) gauge symmetry,
and thus it will certainly modify/destroy a moduli space of a set of connections.
It is mathematically/physically interesting for us to show how a moduli space is modified, 
divided, or ( partially ) disappears/lifted 
by introducing an explicit symmetry breaking parameter. 
Probably this issue will be solved in the framework of Seiberg-Witten theory,
or by a generalization of the Seiberg-Witten-type methods/considerations~[154].

\vspace{2mm}

We have mentioned on a possibility of chaotic behavior caused by an unharmonicity of a valley of phason. 
On this issue, one may use the following formalism:
\begin{eqnarray}
\varrho_{\alpha\beta}(x) &\equiv& \langle {\bf B}(x)|\Psi_{\alpha}(x)\overline{\Psi}_{\beta}(x)|{\bf B}(x)\rangle,  \qquad
( \alpha, \beta = 1,\cdots,8 ),   \\
i\frac{d}{dt}|{\bf B}(x) \rangle &=& \widehat{H}|{\bf B}(x)\rangle,   \\
i\frac{d}{dt}\varrho(x) &=& \langle {\bf B}(x)|[\varrho(x),\widehat{H}]|{\bf B}(x)\rangle,
\end{eqnarray}
namely, one can utilize a time-dependent Hartree(-Fock) formalism for the equation of motion
of the one-body density matrix $\varrho_{\alpha\beta}(x)$.
Here, $|{\bf B}(x)\rangle$ is a vector as an implicit functional of the Schwinger-Dyson mass function
${\bf B}(x)$, and $\widehat{H}$ is a Hamiltonian~[169]. 
Futhermore, one can construct a method of density-matrix-functional theory 
( ${\bf B}(x)$ is the minimizer ) for this purpose~[170,171].
An extension to the SUSY case mentioned above can be done.

\vspace{2mm}

The dynamics of our model has some similarities with axion of the strong-CP problem of QCD
~[235,239].
In our neutrino seesaw model, the Majorana mass terms play the role of an axion potential,
and in fact the dynamics of our model is simpler than those of QCD
( while, our model may have the domain wall problem~[206] ).
We can add the following terms to a Lagrangian,
\begin{eqnarray}
\delta\theta_{\chi}\frac{g^{2}}{16\pi^{2}}G_{\mu\nu}\tilde{G}_{\mu\nu}, \quad {\rm or} \quad 
\delta\Theta\frac{g^{2}}{16\pi^{2}}G_{\mu\nu}\tilde{G}_{\mu\nu}.
\end{eqnarray}
$\delta\theta_{\chi}=0$ or $\delta\Theta=0$ are chosen under the variation principle in our models,
as discussed in Sec. II.
The interaction given above will not modify equations of motion of the theory,
while they may contribute to masses of fluctuating phases similar to the case of $\eta'$ meson~[232]. 
In the massless QCD, $\theta$ is not a physical degree of freedom,
and the Peccei-Quinn type axion theory makes it as physical 
by combining the phase of the Dirac mass term of QCD~[186,187].
We wish to emphasize that the model we have considered here spontaneously choose a CP-conserving vacuum.
Our results they have been shown in this paper are an attempt for understanding
majoron, phason, pion, axion and mixon under a simple/unified point of view.

\vspace{2mm}

The mass relations of neutrino sector may imply us toward both beyond the SM and beyond the BCS-NJL. 
In fact, both of them are based on the gauge principle and spontaneous/dynamical symmetry breakdown,
namely an inconsistency between a local Ward-Takahashi-type relation and a global symmetry,
and some parts of issues of the SM might not be explained by the gauge principle,
while the ( generalized ) Nambu-Goldstone theorem is quite universal,
seems to work well in a large part of model buildings in various energy scales, as shown in this paper.
From our context, our speculation on "beyond the SM" is,
a top-condensation-type gauged-(S)NJL with a set of explicit symmetry breaking 
( in very generic sense ) parameters.
( An example of "beyond the MSSM", see Ref.~[55]. )    
To find a physical solution for the issue "beyond the SM vs beyond the BCS-NJL", 
we wish to hope new experimental results, especially on lepton sector of the SM. 
The author also speculate that a deep mathematical understanding 
on electroweak symmetry breaking and flavor physics~[243] gives the way toward a new physics
( for example, a mathematical investigation of a top-condensation-type gauged-(S)NJL 
with explicit symmetry breaking parameters as an effective theory or its effective theory ).
Our results are generic, some parts of them are not restricted to neutrino seesaw phenomenology:
For example, (S)GUT physics may also have similar phenomena coming from pseudo NG bosons,
and it is an interesting subject for us for the next stage of our investigation:
A cosmological criterion would restrict masses of pNG bosons/fermions of (S)GUT
( breakings of gauge, flavor, SUSY, so on ),
and it may work to give us a key for constructing a (S)GUT.

\vspace{2mm}

We have established that a mass function is a Riemann surface
( more precisely, a (quasi)conformal mapping ),
where lengths and angles are well-defied plane/surface
and a local isothermal coordinates always exists.
( The mass function ${\bf B}(s)$ of the Schwinger-Dyson theory
also be regarded as a "vectorial" ( triplet ) Riemann surface, 
and the theory determines a point on the Riemann surface non-trivially/self-consistently. 
A quantum fluctuation of the neighborhood of the representation point 
on the Riemann surface has a physical meaning. 
Moreover, the Schwinger-Dyson equation we have discussed 
is a coupled non-linear Schr\"{o}dinger equation of the parameter $s$
for the vectorial Riemann surfaces of both holomorphic/anti-holomorphic parts. )
Then, the next issue is to find a method of quantization of a Riemann surface.
In that case, a Riemann surface and its anti-holomorphic counter part will be devided into the following forms:
\begin{eqnarray}
z = z_{c} + \hat{z}, \quad \bar{z} = \bar{z}_{c} + \hat{\bar{z}},
\end{eqnarray}
and the canonical commutation relations are $[\hat{z},\hat{\bar{z}}]\sim\hbar$,
$[\hat{z},\hat{z}]=[\hat{\bar{z}},\hat{\bar{z}}]=0$.
The Wigner distribution function or 
the Heisenberg equation for Riemann surfaces $i\hbar\frac{d}{dt}z=[z,H]$ 
with an appropriately defined Hamiltonian can also be introduced.
For that attempt, one may obtain an interest on a deformation quantization
~[17,117,118,149] 
of Riemann surfaces
~[179,180].
Since (quasi)conformal mappings give collective coordinates/modes of Riemann surfaces,
it is a surprising fact for us that (quasi)conformal mappings and 
quantum algebra of Riemann surfaces can describe not only
( dynamical ) mass functions but also an inflaton 
( if which will be given by a composite=spin-1/2-fermion-bilinear-condensation model )
and its quantum fluctuations, i.e., 
an early history of development of structure/dynamics of our Universe
( {\it Riemann surfaces of the Universe} ).
The results of this paper implies that there is a connection between the generalized NG theorem
and complex structure of a manifold. 
For investigating this connection deeper,
a complexification of Lie group,
the Peter-Weyl theorem over a complex Lie group 
and also the Kodaira-Spencer deformation theory of complex structure
may provide us a useful methods~[115].
For example, it is suggestive fact that a flag manifold $U(N_{f})/{\bf T}_{N_{f}}$ 
is a compact complex manifold, and the CKM/PMNS matrices might be understood 
from a similar context, by utilizing the Kodaira-Spencer theory,
complex analysis of several variables, and the Borel-Weil theory. 
It is also an interesting issue for us to consider an application of our generalized NG theorem 
to a non-commutative field theory
~[8,57,177,204,205,215].

\vspace{2mm}

The origin of spacetime is an attractive, and the most mysterious problem for physics:
Some people may regard it as "the final question for physicists".
However, it seems the case that there are not so much concrete theories for 
this problem, since the question is somewhat abstract.
Especially, approaches from matrix models considered this problem~[85,87].
We can consider an application of our generalized Nambu-Goldstone theorem to this problem,
a possible mechanism of emergence of spacetime. 
We assume a field of "Ur-Raum" with an explicit symmetry breaking parameter, 
and it will spontaneously/dynamically 
get a periodic potential, then a set of discrete vacua arises.
If we assume a set of localized functions of the periodic potential as entries of
Riemannian metric, then spacetime emerges
( first, an infinite-dimensional geometry arises,
and after an appropriate compactification, a finite dimensional manifold
may be obtained ).
Quantum fluctuations around bottoms of valleys might describe quantum gravity
in this mechanism.
A dynamical response of the system may give a Ricci flow.
In other words, a matrix model might be derived under our generalized NG mechanism.

\vspace{2mm}

We cannot make a universe inside a laboratory, 
while only cosmology can answer the reason why
our Universe chooses the SM from infinite number of possible theories.
Hence, each universe might have its proper "SM".
Therefore, ultimately, there is two ways toward "beyond the SM":
Beyond the SM inside, or outside our Universe.
It is interesting for us which is the correct way to study a physical law,
and this also relates to a question "what is physics...".

\appendix

\section{The SU(3) Algebra}

Here, we summarize the $SU(3)$ algebra.
The Gell-Mann matrices are
\begin{eqnarray}
\lambda_{1} &\equiv& \left(
\begin{array}{ccc}
0 & 1 & 0 \\
1 & 0 & 0 \\
0 & 0 & 0 
\end{array}
\right), \quad 
\lambda_{2} \equiv \left(
\begin{array}{ccc}
0 & -i & 0 \\
i & 0 & 0 \\
0 & 0 & 0 
\end{array}
\right), \quad 
\lambda_{3} \equiv \left(
\begin{array}{ccc}
1 & 0 & 0 \\
0 & -1 & 0 \\
0 & 0 & 0 
\end{array}
\right), \quad
\lambda_{4} \equiv \left(
\begin{array}{ccc}
0 & 0 & 1 \\
0 & 0 & 0 \\
1 & 0 & 0 
\end{array}
\right), \nonumber \\
\lambda_{5} &\equiv& \left(
\begin{array}{ccc}
0 & 0 & -i \\
0 & 0 & 0 \\
i & 0 & 0 
\end{array}
\right), \quad 
\lambda_{6} \equiv \left(
\begin{array}{ccc}
0 & 0 & 0 \\
0 & 0 & 1 \\
0 & 1 & 0 
\end{array}
\right), \quad
\lambda_{7} \equiv \left(
\begin{array}{ccc}
0 & 0 & 0 \\
0 & 0 & -i \\
0 & i & 0 
\end{array}
\right), \quad 
\lambda_{8} \equiv \frac{1}{\sqrt{3}}\left(
\begin{array}{ccc}
1 & 0 & 0 \\
0 & 1 & 0 \\
0 & 0 & -2 
\end{array}
\right),
\end{eqnarray}
and the generators are defined as $F_{i}\equiv \lambda_{i}/2$, $[F_{i},F_{j}]=if_{ijk}F_{k}$.
Here, we have chosen our representation as the Cartan subalgebra is given by the set $\lambda_{3},\lambda_{8}$.
The following structure constants $f_{ijk}$ ( and their permutations ) are used for our calculations:
\begin{eqnarray}
& & 
f_{123} = 1, \quad 
f_{147} = \frac{1}{2}, \quad 
f_{156} = -\frac{1}{2}, \quad 
f_{246} = \frac{1}{2}, \quad   
f_{257} = \frac{1}{2},  \nonumber \\
& & 
f_{345} = \frac{1}{2}, \quad 
f_{367} = -\frac{1}{2}, \quad 
f_{458} = \frac{\sqrt{3}}{2}, \quad 
f_{678} = \frac{\sqrt{3}}{2}. 
\end{eqnarray}
The second and third Casimir invariants are given as follows:
\begin{eqnarray}
F^{2} = \sum^{8}_{i=1}F^{2}_{i}, \quad 
F^{3} = 8\sum^{8}_{i=1}\sum^{8}_{j=1}\sum^{8}_{k=1}{\rm tr}\bigl( \{F_{i},F_{j}\}F_{k}\bigr)F_{i}F_{j}F_{k}.
\end{eqnarray}
In the diagonal breaking scheme, $SU(3)\to (U(1))^{2}$,
one should evaluate the following similarity transformation, 
\begin{eqnarray}
& & e^{i(\theta_{1}F_{1}+\theta_{2}F_{2}+\theta_{4}F_{4}+\theta_{5}F_{5}+\theta_{6}F_{6}+\theta_{7}F_{7})}
\widehat{M}_{dyn}
e^{-i(\theta_{1}F_{1}+\theta_{2}F_{2}+\theta_{4}F_{4}+\theta_{5}F_{5}+\theta_{6}F_{6}+\theta_{7}F_{7})},
\end{eqnarray}
with
\begin{eqnarray}
\widehat{M}_{dyn} &=& {\rm diag}(M_{dyn1},M_{dyn2},M_{dyn3})  \nonumber \\
&=& \frac{1}{3}(1-2\sqrt{3}F_{8})M_{dyn1} +
\Bigl[\frac{1}{2}(1+2F_{3}) - \frac{1}{6}(1-2\sqrt{3}F_{8}) \Bigr]M_{dyn2} +
\Bigl[\frac{1}{2}(1-2F_{3}) - \frac{1}{6}(1-2\sqrt{3}F_{8}) \Bigr]M_{dyn3}.
\end{eqnarray}
Of course, the diagonal mass matrix $\widehat{M}_{dyn}$ are expanded by the unit and Cartan subalgebra.
For a consistency, the group elements given in terms of exponential mappings
must not contain any symmetric generators.
The expricit expression of the exponential mapping of a Lie group is known
for low-dimensional cases, $SU(2)$ and $SO(3)$,
while there are still some discussions/investigations on $SU(N)$ ( $N\ge 3$ ) cases.  
In the simpler case, for example, 
\begin{eqnarray}
e^{2i\theta_{7}F_{7}}F_{3}e^{-2i\theta_{7}F_{7}}
&=& F_{3}\cos^{2}\theta_{7} + \frac{F_{3}+\sqrt{3}F_{8}}{2}\sin^{2}\theta_{7} 
+ F_{6}\sin\theta_{7}\cos\theta_{7}.
\end{eqnarray}
In the breaking scheme $SU(3)\to (U(1))^{2}$,
totally six real NG bosons are generated they give three complex boson fields,
as clear from the structure of the Gell-Mann matrices.

\section{The Automorphic and Modular Functions}

A function which is invariant under any linear fractional transformation is called
as an automorphic function~[2]:
\begin{eqnarray}
f\Bigl( \frac{az+b}{cz+d} \Bigr) &=& f(z).
\end{eqnarray} 
A function which is invarinant under a subgroup of the modular group is called 
as a ( elliptic ) modular function, and it will be given by roots of the first derivative
of the Weierstrass function $\wp (z)$.
The definition of the modular group $\Gamma_{0}(L)$ is
\begin{eqnarray}
\Gamma_{0}(L) &=& \left(
\begin{array}{cc}
a & b \\
c & d
\end{array}
\right) \in SL(2,{\bf Z}), \quad c \equiv 0 ({\rm mod} \, L),
\end{eqnarray}
and the modular form of weight $k$ with level $L$ ( called as weakly modular ) is
\begin{eqnarray}
f\Bigl( \frac{az+b}{cz+d} \Bigr) &=& (cz+d)^{k}f(z).
\end{eqnarray}
Here, $k$ and $L$ are positive integers.
Since $f(z+1)=f(z)$, it will have a Fourier series.
The modular function uniquely exists under a full elliptic modular group,
and holomorphic except $z=i\infty$ ( cusp ).
The generic definition of the space of holomorphic modular forms is
\begin{eqnarray}
{\cal M}_{k}(\Gamma) &=& \{f:{\bf H}\to\mathbb{C}|f(\gamma z)=j(\gamma,z)^{k}f(z), \,\,\forall\gamma\in\Gamma\},  \\
f(z) &=& \sum^{\infty}_{n=0}a_{n}q^{n}, \qquad q=e^{2\pi iz},  
\end{eqnarray}
and the case $a_{0}$ is called as a cusp form.
For example, the Laurent expansion of the Dedekind-Klein $j$-function becomes
\begin{eqnarray}
J(q=e^{i\pi\tau}) = \frac{1}{1728}\Bigl( \frac{1}{q^{2}} + 744 + 196884 q^{2} + O(q^{4}) \Bigr). 
\end{eqnarray}
There are similar Laurent serieses of elliptic functions.

\section{The Heun Equation}

Any Fuchsian type differential equation with four sigular points
can be converted into the Heun equation.
The Heun equation wth singular points $z=0,1,a,\infty$ is defined as follows:
\begin{eqnarray}
\Bigg\{ \frac{d^{2}}{dz^{2}} + \Bigl(\frac{\gamma}{z} +\frac{\delta}{z-1} + \frac{\epsilon}{z-a} \Bigr)\frac{d}{dz} + \frac{\alpha\beta z-q}{z(z-1)(z-a)} \Bigg\}U(z) = 0.
\end{eqnarray}
The Fuchs relation 
\begin{eqnarray}
\epsilon &=& \alpha + \beta - \gamma -\delta +1
\end{eqnarray}
must be satisfied if the singular point $z=\infty$ is regular.  
$q$ is called as accessory parameter.
The exponents at $z=0,1,a,\infty$ are obtained as
\begin{eqnarray}
(0,1-\gamma), \quad (0,1-\delta), \quad (0,1-\epsilon), \quad (\alpha,\beta),
\end{eqnarray}
respectively.


\begin{thebibliography}{999}



\bibitem{abada}
A. Abada, A. Le Yaouanc, L. Oliver, O. P\'{e}ne and J.-C. Raynel,
{\it Dynamical Generation of Majorana Masses},
Phys. Rev. {\bf D42}, 1699 (1990).
\bibitem{ahlfors}
L. V. Ahlfors,
{\it Complex Analysis}
( MacGraw-Hill, 1953 ).
\bibitem{alkofer}
R. Alkofer and L. von Smekal,
{\it The Infrared Behaviour of QCD Green's Functions,
Confinement, Dynamical Symmetry Breaking, and Hadrons
as Relativistic Bound States},
Phys. Rep. {\bf 353}, 281 (2001),
R. Alkofer,
{\it QCD Green Functions and Their Application to Hadron Physics},
Braz. J. Phys. {\bf 37}, 144 (2007).
\bibitem{antusch}
S. Antusch, J. Kersten, M. Lindner and M. Ratz,
{\it Dynamical Electroweak Symmetry Breaking by a Neutrino Condensate},
Nucl. Phys. {\bf B658}, 203 (2003).
\bibitem{runneutri}
S. Antusch, J. Kersten, M. Lindner and M. Ratz,
{\it Running Neutrino Masses, Mixings and CP Phases:
Analytical Results and Phenomenological Consequences},
Nucl. Phys. {\bf B674}, 401 (2003).
\bibitem{antusch2}
S. Antusch and E. Fernandez-Martinez,
{\it Signals of CPT Violation and Non-locality in Future Neutrino Oscillation Experiments},
Phys. Lett. {\bf 665}, 190 (2008).
\bibitem{appelq}
T. Appelquist, K. Lane and U. Mahanta,
{\it Ladder Approximation for Spontaneous Chiral-Symmetry Breaking},
Phys. Rev. Lett. {\bf 61}, 1553 (1988).
\bibitem{arebel}
I. Ya. Aref'eva, D. M. Belov, A. A. Giryavets, A. S. Koshelev and P. B. Medvedev,
{\it Noncommutative Field Theories and (Super)string Field Theories},
arXiv:hep-th/0111208.
\bibitem{betamajomass}
F. T. Avignone III, S. R. Elliott and J. Engel,
{\it Double Beta Decay, Majorana Neutrinos, and Neutrino Mass},
Rev. Mod. Phys. {\bf 80}, 481 (2008).
\bibitem{kreimerbos}
G. van Baalen, D. Kreimer, D. Uminsky and K. Yeats,
{\it The QCD $\beta$-Function from Global Solutions to Dyson-Schwinger Equations},
arXiv:0906.1754.
\bibitem{bcs}
J. Bardeen, L. N. Cooper and J. R. Schrieffer,
{\it Theory of Superconductivity},
Phys. Rev. {\bf 108}, 1175 (1957).
\bibitem{top}
W. A. Bardeen, C. T. Hill and M. Lindner,
{\it Minimal Dynamical Symmetry Breaking of the Standard Model},
Phys. Rev. {\bf D41}, 1647 (1990).
\bibitem{barducci}
A. Barducci, R. Casalbuoni, S. De Curtis, D. Dominici and R. Gatto,
{\it Dynamical Chiral-Symmetry Breaking and Determination of the Quark Masses},
Phys. Rev. {\bf D38}, 238 (1988).
\bibitem{cosmocpvio}
V. Barger, D. A. Dicus, H.-J. He and T. Li,
{\it Structure of Cosmological CP-Violation via Neutrino Seesaw},
Phys. Lett. {\bf B583}, 173 (2004).
\bibitem{barger}
A neutrino decay model: 
V. Barger, J. G. Learned, S. Pakvasa and T. J. Weiler,
{\it Neutrino Decay as an Explanation of Atmospheric Neutrino Observations},
Phys. Rev. Lett. {\bf 82}, 2640 (1999).
\bibitem{supergroup}
I. Bars and M. G\"{u}naydin,
{\it Unitary Representations of Non-Compact Supergroups},
Commun. Math. Phys. {\bf 91}, 31 (1983).
\bibitem{defquant}
F. Bayen, M. Flato, C. Fronsdal, A. Lichnerowicz and D. Sternheimer,
{\it Deformation Theory and Quantization I and II},
Ann. Phys. {\bf 111}, 61, 111 (1978).
\bibitem{littleflavon}
F. Bazzocchi, S. Bertolini, M. Fabbrichesi and M. Piai,
{\it Fermion Masses and Mixings in the Little Flavon Model},
Phys. Rev. {\bf D69}, 036002 (2004).
\bibitem{bijne}
J. Bijnens, A. Bramon and F. Cornet,
{\it Chiral Perturbation Theory for Anomalous Processes};
Z. Phys. {\bf C46}, 599 (1990).
\bibitem{muenchen}
M. Blanke, A. J. Buras, A. Poschenrieder, S. Recksiegel, C. Tarantino, S. Uhlig and A. Weiler,
{\it Rare and CP-Violation K and B decays in the Littlest Higgs Models with T-Parity},
JHEP {\bf 0701}, 066 (2007).
\bibitem{schaden}
A. H. Blin, B. Hiller and M. Schaden,
{\it Electromagnetic Form-Factors in the Nambu$-$Jona-Lasinio Model},
Z. Phys. {\bf A331}, 75 (1988).
\bibitem{dark}
A. W. Brookfield, C. van der Bruck, D. F. Mota and D. Tocchini-Valentini,
{\it Cosmology with Massive Neutrinos Coupled to Dark Energy},
Phys. Rev. Lett. {\bf 96}, 061301 (2006). 
\bibitem{bott}
R. Bott,
{\it Lectures on Morse Theory, Old and New},
Bull. Am. Math. Soc. {\bf 7}, 331 (1982).
\bibitem{snjl}
W. Buchm\"{u}ller and S. T. Love, 
{\it Chiral Symmetry and Supersymmetry in the Nambu$-$Jona-Lasinio Model},
Nucl. Phys. {\bf B204}, 213 (1982),
W. Buchm\"{u}ller and U. Ellwanger, 
{\it On the Structure of Composite Goldstone Supermultiplets},
Nucl. Phys. {\bf B245}, 237 (1984).
\bibitem{burges}
C. P. Burgess,
{\it Goldstone and Pseudo-Goldstone Bosons in Nuclear, Particle and Condensed-matter Physics},
Phys. Rep. {\bf 330}, 193 (2000).
\bibitem{cabibbo}
N. Cabibbo,
{\it Unitary Symmetry and Leptonic Decays},
Phys. Rev. Lett. {\bf 10}, 531 (1963).
\bibitem{cantor}
G. Cantor,
{\it Beitr\"{a}ge zur Begr\"{u}ndung der transfiniten Mengenlehre},
( Erster und Zweiter ), 
Mathematische Annalen {\bf 46}, 481 (1895), {\bf 49}, 207 (1897).
\bibitem{snjlmssm}
M. Carena, T. E. Clark, C. E. M. Wagner, W. A. Bardeen and K. Sasaki,
{\it Dynamical Symmetry Breaking and the top Quark Mass in the Minimal Supersymmetric Standard Model},
Nucl. Phys. {\bf B369}, 33 (1992).
\bibitem{carenamssm}
M. S. Carena, J. R. Espinosa, M. Quir\'{o}s and C. E. M. Wagner,
{\it Analytic Expressions for Radiatively Corrected Higgs Masses and Couplings in the MSSM},
Phys. Lett. {\bf B355}, 209 (1995).
\bibitem{carena2}
M. S. Carena, M. Quir\'{o}s and C. E. M. Wagner,
{\it Effective Potential Methods and the Higgs Mass Spectrum in the MSSM},
Nucl. Phys. {\bf B461}, 407 (1996).
\bibitem{susyvarietyscheme}
L. Caston and R. Fioresi,
{\it Mathematical Foundations of Supersymmetry},
arXiv:0710.5742.
\bibitem{family}
M.-C. Chen and K. T. Mahanthappa,
{\it Fermion Masses and Mixing and CP-Violation in $SO(10)$ Models with Family Symmetries},
Int. J. Mod. Phys. {\bf A18}, 5819 (2003). 
\bibitem{finsler}
S. S. Chern and Z. Shen,
{\it Riemann-Finsler Geometry},
( World Scientific, 2005 ).
\bibitem{chikashige}
Y. Chikashige, R. N. Mohapatra and R. D. Peccei,
{\it Are There Real Goldstone Bosons Associated with Broken Lepton Number?},
Phys. Lett. {\bf 98B}, 265 (1981).
\bibitem{mssmreview}
A recent review on MSSM, including CP violation, flavor physics, dark matter, axion and inflation:
D. J. H. Chung, L. L. Everett, G. L. Kane, S. F. King, J. Lykken and L.-T. Wang,
{\it The Soft Supersymmetry-Breaking Lagrangian: Theory and Applications},
Phys. Rep. {\bf 407}, 1 (2005).
\bibitem{stop}
T. E. Clark, S. T. Love and W. A. Bardeen,
{\it The top Quark Mass in a Supersymmetric Standard Model with Dynamical Symmetry Breaking},
Phys. Lett. {\bf B237}, 235 (1990).
\bibitem{cohen}
There are a lot of literatures on quantum kicked rotors, both theories and experiments.
For example,
D. Cohen, 
{\it Quantum Chaos, Dynamical Correlations, and the Effect of Noise on Localization},
Phys. Rev. {\bf A44}, 2292 (1991),
R. Graham, M. Schlautmann and P. Zoller,
{\it Dynamical Localization of Atomic-Beam Deflection by a Modulated Standing Light Wave},
Phys. Rev. {\bf A45}, R19 (1992),
A. Altland and M. R. Zirnbauer,
{\it Field Theory of the Quantum Kicked Rotor},
Phys. Rev. Lett. {\bf 77}, 4536 (199),
H. Ammann, R. Gray, I. Shvarchuck and N. Christensen,
{\it Quantum Delta-Kicked Rotor: Experimental Observation of Decoherence},
Phys. Rev. Lett. {\bf 80}, 4111 (1998),
B. G.Klappauf, W. H. Oskay, D. A. Steck and M. G. Raizen,
{\it Observation of Noise and Disipation Effects on Dynamical Localization}
Phys. Rev. Lett. {\bf 81}, 1203 (1998),
B. Mieck and R. Graham,
{\it Bose-Einstein Condensate of Kicked Rotators with Time-Dependent Interaction},
arXiv:cond-mat/0411648.
\bibitem{cole}
S. Coleman,
{\it Aspects of Symmetry},
( Cambridge University Press, 1985 ).
\bibitem{radiative}
S. Coleman and E. Weinberg,
{\it Radiative Corrections as the Origin of Spontaneous Symmetry Breaking},
Phys. Rev. {\bf D7}, 1888 (1973).
\bibitem{connesckm}
A. Connes,
{\it A Unitary Invariant in Riemannian Geometry},
arXiv:0810.2091.
\bibitem{kreimer}
A. Connes and D. Kreimer,
{\it Renormalization in Quantum Field Theory and the Riemann-Hilbert Problem. 
I: The Hopf Algebra Structure of Graphs and the Main Theorem},
Commun. Math. Phys. {\bf 210}, 249 (2000).
\bibitem{kreimer2}
A. Connes and D. Kreimer,
{\it Renormalization in Quantum Field Theory and the Riemann-Hilbert Problem. 
II: The Beta-function, Diffeomorphisms and the Renormalization Group},
Commun. Math. Phys. {\bf 216}, 215 (2001).
\bibitem{conmar}
A. Connes and M Marcolli,
{\it Quantum Fields and Motives},
J. Geom. Phys. {\bf 56}, 55 (2006).
\bibitem{cdt}
B. Conrad, F. Diamond and R. Taylor,
{\it Modularity of Certain Potentially Barsotti-Tate Galois Representations},
J. Amer. Math. Soc. {\bf 12}, 521 (1999).
\bibitem{experiment}
J. M. Conrad,
{\it Neutrino Experiments},
arXiv:0708.2446.
\bibitem{courant}
R. Courant,
{\it Dirichlet's Principle, Conformal Mapping, and Minimal Surfaces}
( Intersciece, New York, 1950 ).
\bibitem{cranerabin}
L. Crane and J. M. Rabin,
{\it Super Riemann Surfaces: Uniformization and Teichm\"{u}ller Theory},
Commun. Math. Phys. {\bf 113}, 601 (1988).
\bibitem{cvetic}
G. Cveti\v{c},
{\it Top-quark Condensation},
Rev. Mod. Phys. {\bf 71}, 513 (1999).
\bibitem{tanishimuweil}
H. Darmon,
{\it A Proof of the Full Shimura-Taniyama-Weil Conjecture is Announced},
Notices of the AMS, 1397, Dec. 1999.
\bibitem{zetazetazeta}
A. Deitmar,
{\it Panorama of Zeta Functions},
arXiv:math/0210060.
\bibitem{kink}
R. F. Dashen, S-k. Ma and R. Rajaraman,
{\it Finite-Temperature Behavior of a Relativistic Field Theory with Dynamical Symmetry Breaking},
Phys. Rev. {\bf D11}, 1499 (1975).
\bibitem{davies}
P. C. W. Davies,
{\it Scalar Particle Production in Schwarzschild and Rindler Metrics},
J. Phys. {\bf A8}, 609 (1975).
\bibitem{supmani}
B. De Witt,
{\it Supermanifolds}
( Cambridge University Press, 1992 ).
\bibitem{dahr}
A. Dhar and S. R. Wadia,
{\it Nambu$-$Jona-Lasinio Model, An Effective Lagrangian for Quantum Chromodynamics at Intermediate Length Scale},
Phys. Rev. Lett. {\bf 52}, 959 (1984),
A. Dhar, R. Shankar and S. R. Wadia,
{\it Nambu$-$Jona-Lasinio$-$type Effective Lagrangain: Anomalies and Nonlinear Lagrangian of Low-Energy, Large-$N$ QCD},
Phys. Rev. {\bf D31}, 3256 (1985).
\bibitem{dineseibergthomas}
M. Dine, N. Seiberg and S. Thomas,
{\it Higgs Physics as a Window beyond the MSSM},
Phys. Rev. {\bf D76}, 095004 (2007).
\bibitem{littlehiggs}
A. Dobado, L. Tabares-Cheluci and S. Penaranda,
{\it Higgs Effective Potantial in the Littlest Higgs Model at the One-Loop Level},
Phys. Rev. {\bf D75}, 083527 (2007).
\bibitem{dougnek}
M. R. Douglas and N. A. Nekrasov,
{\it Noncommutative Field Theory},
Rev. Mod. Phys. {\bf 73}, 977 (2001).
\bibitem{grogalo}
E. J. Dubuc and C. S. De La Vega,
{\it On The Galois Theory of Grothendieck},
arXiv:math/0009145.
\bibitem{ebert}
D. Ebert, A. A. Belkov, A. V. Lanyov and A. Schaale,
{\it Effective Chiral Lagrangians for Strong, Weak and Electromagnetic-Weak Interactions of Mesons from Quark Flavor Dynamics},
Int. J. Mod. Phys. {\bf A8}, 1313 (1993).
\bibitem{ecker}
G. Ecker,
{\it Chiral Perturbation Theory},
arXiv:hep-ph/9501357.
\bibitem{piondecay}
S. Eidelman et al. ( Particle Deta Group ),
{\it Review of Particle Physics},
Phys. Lett. {\bf B592}, 1 (2004).
\bibitem{supergeometry}
A. M. El Gradechi and L. M. Nieto,
{\it Supercoherent States, SuperK\"{a}hler Geometry and Geometric Quantization},
Commun. Math. Phys. {\bf 175}, 521 (1996).
\bibitem{faltings}
G. Faltings,
{\it The Proof of Fermat's Last Theorem by R. Taylor and A. Wiles},
Notices of the AMS, 743, July 1995.
\bibitem{fischer}
C. S. Fischer,
{\it Infrared Properties of QCD from Dyson-Schwinger Equations},
J. Phys. {\bf G32}, R253 (2006).
\bibitem{frampton}
D. H. Frampton, S. L. Glashow and T. Yanagida,
{\it Cosmological Sign of Neutrino CP Violation},
Phys. Lett. {\bf B548}, 119 (2002).
\bibitem{friedan}
D. Friedan,
{\it Nonlinear models in $2+\epsilon$ Dimensions},
Ann. Phys. {\bf 163}, 318 (1985).
\bibitem{flavon}
C. D. Froggatt and H. B. Nielsen,
{\it Hierachy of Quark Masses, Cabibbo Angles and CP Violation},
Nucl. Phys. {\bf B147}, 277 (1979).
\bibitem{fujishro}
K. Fujikawa and R. Shrock,
{\it The Magnetic Moment of a Massive Neutrino and Neutrino Spin Rotation},
Phys. Rev. Lett. {\bf 45}, 963 (1980).
\bibitem{fukudakugo}
R. Fukuda and T. Kugo,
{\it Schwinger-Dyson Equation for Massless Vector Theory and Absence of Fermion Pole},
Nucl. Phys. {\bf B117}, 250 (1976).
\bibitem{fukugoyangmills}
R. Fukuda and T. Kugo,
{\it Dynamical Theory of the Yang-Mills Field.I},
Prog. Theor. Phys. {\bf 60}, 565 (1978).
This work contains an analytic solution of a matrix Schwinger-Dyson equation,
and all of the essential part of the framework of SD formalism are explained.
\bibitem{fulling}
S. A. Fulling,
{\it Nonuniqueness of Canonical Field Quantization in Riemannian Space-Time},
Phys. Rev. {\bf D7}, 2850 (1973).
\bibitem{fyodweizirn}
Y. V. Fyodorov, Y. Wei and M. R. Zirnbauer,
{\it Hyperbolic Hubbard-Stratonovich Transformation made Rigorous},
J. Math. Phys. {\bf 49}, 053507 (2008).
\bibitem{hyper}
C. F. Gauss, 
{\it Gauss Werke}, dritter Band,
Herausgegeben von der K\"{o}niglichen Gesellschaft der Wissenschaften zu G\"{o}ttingen (1866),
B. Riemann,
{\it Beitr\"{a}ge zur Theorie der durch die Gauss'sche Reihe} $F(\alpha,\beta,\gamma,x)$
{\it darstellbaren Functionen},
Abhandlungen der K\"{o}niglichen Gesellschaft der Wissenschaften zu G\"{o}ttingen (1857).
\bibitem{georgipais}
H. Georgi and A. Pais,
{\it Vacuum Symmetry and the Pseudo-Goldstone Phenomenon},
Phys. Rev. {\bf D12}, 508 (1975).
\bibitem{leutwyler}
J. Gasser and H. Leutwyler,
{\it Chiral Perturbation Theory to One Loop},
Ann. Phys. {\bf 158}, 142 (1984).
\bibitem{gellmannramond}
M. Gell-Mann, P. Ramond and R. Slansky,
{\it Complex Spinors and Unified Theories}, in {\it Supergravity} (P. Niuwenhuizen and D. Z. Freedman. eds.),
( North-Holland, Amsterdam (1979) ), p.315.
\bibitem{gidnel}
S. B. Giddings and P. Nelson,
{\it The Geometry of Super Riemann Surfaces},
Commun. Math. Phys. {\bf 116}, 607 (1988);
{\it Line Bundles on Super Riemann Surfaces},
Commun. Math. Phys. {\bf 118}, 289 (1988).
\bibitem{goert}
O. Goertsches,
{\it Riemannian Supergeometry},
Math. Z. {\bf 260}, 557 (2008).
\bibitem{dcptvio}
M. Gomes, T. Mariz, J. R. Nascimento and A. J. da Silva,
{\it Dynamical Lorentz and CPT Symmetry Breaking in a 4D Four-fermion Model},
Phys. Rev. {\bf D77}, 105002 (2008).
\bibitem{garciamaltoni}
M. C. Gonzalez-Garciz and M. Maltoni,
{\it Phenomenology with Massive Neutrinos},
Phys. Rep. {\bf 460}, 1 (2008).
\bibitem{superstrings}
M. B. Green, J. H. Schwarz and E. Witten,
{\it Superstring Theory: 1 and 2},
( Cambridge University Press, 1987 ).
\bibitem{greenberg}
R. Greenberg,
{\it On p-adic L-functions and Cyclotomic Fields I, II},
Nagoya Math. J. {\bf 56}, 61 (1974), {\bf 67}, 138 (1977). 
\bibitem{grimus}
W. Grimus,
{\it Neutrino Physics $-$ Theory},
arXiv:hep-ph/0307149.
\bibitem{axionsss}
C. Hagmann, H. Murayama, G. G. Raffelt, L. J. Rosenberg and K. van Bibber,
{\it Axions},
Phys. Lett. {\bf B667}, 1 (2008).
( available at Particle Data Group website. )
\bibitem{hanakawakimu}
M. Hanada, H. Kawai and Y. Kimura,
{\it Describing Curved Spaces by Matrices},
Prog. Theor. Phys. {\bf 114}, 1295 (2005).
\bibitem{hartshorne}
R. Hartshorne,
{\it Algebraic Geometry}
( Springer, 1977 ).
\bibitem{jikuhassei}
A. Hashimoto and M. Sethi,
Phys. Rev. Lett. {\bf 89}, 261601 (2002).
\bibitem{hawking}
S. W. Hawking,
{\it Particle Creation by Black Holes},
Commun. Math. Phys. {\bf 43}, 199 (1975).
\bibitem{neumagstar}
A. Heger, A. Friedland, M. Giannotti and V. Cirigliano,
{\it The Impact of Neutrino Magnetic Moments on the Evolution of Massive Stars}, arXiv:0809.4703.
\bibitem{heisenbergeuler}
W. Heisenberg and H. Euler,
{\it Consequences of Dirac's Theory of Positrons},
Z. Phys. {\bf 98}, 714 (1936).
\bibitem{heun}
K. Heun,
{\it Zur Theorie der Riemann'schen Functionen zweiter Ordnung mit vier Verzweigungspunkten},
Mathematische Annalen, {\bf 33}, 161 (1899).
\bibitem{higashij3}
K. Higashijima,
{\it Solutions of the Spinor-Spinor Bethe-Salpeter Equation in the Scalar-Vector Sector},
Prog. Theor. Phys. {\bf 55}, 1591 (1976).
\bibitem{higashijima}
K. Higashijima,
{\it Theory of Dynamical Symmetry Breaking},
Prog. Theor. Phys. Suppl. {\bf 104}, 1 (1991).
\bibitem{hilluty}
C. T. Hill, M. A. Luty and E. A. Paschos,
{\it Electroweak Symmetry Breaking by Fourth-Generation Condensates and the Neutrino Spectrum},
Phys. Rev. {\bf D43}, 3011 (1991).
\bibitem{hillsimon}
C. T. Hill and E. H. Simmons,
{\it Strong Dynamics and Electroweak Symmetry Breaking},
Phys. Rep. {\bf 381}, 235 (2003).
\bibitem{majoronron}
C. T. Hill, I. Mocioiu, E. A. Paschos and U. Sarkar,
{\it Neutrino Phenomenology, Dark Energy and Leptogenesis from Pseudo-Nambu-Goldstone Bosons},
Phys. Lett. {\bf B651}, 188 (2007).
\bibitem{tareknath}
A review mainly on (s)quarks of SUSY models: 
T. Ibrahim and P. Nath,
{\it CP Violation from the Standard Model to Strings},
Rev. Mod. Phys. {\bf 80}, 577 (2008).
\bibitem{teich}
Y. Imayoshi and M. Taniguchi,
{\it An Introduction to Teichm\"{u}ller Spaces}
( Springer Verlag, Berlin und Heidelberg (1992) ).
\bibitem{iwasawa1}
K. Iwasawa, 
{\it On the Theory of Cyclotomic Fields},
Ann. of Math. {\bf 70}, 530 (1959).
\bibitem{iwasawa2}
K. Iwasawa,
{\it On p-adic L-functions},
Ann. of Math. {\bf 89}, 198 (1969).
\bibitem{superdark}
G. Jungman, M. Kamionkowski and K. Griest,
{\it Supersymmetric Dark Matter},
Phys. Rep. {\bf 267}, 195 (1996).
\bibitem{kac}
V. G. Kac,
{\it Lie Superalgebras}, 
Adv. Math. {\bf 26}, 8 (1977),
{\it A Sketch of Lie Superalgebra Theory},
Commun. Math. Phys. {\bf 53}, 31 (1977).
\bibitem{kahana}
D. Kahana and U. Vogl,
{\it Diquark Bosonization of the Nambu Model},
Phys. Lett. {\bf B244}, 10 (1990).
\bibitem{rarerare}
Y. Kahn, M. Schmitt and T. M. P. Tait,
{\it Enhanced Rare Pion Decays from a Model of MeV Dark Matter},
arXiv:0712.0007.
\bibitem{kamkam}
KamLAND Collaboration, K. Eguchi et al.,
{\it First Results from KamLAND: Evidence for Reactor Antineutrino Disappearance},
Phys. Rev. Lett. {\bf 90}, 021802 (2003),
KamLAND Collaboration, I. Shimizu et al.,
{\it KamLAND ( Anti-Neutrino Status )},
J. Phys. Conference Series {\bf 120}, 052022 (2008).
\bibitem{kapnelwei}
D. B. Kaplan, A. E. Nelson and N. Weiner,
{\it Neutrino Oscillations as a Probe of Dark Energy},
Phys. Rev. Lett. {\bf 93}, 091801 (2004).
\bibitem{katotosio}
T. Kato,
{\it Perturbation Theory for Linear Operators}
( Springer, Heidelberg, 1980 ).
\bibitem{kayma}
O. Kaymakcalan and J. Schechter,
{\it Chiral Lagrangian of Pseudoscalars and Vectors},
Phys. Rev. {\bf D31}, 1109 (1985).
\bibitem{axiaxi}
J. E. Kim and G. Carosi,
{\it Axions and the Strong CP Problem},
arXiv:0807.3125.
\bibitem{king}
S. F. King,
{\it Neutrino Mass Models: a road map},
arXiv:0810.0492.
\bibitem{kingmannan}
S. F. King and S. H. Mannan,
{\it The Top Quark Condensate},
Phys. Lett. {\bf B241}, 249 (1990).
\bibitem{klevan}
S. P. Klevansky,
{\it The Nambu$-$Jona-Lasinio Model of Quantum Chromodynamics},
Rev. Mod. Phys. {\bf 64}, 649 (1992).
\bibitem{kobamas}
M. Kobayashi and T. Maskawa,
{\it CP-Violation in the Renormalizable Theory of Weak Interaction},
Prog. Theor. Phys. {\bf 49}, 652 (1973).
\bibitem{kobaoshimlie}
T. Kobayashi and T. Oshima,
{\it Lie Groups and Representations},
( Iwanami, Tokyo, 2005 ),
M. Wakimoto,
{\it Infinite-dimensional Lie Algebra},
( Iwanami, Tokyo, 2008 ),
S. Kumar,
{\it Kac-Moody Groups, Their Flag Varieties and Represenation Theory},
( Birkh\"{a}user, Boston-Basel-Berlin, 2002 ).
\bibitem{kodaira}
K. Kodaira,
{\it Complex Manifolds and Deformation of Complex Structures}
( Springer, Berlin-Heidelberg-New York, 1986 ).
\bibitem{kogutra}
J. B. Kogut, M. A. Stephanov, D. Toublan, J. J. M. Verbaarschot and A. Zhitnitsky,
{\it QCD-like Theories at Finite Baryon Density},
Nucl. Phys. {\bf B582}, 477 (2000).
\bibitem{kontsevich}
M. Kontsevich,
{\it Deformation Quantization of Poisson Manifolds},
Lett. Math. Phys. {\bf 66}, 157 (2003).
\bibitem{kontsevich2}
M. Kontsevich,
{\it Deformation Quantization of Algebraic Varieties},
Lett. Math. Phys. {\bf 56}, 271 (2001).
\bibitem{kostelecky}
The recent overviews of CPT violation,
{\it CPT and Lorentz Symmetry II}, 
edited by V. A. Kostelecky
( World Scientific, Singapore, 2002 ).
\bibitem{sptstri}
V. A. Kostelecky and R. Potting,
{\it CPT and Strings},
Nucl. Phys. {\bf B359}, 545 (1991),
V. A. Kostelecky and R. Potting,
{\it CPT, Strings, and Meson Factories},
Phys. Rev. {\bf D51}, 3923 (1995), 
D. Colladay and V. A. Kostelecky,
{\it CPT Violation and the Standard Model},
Phys. Rev. {\bf D55}, 6760 (1997),
V. A. Kostelecky and M. Mewes,
{\it Lorentz and CPT Violation in Neutrino Sector},
Phys. Rev. {\bf D70}, 031902 (2004).
\bibitem{kreimergauge}
D. Kreimer,
{\it Anatomy of a Gauge Theory},
Ann. Phys. {\bf 321}, 2757 (2006).
\bibitem{kreimersdeq}
D. Kreimer,
{\it Dyson-Schwinger Equations: From Hopf Algebras to Number Theory},
Fields Inst. Commun. {\bf 50}, 225 (2007).
\bibitem{kugoijima}
T. Kugo and I. Ojima,
{\it Local Covariant Operator Formalism of Nonabelian Gauge Theories and Quark Confinement Problem},
Prog. Theor. Phys. Suppl. {\bf 66}, 1 (1979).
\bibitem{kugodyn}
T. Kugo,
{\it Basic Concepts in Dynamical Symmetry Breaking and Bound State Problems},
Preprint KUNS-1086, Jul 1991, ( available at Slac Spires ).
\bibitem{kusnezov}
D. Kusnezov,
{\it Exact Matrix Expansions for Group Elements of $SU(N)$},
J. Math. Phys. {\bf 36}, 898 (1995).
\bibitem{lang}
S. Lang,
{\it Algebraic Number Theory}
( Springer, New York Berlin Heidelberg, 1994 ).
\bibitem{lattanzi}
M. Lattanzi,
{\it Decaying Majoron Dark Matter and Neutrino Masses},
arXiv:0802.3155.
\bibitem{laufer}
A. Laufer,
{\it The Exponential Map of $GL(N)$},
J. Phys. A: Math. Gen. {\bf 30}, 5455 (1997).
\bibitem{lautrup}
B. Lautrup,
Kgl. Danske Videnskab. Selskab. Mat.-fys. Medd. {\bf 35}, No.11, 1 (1967).
\bibitem{project}
C. LeBrun, Y.-S. Poon and R. O. Wells, 
{\it Projective Embeddings of Complex Supermanifolds},
Commun. Math. Phys. {\bf 126}, 433 (1990).
\bibitem{leepakvshr}
B. W. Lee, S. Pakvasa, R. E. Shrock and H. Sugawara,
{\it Muon and Electron Number Nonconservation in a $V-A$ Gauge Model},
Phys. Rev. Lett. {\bf 38}, 937 (1977),
B. W. Lee and R. E. Shrock,
{\it Natural Suppression of Symmetry Violation in Gauge Theories:
Muon- and Electron-Lepton-Number Nonconservation},
Phys. Rev. {\bf D16}, 1444 (1977).
\bibitem{lipman}
J. Lipman,
{\it Notes on Derived Functors and Grothendieck Duality},
Lecture Notes in Math. {\bf 1960}, 1 (2009).
\bibitem{heunheun}
R. S. Maier,
{\it The 192 Solutions of the Heun Equation},
Math. Comp. {\bf 76}, 811 (2007).
\bibitem{makinakasakata}
Z. Maki, M. Nakagawa and S. Sakata,
{\it Remarks on the Unified Model of Elementary Particles},
Prog. Theor. Phys. {\bf 28}, 870 (1962).
\bibitem{threequantgrav}
A. Maloney and E. Witten,
{\it Quantum Gravity Partition Functions in Three Dimensions},
arXiv:0712.0155.
\bibitem{soliton}
S. Mandelstam,
{\it Soliton Operators for the Quantized Sine-Gordon Equation},
Phys. Rev. {\bf D11}, 3026 (1975).
\bibitem{marciano}
W. J. Marciano,
{\it Heavy Top-Quark Mass Predictions},
Phys. Rev. Lett. {\bf 62}, 2793 (1989),
{\it Dynamical Symmetry Breaking and the Top Quark Mass},
Phys. Rev. {\bf D41}, 219 (1990).
\bibitem{marisrob}
P. Maris and C. D. Roberts,
{\it $\pi$- and $K$-Meson Bethe-Salpeter Amplitudes},
Phys. Rev. {\bf C56}, 3369 (1997).
\bibitem{marisrob2}
P. Maris and C. D. Roberts,
{\it Dyson-Schwinger Equations: A Tool for Hadron Physics},
Int. J. Mod. Phys. {\bf E12}, 297 (2003).
\bibitem{schroediffgalois}
I. Marshall and M. Semenov-Tian-Shansky,
{\it Poisson Groups and Differential Galois Theory of Schr\"{o}dinger Equation on the Circle},
arXiv: math:0710.5456.
\bibitem{martin}
S. P. Martin,
{\it Dynamical Electroweak Symmetry Breaking with top-Quark and Neutrino Condensates},
Phys. Rev. {\bf D44}, 2892 (1991).
\bibitem{maskawa}
T. Maskawa and H. Nakajima,
{\it Spontaneous Breaking of Chiral Symmetry in a Vector-Gluon Model},
Prog. Theor. Phys. {\bf 52}, 1326 (1974).
\bibitem{mazurwiles}
B. Mazur and A. Wiles,
{\it Class Fields of Abelian Extensions of {\bf Q}},
Inv. Math. {\bf 76}, 179 (1984).
\bibitem{milnor}
J. Milnor,
{\it Morse Theory}
( Ann. of Math. Studies, no.51. Princeton Univ. Press, Princeton, 1963 ).
\bibitem{theoldestseesaw}
The first paper of the neutrino seesaw mechanism:
P. Minkowski,
{\it mu$\to$e gamma at a Rate of One Out of 1-Billion Muon Decays?},
Phys. Lett. {\bf B67}, 421 (1977).
\bibitem{mirtanyam}
V. A. Miransky, M. Tanabashi and K. Yamawaki,
{\it Dynamical Electroweak Symmetry Breaking with Large Anomalous Dimension and t Quark Condensate},
Phys. Lett. {\bf B221}, 177 (1989).
\bibitem{miransky}
V. A. Miransky,
{\it Dynamical Symmetry Breaking in Quantum Field Theories},
( World Scientific, Singapore, 1993 ).
\bibitem{originsusy}
The first attempt toward supersymmetry ( super-algebra ):
H. Miyazawa,
{\it Spinor Currents and Symmetries of Baryons and Mesons},
Phys. Rev. {\bf 170}, 1586 (1968).
\bibitem{weyltaio}
M. M. Mizrahi,
{\it The Weyl Correspondence and Path Integrals},
J. Math. Phys. {\bf 16}, 2201 (1975).
\bibitem{mochi}
S. Mochizuki,
{\it An Introduction to p-adic Teichm\"{u}ller Theory, cohomologies p-adiques et applications arithm\'{e}tiques I},
Ast\'{e}risque {\bf 278}, 1 (2002).
\bibitem{mohapatra}
R. N. Mohapatra and G. Senjanovi\'{c},
{\it Neutrino Mass and Spontaneous Parity Violation},
Phys. Rev. Lett. {\bf 44}, 912 (1980).
\bibitem{mohamoha}
R. N. Mohapatra et al.,
{\it Theory of Neutrinos: A White Paper},
hep-ph/0510213.
\bibitem{moelmoel}
M. M\"{o}ller,
{\it Teichm\"{u}ller Curves, Galois Action and $\widehat{GT}$-relations},
arXiv:math.AG/0311308.
\bibitem{mooreswinvari}
J. D. Moore,
{\it Lectures on Seiberg-Witten Invariants},
Lect. Note Math. {\bf 1629} ( Springer, 1996 ),
See, especially the Chap. 3, {\it Global Analysis of the Seiberg-Witten Equations}.
\bibitem{moriya}
T. Moriya,
{\it Spin Fluctuations in Itinerant Electron Magnetism}
( Springer, Berlin-Heidelberg-New York, 1985 ).
\bibitem{nahm}
W. Nahm,
{\it Supersymmetries and Their Representations},
Nucl. Phys. {\bf B135}, 149 (1978).
\bibitem{nakahara}
M. Nakahara,
{\it Geometry, Topology and Physics}
( IOP Publishing, 1990 ).
\bibitem{nakanishi}
N. Nakanishi,
Prog. Theor. Phys. {\bf 35}, 1111 (1966),
{\bf 49}, 640 (1973), 
{\bf 52}, 1929 (1974).
\bibitem{nambu}
Y. Nambu,
{\it Quasi-particles and Gauge Invariance in the Theory of Superconductivity},
Phys. Rev. {\bf 117}, 648 (1960).
\bibitem{nambu2}
Y. Nambu and G. Jona-Lasinio,
{\it Dynamical Model of Elementary Particles Based on an Analogy with Superconductivity. I},
Phys. Rev. {\bf 122}, 345 (1961), 
{\it Dynamical Model of Elementary Particles Based on an Analogy with Superconductivity. II},
Phys. Rev. {\bf 124}, 246 (1961).
\bibitem{nambutop}
Y. Nambu,
in {\it New Theories in Physics}, proceedings of the XI International Symposium on Elementary Particle Physics,
Kazimierz, Poland, 1988, edited by Z. Ajduk, S. Pokorski, and A. Trautman ( World Scientific, Singapole, 1989 ).
\bibitem{diffgalois}
K. A. Nguyen and M. van der Put,
{\it Solving Linear Differential Equations},
arXiv:0810.4039.
\bibitem{ninne}
H. Ninnemann,
{\it Deformations of Super Riemann Surfaces},
Comun. Math. Phys. {\bf 150}, 267 (1992).
\bibitem{nishino}
T. Nishino,
{\it Complex Analysis of Several Variables}
( Tokyo University Press, 1996 ).
\bibitem{novikov}
S. P. Novikov, S. V. Manakov, L. B. Pitaevskii and V. E. Zakharov,
{\it Theory of Solitons. The Inverse Scattering Method},
( Plenum Press, New York, 1984 ).
\bibitem{novitai}
S. P. Novikov and I. A. Taimanov,
{\it Modern Geometric Structures and Fields}
( American Mathematical Society, Province, Rhode Insland, 2006 ).
\bibitem{nunokawa}
H. Nunokawa, S. Parke and Jos\'{e} W. F. Valle,
{\it CP Violation and Neutrino Oscillations},
Prog. Part. Nucl. Phys. {\bf 60}, 338 (2008).
\bibitem{nussishro}
S. Nussinov and R. Shrock,
{\it On the $\pi$ and $K$ as $q\bar{q}$ Bound States and Approximate Nambu-Goldstone Bosons},
Phys. Rev. {\bf D79}, 016005 (2009).
\bibitem{ohsaku1}
T. Ohsaku,
{\it Theory of Quantum Electrodynamical Self-consistent Fields},
arXiv:physics/0112087.
\bibitem{ohsaku2}
T. Ohsaku and K. Yamaguchi,
{\it QED-SCF, MCSCF and Coupled-cluster Methods in Quantum Chemistry},
Int. J. Quant. Chem. {\bf 85}, 272 (2001).
\bibitem{ohsaku3}
T. Ohsaku, S. Yamanaka, D. Yamaki and K. Yamaguchi,
{\it Quantum Electrodynamical Density-matrix Functional Theory and Group-theoretical Consideration
of its Solutions},
Int. J. Quant. Chem. {\bf 90}, 273 (2002).
\bibitem{ohsaku4}
T. Ohsaku,
{\it BCS and Generalized BCS Superconductivity in Relativistic Quantum Field Theory: Formulation},
Phys. Rev. {\bf B65}, 024512 (2002).
\bibitem{ohsaku5}
T. Ohsaku,
{\it BCS and Generalized BCS Superconductivity in Relativistic Quantum Field Theory. II. Numerical Calculations},
Phys. Rev. {\bf B66}, 054518 (2002).
\bibitem{tadafumi}
T. Ohsaku,
{\it Dynamical Chiral Symmetry Breaking and its Restoration for an Accelerated Observer},
Phys. Lett. {\bf B599}, 102 (2004).
\bibitem{Ohsaku8}
T. Ohsaku,
{\it Dynamical Chiral Symmetry Breaking and Superconductivity in the Supersymmetric Nambu$-$Jona-Lasinio Model at finite Temperature and Density},
Phys. Lett. {\bf B634}, 285 (2006).
\bibitem{ohsaku9}
T. Ohsaku, 
{\it Dynamical Chiral Symmetry Breaking, Color Superconductivity, and Bose-Einstein Condensation in an $SU(N_{c})\times U(N_{f})_{L}\times U(N_{f})_{R}$-invariant Supersymmetric Nambu$-$Jona-Lasinio Model at finite Temperature and Density},
Nucl. Phys. {\bf B803}, 299 (2008).
\bibitem{ohsakunoncomu}
T. Ohsaku,
{\it Supersymmetric Nambu$-$Jona-Lasinio Model on ${\cal N}=1/2$ Four-Dimensional 
Non(anti)commutative Superspace},
JHEP {\bf 0802}, 021 (2008).
\bibitem{ohsakususy}
T. Ohsaku,
{\it Dynamical Dirac Mass Generation in the Supersymmetric Nambu$-$Jona-Lasinio Model with the Seesawa Mechanism of Neutrinos},
arXiv:0801:1256, submitted for publication.
\bibitem{ohsakuriem}
T. Ohsaku,
{\it Algebra of Noncommutative Riemann Surfaces},
arXiv:math-ph/0606057.
\bibitem{ohsakudef}
T. Ohsaku,
{\it Moyal-Weyl Star-products as Quasiconformal Mappings},
arXiv:math-ph/0610032.
\bibitem{pagels}
H. Pagels,
{\it Dynamical Chiral Symmetry Breaking in Quantum Chromodynamics},
Phys. Rev. {\bf D19}, 3080 (1979),
{\it Models of Dynamically Broken Gauge Theories},
Phys. Rev. {\bf D21}, 2336 (1980).
\bibitem{gravwave}
G. Pagliaroli, F. Vissani, E. Coccia and W. Fulgione,
{\it Neutrinos from Supernovae as a Trigger for Gravitational Wave Search},
Phys. Rev. Lett. {\bf 103}, 031102 (2009).
\bibitem{parke}
S. Parke,
{\it CP Violation in the Neutrino Sector},
arXiv:0807.3311.
\bibitem{Pauli}
W. Pauli, Nuovo Cim. {\bf 6}, 204 (1957), 
G. G\"{u}rsey, Nuovo Cim. {\bf 7}, 411 (1958).
\bibitem{paycudehut}
A. Payez, J. R. Cudell and D. Hutsem\'{e}kers,
{\it Axions and Polarisation of Quasars},
CP1038, Hadronic Physics, Joint Meeting Heidelberg-Li\`{e}ge-Paris-Wroclaw ( HLPW 2008 ).
\bibitem{peccquin}
R. D. Peccei and H. R. Quinn,
{\it Constraints Imposed by CP Conservation in the Presence of Pseudoparticles},
Phys. Rev. {\bf D16}, 1791 (1977).
\bibitem{pecceicp}
R. D. Peccei,
{\it The Strong CP Problem and Axions},
Lect. Notes Phys. {\bf 741}, 3 (2008).
\bibitem{perelman}
G. Perelman,
{\it The Entropy Formula for the Ricci Flow and its Geometric Applications},
arXiv:math/0211159.
\bibitem{lithigs}
M. Perelstein,
{\it Little Higgs Models and Their Phenomenology},
Prog. Part. Nucl. Phys. {\bf 58}, 247 (2007).
\bibitem{pisarsky}
R. D. Pisarsky, T. L. Trueman and M. H. G. Tytgat,
{\it How $\pi^{0}\to\gamma\gamma$ Changes with Temperature},
Phys. Rev. {\bf D56}, 7077 (1997).
\bibitem{politzer}
H. D. Politzer,
{\it Effective Quark Masses in the Chiral Limit},
Nucl. Phys. {\bf B117}, 397 (1976).
\bibitem{ponte}
B. Pontecorvo,
{\it Mesonium and Anti-Mesonium}, 
Sov. Phys. JETP {\bf 6}, 429 (1957),
B. Pontecorvo,
{\it Neutrino Experiments and the Problem of Conservation of Leptonic Charge},
Sov. Phys. JETP {\bf 26}, 984 (1968),
S. M. Bilensky and B. Pontecorvo,
{\it Lepton Mixing and Neutrino Oscillations},
Phys. Rep. {\bf 41}, 225 (1978).
\bibitem{polchin}
J. Polchinski,
{\it String Theory, I and II},
( Cambridge University Press, 1998 ).
\bibitem{putputpu}
M. van der Put and M. Reversat,
{\it A Local-Global Problem for Linear Differential Equations},
arXiv:0711.0815.
\bibitem{putsinger}
M. van der Put and M. F. Singer,
{\it Galois Theory of Linear Differential Equations}
( Grundlehren der mathematischen Wissenschaften, vol. 328, Springer, Heidelberg, 2003 ).
\bibitem{rabinfreund}
J. M. Rabin and P. G. O. Freund,
{\it Supertori are Algebraic Curves},
Commun. Math. Phys. {\bf 114}, 131 (1988).
\bibitem{raffelt}
G. Raffelt and L. Stodolsky,
{\it Mixing of the Photon with Low-mass Particles},
Phys. Rev. {\bf D37}, 1237 (1988).
\bibitem{rajaraman}
R. Rajaraman,
{\it Solitons and Instantons},
( North-Holland, 1982 ).
\bibitem{robewill}
C. D. Roberts and A. G. Williams,
{\it Dyson-Schwinger Equations and Their Applications to Hadron Physics},
hep-ph/9403224,  
C. D. Roberts and S. M. Schmidt,
{\it Dyson-Schwinger Equations: Density, Temperature and Continuum Strong QCD},
Prog. Part. Nucl. Phys. {\bf 45}, S1 (2000).
\bibitem{rodejohann}
W. Rodejohann,
{\it Type II Seesaw Mechanism, Deviations from Bimaximal Neutrino Mixing, and Leptogenesis},
Phys. Rev. {\bf D70}, 073010 (2004).
\bibitem{rogers}
A. Rogers,
{\it Super Lie Groups: Global Topology and Local Structure},
J. Math. Phys. {\bf 22}, 939 (1981).
\bibitem{humihumi}
H. Sato, 
{\it Cosmology}
( Iwanami, Tokyo, 1997 ).
\bibitem{schmithuesen}
G. Schmith\"{u}sen,
{\it An Algorithm for Finding the Veech Group of an Origami},
Experimental Mathematics, {\bf 13}, 459 (2004).
\bibitem{seibergnoncom}
N. Seiberg,
{\it Noncommutative Superspace, ${\cal N}=1/2$ Supersymmetry,
Field Theory and String Theory},
JHEP {\bf 0306}, 010 (2003).
\bibitem{seibwitt}
N. Seiberg and E. Witten,
{\it String Theory and Noncommutative Geometry},
JHEP {\bf 9909}, 032 (1999).
\bibitem{skivie}
P. Sikivie,
{\it Axion Cosmology},
arXiv:astro-ph/0610440.
\bibitem{axionbec}
P. Sikivie and Q. Yang,
{\it Bose-Einstein Condensation of Dark Matter Axions},
arXiv:0901.1106.
\bibitem{singerdiffgalois}
M. F. Singer,
{\it Introduction to the Galois Theory of Linear Differential Equations},
arXiv:0712.4124. 
\bibitem{kolkata}
M. Sinha and B. Mukhopadhyay,
{\it CPT and Lepton Number Violation in the Neutrino Sector:
Modified Mass Matrix and Oscillation due to Gravity},
Phys. Rev. {\bf 77}, 025003 (2008).
\bibitem{studenikin}
A. Studenikin,
{\it Neutrino Magnetic Moment: A Window to New Physics},
arXiv:0812.4716.
\bibitem{sugawara}
H. Sugawara,
{\it Dynamical Calculation of Quark, Lepton, and Gauge-Boson Masses},
Phys. Rev. {\bf D30}, 2396 (1984).
\bibitem{kamioka}
SuperKamiokande Collaboration, Y. Fukuda et al.,
{\it Evidence for Oscillation of Atmospheric Neutrinos},
Phys. Rev. Lett. {\bf 81}, 1562 (1998).
\bibitem{magneneut}
SuperKamiokande Collaboration, D. W. Liu et al.,
{\it Limits on the Neutrino Magnetic Moment using Super-Kamiokande Solar Neutrino Data},
Int. J. Mod. Phys. {\bf A20}, 3110 (2005).
\bibitem{supernova}
Supernova Search Team,
{\it Observational Evidence from Supernovae for an Accelerating Universe and a Cosmological Constant},
Astron. J. {\bf 116}, 1009 (1998).
\bibitem{szabo}
R. J. Szabo,
{\it Quantum Field Theory on Noncommutative Spaces},
Phys. Rep. {\bf 378}, 207 (2003).
\bibitem{takagi}
S. Takagi,
{\it Vacuum Noise and Stress Induced by Uniform Acceleration},
Prog. Theor. Phys. Suppl. {\bf 88}, 1 (1986). 
\bibitem{takeuchi}
T. Takeuchi,
{\it Analytical and Numerical Study of the Schwinger-Dyson Equation with Four-fermion Coupling},
Phys. Rev. {\bf D40}, 2697 (1989). 
\bibitem{taylor}
R.Taylor,
{\it On Galois Representations Associated to Hilbert Modular Forms},
Invent. Math. {\bf 98}, 265 (1989).
\bibitem{taylwiles}
R. Taylor and A. Wiles,
{\it Ring-theoretic Properties of Certain Hecke Algebras},
Ann. of Math. {\bf 141}, 553 (1995).
\bibitem{sugaku}
K. Terasawa, 
{\it Mathematics for Natural Scientists ( Revised and Expanded Version )},
( Iwanami, Tokyo, 1983 ).
\bibitem{terazawa}
H. Terazawa, Y. Chikashige and K. Akama,
{\it Unified Model of the Nambu$-$Jona-Lasinio Type for all Elementary-Particle Forces},
Phys. Rev. {\bf D15}, 480 (1977),
H. Terazawa, 
{\it $t$-Quark Mass Predicted from a Sum Rule for Lepton and Quark Masses},
Phys. Rev. {\bf D22}, 2921 (1980).
\bibitem{thooftmass}
G. 't Hooft, 
{\it Dimensional Regularization and the Renormalization Group},
Nucl. Phys. {\bf B61}, 455 (1973).
\bibitem{unruh}
W. G. Unruh,
{\it Notes on Black Hole Evaporation},
Phys. Rev. {\bf D14}, 870 (1976).
\bibitem{gaugeprinzip}
R. Utiyama,
{\it Invariant Theoretical Interpretation of Interaction},
Phys. Rev. {\bf 101}, 1597 (1956).
\bibitem{vafawitten}
C. Vafa and E. Witten,
{\it Parity Conservation in Quantum Chromodynamics},
Phys. Rev. Lett. {\bf 53}, 535 (1984).
\bibitem{valenzula}
C. Valenzuela,
{\it Spontaneous $CP$ Symmetry Breaking at the Electroweak Scale},
Phys. Rev. {\bf D71}, 095014 (2005).
\bibitem{betadecay}
P. Vogel and A. Piepke,
{\it Neutrinoless Double-$\beta$ Decay},
Particle Data Group, 2008.
\bibitem{complexpote}
E. J. Weinberg and A.-q. Wu,
{\it Understanding Complex Perturbative Effective Potentials},
Phys. Rev. {\bf D36}, 2474 (1987).
\bibitem{smsmsms}
S. Weinberg,
{\it A Model of Leptons},
Phys. Rev. Lett. {\bf 19}, 1264 (1967).
\bibitem{weinbpseud}
S. Weinberg,
{\it Approximate Symmetries and Pseudo-Goldstone Bosons},
Phys. Rev. Lett. {\bf 29}, 1698 (1972).
\bibitem{weinbmass}
S. Weinberg,
{\it New Approach to the Renormalization Group},
Phys. Rev. {\bf D8}, 3497 (1973).
\bibitem{weiberguu}
S. Weinberg,
{\it The $U(1)$ Problem},
Phys. Rev. {\bf D11}, 3583 (1975).
\bibitem{weinbcp}
S. Weinberg,
{\it Gauge Theory of $CP$ Nonconservation}
Phys. Rev. Lett. {\bf 37}, 657 (1976).
\bibitem{weinbperturb}
S. Weinberg,
{\it Implications of Dynamical Symmetry Breaking},
Phys. Rev. {\bf D13}, 974 (1976),
{\it Implications of Dynamical Symmetry Breaking: An Addendum},
Phys. Rev. {\bf D19}, 1277 (1979).
\bibitem{axionmodels}
S. Weinberg,
{\it A New Light Boson?},
Phys. Rev. Lett. {\bf 40}, 223 (1978).
\bibitem{weinbchiral}
S. Weinberg,
{\it Phenomenological Lagrangians},
Physica {\bf 96}, 327 (1979).
\bibitem{weiss}
C. Weiss, A. Buck, R. Alkofer and H. Reinhardt,
{\it Diquark Electromagnetic Form-Factors in a Nambu$-$Jona-Lasinio Model},
Phys. Lett. {\bf B312}, 6 (1993).
\bibitem{susy}
J. Wess and J. Bagger,
{\it Supersymetry and Supergravity},
( Princeton Univ. Press, 1992 ).
\bibitem{wilczek}
F. Wilczek,
{\it Problem of Strong $P$ and $T$ Invariance in the Presence of Instantons}
Phys. Rev. Lett. {\bf 40}, 279 (1978).
\bibitem{wiles1}
A. Wiles,
{\it The Iwasawa Conjecture for Totally Real Fields},
Ann. of Math. {\bf 131}, 493 (1990).
\bibitem{wiles2}
A. Wiles,
{\it Modular Elliptic Curves and Fermat's Last Theorem},
Ann. of Math. {\bf 141}, 443 (1995).
\bibitem{witten}
E. Witten,
{\it Supersymmetry and Morse Theory},
J. Diff. Geom. {\bf 17}, 661 (1982).
\bibitem{edwitten}
E. Witten,
{\it From Superconductors and Four-Manifolds to Weak Interactions},
Bul. Am. Math. Soc. {\bf 44}, 361 (2007).
\bibitem{yamawaki}
K. Yamawaki,
{\it Top Quark Condensate Revisited},
Prog. Theor. Phys. Suppl. {\bf 123}, 19 (1996).
\bibitem{yamawa2}
K. Yamawaki,
{\it Quest for the Dynamical Origin of Mass},
arXiv:0907.5277.
\bibitem{yanagida}
T. Yanagida,
in Proceedings of the Workshop on the Unified Theory and the Baryon Number in the Universe ( O. Sawada and A. Sugamoto, eds. )
( KEK, Tsukuba, Japan (1979) ), p.95.
\bibitem{yangmills}
C.-N. Yang and R. L. Mills,
{\it Conservation of Isotopic Spin and Isotopic Gauge Invariance},
Phys. Rev. {\bf 96}, 191 (1954).
\bibitem{zirndom}
M. R. Zirnbauer,
{\it Symmetry Classes in Random Matrix Theory},
arXiv:math-ph/0404058.
\bibitem{sinegord}
V. E. Zakharov, L. A. Takhtajan and L. D. Faddeev,
{\it Complete Description of Solutions of the "Sine-Gordon" Equation},
Doklady AN SSSR, {\bf 219}, 1334 (1973).
\bibitem{atiyah}
M. F. Atiyah,
Arbeitstagung Bonn-MPIM, Germany,
F. Hirzebruch, J. Schwermer and S. Stuter ed.
( Springer, Berlin-Heidelberg, 1985 ).
\bibitem{matsumura}
H. Matsumura,
{\it Commutative Ring Theory}
( Cambridge University Press, UK, 1989 ).
\bibitem{numtheo}
K. Kato, M. Kurihara, N. Kurokawa, T. Saito, 
{\it Number Theory, 1 and 2}
( Iwanami, Tokyo, 2004 ),
T. Saito, 
{\it The Fermat's Last Theorem}
( Iwanami, Tokyo, 2008 ).
\bibitem{mukais}
S. Mukai,
{\it Moduli Theory, 1 and 2}
( Iwanami, Tokyo, 2008 ).
\bibitem{uenoshim}
K. Ueno and Y. Shimizu,
{\it Deformation of Complex Structures and Periods}
( Iwanami, Tokyo, 2008 ).
\bibitem{hida}
H. Hida,
{\it Modular Forms and Galois Cohomology}
( Cambridge University Press, UK, 2000 ).


\end{thebibliography}
\end{document}